\documentclass[aps,prl,reprint,superscriptaddress,longbibliography,amsmath,amssymb]{revtex4-2}
\usepackage[T1]{fontenc}
\usepackage{lmodern}
\usepackage{bm}
\usepackage{graphicx}
\usepackage{ifpdf}
\ifpdf
  \usepackage{epstopdf}
\fi
\usepackage[hidelinks]{hyperref}
\hypersetup{pdftitle={Analytic Solutions To Fully Quantized Parametric Amplification},
  pdfauthor={Wanli Xing and Timothy C. Ralph},pdfsubject={Revised PRL-style manuscript}}
\newcommand{\avg}[1]{\langle #1\rangle}
\newcommand{\dd}{\mathrm{d}}
\newcommand{\ii}{\mathrm{i}}
\newcommand{\Var}{\operatorname{Var}}

\begin{document}

\title{Analytic Solutions To Fully Quantized Parametric Amplification}
\author{Wanli Xing}
\email{wanli.xing@unsw.edu.au}
\affiliation{UNSW Sydney, Sydney, New South Wales 2033, Australia}
\author{Timothy C. Ralph}
\affiliation{ARC Centre of Excellence for Quantum Computation and Communication Technology,
School of Mathematics and Physics, The University of Queensland,
St Lucia, Queensland 4072, Australia}
\date{November 3, 2026}

\begin{abstract}
We derive analytic expressions for observable dynamics under the fully quantized trilinear Hamiltonian, governing quantum three-wave mixing and parametric amplification. Starting from a coherent pump and vacuum signal and idler, the calculation reorganizes the Heisenberg expansion into resummed depletion and return profiles, yielding directly evaluable expressions for all 1st-, 2nd- and 3rd-order correlations. The construction uses no parametric approximation or fitted parameters. Representative numerical benchmarks show errors of order $10^{-12}$ throughout the entire pump depletion, and below $10^{-3}$ across dozens of depletion-and-back-conversion cycles, up to the first total collapse of the oscillation. The first-depletion treatment additionally provides 4th-order correlations, accommodates bright squeezed-pump inputs and reveals phenomena important to future optical quantum technologies. The framework can be extended to higher moments. 
\end{abstract}
\maketitle

The second-order optical nonlinear interaction converts pump photons into correlated signal and idler photons. It is the underpinning theory of parametric amplification~\cite{LOUISELL_1961,MOLLOW_1967}, single-photon and entangled-pair sources~\cite{HONG_1986,JIN_2014}, squeezed light~\cite{WU_1987,EBERLE_2013}, and continuous-variable quantum technologies~\cite{WEEDBROOK_2012,MENICUCCI_2006}. 
Related three-wave interactions also occur in superconducting microwave circuits. Josephson ring modulators provide nondegenerate parametric amplification, with pump depletion contributing to their finite dynamic range~\cite{ABDO_2013}, and Josephson mixers have generated spatially separated two-mode squeezed microwave fields~\cite{FLURIN_2012}. 

A common approximation is the parametric approximation, which replaces the pump operator by a fixed classical amplitude.  This gives a simple quadratic interaction, but neglects pump depletion, pump fluctuations, and correlations between the pump and the generated fields. Past studies on quantum-pump corrections to squeezing demonstrate the limitation of this approximation~\cite{HILLERY_1984,CROUCH_1988}. Experiments on intense twin beams, high-efficiency squeezing and down-conversion also motivate descriptions that include depletion \cite{ALLEVI_2014, PERINA_2016, FLOREZ_2020}. One can allow the classical pump amplitude to change, which describes energy transfer \cite{BAUMGARTNER_1979}, but a fully quantized treatment is needed to follow its quantum noise and correlations.

Fully quantized three-wave mixing has been studied extensively beyond the parametric approximation. Early quantum calculations treated amplification and frequency conversion directly \cite{WALLS_1970, GAMBINI_1977}, while later work examined mode statistics, entanglement, and pump depletion \cite{DROBNY_1992, DROBNY_1993, DROBNY_1997, BANDILLA_2000}. The same interaction has also been studied as a model of Hawking radiation \cite{NATION_2010}, in coherently seeded down-conversion \cite{BIRRITTELLA_2020}, and in a trapped-ion quantum simulation \cite{DING_2018}. These studies establish that pump depletion is a quantum dynamical problem with structure well beyond a correction to the mean pump amplitude. 

Obtaining readily evaluable, high-precision expressions through strong pump depletion and repeated back-conversion remains challenging. Indeed, Shchesnovich identified the asymptotic limit of large invariant-subspace dimension as an important open problem \cite{SHCHESNOVICH_2026}. Exact spectral and recursive solutions exist \cite{ANDREEV_1992, KARASSIOV_1994,LEE_2010,SHCHESNOVICH_2025, SHCHESNOVICH_2026}, but extracting coherent-pump observables from these representations requires combining many photon-number sectors, whose dimensions grow with the pump population. Exact solvability therefore does not immediately provide compact, readily evaluable expressions for the observable dynamics or physical insight. On the other hand, approximate treatments have specific ranges of validity~\cite{XING_2023,CHINNI_2024,YANAGIMOTO_2022,HOROSHKO_2025,ZHANG_2026}. For example, the recent fifth-order cumulant treatment of Ref. \cite{CHINNI_2024} maintains 1\% relative accuracy for the signal population up to approximately 37\% pump depletion at an initial pump population of $10^5$. This motivates analytic descriptions that retain high precision throughout depletion—and even subsequent conversion cycles.

Here we reorganize the perturbation series according to successive depletion and return passages and resum each contribution. For a coherent pump, this gives a common leading form for five pump moments, systematic corrections, and a description of repeated passages. We also calculate fourth-order correlations and treat bright squeezed pumps through the first depletion only. Together, these results connect the conversion dynamics to experimentally accessible noise, photon statistics, and higher-order correlations. We also recover a fundamental limit on the maximum squeezing possible from the interaction.

\textit{Model and observables.---}
In the interaction picture, the Hamiltonian and evolution operator are
\begin{equation}
\begin{aligned}
H_{\mathrm{int}}&=\ii\hbar g(a^\dagger b^\dagger c-abc^\dagger),\\
\mathcal U(\tau)&=\exp\!\left[\tau(a^\dagger b^\dagger c-abc^\dagger)\right],
\end{aligned}
\label{eq:model}
\end{equation}
where $\tau=gt$, $c$ is the pump mode, and $a,b$ are the signal and idler modes. We first take the initial state to be $\lvert0\rangle_a\lvert0\rangle_b\lvert\alpha\rangle_c$, with real $\alpha>0$, and use $\chi=\alpha\tau$ and $\epsilon=\alpha^{-2}$. Direct diagonalization has polynomial cost within each invariant Fock-space sector, and the relevant sector dimensions grow with the mean pump population $\alpha^2$~\cite{WALLS_1970,SHCHESNOVICH_2026}, making numerical approaches infeasible for large coherent amplitude.

In the present work, we evaluate the normalized moments
\begin{equation}
\begin{gathered}
C=\frac{\avg{c}}{\alpha},\qquad
P=\frac{\avg{c^\dagger c}}{\alpha^2},\qquad
S=\frac{\avg{c^2}}{\alpha^2},\\
U=\frac{\avg{c^\dagger c^2}}{\alpha^3},\qquad
T=\frac{\avg{c^3}}{\alpha^3}.
\end{gathered}
\label{eq:moments}
\end{equation}
Symmetry, conservation laws, and time derivatives then determine all moments of total operator degree up to three.

\textit{Resumming the first depletion.---}
The coherent amplitude illustrates the method. The
Baker--Campbell--Hausdorff expansion of the Heisenberg
operator~\cite{SAKURAI_2011} gives
\begin{equation}
\begin{aligned}
C={}&1-\frac{\tau^2}{2!}+\frac{\tau^4}{4!}(-4\alpha^2+1)\\
&+\frac{\tau^6}{6!}(-16\alpha^4+60\alpha^2-1)\\
&+\frac{\tau^8}{8!}(-64\alpha^6+1552\alpha^4-744\alpha^2+1)\\
&+\frac{\tau^{10}}{10!}\bigl(-256\alpha^8+29632\alpha^6\\
&\hspace{51pt}-92848\alpha^4+8920\alpha^2-1\bigr)+\cdots .
\end{aligned}
\label{eq:bch}
\end{equation}
Writing $\tau=\chi/\alpha$ and regrouping in powers of $\epsilon$ yields
\begin{equation}
C(\chi,\epsilon)\sim\sum_{n=0}^{\infty}(-\epsilon)^nG_n(\chi),
\qquad G_0=1,
\label{eq:epsilon}
\end{equation}
with
\begin{equation}
G_1=\tfrac12\sinh^2\chi
=\tfrac18e^{2\chi}-\tfrac14+\tfrac18e^{-2\chi}.
\label{eq:g1}
\end{equation}
The next coefficient reproduces the pump correction from our previous work~\cite{XING_2023}:
\begin{equation}
\begin{aligned}
G_2={}&\frac{e^{4\chi}+e^{-4\chi}}{64}-\frac{4\chi^2+3}{32}\\
&+\frac{(1-2\chi^2-\chi)e^{2\chi}+(1-2\chi^2+\chi)e^{-2\chi}}{32}.
\end{aligned}
\label{eq:g2}
\end{equation}
All higher coefficients combine  polynomials in $\chi$ with exponentials $e^{2j\chi}$. Keeping the leading exponential from each order gives
\begin{equation}
C_0^{(0)}(z)\sim1+2\sum_{n=1}^{\infty}(-1)^n n!z^n,
\qquad z=\frac{e^{2\chi}}{16\alpha^2}.
\label{eq:factorial}
\end{equation}
The factorial growth suggests Borel summation~\cite{BALSER_1994,ANICETO_2019}. Here the resummed expression is
particularly simple:
\begin{equation}
\begin{aligned}
C_0^{(0)}(z)&=2\int_0^\infty\frac{e^{-v}}{1+zv}\,\dd v-1\\
&=\frac{2e^{1/z}}{z}E_1(1/z)-1,
\end{aligned}
\label{eq:borel}
\end{equation}
where $E_1$ is the exponential integral~\cite[Eq.~6.2.1]{NIST_DLMF}. This leading profile changes from $+1$ to $-1$ across a region centered on $\chi=\log(4\alpha)+O(1)$. During this amplitude passage, the pump population depletes and recovers; the negative final amplitude represents a pump phase reversal. We refer to the window $0\leq\chi\leq2\log(4\alpha)$ as the first depletion throughout this Letter.

Subleading exponential terms generate corrections to Eq.~\eqref{eq:borel}. They also contain terms that become important near the first amplitude return, at $\chi=3\log(4\alpha)+O(1)$. This separation motivates resolving the series into distinct passages. Detailed recurrences, resummations, and numerical benchmarks are given in the Supplemental Material (SM)~\cite{SUPPLEMENTAL}.

\textit{Successive passages and coherent-state clocks.---}
At leading order, the five moments share a family of normalized densities. Define
\begin{equation}
\begin{aligned}
s_m&=\chi-(2m+1)\log(4\alpha),\\
h_m(u)&=\frac{1}{2\pi}\int_{-\infty}^{\infty}H_m(t)e^{\ii tu}\,\dd t,
\qquad \rho_m(u)=|h_m(u)|^2,\\
H_m(t)&=2^{-(2m+1)\ii t/2}
\frac{\Gamma\!\left((1-\ii t)/2\right)^{m+1}}
{\Gamma\!\left((1+\ii t)/2\right)^m}.
\end{aligned}
\label{eq:packets}
\end{equation}
Here $m=0,1,\ldots$ labels successive amplitude passages and $\int\rho_m(u)\,\dd u=1$. Their leading moment profiles are
\begin{equation}
\begin{aligned}
C_m^{(0)}&=(-1)^{m+1}\int\rho_m(u)\tanh(s_m+u)\,\dd u,\\
P_m^{(0)}=S_m^{(0)}&=\int\rho_m(u)\tanh^2(s_m+u)\,\dd u,\\
U_m^{(0)}=T_m^{(0)}&=(-1)^{m+1}\int\rho_m(u)\tanh^3(s_m+u)\,\dd u.
\end{aligned}
\label{eq:profiles}
\end{equation}
All integrals run over the real line. At fixed local time $s_m$, their leading shapes are independent of $\alpha$; the pump amplitude sets the displacement of each passage in time.

For well-separated passages, a first approximation to $C$ is obtained by simply adding the passages:
\begin{equation}
C_{\mathrm{add}}=1+\sum_{m=0}^{M}
\left[C_m^{(0)}-(-1)^m\right].
\label{eq:addition}
\end{equation}
At later times, this addition is improved by retaining interference between passage amplitudes. 

\begin{figure}[t]
\centering
\includegraphics[width=\columnwidth]{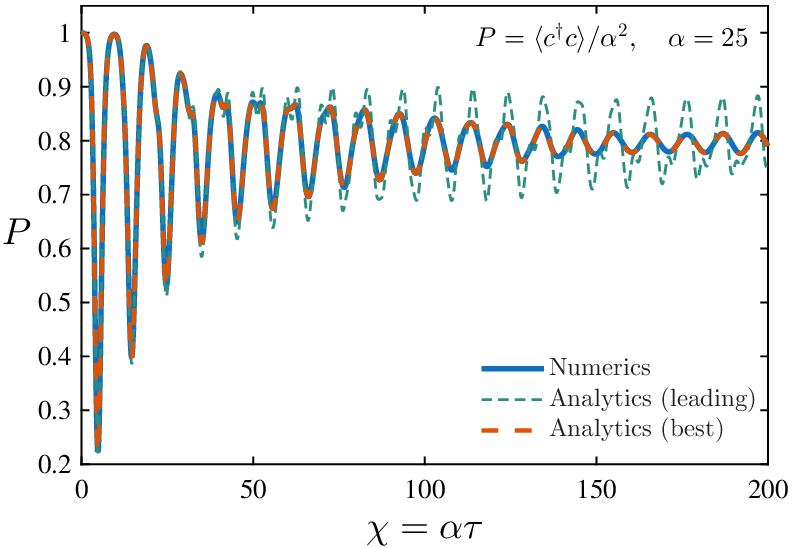}
\caption{Coherent-pump population dynamics at $\alpha=25$. Numerical evolution (blue solid), the leading analytic approximation (green dashed), and the corrected analytic result (orange dashed) are shown through $\chi=200$. The corrected result resolves the repeated conversion cycles and their decaying envelope. Its relative population error, $\Delta P=|P_{\mathrm{num}}-P_{\mathrm{ana}}|/P_{\mathrm{num}}$, is of order $10^{-13}$ during the first depletion and remains below $10^{-3}$ over the plotted interval. The error curves and detailed comparison are given in the SM; see also SM Fig.~3.}
\label{fig:population-benchmark}
\end{figure}

The coherent pump introduces a second effect. Its photon-number distribution is Poissonian~\cite{BACHOR_2019},
\begin{equation}
p_N=e^{-\alpha^2}\frac{\alpha^{2N}}{N!},
\label{eq:poisson}
\end{equation}
but different sectors evolve at slightly different rates. For $C$, the clock prescription uses
\begin{equation}
s_{m,N}=\frac{\chi\sqrt{N+3/2}}{\alpha}
-(2m+1)\log\!\left(4\sqrt{N+3/2}\right).
\label{eq:clocks}
\end{equation}
Averaging with $p_N$ then dephases the oscillations, producing the decaying envelope in Fig. ~\ref{fig:population-benchmark}. Expressions for the other moments are listed in the SM. The Poisson average can be evaluated using fixed-node quadrature~\cite{GOLUB_1969}, so the cost does not grow with the Fock-space sector dimension.

\textit{Numerical comparison.---}
Figure~\ref{fig:population-benchmark} compares the normalized pump population with numerical evolution at $\alpha=25$. The corrected analytic result follows both the repeated oscillations and their decreasing envelope through $\chi=200$. The corresponding relative errors are summarized in the caption and resolved in the SM. For this benchmark we retained terms through $O(\epsilon^{50})$; additional corrections can be incorporated systematically. The accompanying computational package provides direct evaluation of the moments, derived observables, and numerical benchmarks (SM Secs.~A and H.2).

\textit{Quantum noise and higher-order correlations.---}
The leading profiles alone do not resolve every observable. For
$X_c=c+c^\dagger$ and $Y_c=-\ii(c-c^\dagger)$, the coherent-input pump
quadrature variances are
\begin{equation}
\begin{aligned}
V_{X_c}&=1+2\alpha^2(P+S)-4\alpha^2C^2,\\
V_{Y_c}&=1+2\alpha^2(P-S).
\end{aligned}
\label{eq:variance}
\end{equation}
The vacuum variance is one. Although $P^{(0)}=S^{(0)}$, their difference at order $\epsilon$ gives an order-one contribution to $V_{Y_c}$. Likewise, $U^{(0)}=T^{(0)}$ leaves $T-U$ controlled by finite-pump corrections. Accurate mean fields therefore do not, by themselves, establish accurate quantum noise.

The dynamics also fix the signal and idler observables. For example,
\begin{equation}
\avg{n_a}=\avg{n_b}=\alpha^2(1-P),\qquad
\avg{ab}=-\alpha^2\partial_\chi C,
\label{eq:identities}
\end{equation}
where $n_j=j^\dagger j$. Third-order central quadrature moments probe non-Gaussian statistics, since they vanish for a Gaussian state~\cite{WEEDBROOK_2012}. The corrected moments also provide inputs to moment-based entanglement criteria~\cite{SHCHUKIN_2005,HILLERY_2006}.

\begin{figure*}[t]
\centering
\begin{minipage}[t]{0.47\textwidth}
\raggedright
    {\small (a)}\par
\includegraphics[width=\linewidth]{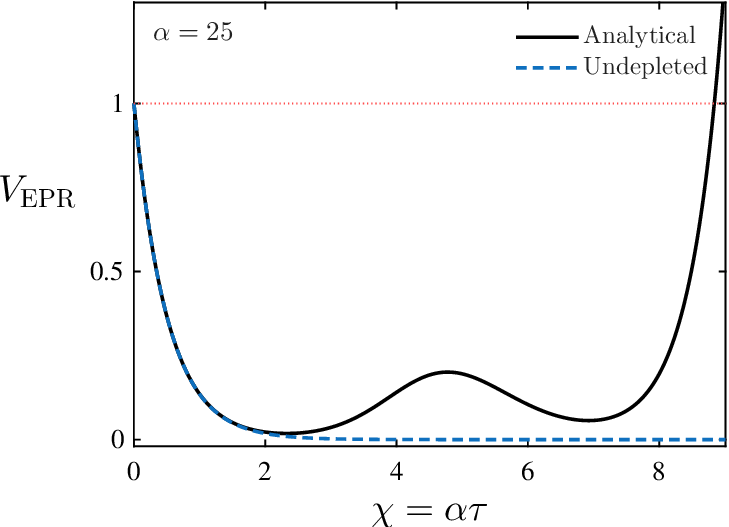}
\end{minipage}\hfill
\begin{minipage}[t]{0.49\textwidth}
\raggedright
    {\small (b)}\par
\includegraphics[width=\linewidth]{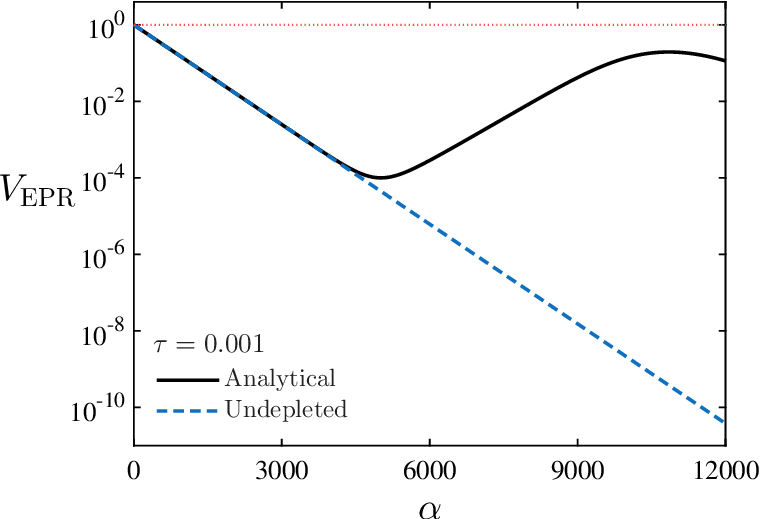}
\end{minipage}
\caption{Two-mode squeezing from a coherent pump, with the signal and
idler initially in vacuum. The normalized joint-quadrature variance
$V_{\mathrm{EPR}}$ is defined in Eq.~\eqref{eq:epr-variance}.
Black solid curves are evaluated from the resummed first-depletion
expressions; blue dashed curves give the undepleted quadratic result
$e^{-2\chi}$. The red dotted line denotes vacuum noise.
(a) Fixed $\alpha=25$, with time varied through
$0\leq\chi\leq2\log(4\alpha)$, on a linear vertical scale.
The first squeezing minimum is finite, and later conversion dynamics
produce nonmonotonic noise.
(b) Fixed $\tau=gt=0.001$, with $0\leq\alpha\leq12000$, on a
logarithmic vertical scale. Increasing $\alpha$ also increases the
gain, revealing the minimum and the subsequent departure from the
monotonically decreasing quadratic prediction.}
\label{fig:two-mode-squeezing}
\end{figure*}

The signal--idler squeezing follows directly from the same two pump moments. With $X_j=j+j^\dagger$ and $Y_j=-\ii(j-j^\dagger)$, define
\begin{equation}
\begin{aligned}
V_{\mathrm{EPR}}
&=\tfrac12\Var(X_a-X_b)
 =\tfrac12\Var(Y_a+Y_b)\\
&=1+2\alpha^2\bigl(1-P+\partial_\chi C\bigr).
\end{aligned}
\label{eq:epr-variance}
\end{equation}
This normalization gives $V_{\mathrm{EPR}}=1$ for the input vacuum. Replacing the pump operator by the constant amplitude $\alpha$ gives $V_{\mathrm{EPR}}^{\mathrm{quad}}=e^{-2\chi}$. The quantized-pump result instead develops a finite first minimum, followed by nonmonotonic evolution through depletion and back-conversion [Fig.~\ref{fig:two-mode-squeezing}(a)].

For a bright coherent pump, the expansion near the first minimum gives $V_{\mathrm{EPR}}\simeq e^{-2\chi}+e^{2\chi}/(16\alpha^2)$. Balancing the decreasing quadratic contribution and the growing
finite-pump correction yields
\begin{equation}
\chi_{\min}\simeq\tfrac12\log(4\alpha),\qquad
V_{\min}\simeq\frac{1}{2\alpha},\qquad
\tau_{\min}=\frac{\chi_{\min}}{\alpha}.
\label{eq:epr-minimum}
\end{equation}
The fixed-$\tau$ scan in Fig.~\ref{fig:two-mode-squeezing}(b) varies both the initial pump population and the gain $\chi=\alpha\tau$, and consequently has a different optimization condition. Pump fluctuations thus affect the squeezed noise even before a large fraction of the pump energy has been converted. For a bright coherent pump, our expansion recovers the established leading finite-pump squeezing limit \cite{CROUCH_1988,KINSLER_1993}.

Through the first depletion, we additionally calculate
\begin{equation}
\begin{gathered}
q_{22}=\frac{\avg{c^{\dagger2}c^2}}{\alpha^4},\qquad
q_{13}=\frac{\avg{c^\dagger c^3}}{\alpha^4},\\
q_{04}=\frac{\avg{c^4}}{\alpha^4},\qquad
R=\frac{\avg{abn_c}}{\alpha^4}.
\end{gathered}
\label{eq:fourth}
\end{equation}
For the coherent input, the three pump moments again share a leading profile:
\begin{equation}
\begin{aligned}
q_{22,0}^{(0)}& =q_{13,0}^{(0)} =q_{04,0}^{(0)}=Q_0^{(0)},\\
Q_0^{(0)}(z)& =\int_0^\infty e^{-v}
\left(\frac{1-zv}{1+zv}\right)^4\,\dd v,
\end{aligned}
\label{eq:fourthprofile}
\end{equation}
while $R_0^{(0)}=P_0^{(0)}-Q_0^{(0)}$. Finite-pump corrections separate the
three pump moments and determine the remaining fourth-order correlations
through the identities in SM Sec.~F. In particular, the pump number variance
and equal-time intensity correlation are
\begin{equation}
\begin{aligned}
\Var(n_c)&=\alpha^4(q_{22}-P^2)+\alpha^2P,\\
g_c^{(2)}(0)&=\frac{q_{22}}{P^2}.
\end{aligned}
\label{eq:numberstatistics}
\end{equation}
The same building blocks give $\avg{X_c^4}$, $\avg{Y_c^4}$, and intermode
intensity correlations. Thus fourth order adds photon-number fluctuations
and quadrature-shape information beyond the mean fields and variances.

The framework also accommodates a bright displaced squeezed pump,
\begin{equation}
\begin{aligned}
\lvert\alpha,\zeta\rangle
&=D_c(\alpha)\mathcal S_c(\zeta)\lvert0\rangle,
\qquad \zeta=re^{\ii\theta},\\
\mathcal S_c(\zeta)
&=\exp\!\left[\tfrac12(\zeta^*c^2-\zeta c^{\dagger2})\right],
\end{aligned}
\label{eq:squeezed}
\end{equation}
where $D_c(\alpha)=\exp(\alpha c^\dagger-\alpha c)$ and $\alpha$ remains real.
Writing $\delta c=c-\alpha$, the input fluctuations are
$\nu=\avg{\delta c^\dagger\delta c}=\sinh^2r$ and
$\eta=\avg{\delta c^2}=-e^{\ii\theta}\sinh r\cosh r$.
The operator BCH coefficients are unchanged. Only their initial-state
expectation values change, obtained from the normal generating
function~\cite{WEEDBROOK_2012,BACHOR_2019}
\begin{equation}
\avg{e^{u c^\dagger}e^{v c}}_0
=\exp\!\left[\alpha(u+v)+\tfrac12(\eta^*u^2+\eta v^2)+\nu uv\right].
\label{eq:generating}
\end{equation}
Derivatives with respect to $u,v$ supply the input moments and hence the squeezing-dependent depletion corrections (SM Sec.~G). Input squeezing provides a direct way to change the noise entering depletion~\cite{COHEN_1995}. The fourth-order and squeezed-pump results reported here are restricted to $\chi\leq2\log(4\alpha)$; their later return passages are not treated.

\begin{figure}[t]
\centering
\includegraphics[width=\columnwidth]{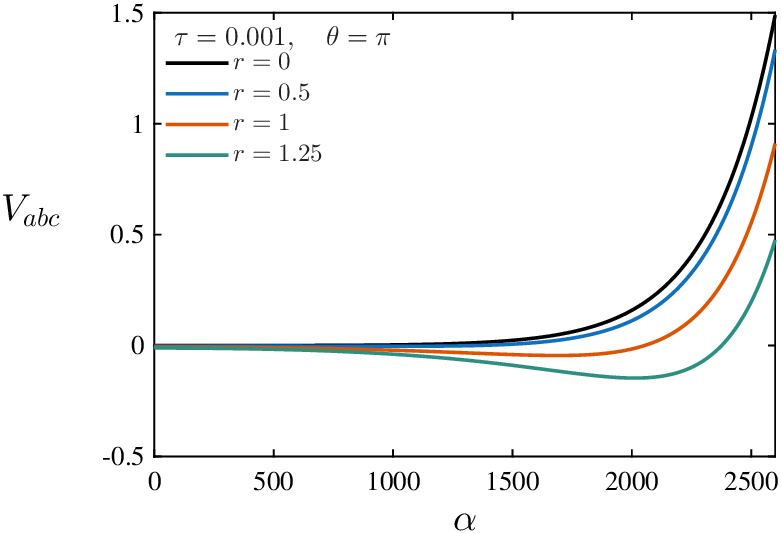}
\caption{Control of a third-order signal--idler--pump correlation by input pump squeezing. The interaction is fixed at $\tau=0.001$, with $\theta=\pi$ and $r=0,0.5,1,1.25$ shown in black, blue, orange, and green, respectively. 
$V_{abc} = -\avg{X_aX_b\delta X_c}$ in the convention of Eq.~\eqref{eq:model}, where $\delta X_c=X_c-\avg{X_c}$. Stronger phase squeezing produces a negative interval followed by a sign reversal as $\alpha$ increases. The nonzero third-order central moment witnesses non-Gaussian correlations; its sign is set by the chosen quadrature phases.}
\label{fig:third-squeezing}
\end{figure}

Figure~\ref{fig:third-squeezing} shows that changing the pump's input noise can alter both the magnitude and the sign of the generated third-order correlation. For $\theta=\pi$, phase squeezing lowers the input phase variance while raising the amplitude variance. At the larger squeezing values, the resulting central moment becomes negative over an intermediate amplitude interval before changing sign again. A zero crossing of this one moment does not establish Gaussianity, but the nonzero values on either side reveal correlations absent from a Gaussian description. The second-order pump-noise scans and their dependence on interaction time are given in the SM.

\textit{Conclusion.---}
Resumming the Heisenberg expansion into depletion and return passages gives explicit expressions for fully quantized parametric down-conversion, valid for all pump amplitudes. A common integral structure connects coherent-pump moments through third order, with finite-pump corrections resolving quantum noise and correlations. Passage interference and sector-clock averaging extend the description to repeated conversion dynamics. Fourth-order correlations and bright squeezed inputs extend the first-depletion calculation to photon statistics and the effect of input quantum noise. These results provide directly evaluable observables without large-sector propagation and a basis for treating more general inputs, losses, and multimode interactions. From the point of view of optical quantum technologies, the most relevant region is that leading up to the first depletion where strong levels of squeezing from highly efficient sources are required for quantum computing \cite{AGHAEE_RAD_2025} and quantum metrology \cite{JIA_2024} applications, amongst others. In this region we recover fundamental analytic limits to squeezing versus efficiency and the emergence of non-Gaussian statistics that may guide future technological developments.

\textit{Acknowledgements.} — We acknowledge the use of ChatGPT (OpenAI; GPT-6 Sol) to assist in identifying patterns and analytic forms for coefficient sequences arising in the perturbative and asymptotic analysis. The resulting expressions were independently verified by the authors, with explicit analytical derivations provided in the Supplemental Material. The authors take full responsibility for the analysis and results presented. We also thank Mark Hillery for drawing our attention to relevant literature.

\ifpdf\else\renewcommand{\href}[2]{#2}\fi
\bibliography{references}

\end{document}


\maketitle

\tableofcontents

\section*{Foreword}
\addcontentsline{toc}{section}{Foreword}
The derivation is long, but only because it is explicit. A significant portion is dedicated to calculating \textbf{higher-order corrections}. However, the part with the most theoretical interest is arguably the derivation of \textbf{leading profiles}. 

For readers who wish to see leading profile derivations, only Appendix \ref{sec:perturbation} and \ref{sec:amplitude} are relevant. Section \ref{subsec:R1} may also be skipped as it's about higher-order corrections. Appendix \ref{sec:book-keeping} is a useful reference including various definitions and raw data. 

\pagebreak
\subsection{Quick references of key results} \label{app:key_results}
Define $\chi = \alpha\tau$ the scaled time, and
\begin{gather}
    s_m = \chi-(2m+1)\log(4\alpha), \qquad \rho_m(u) = \abs*{h_m(u)}^2, \\
    h_m(u) = \frac{1}{2\pi} \int_{-\infty}^\infty H_m(t) e^{itu} dt, \\
    \text{where } \quad H_m(t) = 2^{-(2m+1)it/2} \frac{\Gamma(\frac{1-it}{2})^{m+1}}{\Gamma(\frac{1+it}{2})^m}.
\end{gather}
The $m$-th leading passage of $C = \expval*{c}/\alpha$:
\begin{align}
    \Phi_m^{(0)}(s_m) &= (-1)^{m+1} \int_{-\infty}^\infty \rho_m(u)\tanh(s_m+u) \ du.
\intertext{The $m$-th leading passage of $P=\expval*{c^\dag c}/\alpha^2$ and $S=\expval*{cc}/\alpha^2$: (note: since $P$ and $S$ share the leading profiles, their difference, $V_Y = 1+2\alpha^2(P-S)$, is a major quantity of interest)}
    P_m^{(0)}(s_m) &= \int_{-\infty}^\infty \rho_m(u)\tanh^2(s_m+u) \ du.
\intertext{The $m$-th leading passage of $U=\expval*{c^\dag cc}/\alpha^3$ and $T=\expval*{ccc}/\alpha^3$:}
    U_m^{(0)}(s_m) &= (-1)^{m+1} \int_{-\infty}^\infty \rho_m(u)\tanh^3(s_m+u) \ du.
\end{align}

\begin{table}[H]
    \centering
    \begin{tabular}{c|c|c|c}
        Moment & Correction & Section & Final expression \\\hline
        \multirow{4}{*}{$C$} & 1st correction to depletion & \ref{subsec:correction_c_1} & \eqref{eqn:C1_final} \\
        & 2nd correction to depletion & \ref{subsec:correction_c_2} & \eqref{eqn:C2} \\
        & 1st correction to return & \ref{subsec:R1}& \eqref{eqn:R1_final} \\
        & Poisson averaging & \ref{subsec:poisson_c}& \eqref{eqn:C_poisson} \\\hline
        \multirow{4}{*}{$P$} & 1st correction to depletion & \ref{subsubsec:P_dep_1} & \eqref{eqn:P_dep_1} \\ 
        & 2nd correction to depletion & \ref{subsubsec:P_dep_1} & \eqref{eqn:P_dep_2} \\
        & 1st correction to return & \ref{subsubsec:P_ret_1} & \eqref{eqn:P_ret_1} \\
        & Poisson averaging &  \ref{subsubsec:P_poisson}& \eqref{eqn:P_poisson} \\\hline
        \multirow{4}{*}{$V_Y$} & 1st correction to depletion & \ref{subsubsec:VY_dep_1}&\eqref{eqn:VY_dep_1} \\
        & 2nd correction to depletion & \ref{subsubsec:VY_dep_2} & \eqref{eqn:VY_dep_2} \\
        & 1st correction to return & \ref{subsubsec:VY_dep_2} & \eqref{eqn:VY_ret_1}  \\
        & Poisson averaging & \ref{subsubsec:VY_poisson} & \eqref{eqn:VY_poisson} \\\hline
        \multirow{4}{*}{$U$} & 1st correction to depletion & \ref{subsubsec:U_dep_1} & \eqref{eqn:U_dep_1} \\
        & 2nd correction to depletion & \ref{subsubsec:U_dep_2}&\eqref{eqn:U_dep_2} \\
        & 1st correction to return & \ref{subsubsec:U_ret_1} & \eqref{eqn:U_ret_1} \\
        & Poisson averaging & \ref{subsubsec:U_poisson} & \eqref{eqn:U_poisson} \\\hline
        \multirow{4}{*}{$T$} & 1st correction to depletion & \ref{subsubsec:T_dep_1} & \eqref{eqn:T_dep_1} \\
        & 2nd correction to depletion & \ref{subsubsec:T_dep_2} & \eqref{eqn:T_dep_2} \\
        & 1st correction to return & \ref{subsubsec:T_ret_1} & \eqref{eqn:T1_ret_final}\\
        & Poisson averaging & \ref{subsubsec:T_poisson} & \eqref{eqn:Q_poisson}
    \end{tabular}
    \caption{Locations of higher-order corrections and clock-averaging formulas. The various symbols are defined above. Alternatively, refer to the List of symbols on the next page.} 
    \label{tab:final_expressions}
\end{table}
Furthermore, 4th-order correlations through the first depletion are treated in Section \ref{sec:4th}, and squeezed state is treated in Section \ref{sec:squeezed}.

\pagebreak
\subsection{List of symbols and variables} \label{subsec:list_symbols}

\subsubsection*{Correlation functions}

\begin{list}{}{%
\setlength{\labelwidth}{24mm}
\setlength{\leftmargin}{35mm}}
\item[$a, a^\dag$\dotfill] Signal mode
\item[$b, b^\dag$\dotfill] Idler mode
\item[$c, c^\dag$\dotfill] Pump mode
\item[$F, f$\dotfill] Pump amplitude $\expval c$
\item[$C$\dotfill] Normalized pump amplitude $\frac{\expval c}{\alpha}$
\item[$P$\dotfill] Normalized pump photon number / population $\frac{\expval*{c^\dag c}}{\alpha^2}$
\item[$S$\dotfill] Normalized 2nd-order phase $\frac{\expval{cc}}{\alpha^2}$
\item[$V_Y$\dotfill] Variance $V_Y = 1+2\alpha^2(P-S)$
\item[$U$\dotfill] Normalized 3rd-order mixed correlation $\frac{\expval*{c^\dag cc}}{\alpha^3}$
\item[$T$\dotfill] Normalized 3rd-order phase $\frac{\expval{ccc}}{\alpha^3}$
\item[$Q$\dotfill] 3rd-order phase difference, $Q = C+\alpha^2(U-T)$
\item[$\Gamma$\dotfill] ``Gap'' term in $\expval*{cc}$ calculations, $\Gamma = V_Y-P$
\item[$G$\dotfill] ``Gap'' term in $\expval*{ccc}$ calculations, $G = Q - U$

\end{list}

\subsubsection*{Profiles and their corrections}

\begin{list}{}{%
\setlength{\labelwidth}{24mm}
\setlength{\leftmargin}{35mm}}
\item[$C_0, \mathcal C_0$\dotfill] Leading profile of the first depletion of $C$, equal to the $\mathcal O(\epsilon^0)$ term in $C$. 
\item[$C_1$\dotfill] First correction to the first depletion of $C$
\item[$C_2$\dotfill] Second correction to the first depletion of $C$
\item[$R_0$\dotfill] Leading profile of the first return of $C$
\item[$R_1$\dotfill] First correction to the first return of $C$
\item[$\Phi_m^{(0)}$\dotfill] Leading profile of the $m$-th passage of $C$, $\Phi_0^{(0)} = C_0, \Phi_1^{(0)} = R_0$\dots
\item[$P_0^{(0)}, P_\text{dep}^{(0)}$\dotfill] Leading profile of the first depletion of $P$
\item[$P_0^{(1)},P_\text{dep}^{(1)}$\dotfill] First correction to the first depletion of $P$
\item[$P_1^{(0)},P_\text{ret}^{(0)}$\dotfill] Leading profile of the first return of $P$
\item[$P_1^{(1)},P_\text{ret}^{(1)}$\dotfill] First correction to the first return of $P$
\item[$P_m^{(0)}$\dotfill] Leading profile to the $m$-th passage of $P$ 
\item[\dotfill] Same naming convention holds also for $S,U,T,Q\dots$

\item[$h_m$\dotfill] Leading profile density of the $m$-th passage for all correlation functions
\item[$\rho_m$\dotfill] $\rho_m = \abs{h_m}^2$
\item[$H_m$\dotfill] Inverse Fourier transform of $h_m$ \\

\end{list}

\subsubsection*{Asymptotic and harmonics analysis}
\begin{list}{}{%
\setlength{\labelwidth}{24mm}
\setlength{\leftmargin}{35mm}}
\item[$\alpha$\dotfill] Pump coherent amplitude
\item[$\tau$\dotfill] Time, defined in the time evolution operator as $U(\tau) = e^{\tau(a^\dag b^\dag c - abc^\dag)}$
\item[$\chi,x$\dotfill] Used interchangeably as scaled time, $\chi = \alpha\tau$
\item[$s_m$\dotfill] Shifted (local) time, $s_m = \chi - (2m+1)\log4\alpha$
\item[$\epsilon$\dotfill] Perturbative parameter, $\epsilon=1/\alpha^2$
\item[$z$\dotfill] Perturbative parameter (depletion time scale), $z = e^{2\chi}/16\alpha^2 = \epsilon e^{2\chi}/16$
\item[$w$\dotfill] Perturbative parameter (return time scale), $w=\epsilon^2z = \epsilon^3 e^{2\chi}/16$
\item[$Q$\dotfill] $Q = AB^\dag A$, and $Q^\dag Q$ is the laplacian
\item[$G_N, G_n$\dotfill] $N$-th order perturbative expressions obtained from BCH expansions
\item[$N, n$\dotfill] Used interchangeably as perturbative order in $G_N$
\item[$j$\dotfill] Harmonic order appeared in $e^{2jx}$
\item[$r$\dotfill] $r = N-j$, denotes $r$-th order harmonics 
\item[$H_r$\dotfill] Raw harmonic coefficients
\item[$B_r$\dotfill] Normalized harmonic coefficients
\item[$P_r$\dotfill] Generic polynomial fit of $B_r$, typically $P_r=B_r$ only for large $j$
\item[$E_r$\dotfill] Exceptional terms, $E_r = B_r - P_r$
\item[$D, D_z, D_w$\dotfill] Euler derivative, by default $D = z\frac{d}{dz}$
\item[$q_n, \tilde q_n$\dotfill] Ratios between $E_{2n+1,n}$ and $E_{2n,n}$
\item[$\mathbb E_N$\dotfill] Poisson-weighted average. $N$: Fock state sector here, not perturbative order
\item[$p_N$\dotfill] Poisson distribution function

\end{list}

$Q, H, G$ are re-used but the context should make their meanings clear.

\pagebreak
\appendix

\section{Numerical benchmark comparisons}
\subsection{Coherent amplitude $\expval*{c}$}

\begin{figure}[H]
\centering
\begin{subfigure}{.45\textwidth}
    \centering
    \includegraphics[width=0.95\linewidth]{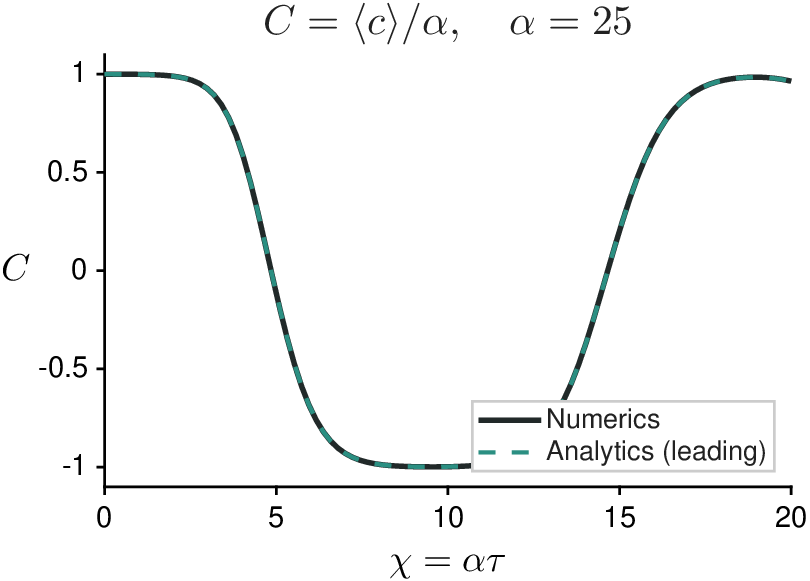}
    \caption{}
    \label{subfig:c_first_passage}
\end{subfigure}
\begin{subfigure}{.45\textwidth}
\centering
    \includegraphics[width=0.95\linewidth]{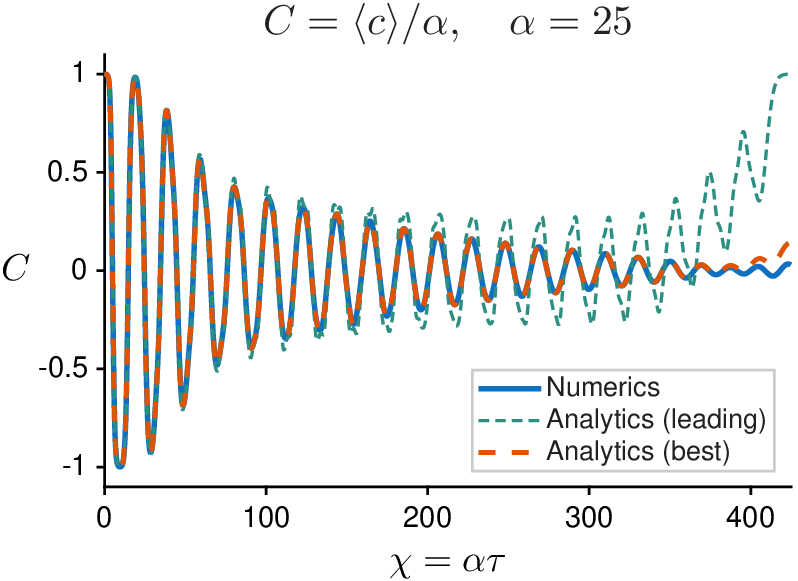}
    \caption{}
    \label{subfig:c_long_time_best}
\end{subfigure}
    \caption{Numerics vs. analytics comparison for the normalized coherent amplitude $C=\expval*{c}/\alpha$, $\alpha=25$. (a) Leading order solution given by equation \eqref{eqn:summ_c}. (b) Ultra long-time behavior of numerics vs. leading order solution vs. best solution. The ``best'' solution has all corrections implemented (higher-order terms, passage interference and Poisson-clocks). The precise errors from numerical benchmark are shown in Figure \ref{fig:c_errors} below.}
    \label{fig:c_comparison}
\end{figure}

\begin{figure}[H]
    \centering
    \begin{subfigure}{.47\textwidth}
    \centering
    \includegraphics[width=0.97\linewidth]{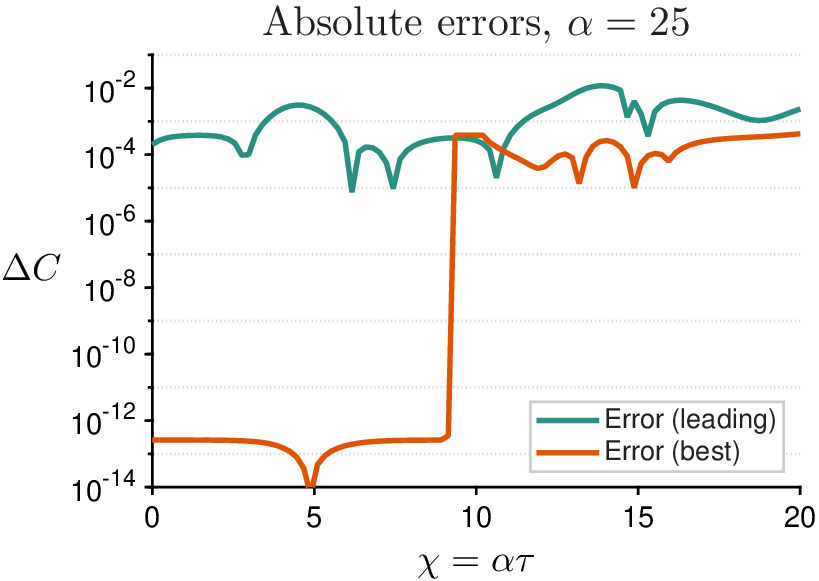}
    \caption{}
    \label{subfig:c_error}
    \end{subfigure}
    \begin{subfigure}{.47\textwidth}
    \centering
    \includegraphics[width=0.97\linewidth]{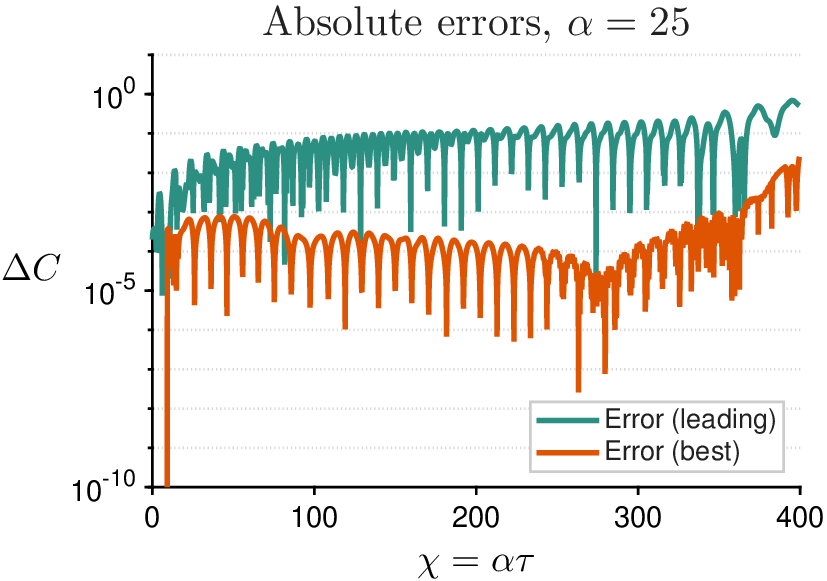}
    \caption{}
    \label{subfig:c_error_long}
    \end{subfigure}
    \caption{Absolute errors in our solutions against numerical benchmark (the numerical results were obtained by directly diagonalizing the Hamiltonian). Error $=\Delta C=\abs{C_\text{numerics} - C_\text{analytics}}$. (a) For the best solution (all corrections applied), the maximum absolute error during the first depletion is $2.6\times10^{-13}$; during the first return it is $3.0\times10^{-4}$. This is a great improvement from the leading profile error, which reaches $0.012$ at maximum. Note the ``jump'' in the blue curve at $\chi\simeq10$ occurs since the depletion profile was calculated to 19 orders, while the return was calculated to 1 order, thus order-by-order matching was not possible. (b) Long-time error: The error in the best solution (with passage interference and Poisson clocks) remains below $10^{-3}$ until the total collapse of oscillation.}
    \label{fig:c_errors}
\end{figure}

The general form of the leading profile is:
\begin{align}
    \Phi_m^{(0)}(s_m) = (-1)^{m+1} \int_{-\infty}^\infty \abs{h_m(u)}^2\tanh(s_m+u) \ du,
\end{align}
where $s_m = \chi - (2m+1)\log(4\alpha)$ is the local time, with $\chi$ the global time; and $h_m(u)$ is a Fourier transformed packet amplitude, 
\begin{gather}
    h_m(u) = \frac{1}{2\pi} \int_{-\infty}^\infty H_m(t) e^{itu} dt, \label{eqn:hm_summary}\\
    \text{where } \quad H_m(t) = 2^{-(2m+1)it/2} \frac{\Gamma(\frac{1-it}{2})^{m+1}}{\Gamma(\frac{1+it}{2})^m}. 
\end{gather}
To get the long-time solution consisting of a total of $M$ passages (meaning $M/2$ round trips), the simplest way is to ``stitch'' the passages together:
\begin{subequations}
\begin{align}
    \Phi_{\text{total,}M}^{(0)} & = 1+\sum_{m=0}^{M-1}\qty[\Phi_m^{(0)}(s_m) - (-1)^m] \\
    & = 1 + \sum_m(-1)^{m+1} \int\abs{h_m(u)}^2 [1+\tanh(s_m+u)] \ du,
\end{align} \label{eqn:summ_c}
\end{subequations}
which already gives great agreement with numerics for several cycles, shown as the orange curves in Figure \ref{fig:c_comparison}. Also, Figure \ref{fig:c_errors} shows that the error in the first depletion-return cycle is at most $10^{-2}$, but becomes more significant later on. 

We demonstrate the leading-order dynamics of the $k$-th order correlator during each passage takes a remarkably universal form as spectral convolutions of $\tanh^k$ soliton profiles. 

\subsubsection{Higher-order corrections}

The leading profiles are useful in describing short-time behavior and in providing a quick, physically intuitive result, but they become unstable as time goes on, as can be seen from Figure \ref{subfig:c_long_time_best}. There are a few ways to improve accuracy:
\begin{enumerate}
    \item The most obvious is to include higher-order corrections. In the present work, we explicitly calculated the 1st- and 2nd-correction to the first depletion in Sections \ref{subsec:correction_c_1} and \ref{subsec:correction_c_2} respectively, and the 1st-correction to the first back-conversion/return in Section \ref{subsec:R1}. Higher-order corrections can be similarly calculated. 
    
    We shall see that for the depletion profile $(m=0)$, higher-order corrections are straightforward to evaluate (summarized in Appendix \ref{app:key_results}). But for each additional passage, the calculations get significantly more complicated. For this reason, we evaluate the depletion profile through $C_{19}$ (19 orders of corrections), and the return profile through $R_1$ (1 order of correction), and use the 0th-order leading profile for every passage afterwards. With the depletion profile, these higher-order corrections reduce the error from $10^{-2}$ to less than $10^{-12}$. 

    As for long-time accuracy, we can make two more improvements:

    \item Passage-interference: Instead of adding passages additively, we add them coherently, taking into account the cross-passage interference. 
    \item Poisson-clocks: Take into account the Poisson distribution of $N$-sectors of the coherent state, as well as their different running clocks. 
    
    The recipes for step 2 and 3 are given in Section \ref{subsec:poisson_c}. 
\end{enumerate}
The late-time agreement becomes almost exact after these two steps. The ``best'' solution including all three corrections are shown in Figure \ref{subfig:c_long_time_best}, the errors are plotted in Figure \ref{fig:c_errors}. 

Our solution eventually appears to deviate from the numerics. At $\alpha=25$, this deviation becomes noticeable at $\chi \approx 800$, after about 40 round trips $(M=80)$. This is likely due to numerical instabilities and can be improved by increasing passage count and Fourier resolution. However, the present comparison does not establish which source dominates the discrepancy near $\chi = 800$, since it is already orders of magnitude outside of the initial down-conversion interval (through the first population minimum at $\chi = \log(4\alpha)+\mathcal O(1)$).

Finally, note that one ``depletion'' of the pump amplitude $\expval*{c}$ is in fact a ``round trip'' of the pump population, $\expval*{c^\dag c}$. So the experimentally-relevant region is most likely only half of the first depletion cycle—or more precisely, at time $\chi \lesssim \log4\alpha$ (for $\alpha=25$, this is about 4.6). 

\begin{table}[H]
    \centering
    \begin{tabular}{c|c|c|c|c|c|c}
        $\alpha$ & dep, leading & dep, best & ret, leading & ret, best & 20 cycles, leading & 20 cycles, best \\\hline
        15 & 0.0072 &$1.6\times10^{-9}$ &  0.025 & 0.0011 & 0.39 & 0.0074 \\
        25 & 0.0031 & $3.6\times10^{-13}$ & 0.0012 & $3.8\times10^{-4}$ & 0.35 & 0.0017\\
        50 & $9.8\times10^{-4}$ & $4.6\times10^{-13}$ & 0.0040 & $9.3\times10^{-5}$ & 0.24 & 0.22 \\
        75 & $4.9\times10^{-4}$ & $3.3\times10^{-12}$ & 0.0021 & $4.4\times10^{-5}$ & 0.11 & 0.0095
    \end{tabular}
    \caption{Maximum error in $C=\expval*{c}/\alpha$ throughout various passages. ``dep''—during first depletion $\chi < 2\log(4\alpha)$; ``ret''—during first return/back-conversion $2\log(4\alpha)<\chi<4\log(4\alpha)$; ``20 cycles''—over 20 depletion-back-conversion cycles. ``leading''—0th-order leading solution; ``best''—the best solution with all corrections applied. We can see the best solution remains highly accurate over 20 cycles.}
    \label{tab:c_error}
\end{table}

\subsection{2nd-order correlations $\expval*{c^\dag c}$ and $\expval*{cc}$}

\subsubsection{Leading profile}
Define the normalized population and phase as
\begin{align}
    P \equiv \frac{\expval*{c^\dag c}}{\alpha^2}, \qquad  S \equiv \frac{\expval*{cc}}{\alpha^2}.
\end{align}
Both $P$ and $S$ have the same leading profiles, described by
\begin{align}
    P_m^{(0)}(s_m) = S_m^{(0)}(s_m) = \int_{-\infty}^\infty \abs*{h_m(u)}^2 \tanh^2(s_m+u) \ du,
\end{align}
with the same $h_m(u)$ defined in equation \eqref{eqn:hm_summary}. Additive stitching gives
\begin{align}
    P_\text{total}^{(0)} = S_\text{total}^{(0)} = 1 + \sum_{m=0}^{M-1} \qty[P_m^{(0)}(s_m) - 1] = 1-\sum_{m=0}^{M-1} \int \abs*{h_m(u)}^2 \sech^2(s_m+u) \ du.
\end{align}
\indent To further improve accuracy, take into account the coherent interference between passages (Section \ref{subsec:stitching_cDc}) and Poisson distribution of internal clocks (Section \ref{subsubsec:P_poisson} for $\expval*{c^\dag c}$ and \ref{subsubsec:VY_poisson} for $\expval*{cc}$). 

\begin{figure}[H]
    \centering
    \begin{subfigure}{.45\textwidth}
    \centering
    \includegraphics[width=0.95\linewidth]{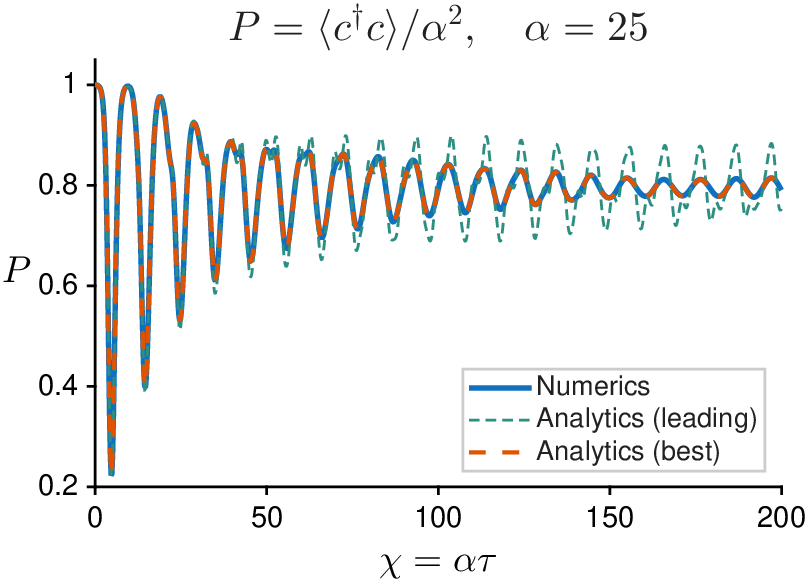}
    \caption{}
    \label{subfig:p_long}
    \end{subfigure}
    \begin{subfigure}{.47\textwidth}
    \centering
    \includegraphics[width=0.97\linewidth]{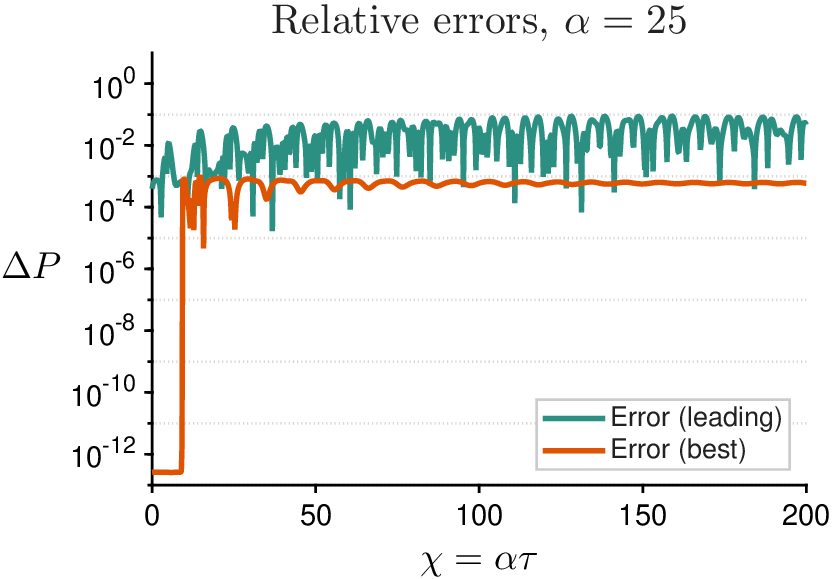}
    \caption{}
    \label{subfig:p_error_long}
    \end{subfigure}
    \caption{(a) Normalized population $P$ comparison. (b) Relative errors against numerical benchmark, $\Delta P = \abs*{P_\text{numerics} - P_\text{analytics}}$. It's safe to use relative errors because there are no zero crossings. For the leading profile, max error reaches about 0.075. As for the best solution with all corrections applied, the max error is $2.6\times10^{-13}$ during the first depletion, and $8.2\times10^{-4}$ afterwards.}
    \label{fig:p_errors}
\end{figure}

\begin{figure}[H]
    \centering
    \begin{subfigure}{.45\textwidth}
    \centering
    \includegraphics[width=0.95\linewidth]{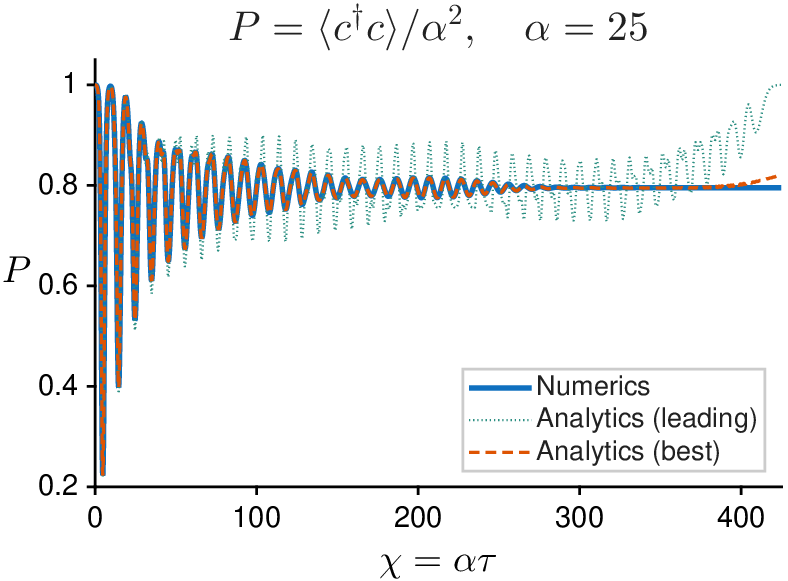}
    \caption{}
    \label{subfig:p_very_long}
    \end{subfigure}
    \begin{subfigure}{.47\textwidth}
    \centering
    \includegraphics[width=0.97\linewidth]{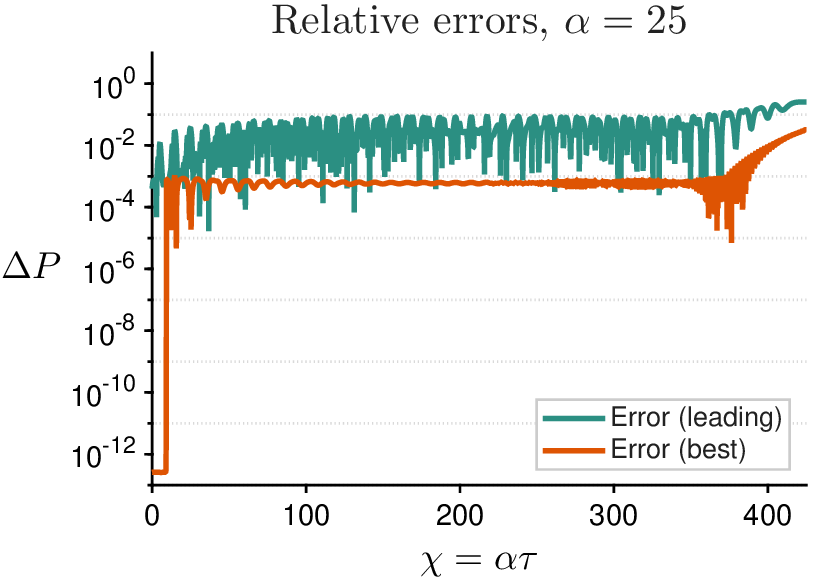}
    \caption{}
    \label{subfig:p_error_very_long}
    \end{subfigure}
    \caption{(a) Normalized population $P$ and (b) absolute errors against numerical benchmark for ultra long time. With the best analytic model, deviation finally occurs at $\chi\approx400$. The accuracy can possibly still be improved by increasing numerical precision and taking into account more passage interference (here, interference only occur between the two nearest neighboring passages), but since $\chi=400$ is already well beyond any experimentally relevant region, this is untested.}
    \label{fig:p_very_long_errors}
\end{figure}

\subsubsection{Higher-order corrections}

To resolve interesting quantum behavior—such as quadrature squeezing and entanglement—we need to compute the $\mathcal O(1/\alpha^2)$ difference between $P$ and $S$. The fact that they have the same leading profile means we must compute higher-order corrections. Define
\begin{subequations}
\begin{gather}
    X = c+c^\dag, \qquad Y = -i(c-c^\dag), \\
    \delta A = A-\expval*{A},  \qquad   V_A \equiv \text{Var}(A) = \expval*{(\delta A)^2}, \qquad \quad A \in \{X,Y\},\\
    \text{explicitly}, \qquad V_X = 1+2\alpha^2(P+S)-4\alpha^2C^2, \qquad V_Y = 1+2\alpha^2(P-S),
\end{gather} \label{eqn:quadratures_intro}
\end{subequations}
then $V_X$ and $V_Y$ are the quadrature variances, plotted in Figure \ref{fig:vx_vy} below.

\begin{figure}[H]
    \centering
    \begin{subfigure}{.49\textwidth}
    \centering
    \includegraphics[width=0.99\linewidth]{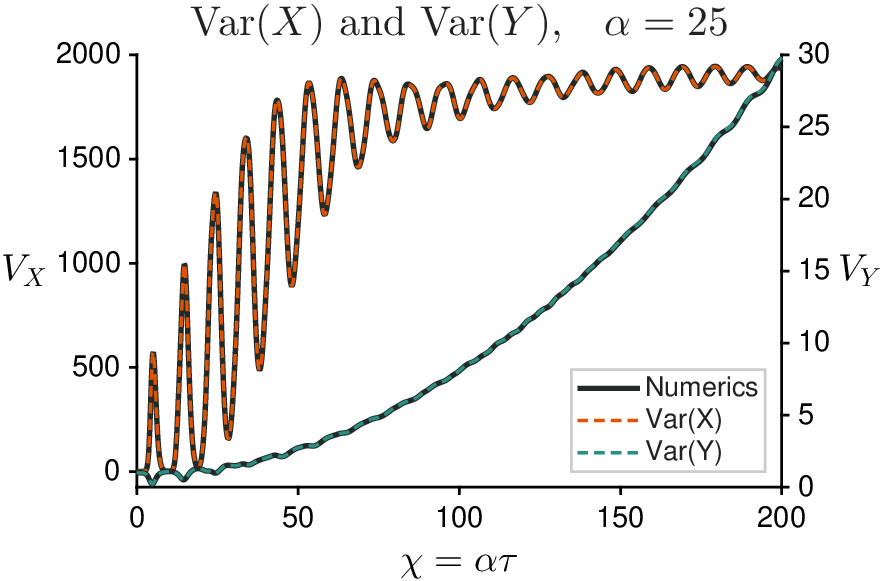}
    \caption{}
    \label{subfig:vx_vy_scaled}
    \end{subfigure}
    \begin{subfigure}{.47\textwidth}
    \centering
    \includegraphics[width=0.97\linewidth]{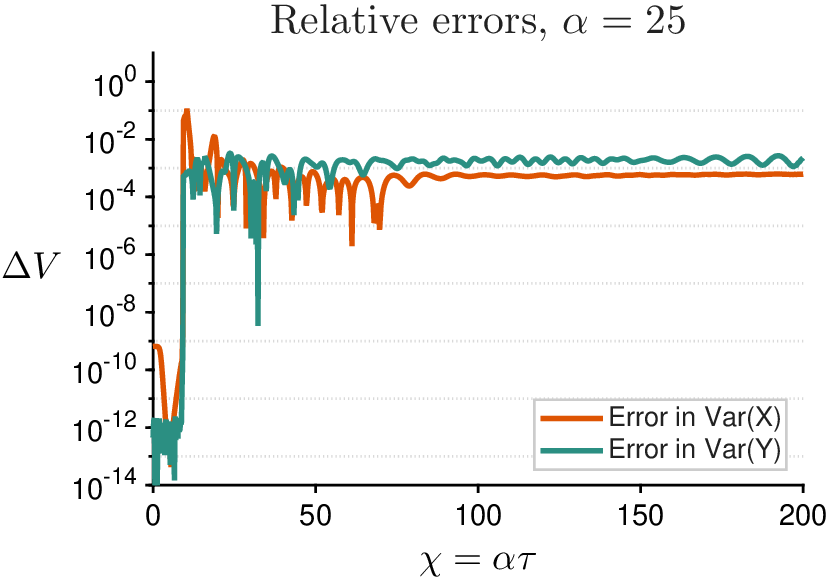}
    \caption{}
    \label{subfig:v_error_long}
    \end{subfigure}
    \caption{(a) Var$(X)$ and Var$(Y)$, numerics vs. analytics comparison. Note $V_X$ and $V_Y$ are plotted on different scales. (b) Relative errors against numerical benchmark, $\Delta V_{X,Y} = \abs*{V_{X,Y,\text{ numerics}} - V_{X,Y,\text{ analytics}}}/V_{X,Y,\text{ numerics}}$. It is safe to use relative errors because there are no zero crossings. During the first depletion, the maximum error in $V_X$ is $6.5\times10^{-10}$, whereas the maximum error in $V_Y$ is $1.4\times10^{-12}$. Note we did not plot the leading-order results here, since to leading order, $V_Y^{(0)} = 1$.}
    \label{fig:vx_vy}
\end{figure}

\begin{table}[H]
    \centering
    \begin{tabular}{c|c|c|c|c|c|c}
        $\alpha$ & $P_\text{dep}$, leading & $P_\text{dep}$, best & $P_\text{ret}$, leading & $P_\text{ret}$, best & 20 cycles, leading & 20 cycles, best \\\hline
        15 & 0.026 &$2.0\times10^{-7}$ &  0.060 & 0.0043 & 0.14 & 0.0084\\
        25 & 0.012 & $2.2\times10^{-12}$ & 0.030 & 0.0011 & 0.10 & 0.0011\\
        50 & 0.0042 & $4.6\times10^{-13}$ & 0.011 & $2.0\times10^{-4}$ & 0.11 & 0.056 \\
        75 & 0.0022 & $3.4\times10^{-12}$ & 0.0058 & $8.9\times10^{-5}$ & 0.083 & 0.011
    \end{tabular}
    \caption{Maximum relative error in normalized population, $P=\expval*{c^\dag c}/\alpha^2$ throughout various passages. Note it is safe to look at relative errors here since $P$ has no zero-crossings.} 
    \label{tab:p_error}
\end{table}

\begin{table}[H]
    \centering
    \begin{tabular}{c|c|c|c|c|c|c}
        $\alpha$ & $V_{X, \text{dep}}$ & $V_{Y, \text{dep}}$ & $V_{X, \text{ret}}$ & $V_{Y,\text{ret}}$ & $V_X$, 20 cycles & $V_Y$, 20 cycles \\\hline
        15 & $7.5\times10^{-5}$ & $2.8\times10^{-7}$ & 0.18  & 0.0077 & 0.18 & 0.10 \\
        25 & $1.8\times10^{-9}$ & $3.0\times10^{-10}$& 0.12 & 0.0025 & 0.12 & 0.010 \\
        50 & $4.6\times10^{-9}$ & $1.7\times10^{-11}$ & 0.12 & $8.6\times10^{-4}$ & 0.12 & 0.23 \\
        75 & $7.5\times10^{-8}$ & $5.1\times10^{-11}$ & 0.10 & $4.7\times10^{-4}$ & 0.10 & 0.035
    \end{tabular}
    \caption{Maximum relative error in $V_X$ and $V_Y$. Notice since $V_X$ and $V_Y$ have no zero crossings, we are safe to use relative errors here. Since leading-order results do not distinguish between $P$ and $S$ and are therefore blind to quadrature squeezing, we only used our best solutions here.}
    \label{tab:vx_vy_error}
\end{table}

In the provided programming package, we include the depletion of $P$ through order 23 and the depletion of $S$ through order 14. The first returns of $P$ and $S$ are included through order 1.

\subsection{3rd-order correlations $\expval*{c^\dag cc}$ and $ \expval*{ccc}$}

\subsubsection{Leading profile}

Define the normalized 3rd-order mixed moment and pure phase as
\begin{align}
    U \equiv \frac{\expval*{c^\dag cc}}{\alpha^3}, \qquad T \equiv \frac{\expval*{ccc}}{\alpha^3}.
\end{align}
Once again, they have the same leading profiles,
\begin{align}
    U_m^{(0)} = T_m^{(0)} = (-1)^{m+1} \int_{-\infty}^\infty \abs*{h_m(u)}^2 \tanh^3(s_m+u) \ du,
\end{align}
with the same $h_m(u)$ defined in equation \eqref{eqn:hm_summary}.

\begin{figure}[H]
    \centering
    \begin{subfigure}{.45\textwidth}
    \centering
    \includegraphics[width=0.95\linewidth]{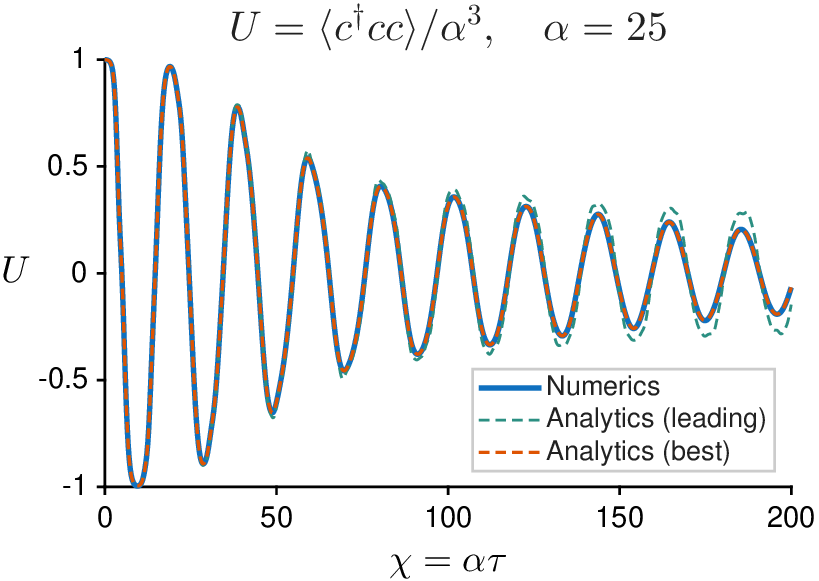}
    \caption{}
    \label{subfig:u}
    \end{subfigure}
    \begin{subfigure}{.47\textwidth}
    \centering
    \includegraphics[width=0.97\linewidth]{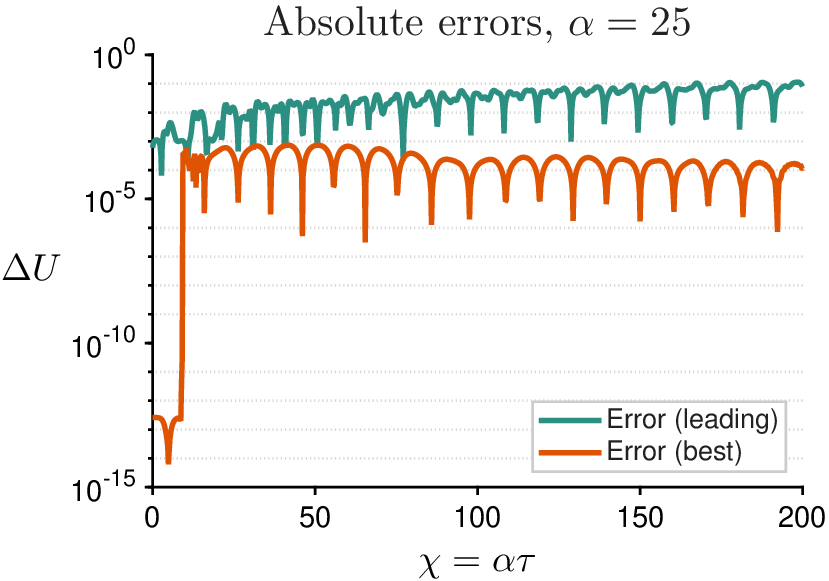}
    \caption{}
    \label{subfig:u_error}
    \end{subfigure}
    \caption{(a) Comparison of the normalized mixed moments $U$. (b) Absolute errors against numerical benchmark, $\Delta U = \abs*{U_\text{numerics} - U_\text{analytics}}$. For the leading profile, the maximum error is 0.011. For the best solution, the maximum error is $2.6\times10^{-13}$ during the first depletion, and $7.5\times10^{-4}$ over the remainder of the interval shown in panel (b).}
    \label{fig:u}
\end{figure}

\begin{table}[H]
    \centering
    \begin{tabular}{c|c|c|c|c|c|c}
        $\alpha$ & $U_\text{dep}$, leading & $U_\text{dep}$, best & $U_\text{ret}$, leading & $U_\text{ret}$, best & 20 cycles, leading & 20 cycles, best \\\hline
        15 & 0.011 & $3.9\times10^{-8}$ & 0.025 &0.0020 & 0.36 & 0.011 \\
        25 & 0.0045 & $2.7\times10^{-11}$ & 0.011 & $5.7\times10^{-4}$ & 0.33 & 0.0034\\
        50 & 0.0013 & $4.6\times10^{-13}$ & 0.0036 & $1.9\times10^{-4}$ & 0.26 & 0.22 \\
        75 & $6.5\times10^{-4}$ & $3.3\times10^{-12}$ & 0.0019 & $7.6\times10^{-5}$ & 0.086 & 0.0095
    \end{tabular}
    \caption{Maximum absolute error in normalized mixed moment, $U=\expval*{c^\dag cc}/\alpha^3$ throughout various passages.}
    \label{tab:u_error}
\end{table}

\subsubsection{Higher-order corrections}

We will compare the normalized three-way correlation, $\tilde V_{abc}$, which is representative of 3rd-order correlations. It is defined as
\begin{align}
     \tilde V_{abc} = \frac{V_{abc}}{\alpha^3}, \qquad V_{abc} = \expval*{X_aY_b\delta X_c},
\end{align}
with the quadratures defined in equation \eqref{eqn:quadratures_intro} but adopted for $a$ and $b$ modes as well.

\begin{figure}[H]
    \centering
    \begin{subfigure}{.45\textwidth}
    \centering
    \includegraphics[width=0.95\linewidth]{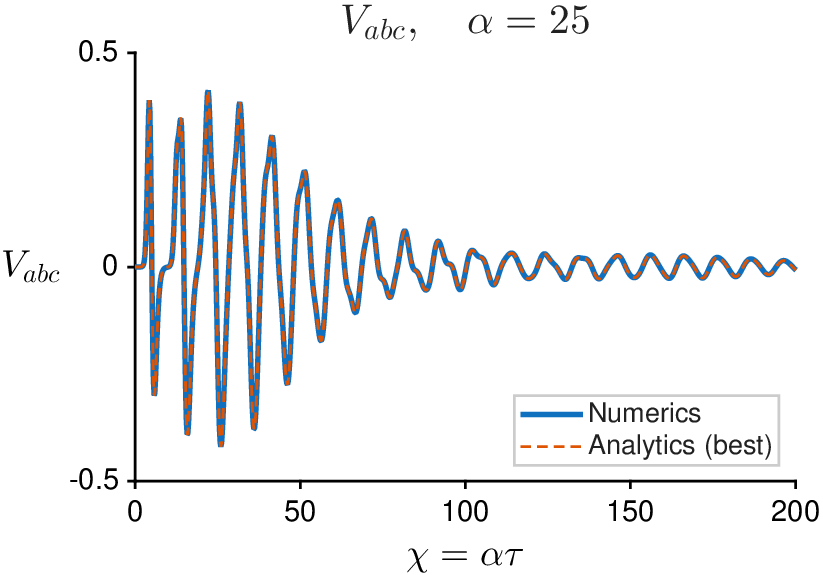}
    \caption{}
    \label{subfig:v_abc}
    \end{subfigure}
    \begin{subfigure}{.47\textwidth}
    \centering
    \includegraphics[width=0.97\linewidth]{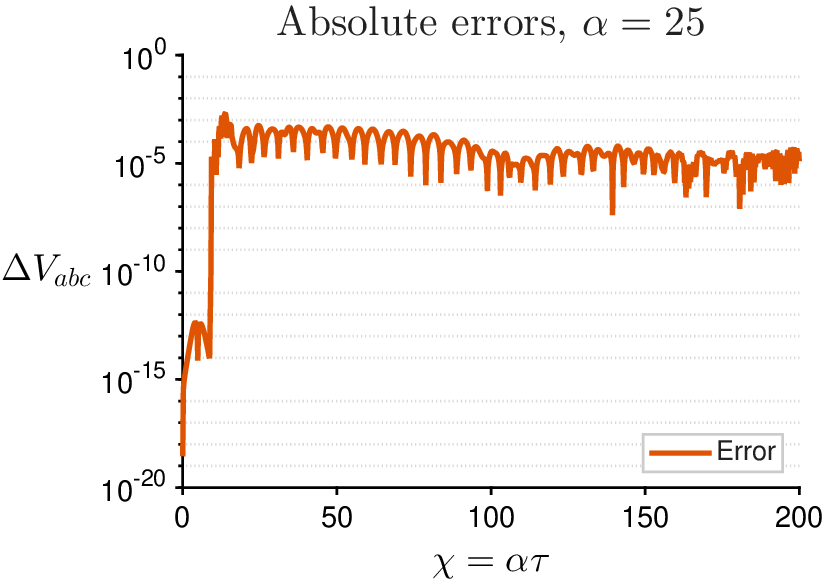}
    \caption{}
    \label{subfig:v_abc_error}
    \end{subfigure}
    \caption{(a) $V_{abc}/\alpha^3$ comparison between numerics and analytics. (b) Absolute errors against numerical benchmark, $\Delta V_{abc} = \abs*{V_{abc,\text{numerics}} - V_{abc,\text{analytics}}}$.}
    \label{fig:v_abc}
\end{figure}

\begin{table}[H]
    \centering
    \begin{tabular}{c|c|c|c}
        $\alpha$ & $\Delta V_{abc,\text{dep}}$  &$\Delta V_{abc,\text{ret}}$ & $\Delta V_{abc}$, 20 cycles \\\hline
        15 & $8.6\times10^{-6}$ & 0.0083 & 0.041  \\
        25 & $7.8\times10^{-11}$ & 0.0021 & 0.0043\\
        50 & $7.5\times10^{-13}$  & $2.8\times10^{-4}$ & 0.042  \\
        75 & $5.6\times10^{-12}$ & $8.0\times10^{-5}$ & 0.0032
    \end{tabular}
    \caption{Maximum absolute error in $V_{abc}$ throughout various passages.}
    \label{tab:v_abc_error}
\end{table}

In the provided programming package, the first depletion profiles of $T$ and $U$ are calculated to order 13. The first return profiles of $T$ and $U$ are calculated to order 1.

\subsection{4th-order correlations (first depletion only)}
We calculated 4th-order correlations for the first depletion only, to $\chi\leqslant2\log(4\alpha)$. The key building blocks are the normalized correlation functions,
\begin{align}
    q_{22} = \frac{\expval*{c^{\dag2}c^2}}{\alpha^4}, \quad q_{13} = \frac{\expval*{c^\dag c^3}}{\alpha^4}, \quad q_{04} = \frac{\expval*{c^4}}{\alpha^4}, \quad R = \frac{\expval*{ab\,n_c}}{\alpha^4},
\end{align}
from which all other correlation functions can be calculated, and they are automatically calculated in the numerical package. $q_{22}, q_{13}$ and $q_{04}$ share the leading profile of
\begin{align}
    Q_0^{(0)} = \int_0^\infty e^{-u} \qty(\frac{1-zu}{1+zu})^4 \ du,
\end{align}
where $z = e^{2s}$ and $s = \chi - \log(4\alpha)$. Higher-order corrections are automatically included in the numerical package, and briefly treated in Section \ref{subsec:4th_corrections}.

We plotted the photon number variances, $\expval*{n_c^2} = \expval*{n_c^2} - \expval*{n_c}^2$, its comparison with numerics and its errors, in Figure \ref{fig:varn} below. The moment used here is $q_{22}$, which is calculated to the 23rd order. The other moments are calculated to the 19th order.

\begin{figure}[H]
\centering
\begin{subfigure}{.45\textwidth}
    \centering
    \includegraphics[width=0.95\linewidth]{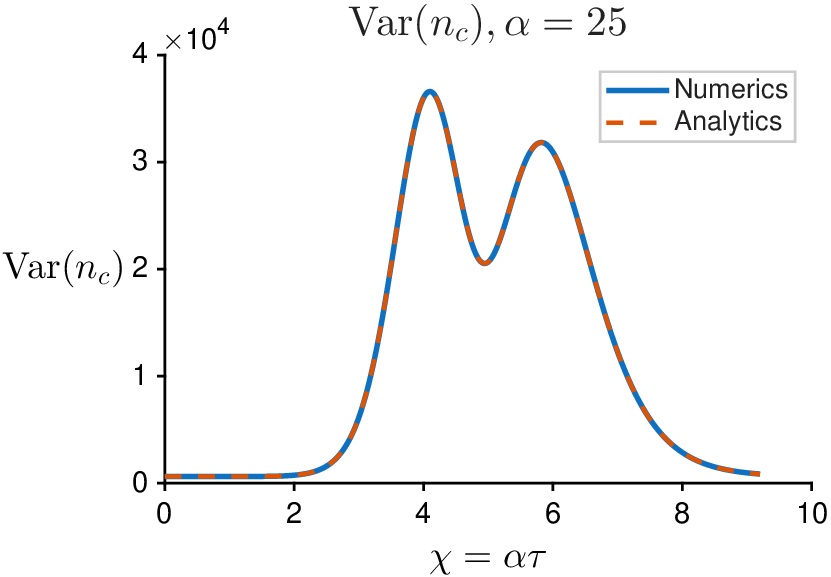}
    \caption{}
    \label{subfig:varn}
\end{subfigure}
\begin{subfigure}{.45\textwidth}
\centering
    \includegraphics[width=0.95\linewidth]{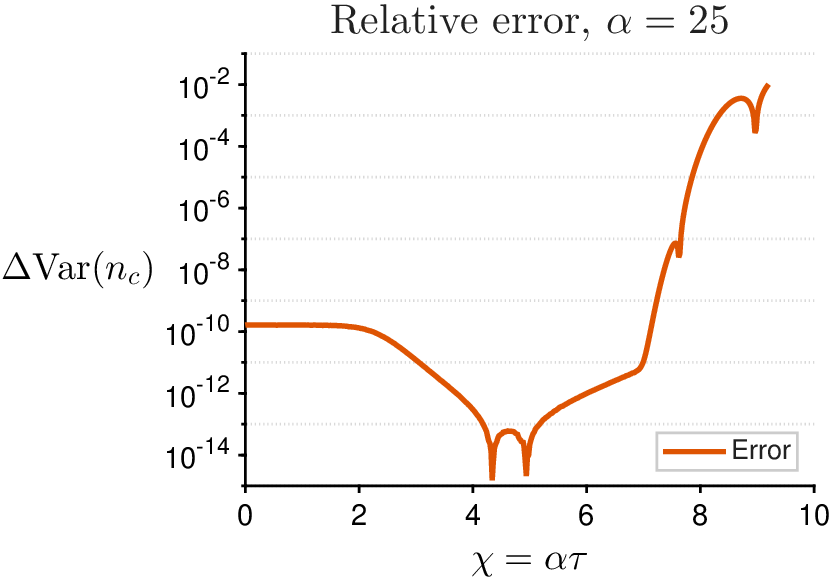}
    \caption{}
    \label{subfig:varn_error}
\end{subfigure}
    \caption{Numerics vs. analytics comparison for the pump photon variance $\expval*{n_c^2} = \expval*{n_c^2} - \expval*{n_c}^2$, analytic result calculated to the 23rd order.
    (a) Direct comparison. Note a variance of 40,000 corresponds to a photon number standard deviation of about 200. (b) Relative error comparison, it remains extremely low until near $2\log4\alpha$.}
    \label{fig:varn}
\end{figure}

In the supplementary programming package, $q_{22}$ is supplied to the 23rd order ($r=0,\dots,23)$ for the first depletion, whereas $q_{13}, q_{04}$ and $R$ are calculated to 19th order ($r=0,\dots,19$). Maximum error occurred across all four moments $(q_{22}, q_{13}, q_{04}, R)$ for various $\alpha$ and correction orders are given in Table \ref{tab:order_4_errors} below.
\begin{table}[H]
    \centering
    \begin{tabular}{c|c|c|c|c|c}
        $\alpha$ & Leading & Through $r=1$ & Through $r=3$ & Through $r=6$ & Through $r=10$ \\\hline
        5 & $8.3\times10^{-2}$ & $3.5\times10^{-2}$ & $1.8\times10^{-3}$ & $4.7\times10^{-5}$ & $1.7\times10^{-5}$ \\
        10 & $2.8\times10^{-2}$ & $3.1\times10^{-3}$ & $1.1\times10^{-5}$ & $1.1\times10^{-8}$ & $6.6\times10^{-12}$ \\
        25 & $5.9\times10^{-3}$ & $1.1\times10^{-4}$ & $1.4\times10^{-8}$ & $6.2\times10^{-14}$ & $8.0\times10^{-15}$ \\
        50 & $1.8\times10^{-3}$ & $8.8\times10^{-6}$ & $8.3\times10^{-11}$ & $9.5\times10^{-15}$ & $9.5\times10^{-15}$ 
    \end{tabular}
    \caption{Maximum absolute errors against numerical benchmarks across all 4 moments for various $\alpha$ and correction orders.}
    \label{tab:order_4_errors}
\end{table}

\subsection{Squeezed pump (first depletion only)}
What if the pump starts not in a coherent state, but a squeezed coherent pump? We calculated the various correlations for the first depletion only, to $\chi\leqslant2\log(4\alpha)$. 

We found due to its smallness, one quantity where squeezing has the most effect on is $V_Y$. Another quantity representative of squeezing is Var$(n_c)$, connecting to 4th-order results. We plot them in Figure \ref{fig:squeezing_VY} below.

\begin{figure}[H]
\centering
\begin{subfigure}{.45\textwidth}
    \centering
    \includegraphics[width=0.95\linewidth]{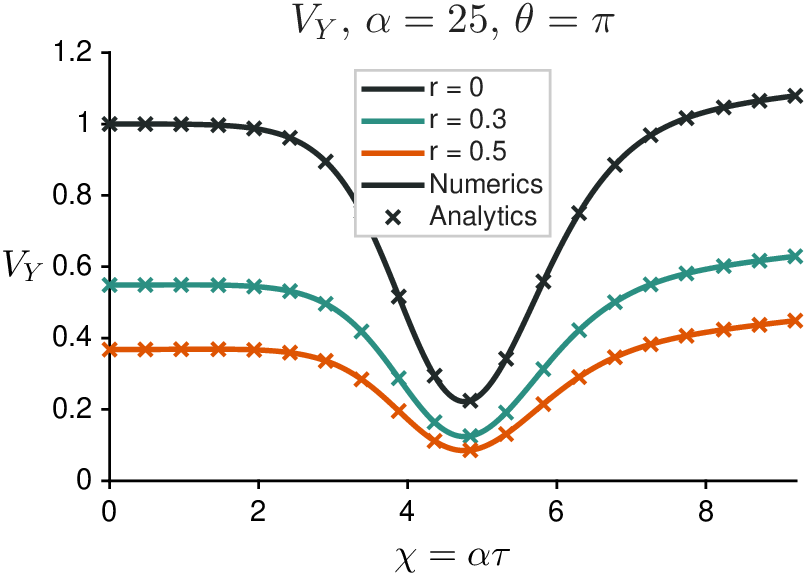}
    \caption{}
    \label{subfig:squeezing_VY}
\end{subfigure}
\begin{subfigure}{.45\textwidth}
\centering
    \includegraphics[width=0.95\linewidth]{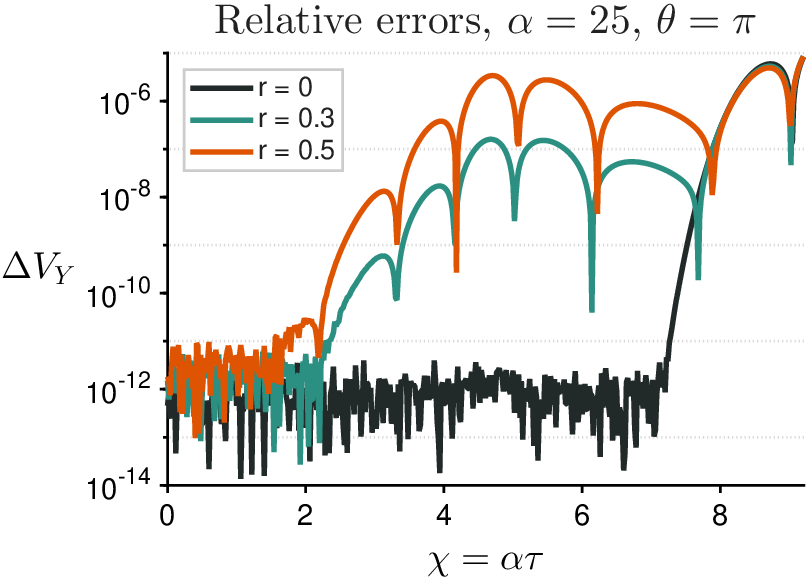}
    \caption{}
    \label{subfig:squeezing_VY_error}
\end{subfigure}
    \caption{Numerics vs. analytics comparison for $V_Y$, now with various degrees of squeezing $r$.}
    \label{fig:squeezing_VY}
\end{figure}

\begin{figure}[H]
\centering
\begin{subfigure}{.45\textwidth}
    \centering
    \includegraphics[width=0.95\linewidth]{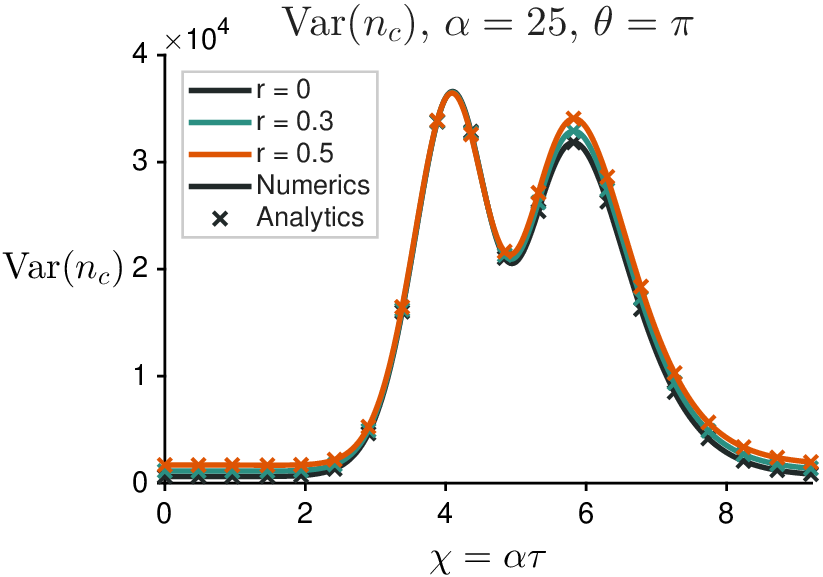}
    \caption{}
    \label{subfig:squeezing_varNc}
\end{subfigure}
\begin{subfigure}{.45\textwidth}
\centering
    \includegraphics[width=0.95\linewidth]{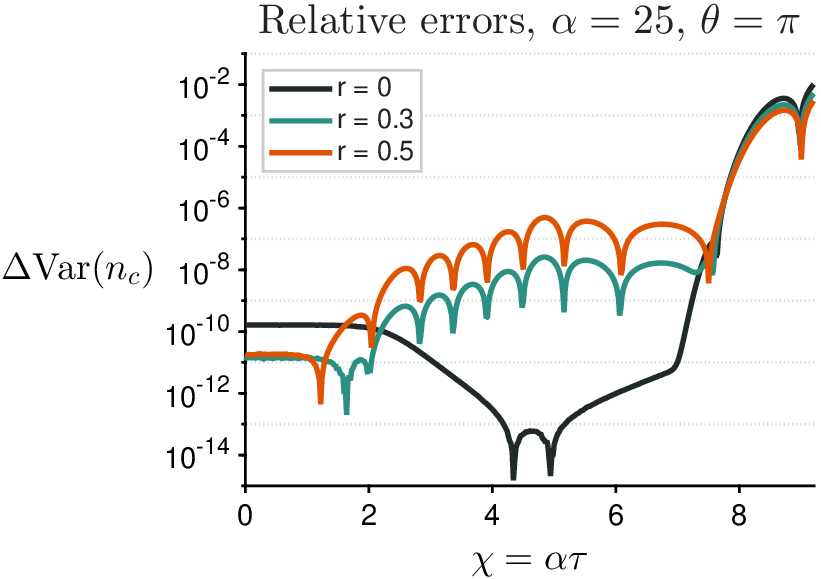}
    \caption{}
    \label{subfig:squeezing_varNc_error}
\end{subfigure}
    \caption{Numerics vs. analytics comparison for Var$(n_c)$, now with various degrees of squeezing $r$.}
    \label{fig:squeezing_varNc}
\end{figure}

It is possible to include higher squeezings. We found that the maximum error scales with both $r$ and $\alpha$, and the error remains small if the ratio 
\begin{align}
    \varepsilon = \frac{e^r\log \alpha}{\alpha} \ll 1.
\end{align}
The precise scaling is shown in Figure \ref{fig:error_squeeze} below. So increasing $\alpha$ would slowly reduce the errors; while increasing squeezing would increase the error exponentially. The empirical fit gives Error $\approx0.0129\varepsilon^{9.70}$, meaning at $r=2$, for example, we need $\alpha \approx 35$ for $10^{-3}$ error, and $\alpha\approx200$ for $10^{-8}$ error.

\begin{figure}[H]
\centering
\begin{subfigure}{.45\textwidth}
\centering
    \includegraphics[width=0.95\linewidth]{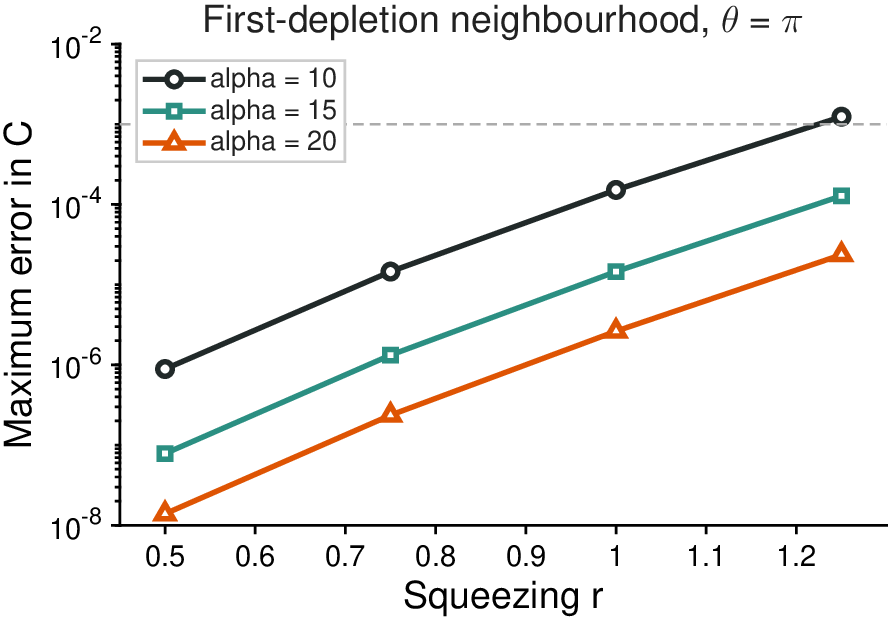}
    \caption{}
    \label{subfig:error_vs_r}
\end{subfigure}
\begin{subfigure}{.45\textwidth}
    \centering
    \includegraphics[width=0.95\linewidth]{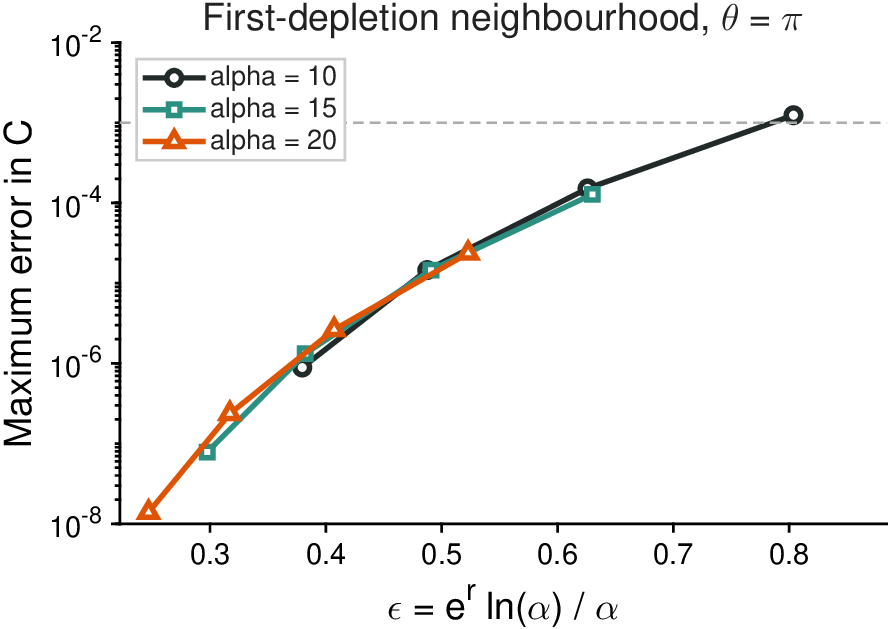}
    \caption{}
    \label{subfig:error_vs_epsilon}
\end{subfigure}
    \caption{Error scaling for squeezed states. The most natural variable is the ratio $e^r \log \alpha / \alpha$.}
    \label{fig:error_squeeze}
\end{figure}

\subsection{Error scaling—general} \label{subsec:error_scaling}
There is a clear trend as we increase $\alpha$: If we focus on the first depletion window, we see that the leading-order results definitively get more and more accurate, and this trend holds across all observables, as shown in Table \ref{tab:alpha_full_normalized_errors} below. Furthermore, the error clearly scales with $\alpha$ as
\begin{align}
    E = A\alpha^{-p}, \qquad \Rightarrow \qquad \log E = \log A - p\log \alpha,
\end{align}
where $p \approx 2$ from the log-log fit. This implies absolute error would go below $10^{-8}$ at around $\alpha\approx10^4$ or so.

\begin{table}[H]
\centering
\small
\begin{tabular}{rrrrrr}
\toprule
$\alpha$ & $E_C$ & $E_P$ & $E_U$ & $E_S$ & $E_T$ \\
\midrule
15 & $7.1885\times10^{-3}$ & $7.7907\times10^{-3}$ & $1.1014\times10^{-2}$ & $2.7629\times10^{-3}$ & $4.4203\times10^{-3}$ \\
20 & $4.495\times10^{-3}$ & $4.9015\times10^{-3}$ & $6.7139\times10^{-3}$ & $1.4574\times10^{-3}$ & $2.2905\times10^{-3}$ \\
25 & $3.1122\times10^{-3}$ & $3.4117\times10^{-3}$ & $4.5594\times10^{-3}$ & $9.0512\times10^{-4}$ & $1.3935\times10^{-3}$ \\
30 & $2.2997\times10^{-3}$ & $2.5333\times10^{-3}$ & $3.3181\times10^{-3}$ & $6.2643\times10^{-4}$ & $9.4282\times10^{-4}$ \\
35 & $1.7782\times10^{-3}$ & $1.9673\times10^{-3}$ & $2.5335\times10^{-3}$ & $4.6748\times10^{-4}$ & $6.8874\times10^{-4}$ \\
40 & $1.4217\times10^{-3}$ & $1.5792\times10^{-3}$ & $2.0039\times10^{-3}$ & $3.6793\times10^{-4}$ & $5.3242\times10^{-4}$ \\
45 & $1.1663\times10^{-3}$ & $1.2999\times10^{-3}$ & $1.6286\times10^{-3}$ & $3.0085\times10^{-4}$ & $4.2924\times10^{-4}$ \\
50 & $9.7631\times10^{-4}$ & $1.0919\times10^{-3}$ & $1.3523\times10^{-3}$ & $2.5295\times10^{-4}$ & $3.571\times10^{-4}$ \\
\bottomrule
\end{tabular}
\caption{Maximum absolute errors of the normalized moments over the first depletion $0\leqslant\chi\leqslant2\log(4\alpha)$, for the moments $C, P, U, S, T$. The expressions for $C,P,U$ are leading order; those for $S,T$ include the first correction. Errors are $E_M=\max_\chi|M_{\mathrm{anal}}-M_{\mathrm{num}}|$.}
\label{tab:alpha_full_normalized_errors}
\end{table}

\begin{figure}[H]
\centering
\begin{subfigure}{.45\textwidth}
    \centering
    \includegraphics[width=0.95\linewidth]{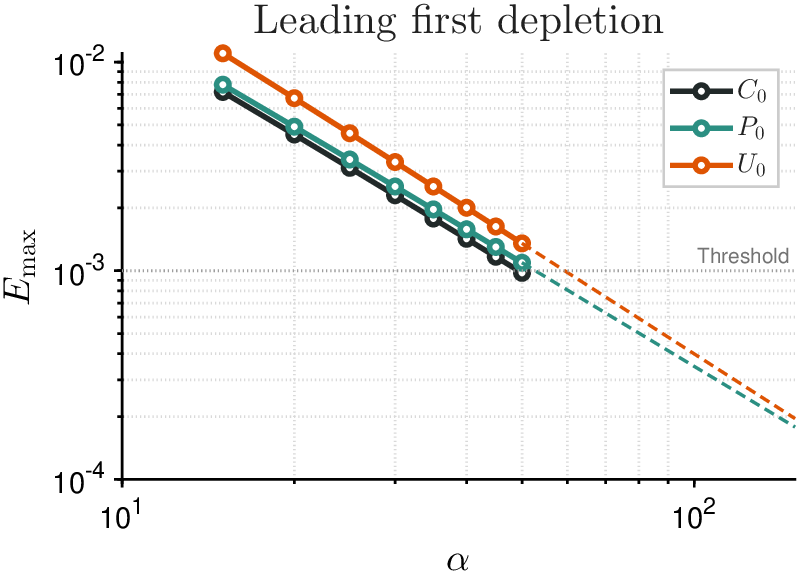}
    \caption{}
    \label{subfig:leading_max_error}
\end{subfigure}
\begin{subfigure}{.45\textwidth}
\centering
    \includegraphics[width=0.95\linewidth]{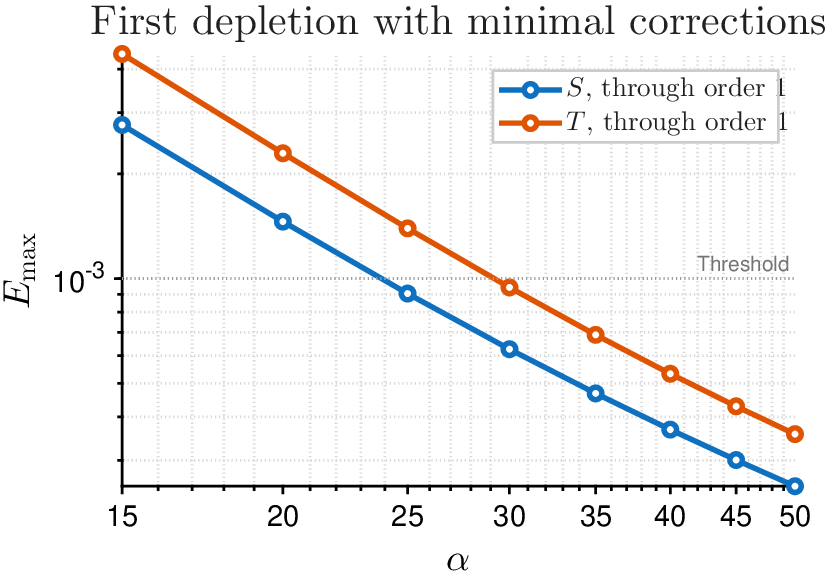}
    \caption{}
    \label{subfig:corrected_max_error}
\end{subfigure}
    \caption{Max error during first depletion $0\leqslant\chi\leqslant2\log(4\alpha)$. (a) For the leading profiles $C_0, P_0, U_0$. (b) For $S$ and $T$ through the first correction order.}
    \label{fig:max_error_scaling}
\end{figure}

However, after we apply all higher-order corrections (for example, through to the 19th order for $C$), the results do not necessarily get more accurate (it went from order $10^{-13}$ to $10^{-12}$ as we increase $\alpha$ to 75). 

It is unclear if the errors in the maximally-corrected case approach a limit—and due to hardware limit, it is not feasible to calculate numerical benchmark for arbitrary large $\alpha$. But the behavior is consistent with what we understand about asymptotic series—increasing order does not necessarily improve precision, and there is typically an optimal truncation point, beyond which the higher-order corrections may even diverge \cite{BENDER_1999}. 

For readers who seek extreme precision, there are two simple solutions: One could decrease the correction order to test convergence; another way is simply perform a Pad\'e approximation on the higher-order corrections, which should both force convergence and increase accuracy. Pad\'e approximants are discussed in Section \ref{subsec:borel_pade} and \ref{subsec:pade}.

\pagebreak
\section{Start of detailed derivations: asymptotic perturbation theory} \label{sec:perturbation}
\subsection{BCH expansion}
\textit{System setup. BCH expansion. Naive summation fails as pump starts to deplete. Borel-Pad\'e accurately describes the first depletion but fails at the first return.}

Recall we will work with $U(\tau) = \exp(\tau(a^\dag b^\dag c - abc^\dag))$, where $\chi = \alpha\tau$ is the scaled time. The $c$ mode starts in coherent state $|\alpha\rangle$, $\alpha>0$ and large, and $a, b$ are in vacuum states $|0\rangle$. 

We shall focus on $\langle c \rangle$. Using the BCH expansion, we can expand $c$ to arbitrary order \cite{SAKURAI_2011}, and we find
\begin{subequations}
\begin{align}
    \langle c \rangle & = \alpha + \frac{\tau^2}{2!}(-\alpha) + \frac{\tau^4}{4!}(-4\alpha^3+\alpha) + \frac{\tau^6}{6!}(-16\alpha^5+60\alpha^3-\alpha) \\
    & +\frac{\tau^8}{8!}(-64\alpha^7+1552\alpha^5-744\alpha^3+\alpha) \\
    & + \frac{\tau^{10}}{10!}(-256\alpha^9+29632\alpha^7-92848\alpha^5+8920\alpha^3-\alpha)+...
\end{align}
\end{subequations}
Instead of grouping terms by powers of $\tau$, we can group them by powers of $1/\alpha$ instead. Recall $\chi=\alpha\tau$, so $\tau = \chi/\alpha$, and we can re-arrange the terms as
\begin{subequations}
\begin{align}
    \langle c\rangle & = \alpha + \frac{1}{\alpha}(-\frac{\chi^2}{2!}-\frac{4\chi^4}{4!}-\frac{16\chi^6}{6!}-\frac{64\chi^8}{8!}-...) \\
    & + \frac{1}{\alpha^3} (\frac{\chi^4}{4!}+\frac{60\chi^6}{6!}+\frac{1552\chi^8}{8!}+\frac{29632\chi^{10}}{10!}+...) \\
    & + \frac{1}{\alpha^5}(-\frac{\chi^6}{6!}-\frac{744\chi^8}{8!}-\frac{92848\chi^{10}}{10!}-...) + \dots
\end{align}
\end{subequations}
Let's look at the first infinite sum (the $1/\alpha^1$ terms). The coefficients $\{1, 4, 16, 64...\}$ are clearly $4^n$, and the sum evaluates to
\begin{align}
    G_1(\chi) \equiv \sum_{n=0}^\infty p_1(n)\frac{\chi^{2n+2}}{(2n+2)!} = \frac{1}{4}(-1+\cosh(2\chi)),
\end{align}
with $p_1(n) = 4^n$. 

It turns out all the higher-order sums evidently also have closed form expressions. Let's go to the next order, $1/\alpha^3$: Once we know the coefficients $\{1,60,1552,29632...\}$, we can find the function that generates them---in this case it's an exponential polynomial, call it $p_2(n)$:
\begin{align}
    p_2(n) = 2^{3+4n}-4^n(2n^2+8n+7), \label{eqn:p2}
\end{align}
then the $(1/\alpha)^3$ sum evaluates to
\begin{align}
    G_2(\chi) = \sum_{n=0}^\infty p_2(n) \frac{\chi^{2n+4}}{(2n+4)!} = \frac{-3-4\chi^2+2\cosh(2\chi)-4\chi^2\cosh(2\chi)+\cosh(4\chi)-2\chi\sinh(2\chi)}{32}.
\end{align}
The result of our previous work \cite{XING_2023}. 

For the next order, we have
\begin{subequations}
\begin{align}
    G_3(\chi) &= \sum_{n=0} p_3(n) \frac{\chi^{2n+6}}{(2n+6)!}  = \frac{1}{1536} \Big(-108-240 \chi^2+3 (45-52 \chi^2+16 \chi^4) \cosh(2 \chi) + \\
    & \quad \dots -12 (3+8 \chi^2) \cosh(4 \chi) + 9 \cosh(6 \chi)+30 \chi \sinh(2 \chi)+16 \chi^3 \sinh(2 \chi)+72 \chi \sinh(4 \chi)\Big),
\end{align}
\end{subequations}
where the coefficients $\{1, 744, 92848...\}$ are found to be generated by the polynomial
\begin{align}
    p_3(n) = \frac{1}{3}\times 2^{-3+2n}(375-27\times 4^{4+n}+9^{4+n}+16 n (95+87 n+28 n^2+3 n^3-3 \times 2^{5+2 n} (4+n))). \label{eqn:p3}
\end{align}

\subsection{(Optional) fast algorithm for calculating BCH coefficients}

\textit{We give a very fast algorithm to calculate BCH expansion to thousands of orders.}

Recall BCH expansion: 
\begin{align}
    c(t) = e^{-\tau K} c e^{\tau K} = c + \tau[c,K] + \frac{\tau^2}{2!}[[c,K],K]+\frac{\tau^3}{3!}[[[c,K],K],K]+\dots
\end{align}
\indent In our current work, we were only able to perform so much analysis thanks to the large number of BCH coefficients we collected, which allowed us to extract closed-form expressions in terms of cosh and sinh to a high order, which then enabled all the other harmonic analysis. 

To find each closed form expression $G_n$ (order $\epsilon^n)$, it turns out we need about $n^2$ coefficients (this is to fit polynomial expressions like equation \eqref{eqn:p2} and \eqref{eqn:p3}), obtained by reaching $2n^2$ BCH order. For example, to discover the tower of exceptional terms we needed to calculate at least up to $G_{15}$, which required us to go to about 450 BCH orders. And everything in our current work is verified through $\epsilon^{50}G_{50}$, meaning order $\sim5,000$ in BCH expansion. 

\begin{figure}[h!]
  \centering
  \resizebox{\linewidth}{!}{\input{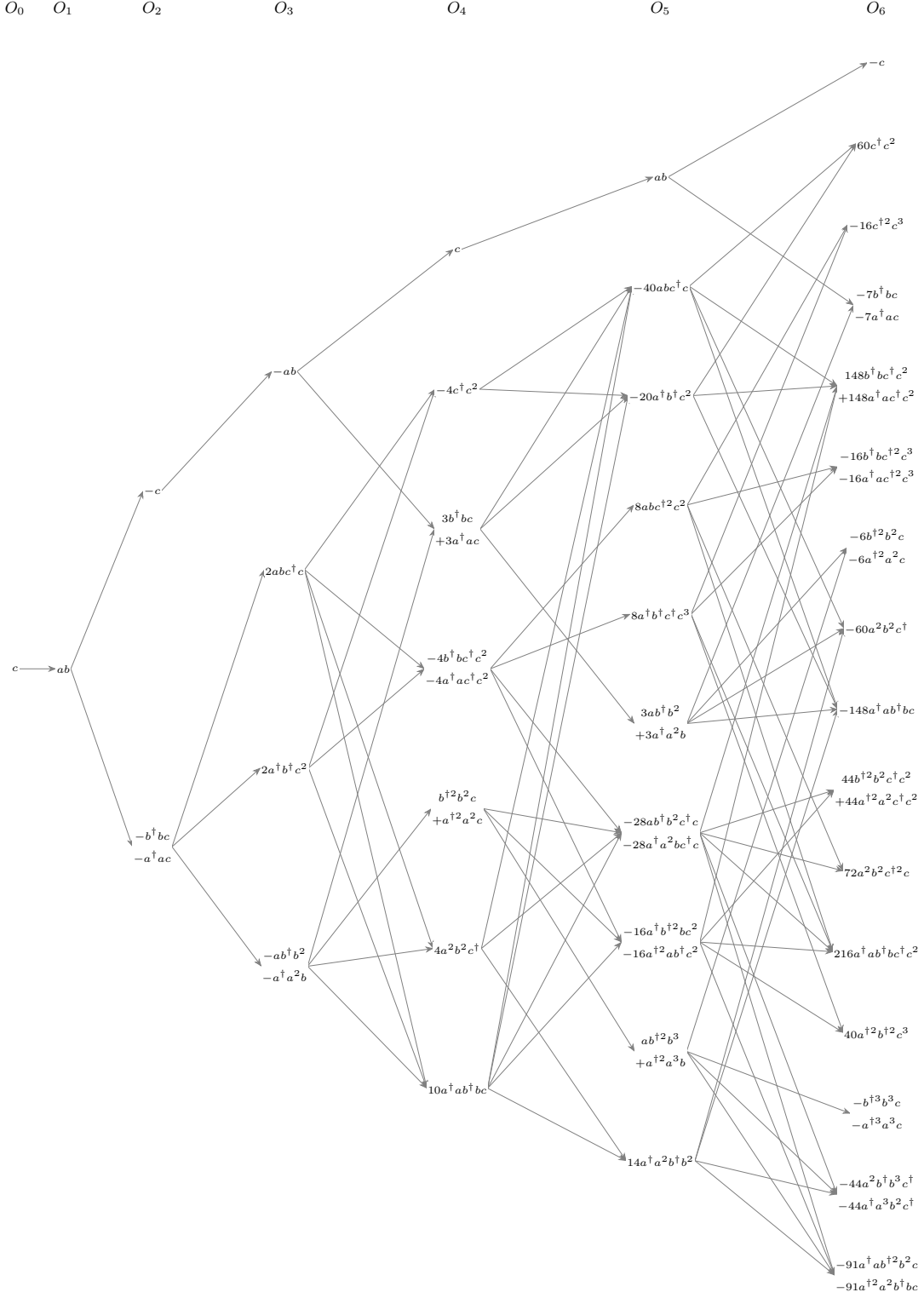}}
  \caption{Nested-commutator cascade of the pump operator.}
  \label{fig:cascade}
\end{figure}

Typical evaluation of BCH commutators using symbolic manipulation would not get us past order 30. This is because the number of operators cascade and quickly become untractable—a visual representation of the first few orders of BCH expansions is shown in Figure \ref{fig:cascade} on the next page, which gives an idea of what the cascade looks like.

Here we give a very fast way to generate these coefficients. It allowed us to get to $\epsilon^{50}$/BCH order 5,000 in any expectation value in a couple of hours of computation on a personal laptop.

First of all, all commutators we encounter in the BCH expansion fall into two possibilities: commuting with $a^\dag b^\dag c$:
\begin{subequations}
\begin{align}
    [(a^\dag)^{p_1} a^{p_2} (b^\dag)^{p_3} b^{p_4} (c^\dag)^{p_5} c^{p_6}, a^\dag b^\dag c] & = p_2p_4 (a^\dag)^{p_1} a^{p_2-1} (b^\dag)^{p_3} b^{p_4-1} (c^\dag)^{p_5} c^{p_6+1} \\
    & + p_2 (a^\dag)^{p_1} a^{p_2-1} (b^\dag)^{p_3+1} b^{p_4} (c^\dag)^{p_5} c^{p_6+1} \\
    & + p_4 (a^\dag)^{p_1+1} a^{p_2} (b^\dag)^{p_3} b^{p_4-1} (c^\dag)^{p_5} c^{p_6+1} \\
    & - p_5 (a^\dag)^{p_1+1} a^{p_2} (b^\dag)^{p_3+1} b^{p_4} (c^\dag)^{p_5-1} c^{p_6}, 
\end{align}
\end{subequations}
or commuting with $abc^\dag$:
\begin{subequations}
\begin{align}
    [(a^\dag)^{p_1} a^{p_2} (b^\dag)^{p_3} b^{p_4} (c^\dag)^{p_5} c^{p_6}, abc^\dag] & = -(p_1p_3) (a^\dag)^{p_1-1} a^{p_2} (b^\dag)^{p_3-1} b^{p_4} (c^\dag)^{p_5+1} c^{p_6} \\
    & - p_1 (a^\dag)^{p_1-1} a^{p_2} (b^\dag)^{p_3} b^{p_4+1} (c^\dag)^{p_5+1} c^{p_6} \\
    & - p_3 (a^\dag)^{p_1} a^{p_2+1} (b^\dag)^{p_3-1} b^{p_4} (c^\dag)^{p_5+1} c^{p_6} \\
    & + p_6 (a^\dag)^{p_1} a^{p_2+1} (b^\dag)^{p_3} b^{p_4+1} (c^\dag)^{p_5} c^{p_6-1}.
\end{align}
\end{subequations}
So instead of evaluating commutators, we can use an algebraic dictionary approach. Let a term $(a^\dag)^{p_1} a^{p_2} (b^\dag)^{p_3} b^{p_4} (c^\dag)^{p_5} c^{p_6}$ be represented as a tuple $(p_1, p_2, p_3, p_4, p_5, p_6)$. Then under the action of the commutator with $a^\dag b^\dag c$:
\begin{subequations}
\begin{align}
    \mqty(p_1\\ p_2\\ p_3\\ p_4\\ p_5\\ p_6) & \rightarrow (p_2p_4) \mqty(p_1\quad \ \ \\ p_2-1\\ p_3\quad \ \ \\ p_4-1\\ p_5\quad \ \ \\ p_6+1) + p_2\mqty(p_1\quad \ \ \\ p_2-1\\ p_3+1\\ p_4\quad \ \ \\ p_5\quad \ \ \\ p_6+1) + p_4\mqty(p_1+1\\ p_2\quad \ \ \\p_3\quad \ \ \\ p_4-1\\ p_5\quad \ \ \\ p_6+1) - p_5\mqty(p_1+1\\ p_2\quad \ \ \\ p_3+1\\ p_4\quad \ \ \\ p_5-1\\ p_6\quad \ \ ),
\end{align}
\end{subequations}
under commutation with $abc^\dag$: 
\begin{subequations}
\begin{align}
    \mqty(p_1\\ p_2\\ p_3\\ p_4\\ p_5\\ p_6) & \rightarrow - (p_1p_3) \mqty(p_1-1\\p_2\quad \ \ \\p_3-1\\p_4\quad \ \ \\p_5+1\\p_6) - p_1\mqty(p_1-1\\p_2\quad \ \ \\p_3\quad \ \ \\p_4+1\\p_5+1\\p_6\quad \ \ ) - p_3\mqty(p_1\quad \ \ \\p_2+1\\p_3-1\\p_4\quad \ \ \\p_5+1\\p_6\quad \ \ ) + p_6\mqty(p_1\quad \ \ \\ p_2+1\\ p_3\quad \ \ \\ p_4+1\\ p_5\quad \ \ \\ p_6-1).
\end{align}
\end{subequations}
This map allows us to evaluate commutators much more efficiently than the symbolic approach (the speed can be further improved by using $su(1,1)$ representation \cite{LEE_2010} for $a$ and $b$ modes, so it's a 4-tuple instead of a 6-tuple).

\begin{figure}[h!]
  \centering
  \resizebox{\linewidth}{!}{\input{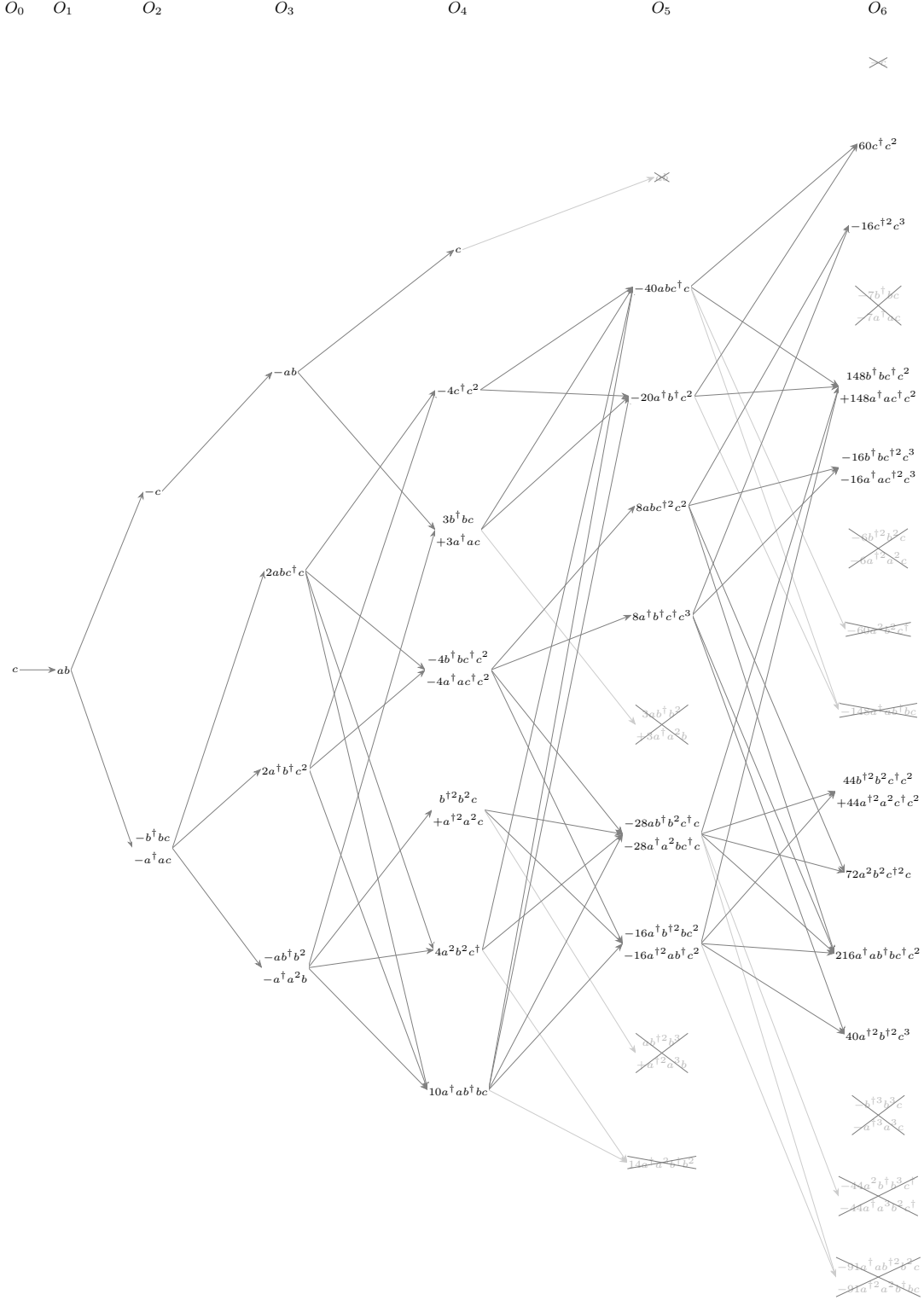}}
  \caption{Crossed-out terms cannot contribute to the $\alpha^{-3}$ sector. To $O(\alpha^{-3})$, we may neglect them and restrict to a much smaller subspace of operators. In this case the cascade stops, and the series become manageable.}
  \label{fig:cascade_chi3}
\end{figure}

The second and more important simplification is this: normally when doing BCH expansion, the number of operators cascade, as shown in Figure \ref{fig:cascade}. However, the key finding is that if we take the vacuum expectation value of $a,b$, while restricting to a certain $(1/\alpha)^N$ order, the cascade stops! For example, the only terms that contribute to $\expval*{c}$ at $1/\alpha^3$ occur at even powers of $\tau$, and have the form 
\begin{align}
    p_2(n-2) \frac{\tau^{2n}}{(2n)!}(c^\dag)^{n-2} c^{n-1},
\end{align}
because if we take the vev of such terms, we have
\begin{align}
    \frac{\expval*{(c^\dag)^{n-2} c^{n-1}}}{\alpha^{2n}} = \frac{\alpha^{2n-3}}{\alpha^{2n}}  = \frac{1}{\alpha^3}.
\end{align}
To find its coefficients, we need to look at who contributes to this term. Only 2 terms contribute to $(c^\dag)^{n-2} c^{n-1}$ (or $(0,0,0,0,n-2,n-1)$ in the tuple representation), occurring at odd powers of $\tau$: 
\begin{itemize}
    \item $(0,1,0,1,n-2,n-2)\rightarrow(0,0,0,0,n-2,n-1)$ via contraction with $a^\dag b^\dag c$ in the Hamiltonian,
    \item $(1,0,1,0,n-3,n-1)\rightarrow(0,0,0,0,n-2,n-1)$ via contraction with $abc^\dag$ in the Hamiltonian.
\end{itemize}
\indent Then there are 6 terms (at even powers of $\tau$) that contribute to these two terms (shift $n\rightarrow n+1$):
\begin{itemize}
    \item $(0,0,0,0,n-1,n) \mapsto n \times (0,1,0,1,n-1,n-1) + (n-1)\times (1,0,1,0,n-2,n)$, 
    \item $ (1,1,0,0,n-2,n-1) \mapsto (0,1,0,1,n-1,n-1) + (1,0,1,0,n-2,n),$
    \item $(0,0,1,1,n-2,n-1) \mapsto (0,1,0,1,n-1,n-1) + (1,0,1,0,n-2,n)$,
    \item $(1,1,1,1,n-2,n-1) \mapsto (0,1,0,1,n-1,n-1) + (1,0,1,0,n-2,n)$,
    \item $(0,2,0,2,n-1,n-2) \mapsto 4 (0,1,0,1,n-1,n-1)$,
    \item $(2,0,2,0,n-3,n) \mapsto 4(1,0,1,0,n-2,n)$.
\end{itemize}
Note this can be represented by a matrix multiplication 
\begin{align}
    \left(\begin{smallmatrix}(0,1,0,1,n-1,n-1) \\ (1,0,1,0,n-2,n))\end{smallmatrix}\right) = \left(\begin{smallmatrix}
    n & 1 & 1 & 1 & 4 & 0 \\ n-1 & 1 & 1 & 1 & 0 & 4
\end{smallmatrix}\right) \left( \begin{smallmatrix}
    (0,0,0,0,n-1,n) \\ (1,1,0,0,n-2,n-1) \\ (0,0,1,1,n-2,n-2) \\ (1,1,1,1,n-2,n-1)\\(0,2,0,2,n-1,n-2)\\(2,0,2,0,n-3,n)
\end{smallmatrix}\right).
\end{align}
\indent Then there are 7 terms (at odd powers of $\tau$) that contribute to these 6 terms... It seems like it starts to cascade again, but crucially, the cascade stops at some point. A visual representation is shown in Figure \ref{fig:cascade_chi3} on the next page, where we see many terms can be neglected as they never contribute to $\expval*{c}$ at the $1/\alpha^{3}$ level.

In the end, there are only 19 relevant terms, with a clear linear map (matrix) between them. So instead of evaluating commutators, if we want to find the $(1/\alpha)^3$ coefficients to $k$th order, all we need is to find the relevant matrix, then do the matrix multiplication $k$ times. 

This trend appears to generalize to arbitrary order. The number of relevant terms increases (for example, at order $(1/\alpha)^5$, there are 46 relevant terms), but there is always a simple linear map describing their relations. The process of finding the linear map and doing matrix multiplication can be fully automated. On a personal computer, it takes seconds to get to order $(1/\alpha)^{30}$, minutes to get to $(1/\alpha)^{60}$, and hours to get to $(1/\alpha)^{120}$. Finally, all VEVs can be calculated this way, not just $\expval{c}$. This means that, even without further analysis, one can simply take the calculated BCH coefficients, apply Borel-Pad\'e summation, and get a highly accurate description of arbitrary correlation function, at least through to first depletion. This is treated in the next section. 

\subsection{Na\"ive summation vs. Borel-Pad\'e} \label{subsec:borel_pade}

\textit{A Pad\'e approximation to the Borel transform is immediately useful for the first-depletion regime, without requiring additional harmonic analysis.}

\indent Back to our observable $\expval{c}$, define the scaled coherent amplitude as $C = \expval{c}/\alpha$, $-1\leqslant C\leqslant1$, then the ``perturbation parameter'' naturally emerges as 
\begin{align}
    \epsilon = \frac{1}{\alpha^2},
\end{align}
and we have
\begin{align}
    C(\alpha,\chi) & = \frac{\expval{c}}{\alpha} \sim \sum_{n=0}^\infty (-1)^n \frac{G_{n}(\chi)}{\alpha^{2n}} \equiv \sum_{n=0}^\infty C_n(\chi)\epsilon^n,
\end{align}
where the $\sim$ sign indicates this is a formal power series and not to be summed naively—if we naively sum the terms as they are, we find good agreement at shorter times, before $C$ quickly diverges to $\pm\infty$ as the pump starts to deplete at scaled time $\chi\sim\log\alpha$ (to be derived later).

\begin{figure}[H]
    \centering
    \includegraphics[width=0.5\linewidth]{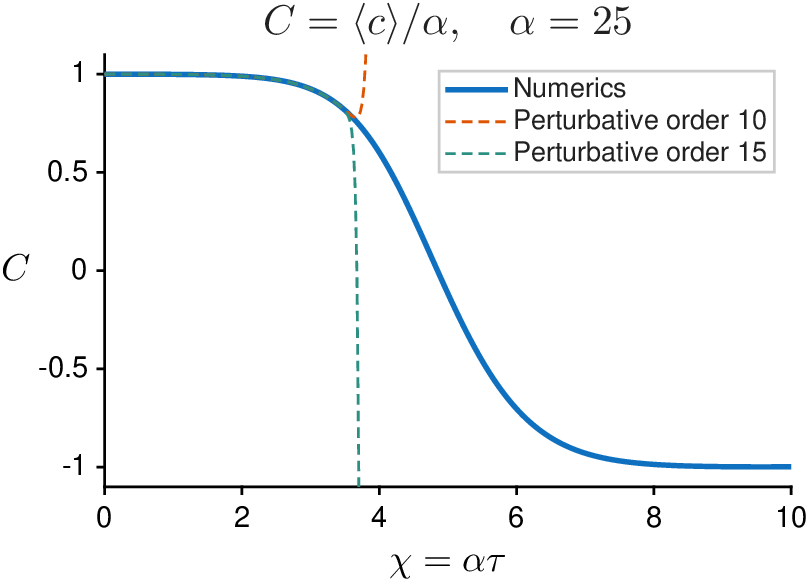}
    \caption{Naively adding perturbative terms does not give good results. In fact, the result gets worse the more orders we add. (For example, if we include corrections through to order 15, the result diverges faster than through to order 10.)}
    \label{fig:perturbative}
\end{figure}

This behavior is the hallmark of a divergent series that grows factorially as $n!$—meaning no matter how small $\epsilon$ is (how large $\alpha$ is), $n\epsilon$ becomes large at some point, and the infinite series would diverge \cite{BENDER_1999}. Such divergent series are ubiquitous in quantum mechanical perturbative expansions. The typical way to deal with factorial growth is via a Borel summation \cite{ANICETO_2019}: Given a divergent formal series
\begin{align}
    C(\epsilon,\chi) \sim \sum_{n=0}^\infty f_n(\chi) \epsilon^n, 
\end{align}
for some small perturbation parameter $\epsilon$, and $f_n(\chi)\sim n!$ growing factorially, we can define the Borel transformation: 
\begin{align}
    \hat C (\epsilon t,\chi) = \sum_{n=0}^\infty \frac{f_n(\chi)\epsilon^n}{n!}t^n. \label{eqn:borel_transform}
\end{align}
Now define Borel summation to be
\begin{align}
    \mathcal{C}(\epsilon,\chi) = \int_0^\infty e^{-t} \hat C(\epsilon t,\chi) \ dt.\label{eqn:borel_sum}
\end{align}
One can easily verify the Borel summation of a Borel transform returns the original formal series $C$ (by exchanging the sum and the integral). The idea of a Borel summation is to remove the factorial growth by first dividing by $n!$, then later ``putting it back'' via the integral which evaluates to $\Gamma(n+1)$. This procedure can often eliminate the divergent behavior, as we shall see later. 

Generally, if a series is growing as $(m!)^s$ (called ``Gevrey-$s$'') for some integer $s$, we simply do a higher-order Borel sum. While it is quite rare in physics to have Gevrey class higher than 1, we shall in fact encounter a $(m!)^3$ divergence later. In such cases, we manage the series by dividing it by $(m!)^3$ to obtain the Borel transformation, and then putting the transformed expression into a triple integral as the Borel summation \cite{CAVALCANTI_2021, CREW_2024}. 

So now we know our series is factorially divergent. But what term exactly is contributing to the divergence? We need to expand our functions $G_n$ into a more elementary form. 

If we are only interested in the first depletion profile, we can in fact immediately get a highly accurate answer without any additional analysis, simply by performing a Pad\'e approximation (a tool that inexplicably but reliably works, see \cite[Sec.~8.3]{BENDER_1999}. A brief tutorial is included in the next Section). This allows us to achieve extreme precision, shown in Table \ref{fig:borel_pade} below.

\begin{figure}[H]
\centering
\begin{subfigure}{.45\textwidth}
    \centering
\includegraphics[width=0.95\linewidth]{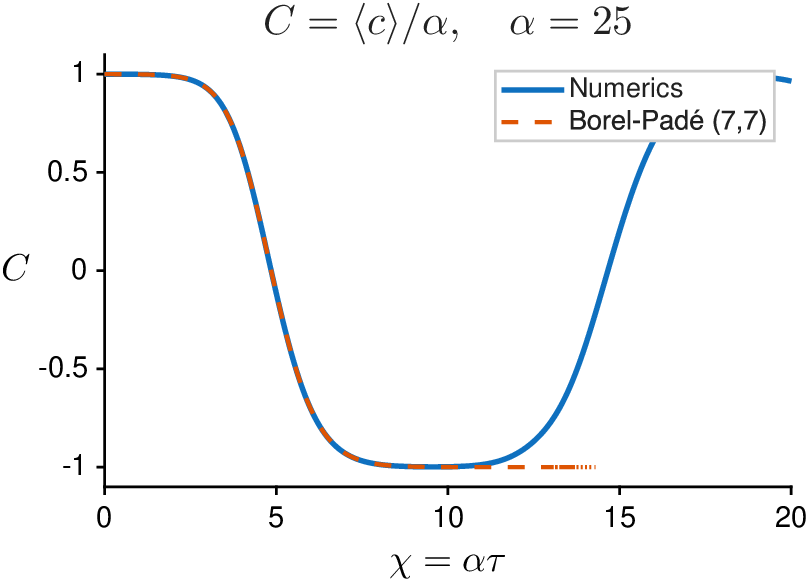}
    \caption{}
    \label{subfig:borelpade}
\end{subfigure}
\begin{subfigure}{.45\textwidth}
\centering
\includegraphics[width=0.95\linewidth]{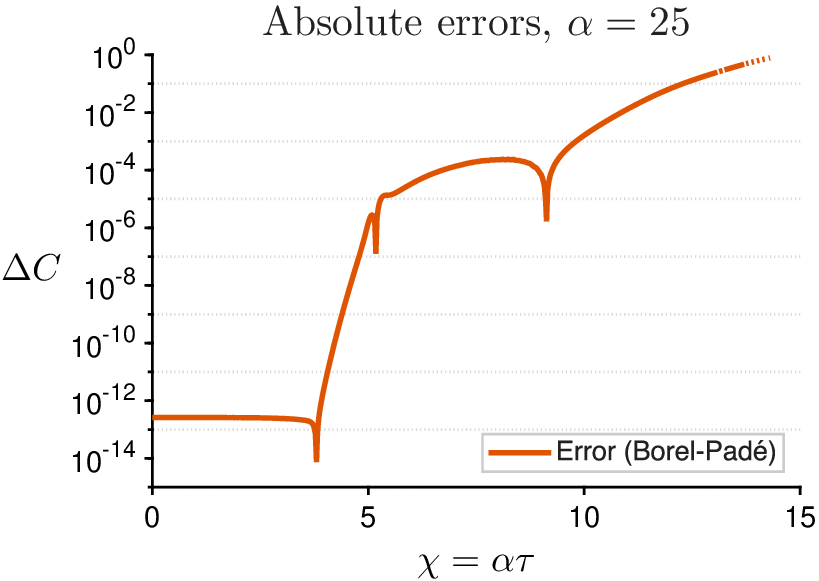}
    \caption{}
    \label{subfig:borel_pade_error}
\end{subfigure}
    \caption{Borel-Pad\'e summation immediately improves the accuracy for the first depletion. Here we used Pad\'e approximants summed up to order 15, of type [7,7], defined in Section \ref{subsec:pade} below. Note our definition of ``depletion'' does not mean the pump has lost its coherence; it instead means the coherence has crossed sign and reached $C\approx -1$. As $C$ changes from $+1$ to $-1$, the pump population already has depleted and then recovered.}
    \label{fig:borel_pade}
\end{figure}

However, while the Borel-Pad\'e approach is able to stabilize the divergence and accurately simulate the first full depletion to extreme precision—it does not provide a physical understanding, and utterly fails to simulate the ``return''—or back-conversion—of the coherent amplitude, no matter how many orders we sum up to. Later we shall see why: There are different terms in our infinite series that contribute to different collapse/return cycles. However, Borel-Pad\'e does not distinguish them and tame all of their growths. In reality, each cycle should be resummed separately. We shall see a first hint of this in Section \ref{subsec:first_return_hint}. 

Nevertheless, if we're only interested in the short time regime ($\chi \lesssim 2\log4\alpha$), then applying Pad\'e approximants is still a great approach that provides incredible accuracy. Also, this method works for any moment, not just $\expval*{c}$. While the computation time is much longer than the analytical expressions we will derive, it is still faster than traditional numerical diagonalization, since the cost does not scale with $\alpha$. So many readers may even choose to stop here. But in the spirit of finding an analytic solution, we will press on.

\subsection{(Optional) Padé approximants} \label{subsec:pade}

For the normalized pump amplitude, write
\begin{align}
    C(\epsilon,\chi)=\frac{\langle c\rangle}{\alpha} \sim\sum_{n=0}^{\infty}a_n(\chi)\epsilon^n, \qquad \epsilon=\alpha^{-2},
\end{align}
where
\begin{align}
    a_0=1,\qquad a_n(\chi)=(-1)^nG_{2n-1}(\chi).
\end{align}
Padé approximation can then be constructed at each fixed $\chi$, using the original coefficient functions $G_{2n-1}$ involving cosh and sinh. 

A Padé approximant of type $[L,M]$ is defined as
\begin{align}
    R_{L,M}(\epsilon)=
\frac{p_0+p_1\epsilon+\cdots+p_L\epsilon^L}
{1+q_1\epsilon+\cdots+q_M\epsilon^M},
\end{align}
where the coefficients are chosen so that the expansion of $R_{L,M}$ agrees with the supplied series through $\epsilon^{L+M}$. So $L+M$ correction coefficients together with $a_0$ are required.

Here's an explicit example: given
\begin{align}
    C=1+a_1\epsilon+a_2\epsilon^2+\cdots,
\end{align}
set
\begin{align}
    R_{1,1}=\frac{1+p_1\epsilon}{1+q_1\epsilon},
\end{align}
expanding the denominator gives
\begin{align}
    R_{1,1}
=1+(p_1-q_1)\epsilon+(q_1^2-p_1q_1)\epsilon^2+\cdots.
\end{align}
We can then match coefficients to find, for $a_1\neq0$,
\begin{align}
    q_1=-\frac{a_2}{a_1},
\qquad p_1=a_1-\frac{a_2}{a_1},
\end{align}
so we have
\begin{align}
    R_{1,1}(\epsilon)=
\frac{1+(a_1-a_2/a_1)\epsilon}
{1-(a_2/a_1)\epsilon}.
\end{align}
For the coefficients in our model, $a_1=-1, a_2=2$, we then have
\begin{align}
    R_{1,1}=\frac{1+\epsilon}{1+2\epsilon}
=1-\epsilon+2\epsilon^2-4\epsilon^3+\cdots,
\end{align}
matching the actual coefficients. The subsequent coefficients are predictions of the Pad\'e approximation. Note for factorially growing coefficients, we may instead apply Pad\'e approximants to the Borel transform.

\pagebreak
\section{Coherent amplitude calculation} \label{sec:amplitude}

\subsection{The first depletion}

\textit{By taking the leading-order harmonics, we derive equation \eqref{eqn:C0_tanh} in the form of $\mathcal C_0 = -\int \abs{h}^2 \tanh$ which accurately describes the first pump depletion.}

Let's expand our closed-form expressions containing $\sinh$ and $\cosh$ into their exponential (harmonics) forms (and we will use $x$ and $\chi$ interchangeably): 
\begin{subequations}
\begin{align}
    G_1(x) & = \frac{1}{8}e^{2x}  - \frac{1}{4} + \frac{1}{8}e^{-2x}, \\
    G_2(x) & = \frac{1}{64}e^{4x} \\
    &  + \frac{1}{32}\qty(-2x^2-x+1)e^{2x} \\
    &  + \frac{1}{32} \qty(-4x^2-3) \\
    &  + \frac{1}{32}\qty(-2x^2+x+1)e^{-2x}\\
    &  + \frac{1}{64}e^{-4x}, \\
    G_3(x) & = \frac{3}{1024}e^{6x} \\
    &  + \frac{1}{256}\qty(-8x^2+6x-3)e^{4x}\\
    &  + \frac{1}{3072}\qty(48x^4+16x^3-156x^2+30x+135) e^{2x} \\
    &  + \frac{1}{128}\qty(-20x^2-9)\\
    &  + \frac{1}{3072}\qty(48x^4-16x^3-156x^2-30x+135) e^{-2x} \\
    &  + \frac{1}{256}\qty(-8x^2-6x-3)e^{-4x}\\
    &  + \frac{3}{1024} e^{-6x}, \\
    G_4(x) & = \frac{3}{4096}e^{8x} \\
    & + \frac{3}{4096} (-18x^2+21x-13) e^{6x} \\
    & + \frac{1}{6144} (192x^4 - 352x^3 + 384x^2 - 369x + 348) e^{4x}, \\
    & + \frac{1}{12288}(-32x^6 + 16x^5 + 232x^4 - 376x^3 - 696x^2 - 426x + 45)e^{2x} \\
    & + \frac{1}{6144}(80x^4 - 1080x^2 - 633) \\
    & + \frac{1}{12288}(-32x^6 - 16x^5 + 232x^4 + 376x^3 - 696x^2 + 426x + 45)e^{-2x} \\
    & + \frac{1}{6144} (192x^4 + 352x^3 + 384x^2 + 369x + 348) e^{-4x} \\
    & + \frac{3}{4096} (-18x^2-21x-13) e^{-6x} \\
    & + \frac{3}{4096}e^{-8x}, \\
    \dots
\end{align}
\label{eqn:harmonics_full}
\end{subequations}
note the negative harmonics are related to the positive ones by $x\mapsto -x$. 

Now let's look at the leading harmonics (the largest contribution in each $G_n(x)$), their coefficients have the simple forms of
\begin{align}
    [e^{2nx}]G_{n} = \frac{2n!}{16^n} \quad \text{ for } \quad n\geqslant1.
\end{align}
This also suggests a natural perturbation parameter 
\begin{align}
    z \equiv \frac{e^{2\chi}}{16\alpha^2} = \frac{\epsilon e^{2\chi}}{16}.
\end{align}
This is a convenient normalization, but later we shall see $z$ is really a measure of time, and it sets the time scale of the transition. Furthermore, we now have an alternative way to express our formal series. Instead of in powers of $\epsilon$, we can do it in powers of $e^{2\chi}$, that is, in terms of $z$. Write
\begin{align}
    C \sim \sum_{n\geqslant0} (-1)^n \epsilon^n G_{n}(\chi) \sim \sum_{r\geqslant0} \epsilon^r C_r(z) = C_0(z) + \epsilon C_1(z) + \epsilon^2C_2(z) + \dots,
\end{align}
where $G_n(\chi)$ are the cosh/sinh expressions in terms of scaled time $\chi$, and $G_0(\chi) = 1$; whereas $C_r(z)$ is the expression in terms of the $r$-th leading harmonics in each $G_n$, and $C_0(z)\neq1$.

\indent Let's look at $C_0(z)$ explicitly:  substitute the leading harmonics into our formal power series, we have 
\begin{subequations}
\begin{align}
    G_0: && & \quad +1, \\
    -\epsilon G_1: && -\epsilon\frac{1}{8}e^{2x} &= -2z, \\
    \epsilon^2 G_2: && + \epsilon^2 \frac{1}{64}e^{4x} &= +4z^2, \\
    -\epsilon^3 G_3: && -\epsilon^3 \frac{3}{1024}e^{6x} &= -12z^3, \\
    \epsilon^4 G_4: && +\epsilon^4 \frac{3}{4096}e^{8x} &= +48z^4,
\end{align}
\end{subequations}
giving
\begin{align}
    C_0(z) = 1-2z+4z^2-12z^3+48z^4-\dots = 1+2\sum_{j=1}^\infty(-1)^j j! z^j, \label{eqn:C0_coeff}
\end{align}
which once again diverges factorially. Performing Borel summation as in equation \eqref{eqn:borel_transform} to eliminate the $j!$ growth, we have (and note the Borel transform of 1 is 1)
\begin{align}
    \hat C_0(t) = 1 + 2\sum_{j=1}^\infty (-1)^j t^j = 1- \frac{2t}{1+t} = \frac{1-t}{1+t} = \frac{2}{1+t}-1,
\end{align}
the Borel sum is
\begin{align}
    \Aboxed{\mathcal{C}_0(z) & = -1 + 2\int_0^\infty \frac{e^{- t}}{1+z t} dt = 2\times \frac{e^{1/z}}{z}E_1(1/z) - 1}, \label{eqn:C0_E1}
\end{align}
where $E_1(z) = \int_z^\infty \frac{e^{-w}}{w} dw $ is the exponential integral; and to see the last equality, make two changes of variables: let $u = 1+zt$, then let $w = u/z$. This simple formula already describes the first depletion almost perfectly. 
\begin{figure}[H]
    \centering
    \includegraphics[width=0.5\linewidth]{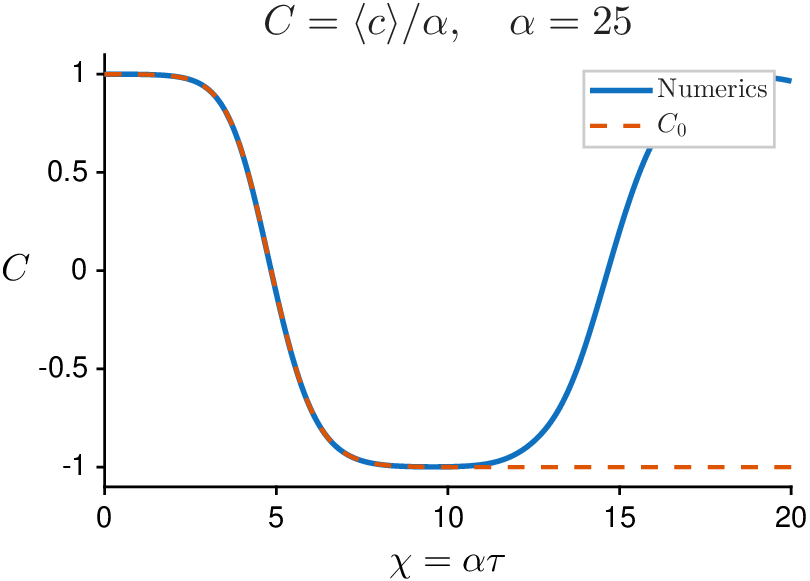}
    \caption{Borel summed $\mathcal C_0$, the simple expression accurately describes the first depletion.}
    \label{fig:c0}
\end{figure}
\indent To get a more physical picture, consider what $z = e^{2\chi}/16\alpha^2$ does physically—it's the time scale of the problem. As $z\rightarrow0$, $\chi\rightarrow-\infty$ and the pump is almost undepleted. As $z\rightarrow\infty$, the pump has reached the negative plateau. That means the depletion happens at $z\approx1$, that is, when
\begin{align}
    z = \frac{e^{2\chi}}{16\alpha^2} = e^{2\chi} e^{-\log(16\alpha^2)} = e^{2(\chi-\log4\alpha)} \rightarrow 1, \quad \Rightarrow \quad \chi\approx\log4\alpha,
\end{align}
so the first depletion happens at $\chi\approx\log4\alpha$. 

Now, first note that the Borel sum can be written as
\begin{align}
    \mathcal C_0 = \int_0^\infty e^{-t}\frac{1-zt}{1+zt} dt = -\int_0^\infty e^{-t} \tanh(\frac{1}{2}\log(zt)) \ dt, \label{eqn:C0_fraction}
\end{align}
then define the shifted depletion time $s_0 = \chi - \log(4\alpha)$, then make the change of variable $t = e^{2u}$, where $u$ also represents time, we have
\begin{align}
    \Aboxed{ \mathcal{C}_0(s_0) = -\int_{-\infty}^\infty 2e^{2u}e^{-e^{2u}} \tanh(s_0+u) \ du}, \label{eqn:C0_tanh}
\end{align}
which looks like an soliton ($\tanh$) profile convoluted with a normalized weight (note $\int du\ 2e^{2u}e^{-e^{2u}} = 1$), the result is a smoothed out soliton profile as in Figure \ref{fig:c0}. 

The most obvious improvement one can do is to take into account sub-leading harmonics $C_1(z)$ and possibly $C_2(z)$. We're going to do just that and compute the first two corrections $C_1(z)$ and $C_2(z)$ explicitly. But we shall see this computation turns out to be much more important than just marginally improving accuracy, because in fact, the first return/back-conversion profile naturally emerges starting from $C_2(z)$!

\subsection{First correction to the first depletion} \label{subsec:correction_c_1}

\textit{The sub-leading harmonics give the first correction to the first depletion, in the form of a generic polynomial and derivatives acting on the 0th order term.}

Recall $C \sim C_0 + \epsilon C_1 + \epsilon^2 C_2 + \dots $ (in terms of harmonics) and $C \sim 1-\epsilon G_1 + \epsilon^2 G_2 - \epsilon^3 G_3 \dots  $ (in terms of cosh and sinh). Extract the subleading harmonics from each $G_n(x)$ (so the $e^{0}$ term in $G_1(x)$, plus the $e^{2x}$ term in $G_2(x)$, the $e^{4x}$ term in $G_3(x)$, etc.), we have
\begin{subequations}
\begin{align}
    \epsilon C_1 & = \epsilon\qty[\frac{1}{4} + \frac{\epsilon}{32}(-2x^2-x+1)e^{2x} - \frac{\epsilon^2}{256}(-8x^2+6x-3)e^{4x} + \frac{3\epsilon^3}{4096} (-18x^2+21x-13)e^{6x} + \dots ] \\
    & = \epsilon\qty[\frac{1}{4} + \frac{1}{2}(-2x^2-x+1)z-(-8x^2+6x-3)z^2+3(-18x^2+21x-13)z^3 + \dots],
\end{align}
\end{subequations}
where we used $16z = \epsilon e^{2x}$, and the bracketed expression is $C_1(z,x)$. Compare this with 
\begin{align}
    C_0 = 1-2z+4z^2-12z^3+\dots,
\end{align}
and change the normalization of all the $z$ terms to match the coefficients of $C_0$:
\begin{align}
    C_1 = \frac{1}{4} ( 1) + \qty(\frac{1}{2}x^2+\frac{1}{4}x - \frac{1}{4})(-2z) + \qty(2x^2-\frac{3}{2}x+\frac{3}{4})(4z^2)  + \qty(\frac{9}{2}x^2 - \frac{21}{4}x + \frac{13}{4})(-12z^3) + \dots 
\end{align}
This normalization is achieved by multiplying by $-\frac{16^j}{2j!}$. So let's define the ``normalized coefficients'' $B_1(j,x)$ as
\begin{align}
    B_1(j,x) = -\frac{16^j}{2j!}[e^{2jx}]G_{j+1}(x), \qquad j\geqslant1, \label{eqn:B1_normalization}
\end{align}
where $[e^{2jx}]G_{j+1}(x)$ are the coefficients of the sub-leading harmonics in equation \eqref{eqn:harmonics_full}.
Then
\begin{align}
    C_1(z,x) = B_1(0,x)(1) + B_1(1,x)(-2z) + B_1(2,x)(4z^2) + B_1(3,x)(-12z^3) + \dots,  \label{eqn:C1_series}
\end{align}
where
\begin{subequations}
\begin{align}
    B_1(1,x) & = -\frac{16}{2} \times \frac{1}{32}(-2x^2-x+1) = \frac{1}{2}x^2 +\frac{1}{4}x - \frac{1}{4}, \\
    B_1(2,x) & = -\frac{256}{4} \times \frac{1}{256}(-8x^2+6x-3) = 2x^2 - \frac{3}{2}x +\frac{3}{4}, \\
    B_1(3,x) & = -\frac{4096}{12} \times \frac{3}{4096}(-18x^2+21x-13) = \frac{9}{2}x^2 - \frac{21}{4}x + \frac{13}{4}.
\end{align}
\end{subequations}
Look at the coefficients more closely:
\begin{table}[H]
    \centering
    \begin{tabular}{c|ccc}
        $j$ & $[x^2]B_1$ &$ [x] B_1$ & $[1]B_1$ \\ \hline
        1 & 1/2 & 1/4 & $-$1/4 \\
        2 & 2 & $-$3/2& 3/4 \\
        3 & 9/2& $-$21/4 & 13/4,
    \end{tabular}
    \caption{First few coefficients of the normalized sub-leading harmonics in $C_1$. Note: in the supplementary ``Amplitude Reference'' table, these coefficients are found in the ``2 Normalized'' tab, under the $r=1$ column where $r=N-j$.}
    \label{tab:C1_j}
\end{table}
\noindent we see each column is a quadratic polynomial in $j$:
\begin{subequations}
\begin{align}
    [x^2]B_1 & = \frac{j^2}{2}, \\
    [x]B_1 & = -j^2 + \frac{5}{4}j, \\
    [1]B_1 & = \frac{3j^2-5j+1}{4},
\end{align}
\end{subequations}
combining, we have
\begin{align}
    B_1(j,x) = \frac{1}{4}[(2x^2-4x+3)j^2 + 5(x-1)j+1], \label{eqn:B1}
\end{align}
this pattern is verified for every available $j$ to about $j=50$. Now, plug this back into the expression for $C_1(z,x)$ (equation \eqref{eqn:C1_series}), we have
\begin{align}
    C_1(z,x) = \frac{1}{4} \qty[(2x^2-4x+3) \sum_{j=1}^\infty j^2\times2(-1)^jj!z^j + 5(x-1)\sum_{j=1}^\infty j\times2(-1)^jj!z^j + C_0],
\end{align}
i.e. the $j$th term in $B_1$ multiplies the $j$th term in $C_0$ (this becomes clear by writing out the first few terms explicitly). This motivates us to define the Euler derivative at fixed $x$,
\begin{align}
    D = z\frac{d}{dz}, \qquad \Rightarrow \qquad Dz^j = jz^j,
\end{align}
this way, we may replace $j$ by $D$, and we obtain
\begin{align}
    \Aboxed{C_1(z,x) = \frac{1}{4} \qty[(2x^2-4x+3)D^2 + 5(x-1)D + 1]C_0(z)}. \label{eqn:C1_final}
\end{align}
The derivatives can be very easily computed by using the integral form of $\mathcal C_0$. In fact, since $\mathcal C_0$ is already Borel summed, $C_1$ calculated this way is also automatically Borel summed.

A minor point that is interesting: note equation \eqref{eqn:B1_normalization} that defines the normalisation of $B_1(j,x)$ works for all $j>0$, but is off by a factor of 2 for $j=0$. The correct unified formula is
\begin{align}
    B_1(j,x) = -\frac{16^j}{(2-\delta_{j0})j!}[e^{2jx}]G_{j+1}(x).
\end{align}
However, ultimately this did not matter because the polynomial expression, equation \eqref{eqn:B1}, still works! But this gives us a first hint to the ``boundary terms'' that will appear in low-$j$-order terms that will become important starting from $C_2$.

\subsection{Second correction to the first depletion} \label{subsec:correction_c_2}

\textit{The next-order harmonics are supposed to give the 2nd-order correction to the first depletion. However, we identify three ``exceptional'' terms that do not fit into the generic polynomial-derivative form we established in the previous section.}

We can try the same procedure with the second sub-leading term $C_2(z)$, i.e. write it as 
\begin{align}
    C_2(z,x) = P_2(D,x)C_0(z),
\end{align}
where $P_2$ is written as a polynomial expression of $x$ multiplied by the Euler derivative $D$. 

However, here things go wrong starting from the very first term, 
\begin{align}
    -\epsilon G_1(x) = \dots -\frac{\epsilon}{8}e^{-2x} = \dots -\frac{\epsilon^2}{128z},
\end{align}
its contribution to $C_2$ is therefore $-\frac{1}{128z}$. But the derivative $D$ leaves powers of $z$ unchanged, so the generating series $C_0 = 1-2z+4z^2-...$ could not possibly produce a $1/z$ term! 

Furthermore, we found that there seems to be no polynomial sequence one can cleanly extract, unlike the quadratic sequence in the previous case.

Astonishingly, after examining the whole sequence, we found that there is indeed an exact, quartic sequence which predicts all coefficients starting from $G_4$ to $G_{50}$ (meaning it works for all other $G_n$ except $G_1, G_2$ or $G_3$)! And the discrepancies contain crucial information.

Here we skim over the derivation and simply state key results. The normalized coefficients are defined as
\begin{align}
    B_2(j,x) & = \frac{16^j}{2j!}[e^{2jx}]G_{2+j}(x), \\
    C_2(z,x) & = \sum_{j\geqslant1}2(-1)^jj!B_2(j,x)z^j,
\end{align}
as before. Now we pick the third highest harmonics from each $G_n$ starting at $G_4$ (so the $e^{4x}$ term in $G_4$, $e^{6x}$ in $G_5$, etc.), and write 
\begin{align}
    B_2(j,x) = b_0(j) + b_1(j)x + b_2(j)x^2+ b_3(j)x^3 + b_4(j)x^4, 
\end{align}
we have the table
\begin{table}[H]
    \centering
    \begin{tabular}{c|ccccc}
         j & $b_0(j)$ & $b_1(j)$ & $b_2(j)$ & $b_3(j)$ & $b_4(j)$  \\\hline 
         2 & 29/8 & $-$123/32 & 4 & $-$11/3 & 2 \\
         3 & 1049/64 & $-$1949/64 & 1163/32 & $-$207/8 & 81/8 \\
         4 & 1675/32 & $-$1867/16 & 1127/8 & $-$280/3 & 32 \\ 
         5 & 8443/64 & $-$20319/64 & 12251/32 & $-$5875/24 & 625/8 \\
         6 & 4517/16 & $-$22585/32 & 1697/2 & $-$531 & 162
    \end{tabular}
    \caption{Polynomial coefficients in $B_2$ necessary for a quartic fit. Note: In the supplementary ``Amplitude Reference'' table, these are found in the ``2 Normalized'' tab, under the $r=2$ column where $r=N-j$, and in the $G_4 (N=4)$ to $G_8 (N=8)$ rows.}
    \label{tab:B2_coeff}
\end{table}
\noindent these can be fitted to obtain the following quartic expressions,
\begin{subequations}
\begin{align}
    b_0(j) & = -\frac{11}{32} + \frac{3}{64}j + \frac{29}{32}j^2 - \frac{17}{32}j^3 + \frac{9}{32}j^4, \\
    b_1(j) & = \frac{1}{4} - \frac{7}{64}j - \frac{21}{32}j^2 + \frac{43}{32}j^3 - \frac{3}{4}j^4, \\
    b_2(j) & = -\frac{1}{8} - \frac{1}{8}j + \frac{11}{32}j^2 - \frac{11}{8}j^3 + \frac{7}{8}j^4, \\
    b_3(j) & = \frac{13}{24}j^3 - \frac{1}{2}j^4, \\
    b_4(j) & = \frac{1}{8}j^4. 
\end{align}
\end{subequations}
Adding them up, we can then try to express $B_2$ as polynomials of $x$ multiplied by powers of $j$, as before:
\begin{align}
    B_2(j,x)\bigg\vert_{j\geqslant2} = p_0(x) + p_1(x)j + p_2(x)j^2 + p_3(x)j^3 + p_4(x)j^4, 
\end{align}
where
\begin{subequations}
\begin{align}
    p_0(x) & = -\frac{11}{32} + \frac{x}{4} - \frac{x^2}{8}, \\
    p_1(x) & = \frac{3}{64} - \frac{7x}{64} - \frac{x^2}{8}, \\
    p_2(x) & = \frac{29}{32} - \frac{21x}{32} + \frac{11x^2}{32}, \\
    p_3(x) & = -\frac{17}{32} + \frac{43x}{32} - \frac{11x^2}{8} + \frac{13x^3}{24}, \\
    p_4(x) & = \frac{9}{32} - \frac{3x}{4} + \frac{7x^2}{8} - \frac{x^3}{2} + \frac{x^4}{8}, 
\end{align}\label{eqn:P2_polynomials}
\end{subequations}
this relation holds for all $j$ starting from $j=2$ (corresponding to $G_4$). And finally, we may replace $j\rightarrow D$ to obtain the expression
\begin{align}
    P_2(D,x) & = p_0(x) + p_1(x)D + p_2(x)D^2 + p_3(x)D^3 + p_4(x)D^4, \label{eqn:P2}  \\
    C_2 & \stackrel{?}{=} P_2(D,x) \ C_0.
\end{align}
This is not exactly right, because while it gives us the contribution to $C_2$ starting from $G_4$, it does not yet include $G_1, G_2, G_3$. We have already seen $G_1$ contributes $-\frac{1}{128z}$ which cannot be included in the polynomial framework. Now look at the $G_2$ term,
\begin{align}
    G_2(x) = \frac{1}{64}e^{4x} + \dots + \frac{1}{32}(-4x^2 - 3) + \dots, 
\end{align}
the zero harmonic contribution $\frac{1}{32}(-4x^2-3)$ should be the actual $j=0$ term in $C_2$. But our generic quartic expression gives
\begin{align}
    p_0(x) = -\frac{11}{32} + \frac{x}{4} - \frac{x^2}{8},
\end{align}
the difference is
\begin{align}
    \frac{1}{32}(-4x^2-3) - \qty(-\frac{11}{32} + \frac{x}{4} - \frac{x^2}{8}) = \frac{1-x}{4}, 
\end{align}
therefore the generic polynomial must be corrected by $+\frac{1-x}{4}$. 

Now look at $G_3$ term: the relevant harmonic is
\begin{subequations}
\begin{align}
    G_3 & = \dots + \frac{1}{3072}\qty(48x^4 + 16x^3 - 156x^2 + 30x + 135) e^{2x} \\
    & = \dots + \qty(\frac{x^4}{64} + \frac{x^3}{192} - \frac{13x^2}{256} + \frac{5x}{512} + \frac{45}{1024})e^{2x}, 
\end{align}
\end{subequations}
converting to the normalized $B_2$, we have
\begin{subequations}
\begin{align}
    B_2(j=1,x) & = \frac{16^j}{2j!}[e^{2jx}]G_3(x) = \frac{16^1}{2\cdot 1!}\qty(\frac{x^4}{64} + \frac{x^3}{192} - \frac{13x^2}{256} + \frac{5x}{512} + \frac{45}{1024}) \\
    & = \frac{x^4}{8} + \frac{x^3}{24} - \frac{13x^2}{32} + \frac{5x}{64} + \frac{45}{128}.
\end{align}
\end{subequations}
whereas the polynomial fit predicts
\begin{align}
    P_2(1,x) & = \frac{x^4}{8} + \frac{x^3}{24} - \frac{13x^2}{32} + \frac{5x}{64} + \underbracket{\frac{46}{128}},
\end{align}
we see there is a tiny difference, of $1/128$, between the two expressions! In the normalized $B_2$ notation,
\begin{align}
    B_2(1,x) = P_2(1,x) - \frac{1}{128},
\end{align}
and the $G_3$ term in $C$ is
\begin{align}
    -\epsilon^3G_3\bigg\vert_{e^{2x}} & = -\epsilon^3 e^{2x} \qty[\frac{1}{8}P_2(1,x) - \frac{1}{1024}] = \epsilon^2 \qty[-2P_2(1,x) + \frac{1}{64}]z,
\end{align}
the $-2P_2z$ term comes from taking the Euler derivative of the $-2z$ term in $C_0$. The second term, $\frac{z}{64}$, is a genuine extra correction. 

This means we have 3 additional (``exceptional'') terms. The full answer is
\begin{align}
    \Aboxed{C_2(z,x) = P_2(D,x) C_0(z) + \frac{1-x}{4} + \frac{z}{64} - \frac{1}{128z}}, \label{eqn:C2}
\end{align}
with the polynomials in $P_2$ given by equation \eqref{eqn:P2}.

\subsection{The first return: A hint} \label{subsec:first_return_hint}
\textit{From the exceptional terms, we derive the leading contribution to the first return profile, expressed in a triple-Borel sum.}

The exceptional term, $\frac{z}{64}$, is of particular interest. In fact it contains the information of the first return profile! In the full expression of $C$, it's in fact $\frac{\epsilon^2z}{64}$. Let's define a new variable,
\begin{align}
    w = \epsilon^2z = \frac{e^{2x}}{16\alpha^6}, 
\end{align}
then our exceptional term becomes $\frac{w}{64}$. This term determines the time scale of the first return, $\chi = 3\log\alpha +\mathcal O(1)$. (Later, with more careful analysis, we shall see that the first return occurs more specifically at $3\log4\alpha$, which is confirmed by numerical results.)

Note we only care about this term because, at this time scale, the two other exceptional terms in equation \eqref{eqn:C2} all go to 0, whereas the $P_2C_0$ term is already Borel-summed and would've vanished by this point. More specifically, recall
\begin{align}
    C = C_0 + \epsilon C_1 + \epsilon^2 C_2 + \dots, 
\end{align}
so the contributions of all terms are
\begin{subequations}
\begin{gather}
    P_2(D,x)C_0 \longrightarrow -p_0(x) = \frac{11}{32}-\frac{x}{4}+\frac{x^2}{8}, \qquad \epsilon^2 P_2(D,x)C_0\longrightarrow0 \quad \text{ as } \quad \epsilon\rightarrow0, \\ 
    \frac{\epsilon^2(1-x)}{4} + \frac{w}{64} - \frac{\epsilon^4}{128w} \quad \longrightarrow \quad \frac{w}{64} \quad \text{ as } \quad \epsilon\rightarrow0.
\end{gather}
\end{subequations}
Only $\frac{w}{64}$ remains finite. 

\indent In fact, $\frac{w}{64}$ is the first of many such exceptional terms. We found this pattern continues, in the sense that: 
\begin{itemize}
    \item The $C_4$, $j=2$ harmonic contains an exceptional $z^2$ term, 
    \item The $C_6$, $j=3$ harmonic contains an exceptional $z^3$ term, 
    \item The $C_8$, $j=4$ harmonic contains an exceptional $z^4$ term, 
    \item \dots
\end{itemize}
Expressed in terms of $G_k$: starting from $G_3$, at each $G_{3n}$, there is an exceptional return term in powers of $w$ that doesn't fit into the polynomials:
\begin{table}[H]
    \centering
    \begin{tabular}{c|c}
        \text{Source} & \text{Exceptional return term} \\ \hline
        $G_3$ & $+\frac{w}{64}$ \\[0.25em]
        $G_{6}$ & $-\frac{5w^2}{2048}$ \\[0.25em]
        $G_{9}$ & $+\frac{189w^3}{131072}$ \\[0.25em]
        $G_{12}$ & $-\frac{4671w^4}{2097152}$ \\[0.25em]
        $G_{15}$ & $+\frac{1900125w^5}{268435456}$ \\
        \dots & \dots 
    \end{tabular}
    \caption{The first few exceptional terms, contributing to $R_0$. Note these are most apparent in the attached ``Amplitude Reference'' table, under the ``4 Corrections'' tab, and these coefficients are the green terms, $-1/128$, $-5/8192$, $-63/524288$, $-1557/33554432$, $-126675/4294967296...$. To convert them into observables, refer to the ``5 Index Guide'' tab, and we see they should be multiplied, respectively, by $-2z$, $4z^2$, $-12z^3$, $48z^4$, $-240z^5\dots$.}
    \label{tab:return_terms}
\end{table}
We will shift the series by $-1$, because the pump has already reached the negative plateau at this point, so $C \approx -1$ at the start. 
\begin{align}
    R_0(w) \sim -1 + \frac{w}{64} - \frac{5w^2}{2048} + \frac{189w^3}{131072} -\frac{4671w^4}{2097152} + \dots.
\end{align}
The coefficients can be expressed as
\begin{align}
    R_0(w) \sim -1 + 2\sum_{m=1}^\infty (-1)^{m+1} (m!)^3 \qty(\frac{w}{256})^mF_m, \label{eqn:R0_franel}
\end{align}
where  
\begin{align}
    F_m = \sum_{k=0}^m \mqty(m \\ k)^3,
\end{align}
are the Franel numbers \cite{OEIS_A000172}, 
\begin{align}
    F_1 = 2, \qquad F_2 = 10, \qquad F_3 = 56, \qquad F_4 = 346, \dots 
\end{align}
$R_0$ has a $(m!)^3$ growth (Gevrey-3), so we divide it by three factorials to perform a Borel transform:
\begin{align}
    \hat R_0(s) = -1 + 2\sum_{m=1}^\infty (-1)^{m+1} F_m \frac{s^m}{256^m} = -1-2\sum_{m=1}^\infty F_m\frac{(-s)^m}{256^m} \equiv 1-2\mathcal F(-s/256),
\end{align}
where
\begin{align}
    \mathcal F(z) = \sum_{m=0}^\infty F_mz^m. 
\end{align}
and we used the fact that $F_0 = 1$ in changing the sum from $\sum_{m\geqslant1}$ to $\sum_{m\geqslant0}$. 

$\mathcal F(x)$ is a known generating function with the following properties \cite{OEIS_A000172}: 
\begin{align}
    \mathcal F(x) & = \frac{1}{1-2x} \ _2F_1\qty(\frac{1}{3},\frac{2}{3};1;\frac{27x^2}{(1-2x)^3}) = \sum_{n=0}^\infty \frac{(3n)!}{(n!)^3} \frac{x^{2n}}{(1-2x)^{3n+1}},
\end{align}
expanding the denominator for the $n=0$ and $n=1$ terms gives 
\begin{align}
    \mathcal F(z) = 1+2z+10z^2 + 56z^3 + \dots
\end{align}
as expected. And $ _2F_1(a,b,c;z)$ is the Gauss hypergeometric function, which is known to be singular at $z=1$ and $\infty$, i.e. when $\frac{27x^2}{(1-2x)^3} = 1$ or $\infty$ \cite[Sec.~15.10]{NIST_DLMF}. The $z=1$ case applies here, and it solves to $x = -1$ and $x=\frac{1}{8}$. And since $x = -\frac{s}{256}$, these correspond to $s=256$ and $s=-32$. 

The explicit triple Borel sum is
\begin{align}
    \iiint_0^\infty e^{-(t_1+t_2+t_3)}\hat R_0(wt_1t_2t_3) \  dt_1 \, dt_2 \, dt_3.
\end{align}
Without evaluating the integral, we already see that the integration path $[0,\infty)$ will pass through the singularity at $x = -1$ (in fact a logarithmic branch cut), and therefore we must deform our contour around it either above or below the cut (or in fact, infinitely many other deformations by going to other Riemann surfaces). Procedures involving deforming over contours such as this is typical of the Resurgence Theory \cite{ANICETO_2019}. The simplest way is to just take the principal value of the integration, and numerically we found this does give the sensible result. However, this is not a formal proof that it is guaranteed to work.

\subsection{The resolvent shortcut} \label{subsec:resolvent}

\textit{Abandoning the triple Borel sum, we show the return coefficients are generated by a simple resolvent operator proportional to the laplacian.}

In short, we are not going to work with the awkward triple integral derived in the previous section. Our goal remains to generate the series
\begin{align}
    R_0(w) \sim -1 + \frac{w}{64} - \frac{5w^2}{2048} + \frac{189w^3}{131072} -\frac{4671w^4}{2097152} + \dots.,
\end{align}
but we shall take a different approach, which we shall see leads us to a more physically intuitive and unambiguous answer, and without heavy calculus. But even so, this detour to derive \eqref{eqn:R0_franel} has been important, because it teaches us three things:
\begin{enumerate}
    \item Recall how $z = e^{2\chi}/16\alpha^2$ was defined, especially the factor of 16—it's the convenient normalization that captures how $C_0$ grows. From equation \eqref{eqn:R0_franel}, we see the corresponding factor for $w$ is really 256, since we're technically doing expansion not just in $w^n$, but in $(w/256)^n$. The $w/256\rightarrow0$ region represents when the return hasn't started yet, and $w/256\rightarrow\infty$ represents when the return has finished. The important part is when $w/256\approx1$. Write
    \begin{align}
        \frac{w}{256} = \frac{\epsilon^2z}{256} = \frac{e^{2\chi}}{16\times256\times\alpha^6} = e^{2\chi} e^{-\log(16\times256\times\alpha^6)} = e^{2\chi - 6\log4\alpha} = e^{2(\chi-3\log4\alpha)},
    \end{align}
    this sets the relevant time scale to be
    \begin{align}
        \chi \approx 3\log4\alpha.
    \end{align}
    In contrast to the depletion time scale, $\chi\approx\log4\alpha$. All this is very consistent with numerical results, where the coherent amplitude is half-depleted almost precisely at time equals $\log4\alpha$, and half is converted back almost exactly at $3\log4\alpha$. 
    
    \item The fact the elements in $R_0(w)$ are proportional to the binomial coefficients (the Franel numbers) indicates the series comes from some binomial expansion. 
    
    \item Finally, the Gevrey-3 growth is interesting as it is not often seen in physics \cite{CAVALCANTI_2021}.
    It suggests the expansion of some highly nonlinear combination of operators. This is because operators acting on states in Hilbert space naturally accumulates factorial growth. For example,
\begin{gather}
    \norm{((a^\dag)^p)^n\ket0}^2 \approx (pn)!, \qquad     \text{Ratio test: } \quad \frac{(p(n+1))!}{(pn)!} \sim (pn)^p.
\end{gather}
so to achieve $(n!)^3$ divergence, we need at least 3 operators.
\end{enumerate}
So, in summary, we should be able to obtain the series $R_0(w)$ by the binomial expansion of a product of at least 3 operators. 

Indeed, there is a combination of 3 operators we can define that satisfies this condition and whose perturbative expansion gives the precise answer we want! We found that in fact, these coefficients are generated by the following resolvent operator:
\begin{align}
    R_0(w) = 1-2\expval{\qty(1+\frac{w}{4}Q^\dag Q)^{-1}}, \label{eqn:R0_formal}
\end{align}
where 
\begin{subequations}
\begin{gather}
    Q = AB^\dag A , \qquad  A = \frac{a+b^\dag}{2}, \qquad B = \frac{a-b^\dag}{2}, \\
    [A,B^\dag] = \frac{1}{2}, \qquad [Q,Q^\dag] = 0,
\end{gather}
\end{subequations}
and all other commutators involving $A$ and $B$ vanish. Let's compute the first few coefficients to show they match our desired results. This expression for $R_0(w)$ generates the geometric series via expanding the bracket:
\begin{align}
    \expval{\qty(1+\frac{w}{4}Q^\dag Q)^{-1}} = 1-\frac{w}{4}\expval*{Q^\dag Q} + \frac{w^2}{16} \expval*{(Q^\dag Q)^2} - \frac{w^3}{64} \expval*{(Q^\dag Q)^3} + \dots . \label{eqn:R0_vev}
\end{align}
One can check that, for example,
\begin{align}
    \expval*{Q^\dag Q} = \norm{\frac{1}{4\sqrt2}\ket{12}}^2 = \frac{1}{32}, \quad \expval*{(Q^\dag Q)^2} = \frac{1}{32}\norm{Q^\dag\ket{12}}^2 = \frac{5}{256}, \quad \expval*{(Q^\dag Q)^3} = \frac{189}{4096} \dots 
\end{align}
Substituting back to equation \eqref{eqn:R0_formal} and \eqref{eqn:R0_vev}, we retrieve the desired results. 

Since the calculation matches their formal return series, our next goal is to express the resolvent in $R_0(w)$ in a more manageable form. In fact, we will see that in an appropriate basis, the seemingly complicated $Q^\dag Q$ is simply the laplacian. 

There are likely a number of slicker approaches to show this—like using Green's function or heat kernel, to name a few. Here, in the spirit of quantum optics, we start by re-writing the operators in the position and momentum quadratures. Define:
\begin{subequations}
\begin{gather}
    a = \frac{x_a + ip_a}{\sqrt2}, \qquad b = \frac{x_b+ip_b}{\sqrt2}, \qquad [x_i, p_j] = i\delta_{ij}, \\
    \Rightarrow \ A = \frac{a+b^\dag}{2} = \frac{x_a+x_b+i(p_a-p_b)}{2\sqrt2}, \qquad B = \frac{a-b^\dag}{2} = \frac{x_a-x_b+i(p_a+p_b)}{2\sqrt2}.
\end{gather}
\end{subequations}
Now define a new set of coordinates and their conjugate momentum:
\begin{align}
    u = \frac{x_a+x_b}{\sqrt2}, \qquad v = \frac{p_a-p_b}{\sqrt2}, \qquad 
    p_u = \frac{p_a+p_b}{\sqrt2}, \qquad p_v = -\frac{x_a-x_b}{\sqrt2}, 
\end{align}
one can check the canonical commutation relations are satisfied. Then our original operators become
\begin{align}
    A = \frac{1}{2}(u+iv), \qquad B = \frac{1}{2}(-p_v + ip_u), 
\end{align}
now introduce complex coordinates,
\begin{gather}
    z = u+iv, \qquad \bar z = u-iv, \qquad 
    \Rightarrow \qquad A = \frac{z}{2}, \qquad A^\dag = \frac{\bar z}{2}, 
\end{gather}
and apply the standard momentum substitution:
\begin{subequations}
\begin{gather}
    p_u = -i\partial_u, \qquad  p_v = -i\partial_v, \\
    \Rightarrow \ B = \frac{1}{2}(i\partial_v + \partial_u) = \partial_{\bar z}, \qquad B^\dag = - \frac{1}{2}(\partial_u - i\partial_v) = -\partial_z, \\
    \Rightarrow \ [A,B^\dag] = \qty[\frac{z}{2}, -\partial_z] = \frac{1}{2}.
\end{gather}
\end{subequations}
Using the commutation relations of $A$ and $B$, we can write $Q^\dag Q$ as
\begin{align}
    Q^\dag Q & = A^\dag B A^\dag AB^\dag A = A^\dag BAA^\dag B^\dag A = A^\dag ABB^\dag A^\dag A = (A^\dag A) (B^\dag B) (A^\dag A) = -\frac{\abs{z}^2}{4} \partial_{\bar z}\partial_z \frac{\abs{z}^2}{4},
\end{align}
note that
\begin{gather}
    \partial_{\bar z}\partial_z = \frac{1}{4}\laplacian \equiv \frac{1}{4}(\partial_u^2+\partial_v^2), \quad 
    \Rightarrow \quad  Q^\dag Q = -\frac{1}{64}\abs{z}^2\laplacian\abs{z}^2.
\end{gather}
This expression is to be understood as
\begin{align}
    Q^\dag Qf = -\frac{1}{64} \abs{z}^2 \laplacian\qty(\abs{z}^2f) = -\frac{1}{64}\abs{z}^2\qty(\abs{z}^2 \laplacian f + 4(u\partial_u + v\partial_v)f + 4f).
\end{align}
We have almost gotten $Q^\dag Q$ into a neat form, but not quite. We can eliminate the $\abs{z}^2$ factors by exploiting the conformal invariance of the 2D laplacian. Under a conformal transformation, $z\rightarrow \eta$, and 
\begin{align}
    \laplacian_z = \abs{\frac{d\eta}{dz}}^2 \laplacian_\eta, 
\end{align}
let $\eta = 2/z$, the factor of 2 is simply a convenient normalization. Then $d\eta/dz = -2/z^2 = -\eta^2/2$. Note that
\begin{align}
    \expval{Q^\dag Q}{\phi_0} & = \expval{U^{-1}U (Q^\dag Q) U^{-1}U}{\phi_0} = \expval{U(Q^\dag Q)U^{-1}}{\tilde\phi_0}.
\end{align}
We're interested in the transformed operator $U(Q^\dag Q)U^{-1}$, and $U$ acts on an arbitrary function as
\begin{align}
    (Uf)(\eta) = \frac{2}{\abs{\eta}^2}f\qty(\frac{2}{\eta}), \qquad (U^{-1}f)(z) = \frac{2}{\abs{z}^2}f\qty(\frac{2}{z}),
\end{align}
then 
\begin{subequations}
\begin{align}
    U(Q^\dag Q) U^{-1}f & = U(Q^\dag Q) \frac{2}{\abs{z}^2}f\qty(\frac{2}{z}) = U \qty(-\frac{1}{64}\abs{z}^2 \laplacian_z 2f\qty(\frac{2}{z})) \\
    & = -\frac{1}{32} \frac{2}{\abs{\eta}^2} \qty(\frac{4}{\abs{\eta}^2} \cdot\frac{\abs{\eta}^4}{4}\laplacian_\eta f(\eta)) = -\frac{1}{16}\laplacian_\eta f(\eta),
\end{align}
\end{subequations}
so in these new coordinates, we have a very simple form for $Q^\dag Q$:
\begin{align}
    U(Q^\dag Q)U^{-1} = -\frac{1}{16}\laplacian_\eta.
\end{align}
The fact that this highly nonlinear combination of operators somehow reduce to a laplacian is interesting in its own right. Just from the fact $Q^\dag Q$ is hermitian and positive we know it's a ``Hamiltonian'', but now we see it also has a continuous spectrum. 

Now note the vacuum has transformed too. Before the conformal transformation, the vacuum wavefunction is uniquely determined by 
\begin{subequations}
\begin{gather}
    a\phi_0 = (A+B)\phi_0 = \qty[\frac{u+iv}{2} + \frac{\partial_u + i\partial_v}{2}]\phi_0(u,v) = 0, \\
    b\phi_0 = (A^\dag - B^\dag)\phi_0 = \qty[\frac{u-iv}{2} + \frac{\partial_u - i\partial_v}{2}]\phi_0(u,v) = 0, \\
    \Rightarrow \ \phi_0(u,v) = \frac{1}{\sqrt\pi} e^{-(u^2+v^2)/2} = \frac{1}{\sqrt\pi} e^{-\abs{z}^2/2}.
\end{gather}
\end{subequations}
Transforming:
\begin{align}
\tilde\phi_0 = U\phi_0 = \frac{2}{\abs{\eta}^2} \phi_0\qty(\frac{2}{\eta}) =  \frac{2}{\sqrt\pi\abs{\eta}^2} e^{-2/\abs{\eta}^2}.
\end{align}
Put our $Q^\dag Q$ back into the original expression \eqref{eqn:R0_formal}, we have
\begin{align}
    \Aboxed{R_0(w) = 1-2\expval{\qty(1-\frac{w}{64}\laplacian_\eta)^{-1}}{\tilde\phi_0}}.
\end{align}

\subsection{The first return profile}
\textit{We derive the first return profile which has the form $R_0 = \int \abs{h}^2 \tanh$ for a weight function $h$.}

This expression for $R_0(w)$ can now be very easily Fourier transformed, with $-\laplacian \rightarrow k^2$:
\begin{align}
    R_0(w) = 1-2\int \frac{\abs*{\hat\phi(k)}^2}{1+wk^2/64} d^2k. \label{eqn:R0_FT}
\end{align}
Define 
\begin{align}
    p=\log k \ \Rightarrow \ dp = dk/k, \qquad h(p) = \sqrt{2\pi} k \hat\phi(k),
\end{align}
where $\hat\phi(k)$ is the Fourier transform of (the already conformally transformed) $\tilde\phi_0$. One can then check that its normalization condition implies
\begin{align}
    \int  \abs*{\hat\phi(k)}^2 d^2k = 1 \quad \Rightarrow \quad \int_{-\infty}^\infty \abs{h(p)}^2 dp = 1, 
\end{align}
and
\begin{subequations}
\begin{align}
    R_0(w) & = 1- 2\int \frac{\abs{h(p)}^2}{2\pi k^2} \frac{1}{1+wk^2/64} k \ dk \ d\theta  \\
    & = \int \abs{h(p)}^2 \qty(1-\frac{2}{1+wk^2/64}) \ dp \\
    & = \int \abs{h(p)}^2 \tanh(\frac{1}{2}\log(\frac{wk^2}{64})) dp.
\end{align}
\end{subequations}
Here we directly jumped to the final answer because we recognized the second line as having the exact same form as equation \eqref{eqn:C0_E1} for $C_0$!  

Recall how our parameters are defined. The return time is now centered at $\chi\sim 3\log(4\alpha)$, so define the shifted time
\begin{align}
    s_1 = \chi - 3\log(4\alpha),
\end{align}
and recall 
\begin{gather}
    w = \frac{e^{2\chi}}{16\alpha^6} = 256 e^{2s_1}, \quad    \Rightarrow \quad\frac{1}{2}\log(\frac{wk^2}{64}) = s_1 + p + \log2,
\end{gather}
define $u = p+\log2$, we have
\begin{align}
    R_0(s_1) = \int\abs{h_1(u)}^2 \tanh(s_1+u) \ du,
\end{align}
To calculate $h(u)$, first note the Fourier transform of the vacuum wavefunction is
\begin{align}
    \hat\phi(k) = \frac{1}{2\pi} \int \tilde\phi_0(r)e^{-ikr\cos\theta} \ r \, dr\, d\theta = \int_0^\infty r\tilde\phi_0(r) J_0(kr) \,dr,
\end{align}
where $J_0$ is the Bessel function. While we can in principle perform the integral numerically, $J_0(kr)$ is highly oscillatory at large $k$. It is more convenient to go into the ``log momentum space'' again—by simply taking the Fourier transform of $h(p)$:
\begin{subequations}
\begin{align}
    H(t) & = \int_{-\infty}^\infty h(p) e^{-itp} dp = \sqrt{2\pi} \int_0^\infty k\hat\phi(k)e^{-it\log k} \frac{dk}{k} \\
    & = \sqrt{2\pi} \int_0^\infty dk \ k^{-it} \int_0^\infty dr \ r \ \tilde\phi_0(r) J_0(kr) \\
    & = \sqrt{2\pi} \int_0^\infty dr \ r \tilde\phi_0(r) \underbracket{\int_0^\infty dk \ k^{-it} J_0(kr)} \\
    & = \sqrt{2\pi} \int_0^\infty dr \ r \tilde\phi_0(r) \times \frac{2^{-it}r^{it-1}\Gamma(\frac{1-it}{2})}{\Gamma(\frac{1+it}{2})} \\
    & = \sqrt{2\pi}2^{-it} \frac{\Gamma(\frac{1-it}{2})}{\Gamma(\frac{1+it}{2})} \underbracket{\int_0^\infty dr \  r^{it} \frac{2}{\sqrt\pi r^2} e^{-2/r^2}}  \\
    & = \sqrt{2\pi} 2^{-it} \frac{\Gamma(\frac{1-it}{2})}{\Gamma(\frac{1+it}{2})} \times \frac{2^{1-\frac{3}{2}+\frac{it}{2}}\Gamma(\frac{1-it}{2})}{\sqrt\pi} \\
    \Rightarrow \ \Aboxed{ H(t) & = 2^{-it/2} \frac{\Gamma(\frac{1-it}{2})^2}{\Gamma(\frac{1+it}{2})}}.
\end{align}
\end{subequations}
Numerically this is a much more manageable form; analytically, we can see that $\abs{H(t)}$ decays in $t$: 
\begin{subequations}
\begin{gather}
    \Gamma\qty(\frac{1-it}{2})^* = \Gamma\qty(\frac{1+it}{2}) \qquad \Rightarrow \qquad \abs{\Gamma\qty(\frac{1-it}{2})} = \abs{\Gamma\qty(\frac{1+it}{2})}, \\
    \Rightarrow \ \abs{H(t)}^2 = \abs{\Gamma\qty(\frac{1+it}{2})}^2 = \frac{\pi}{\cosh(\frac{\pi t}{2})} \rightarrow 2\pi e^{-\pi t/2}.
\end{gather}
\end{subequations}
\indent So, in summary, we have two working formulas for $R_0$:
\begin{align}
    \Aboxed{R_0(w) & = \int \abs{h(p)}^2 \tanh(\frac{1}{2}\log(\frac{wk^2}{64})) dp}, \\
    \text{or } \ \Aboxed{R_0(s_1) & =\int_{-\infty}^\infty \abs{h_1(u)}^2 \tanh(s_1+u) \ du },
\end{align}
with $s_1 = \chi-3\log(4\alpha)$, $u = p+\log 2$. And $h(p)$ is obtained via the inverse Fourier transform,
\begin{align}
    h(p) & = \frac{1}{2\pi} \int_{-\infty}^\infty H(t)e^{itp} dt \\
    \Rightarrow \ \Aboxed{h(p) & = \frac{1}{2\pi} \int_{-\infty}^\infty 2^{-it/2}\frac{\Gamma(\frac{1-it}{2})^2}{\Gamma(\frac{1+it}{2})} e^{itp} \ dt}.
\end{align}
Or if we want $h_1(u)$, it is simply 
\begin{align}
    h_1(u) \equiv h(u-\log 2). \label{eqn:h1u}
\end{align}
The Fourier transform of $h_1(u)$ is slightly different, 
\begin{gather}
    H_1(t) = \int h_1(u) e^{-itu} du = e^{-it \log 2} H(t) = 2^{-it} H(t), \\
    \Rightarrow \ \boxed{H_1(t) = 2^{-3it/2} \frac{\Gamma(\frac{1-it}{2})^2}{\Gamma(\frac{1+it}{2})}}, \label{eqn:H1t}
\end{gather}
meaning (either from equation \eqref{eqn:h1u} or equation \eqref{eqn:H1t})
\begin{align}
    \Aboxed{h_1(u) = \frac{1}{2\pi} \int_{-\infty}^\infty 2^{-3it/2} \frac{\Gamma(\frac{1-it}{2})^2}{\Gamma(\frac{1+it}{2})} e^{itu}dt}.
\end{align}

\begin{figure}[H]
    \centering
    \includegraphics[width=0.5\linewidth]{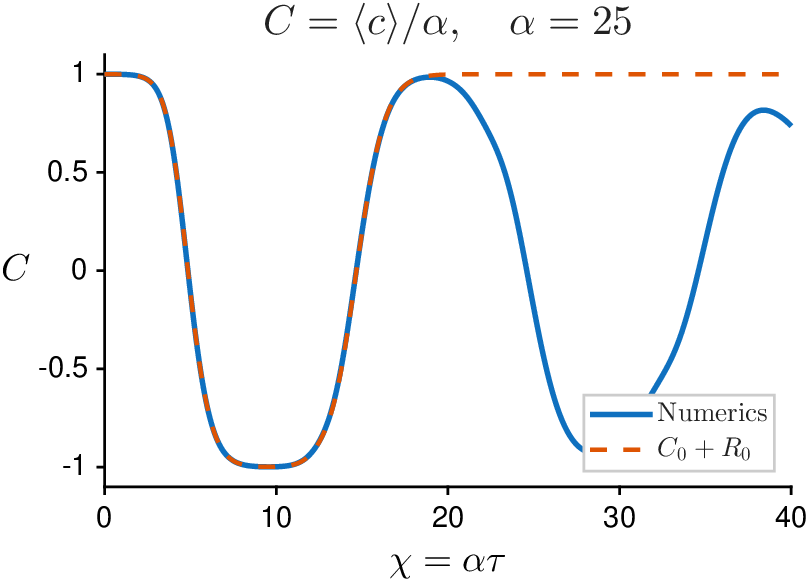}
    \caption{The matched leading profile: $\mathcal C_0$ combined with $R_0$.}
    \label{fig:r0}
\end{figure}

Finally, the normalization of $H(t)$ (or $H_1(t)$) is 
\begin{align}
    \frac{1}{2\pi}\int_{-\infty}^\infty \abs{H(t)}^2 dt = 1.
\end{align}
\indent Comparison with numerics is shown in Figure \ref{fig:r0}, we see it accurately predicts the first return.

\subsection{The $m$-th depletion and return}
\textit{Recognizing a clear pattern, we postulate a general leading-order profile for the $m$-th passage.}

\begin{figure}[H]
    \centering
    \includegraphics[width=0.5\linewidth]{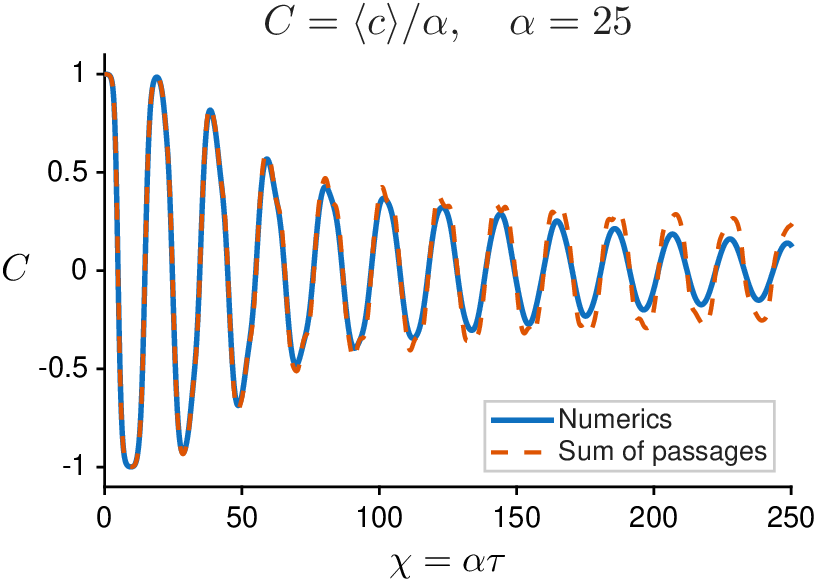}
    \caption{The additive stitched leading profile of equation \eqref{eqn:additive}.}
    \label{fig:scattering}
\end{figure}

Let's recall our first depletion profile:
\begin{align}
    \mathcal{C}_0 = -\int_{-\infty}^\infty 2e^{2u}e^{-e^{2u}} \tanh(s_0+u) \ du \equiv -\int_{-\infty}^\infty \abs{h_0(u)}^2 \tanh(s_0+u) \ du.
\end{align}
We see it takes the same form as our first return profile, $R_0$. The similarity goes even further: Let's perform a Fourier transform on $h_0(u)$,
\begin{align}
    H_0(t) = \int du \ h_0(u) e^{-itu} =  \int du \ \sqrt2e^{u} e^{-e^{2u}/2} e^{-itu} = 2^{-it/2}\Gamma\qty(\frac{1-it}{2}).
\end{align}
Compare with $H_1$:
\begin{align}
    H_1(t) = 2^{-3it/2} \frac{\Gamma\qty(\frac{1-it}{2})^2}{\Gamma\qty(\frac{1+it}{2})}.
\end{align}
Recognizing the clear similarity between $H_0$ and $H_1$, we make a bold postulation that, for the $m$-th passage, the weight function has the Fourier transform:
\begin{align}
    \Aboxed{H_m(t) = 2^{-(2m+1)it/2} \frac{\Gamma((1-it)/2)^{m+1}}{\Gamma((1+it)/2)^m}}. 
\end{align}
And from $s_0 = \chi - \log(4\alpha)$ and $s_1 = \chi - 3\log(4\alpha)$—and the fact that numerically we observe the period of oscillation is remarkably consistent—we predict the shifted time variable for the $m$-th passage is
\begin{align}
    \Aboxed{s_m = \chi - (2m+1)\log(4\alpha)}.
\end{align}
Define the inverse Fourier transform,
\begin{align}
    \Aboxed{h_m(u) = \frac{1}{2\pi} \int_{-\infty}^\infty H_m(t) e^{itu} dt}, 
\end{align}
and for the $m$-th passage, the leading profile is
\begin{align}
    \Aboxed{\Phi^{(0)}_m(s_m) = (-1)^{m+1} \int_{-\infty}^\infty \abs{h_m(u)}^2\tanh(s_m+u) \ du}. \label{eqn:Phi_m0}
\end{align}
It turns out these guesses must have some truth in them, because we observe strong agreement with the numerics, shown in Figure \ref{fig:scattering}. 

Numerical recipe: Evaluate 
\begin{align}
    \log H_m (t_k) = -\frac{(2m+1)it_k}{2}\log2 + (m+1) \log\Gamma\qty(\frac{1-it_k}{2})-m\log\Gamma\qty(\frac{1+it_k}{2})
\end{align}
on a Fourier grid. Exponentiate and inverse FFT to obtain the weight $h_m(u)$, then calculate $\rho_{m,j} = \abs{h_m(u_j)}^2 \ du$, and evaluate
\begin{align}
    \Phi^{(0)}_m(s) \approx (-1)^{m+1} \sum_j \rho_{m,j} \tanh(s+u_j). 
\end{align}
\indent The computational cost is negligible and independent of $\alpha$. In fact, the accuracy improves as $\alpha$ increases! This makes sense as $1/\alpha^2$ is the ``small parameter'' in our perturbation expansion, and also because a large $\alpha$ increases the time interval between successive passages, making the system resemble more of a ``dilute gas''. 

One might ask how to combine different passages to create a unified profile. The easiest and most obvious way is to simply add them. If we want to describe a total of $M$ passages ($M/2$ cycles):
\begin{subequations}
\begin{align}
    \Phi_\text{total}^{(0)} & = 1+\sum_{m=0}^{M-1}\qty[\Phi_m^{(0)}(s_m) - (-1)^m] \\
    & = 1 + \sum_m(-1)^{m+1} \int\abs{h_m(u)}^2 [1+\tanh(s_m+u)] \ du,
\end{align} \label{eqn:additive}
\end{subequations}
which produced Figure \ref{fig:scattering}. Here $s_m = \chi - (2m+1)\log(4\alpha)$ is the local passage time. This means successive passage centers are
separated by $2\log(4\alpha)$; the round-trip time is $4\log(4\alpha)$. 

Now, if we look at our Figure \ref{fig:scattering}, we see great agreement for short time, but the agreement worsens as time goes on. Is this the limit of our model? Not quite. There are two improvements we can implement which substantially improve late-time agreement: Taking into account the coherent interference between passages, and the Poissonian distribution of $N$-sectors of a coherent state. In our study, these effects were discovered by considering the scattering interpretation of $\Phi_\text{total}^{(0)}$, presented in the section below. For readers who only wish to see the final answers, we direct them to Section \ref{subsec:poisson_c}.

\subsection{(Optional) the scattering interpretation of $C$} 

\textit{We deduce the more complete and more consistent form of the solution, including coherent interference between passages, and Poisson-clock averaging.}

Let's examine our solution and try to figure out what everything means. Start by shifting to a single time coordinate, $y = s_m+u$. Now define
\begin{align}
    \psi_m(y) = h_m(y-s_m),
\end{align}
exchange the sum and integral,
\begin{subequations}
\begin{align}
    \Phi_\text{total}^{(0)} & = 1 + \int_{-\infty}^\infty\sum_m(-1)^{m+1} \abs*{\psi_m(y)}^2 (1+\tanh y) \ dy \\
    & = 1 - \int\qty[\abs{\psi_0}^2 - \abs{\psi_1}^2 + \abs{\psi_2}^2 - \abs{\psi_3}^2 + \dots ] (1+\tanh y) \ dy \\
    & = 1 - \int \qty[\sum_{m \text{ even}} \abs{\psi_m}^2  - \sum_{m \text{ odd}} \abs{\psi_m}^2] (1+\tanh y) \ dy \\
    & = 1- \qty[\sum_{m \text{ even}} \expval*{W(y)}{\psi_m} - \sum_{m\text{ odd}}\expval*{W(y)}{\psi_m}],
\end{align}
\end{subequations}
where we defined
\begin{align}
    W(y) \equiv 1+\tanh y.
\end{align}
Previously, we interpreted the profile as successive soliton transitions. In the current picture, we don't have an infinite number of solitons. At first sight, we simply have a single ``interaction potential'' $W=(1+\tanh y)$, and many competing ``wavefunctions'' $\psi$'s, and we're simply calculating the ``expectation value of potential energy''. 

Here we take the analogy further, and treat $\psi_m$ as literal wavefunctions. And we shall show that, in fact, this is equivalent to a scattering problem. Define
\begin{align}
    \Gamma_\pm = \Gamma\qty(\frac{1\pm it}{2}), 
\end{align}
we have
\begin{subequations}
\begin{align}
    H_m(t) & = 2^{-(2m+1)it/2} \frac{\Gamma_-^{m+1}}{\Gamma_+^m} = 2^{-it/2} \Gamma_- \qty[2^{-it}\frac{\Gamma_-}{\Gamma_+}]^m, \\
    \Rightarrow \psi_m(y) & = \frac{1}{2\pi} \int 2^{-it/2}\Gamma_- \qty[2^{-it}\frac{\Gamma_-}{\Gamma_+}]^m e^{it(y-x+(2m+1)\log(4\alpha))} dt \\
    & = \frac{1}{2\pi} \int 2^{-it/2}\cdot 2^{2it}  \alpha^{it} \Gamma_- \qty[2^{-it} \cdot 2^{4it}\alpha^{2it} \frac{\Gamma_-}{\Gamma_+}]^m e^{it(y-x)} dt \\
    & = \frac{1}{2\pi} \int (2\sqrt2\alpha)^{it}\Gamma_- \qty[(8\alpha^2)^{it}\frac{\Gamma_-}{\Gamma_+}]^m e^{it(y-x)} dt. 
\end{align}
\end{subequations}
Let 
\begin{align}
    \tilde \psi_0(t) = (2\sqrt2\alpha)^{it}\Gamma_-, \qquad \hat R(t) = (8\alpha^2)^{it}\frac{\Gamma_-}{\Gamma_+},
\end{align}
then
\begin{align}
    \psi_m(y) = \frac{1}{2\pi} \int \tilde\psi_0(t) \hat R(t)^m \ e^{it(y-x)} dt.
\end{align}
Note that since
\begin{align}
    \abs{(8\alpha^2)^{it}} = \abs{\Gamma_-/\Gamma_+} = 1, \qquad \Rightarrow \qquad \abs*{\hat R} = 1.
\end{align}
$\hat R$ is simply a phase. The scattering changes the phase, but not the Fourier-space magnitude. 
Let the Fourier transform of $\psi_m$ be $\tilde \psi_m$ (this was called $H_m$ previously), then we have
\begin{align}
    \tilde \psi_m = \hat R^m\tilde \psi_0. 
\end{align}
This means we can always define an operator $R$ in the real space ($y$-space):  satisfying
\begin{align}
    R = \mathcal F^{-1}(\hat R\mathcal F(\cdot)) \qquad \Rightarrow \qquad \psi_m = R^m \psi_0.
\end{align}
This really is a repeated-scattering problem! And there is a natural interpretation of the sum of the wavefunctions, 
\begin{align}
    \Psi_\text{out} = \psi_0 + R\psi_0 + R^2\psi_0 + R^3\psi_0 + \dots = \sum_m \psi_m,
\end{align}
it simply represents the amplitude of 
\begin{align}
    \text{free wave} + \text{wave scattered once} + \text{wave scattered twice} + ...,
\end{align}
in other words, it's the standard expansion used to calculate the final state amplitude.

However, there is no interpretation for the diagonal sum $\sum_m\abs{\psi_m}^2$, and seeing wavefunctions added diagonally and disjointed like that should worry us. Since all physical observables are proportional to $\abs{\Psi_\text{out}}^2$, we postulate that the diagonal sum is simply the first-order approximation of a coherent sum,
\begin{align}
    \sum_m \abs{\psi_m}^2 \longrightarrow \abs{\sum_m \psi_m}^2 = \sum_m\abs{\psi_m}^2 + \sum_{n< m}2\Re(\psi_n^*\psi_m).
\end{align}
Now, to be precise, we had two disjointed sums representing the even and odd terms. So the scattering operator really is $R^2$, and each ``scattering'' event is in fact a round-trip:
\begin{subequations}
\begin{align}
    \Psi_\text{even} &= \psi_0 + R^2 \psi_0 + R^4 \psi_0 + \dots , \\
    \Psi_\text{odd} & = R(\psi_0 + R^2 \psi_0 + R^4\psi_0 + \dots),
\end{align}
\end{subequations}
but this does not change any of the discussions above. Also, only wavefunctions of the same ``parity'' interfere with each other: the even wavefunctions interfere among themselves, and the odd wavefunctions interfere among themselves, but the two do not match. This is motivated by the following observations:
\begin{itemize}
    \item Combining even and odd sector cannot work. $\abs{\sum\psi_m}^2$ can never produce $\abs{\psi_i}^2 - \abs{\psi_j}^2$ terms. Therefore, the two sectors must be disjoint.
    \item Thankfully, numerically confirms this. Table \ref{tab:even_odd_overlap} below shows the largest overlap between even and odd passages is of the order $10^{-15}$. 
\end{itemize}
\begin{table}[H]
    \centering
    \begin{tabular}{c|c}
        $\alpha$ & Largest $\expval*{\psi_m, \psi_n}$ \\\hline
        5 & $1.3\times10^{-15}$ \\
        10 & $3.4\times10^{-15}$ \\
        25 & $1.4\times10^{-15}$ \\
        50 & $1.9\times10^{-15}$
    \end{tabular}
    \caption{Largest numerical even-odd overlap between $\psi_m$ and $\psi_n$ for various $\alpha$, where $m$ is even and $n$ is odd, $0\leqslant m,n \leqslant79$, and $\expval*{.,.}$ denotes ordinary inner product.}
    \label{tab:even_odd_overlap}
\end{table}
A few notes:
\begin{enumerate}
    \item Looking at these series, it is very tempting to attempt a geometric sum:
    \begin{align}
        \Psi_\text{even} = (1+R^2+R^4+\dots)\psi_0 \stackrel{?}{=} 
        \frac{1}{1-R^2}\psi_0,
    \end{align}
    but this doesn't work because $\abs*{R^2}=1$, unfortunately.
    \item On $W = 1+\tanh y$: We now see that it is in fact not the interaction energy or scattering potential. The scattering was done purely by $R$. But $W$ may be interpreted as some ``measuring window''.
    \item On numerical implementation: There is no need to include all combinations of $\psi_n^*\psi_m$. Instead, simply take into account the 2 nearest neighbors already gives excellent precision, and this is the procedure we used to produce the ``best'' numerical results in our comparison figures—for example, in Figure \ref{fig:c_comparison}.
\end{enumerate}
\indent Now, did this improvement help with fixing the late-time errors? Yes, but only marginally. In fact, since the passages are very well separated for large $\alpha$ (by $\sim2\log4\alpha$), they do not interfere very much. It is only when $\alpha$ is small that coherence has a larger effect. 

So at first sight, all this seems a great exercise in futility. However, that's not quite true. In fact, this scattering interpretation did help us uncover the true culprit. Let's look at the quantity
\begin{align}
    C(x) = 1- \int \qty[\abs{\Psi_\text{even}(y)}^2 - \abs{\Psi_\text{odd}(y)}^2] W(y) \ dy
\end{align}
should also make us feel uneasy for two reasons:
\begin{enumerate}
    \item The suspicious minus sign. 
    \item No interference between even and odd terms.
\end{enumerate}
A typical quantum mechanical amplitude,
\begin{gather}
    \Psi = \Psi_\text{even} + \Psi_\text{odd} = \psi_0 + \psi_1 + \psi_2 + \psi_3 + \dots, \\
    \Rightarrow \ \abs*{\Psi}^2 = \sum_m \abs*{\psi_m}^2 + \sum_{m\neq n}\psi_m^*\psi_n,
\end{gather}
would not have a relative minus sign, but would have interference between all passages. So what happened? In other words, what hidden structure could preserve interference when $m-n$ is even, but destroy it when $m-n$ is odd?

Upon more careful consideration, we realize there is an obvious solution that perfectly explains both anomalies. First of all, the alternating sign must be caused by a sector-dependent phase, one possibility is $(-1)^{Nm}$ (this can be verified by considering the action of the scattering operator $\hat R$ on even- and odd-$N$ subspace, so the model is internally consistent. But for now, we simply assume such a phase exists).

Secondly, we need to re-instate the statistical distribution of our coherent state. Recall evaluating expectation values of an operator $\hat A$ (that connects sectors $\ket N$ to $\ket{N-m}$) on a coherent state really relies on a Poisson-weighted sum \cite{WALLS_2008}, i.e., 
\begin{subequations}
\begin{gather}
    \expval*{\hat A} = \alpha^m \sum_{N=0}^\infty p_N \sqrt{\frac{N!}{(N+m)!}}\mel*{N}{\hat A}{N+m},\\
    \text{where } \ p_N = e^{-\abs{\alpha}^2} \frac{\abs{\alpha}^{2N}}{N!}, \\
    \text{and } \ \frac{\expval*{\hat A}}{\alpha^m}  \equiv \mathbb E_N \qty[\sqrt{\frac{N!}{(N+m)!}}\mel*{N}{\hat A}{N+m}],
\end{gather}
\end{subequations}
where we defined Poisson sum $\mathbb E_N$ to be sum weighted by $p_N$, that is, $\mathbb E_N \dots = \sum p_N \dots$. 

So far, we have neglected the statistical distribution of $N$-sectors in time evolution—we have pretended we only had one state with $\alpha^2$ photons. In reality, the entire Hilbert space is built out of many fixed-$N$ sub-spaces, where $N$ represents the maximum pump photon number in the subspace. So more precisely, we should have written
\begin{align}
    \ket{\Psi} \approx \sum_{N,m} \sqrt{p_N} \ket{N \text{ sector}}\otimes(-1)^{Nm} \ket{\psi_{m|N}}.
\end{align}
where we set our phase to be $(-1)^{Nm}$. We shall see this is the right choice. Now let's calculate $\expval*{c}$. First note that, since $c$ lowers $\ket N$ to $\ket{N-1}$, only terms of the form $\bra N c \ket{N+1}$ survive. Now $C=\expval*{c}/\alpha$ becomes:
\begin{align}
    C & \sim 1 - \sum_{N,m,n} p_N (-1)^{N(m-n)} (-1)^n \mel*{\psi_{m|N}}{W(y)}{\psi_{n|N+1}}.
\end{align}
$\sum_N p_N(-1)^{N(m-n)}$ has an interesting property:
\begin{align}
    \sum_N p_N(-1)^{N(m-n)} = \begin{cases}
        1, & m-n \text{ even}, \\
        e^{-2\alpha^2}, & m-n \text{ odd},
    \end{cases}
\end{align}
as can be readily checked. But this means only terms with the same parity survive, whereas terms with opposite parities are suppressed by $e^{-2\alpha^2}$ which completely disappear when $\alpha$ is large! This explains why $\Psi_\text{even}$ and $\Psi_\text{odd}$ do not mix. 

But there's more. We still have a factor of $(-1)^n$. Since we established $m-n$ is even, it means that there are two possibilities:
\begin{align}
    (-1)^n = \begin{cases}
        +1, & m, n \text{ both even},\\
        -1, & m, n \text{ both odd}.
    \end{cases}
\end{align}
This explains the minus sign! So in the end, we have
\begin{align}
    C \simeq 1-\expval*{W}{\Psi_\text{even}} + \expval*{W}{\Psi_\text{odd}},
\end{align}
exactly as desired. This is why the Poisson distribution of $N$-states must be taken into account. 

Now, back to the equation for $\expval*{\hat A}$ for a moment—note that the observable (in our case $\hat A = C = \expval*{c}/\alpha$) must have $N$-dependence. Otherwise, Poisson averaging does nothing. To introduce $N$-dependence in our solution, we must look at the time variable. Recall in our definitions:
\begin{align}
    \chi = \alpha \tau, \quad \Rightarrow \quad \tau = \frac{\chi}{\alpha},
\end{align}
but in doing so, we have assumed every sector evolves in time at the same rate, determined by the mean-field amplitude $\alpha$. In reality each sector should evolve approximately according to
\begin{align}
    \chi_N \equiv \tau \sqrt N  = \chi \frac{\sqrt N}{\alpha},
\end{align}
meaning sectors with $N>\alpha^2$ run faster, whereas sectors with $N<\alpha^2$ run slower. This effect decreases with $\alpha$, but accumulates with $\chi$. So instead of
\begin{align}
    s_m = \chi-(2m+1)\log4\alpha,
\end{align}
we really should assign a different local time coordinate to each sector,
\begin{align}
    s_{m,N} \approx \chi \frac{\sqrt N}{\alpha} - (2m+1) \log(4\sqrt{N}).
\end{align}
In fact, we found $N\rightarrow N+\frac{3}{2}$ gives the most accurate predictions. The matrix element connects sector $N$ and $N+1$. However, sector $N$ has length $N+1$, and we found that normalizing the sector Hamiltonian according to the sector length is a better choice than normalizing according to maximum photon number. That is, $H_N \rightarrow \sqrt{N+1} \bar H_N$ provides the most natural choice where $\bar H_N$ transforms in a nice way (to be treated in a future work). This means the most natural sector time scaling is $\sqrt{N+1}\tau$ as opposed to $\sqrt N\tau$. So the squared clocks of the two sectors are $N+1$ and $N+2$, whose midpoint is $N+3/2$. And we shall replace $N\rightarrow N+\frac{3}{2}$ in the above expression, make the replacement $s_m\rightarrow s_{m,N}$, then perform a Poisson sum over different $N$-sectors, and we arrive at our final expression, to be listed explicitly below.

\subsection{Coherent interference and Poisson averaging of clocks} \label{subsec:poisson_c}

\textit{Taking into account the interference between passages, as well as the statistical distribution of coherent state, we drastically improve late-time numerical agreement.}

In short, we studied the strange relative minus sign between even- and odd-wavefunctions, and the fact that states of  opposite parity do not mix. We see these particularities can be explained by taking into account the Poisson distribution of different $N$-sectors of a coherent state. Explicitly: For every sector $N$, we construct the even and odd wave packet coherent sum using the sector's own clocks:
\begin{subequations}
\begin{align}
    \Psi_\text{even, N}(y, \chi) & = \sum_{\substack{m \text{ even} \\ m=0}}^{M-1} h_m(y-s_{m,N}(\chi)), \\
    \Psi_\text{odd, N}(y, \chi) & = \sum_{\substack{m \text{ odd} \\ m=0}}^{M-1} h_m(y-s_{m,N}(\chi)), \\
    \text{where } \ 
    s_{m,N} & \simeq \chi \frac{\sqrt{N+\frac{3}{2}}}{\alpha} - (2m+1) \log(4\sqrt{N+\frac{3}{2}}).
\end{align}
Then, for each sector $N$, we have
\begin{gather}
    C_N(\chi) = 1-\int_{-\infty}^\infty\qty[\abs{\Psi_\text{even, N}(y, \chi)}^2 - \abs{\Psi_\text{odd, N}(y, \chi)}^2] W(y) \ dy,\\
    \text{averaging, } \quad C_\text{final}(\chi) = \sum_N p_N C_N(\chi), \quad \text{where }\quad  p_N = e^{-\abs{\alpha}^2} \frac{\abs{\alpha}^{2N}}{N!}.
\end{gather}
\end{subequations}
So we have arrived at the final expression. More compactly, this is written as
\begin{align}
    \Aboxed{C_\text{final} = \mathbb E_{N} \qty[1-\int \qty(\abs{\Psi_{\text{even}, N}}^2 - \abs{\Psi_{\text{odd}, N}}^2) W \ dy]}, \label{eqn:C_poisson}
\end{align}
where $\mathbb E_N$ is understood as a Poisson sum, $\mathbb E_N \dots = \sum p_N \dots$. 

So now, we finally have a close agreement, even for ultra-long-time! 

\begin{figure}[H]
    \centering
    \includegraphics[width=0.5\linewidth]{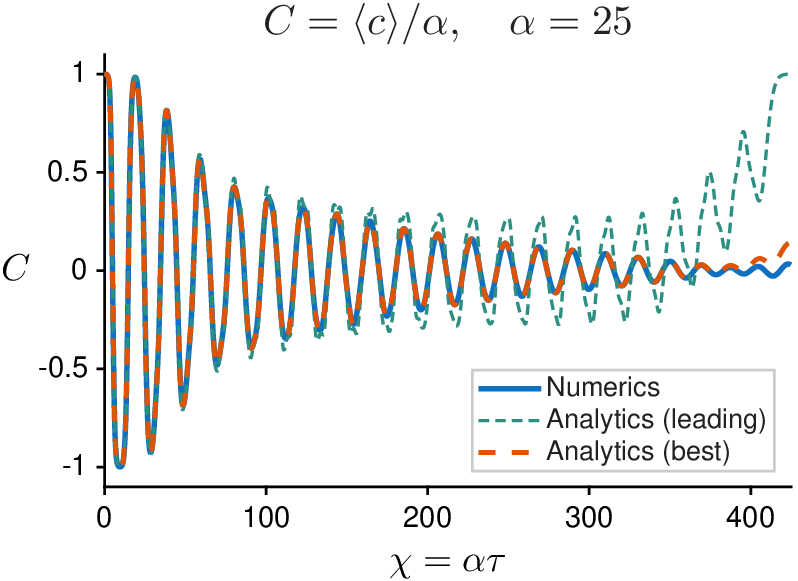}
    \caption{Long time agreement, reproduced from Figure \ref{subfig:c_long_time_best}}
    \label{fig:c_long_time_best_2}
\end{figure}
Physically, having different passages occur at slightly different times dephases the oscillations, which produces the desired decaying envelope. And they occur at different times because different photon sectors oscillate at different frequencies, analogous to how Rabi oscillation collapses in the Jaynes-Cummings model \cite{EBERLY_1980}. 

A note on numerical evaluations: 
\begin{itemize}
    \item The infinite Poisson sum need not slow down the computation. In our numerical algorithms, we in fact use Gauss quadratures for the Poisson sum, \cite{GOLUB_1969}
    \begin{align}
        \sum_{N=0}^\infty p_N f(N) \approx \sum_{q=1}^Q\omega_qf(N_q),
    \end{align}
    and we used $Q = 97$ nodes in the reported calculations. This is equivalent to evaluating 97 sectors, regardless of how large $\alpha$ is. But even if we do not use this simplification, the time scales linearly as $\mathcal O(\alpha)$.
    \item For most practical purposes, we are likely only studying the very first depletion profile. In this case, we recommend ignoring passage interference, while only taking into account the Poisson sum.
\end{itemize}

\subsection{The first return correction} \label{subsec:R1}

\textit{We derive the first correction to the first return.}

Now let's derive the correction to the first return profile. This might not seem necessary since we already have great numerical agreements, but we include the derivation because the result in this section becomes absolutely necessary in the 2nd- and 3rd-order correlation calculations. 

The various indices can be confusing. For readers who wish to follow this derivation (and every 2nd- and 3rd-order derivation), we highly recommend they have the corresponding Reference Table at hand, especially the ``4 Corrections'' tab, which would immediately render the indices almost trivial to understand. 

Recall that the very first exceptional tower we discovered leads to the first return terms,
\begin{align}
    R_0(w) = -1 + \sum_{n=1}^\infty a_nw^n = -1 + \frac{w}{64} - \frac{5w^2}{2048} + \frac{189w^3}{131072} - \dots, \label{eqn:R0_formal_an}
\end{align} 
corresponding to the tower of green terms in the Reference Table (multiplied by the multiplicative factors in the ``5 Indices'' tab). 

Let's use $E_{r,j}(x)$ to denote the exceptional term in the $r$-th harmonics, where $r=N-j$. Then these exceptional terms contribute
\begin{align}
    C_r^\text{exc} = 2(-z)^j\, j!\, E_{r,j}(x), \qquad \text{ and } \qquad a_n = 2(-1)^n \, n! \, E_{2n,n}
\end{align}
We also notice that the relevant leading terms only occur at $r=2n$, that is, $r=2, 4, 6\dots$, and these terms contribute to $C_{2n}$. For example,
\begin{align}
    E_{2,1} = -\frac{1}{128}, \quad E_{4,2} = -\frac{5}{8192}, \quad E_{6,3} = -\frac{63}{524288}, \dots
\end{align}
these are simply the green terms in the Table. 

Now note that immediately after $C_{2n}$, the next-order corrections come from $C_{2n+1}$. The exceptional term contribution is
\begin{align}
    C_{2n+1}^\text{exc} = 2(-z)^n\, n!\, E_{2n+1,n}(x),
\end{align}
these are the blue terms in the Reference Table, whose contribution to the full amplitude is $\epsilon^{2n+1}C_{2n+1}^\text{exc}$. 

It's a convenient book-keeping device to define the ratio,
\begin{align}
    q_n(x) = \frac{E_{2n+1,n}(x)}{E_{2n,n}}.
\end{align}
Since 
\begin{align}
    a_n = 2(-1)^n\,n!\,E_{2n,n},
\end{align}
the odd exception becomes
\begin{align}
    \epsilon^{2n+1}C_{2n+1}^\text{exc} = \epsilon a_n q_n(x)w^n,
\end{align}
and the first return correction thus has the structure
\begin{align}
    R_1(w,x) = c + \sum_{n=1}^\infty a_nq_n(x)w^n, \label{eqn:R1_formal}
\end{align}
for some constant $c$. 

Let's work through some explicit examples, both to see the ratio $q_n$ in action, and to verify the expansion in equation \eqref{eqn:R1_formal} is correct. For $n=1$, we can read from the table 
\begin{subequations}
\begin{gather}
    E_{2,1} = -\frac{1}{128}, \qquad E_{3,1}(x) = -\frac{1}{512} + \frac{7x}{512} - \frac{x^2}{256}, \\
    \Rightarrow \ q_1(x) = \frac{E_{3,1}(x)}{E_{2,1}} = \frac{x^2}{2}-\frac{7x}{4}+\frac{1}{4},
\end{gather}
\end{subequations}
its contribution to amplitude is
\begin{subequations}
\begin{align}
    \epsilon^3 C_3^\text{exc} & = \epsilon^3 (-2z) E_{3,1}(x) \\
    & = \epsilon^3(-2z) \qty(-\frac{1}{512} + \frac{7x}{512}- \frac{x^2}{256}) \\
    & = \epsilon^3 z\qty(\frac{1}{256}-\frac{7x}{256}+\frac{x^2}{128}) \\
    & = \epsilon w \qty(\frac{1}{256}-\frac{7x}{256}+\frac{x^2}{128}),
\intertext{and since $a_1 = 1/64$ (read off equation \eqref{eqn:R0_formal_an}),}
    \Rightarrow \ \epsilon^3 C_3^\text{exc} & = \epsilon \frac{w}{64} \qty(\frac{1}{4} - \frac{7x}{4} + \frac{x^2}{2}) \\
    & = \epsilon a_1 q_1(x) w,
\end{align}
\end{subequations}
which is exactly the format of equation \eqref{eqn:R1_formal}! We've verified it to the first order. 

We can verify it to one more order. Read off the next green and blue terms, we have
\begin{gather}
    E_{4,2} = -\frac{5}{8192}, \qquad E_{5,2}(x) = -\frac{171}{32768} + \frac{95x}{16384} - \frac{5x^2}{4096}. \\
    \Rightarrow \ q_2(x) = \frac{E_{5,2}(x)}{E_{4,2}} = 2x^2 - \frac{19x}{2} + \frac{171}{20}.
\end{gather}
And since $a_2 =  -\frac{5}{2048}$, the $n=2$ contribution to $R_1$ is
\begin{align}
    -\frac{5w^2}{2048}\qty(2x^2 -\frac{19x}{2}+\frac{171}{20}). 
\end{align}
The first few $a_n$ and $q_n$ are:
\begin{table}[H]
    \centering
    \begin{tabular}{c|c|c}
        $n$ & $a_n$ in $R_0$ & $q_n(x)$ \\\hline
        1 & 1/64 & $\frac{1}{2}x^2 -\frac{7}{4}x+\frac{1}{4}$ \\[0.5em]
        2 & $-5/2048$ & $2x^2 - \frac{19}{2}x + \frac{171}{20}$ \\[0.5em]
        3 & 189/131072 & $\frac{9}{2}x^2 - \frac{93}{4}x + \frac{691}{28} $ \\[0.5em]
        4 & $-4671/2097152$ & $8x^2 - 43x + \frac{33809}{692}$ \\[0.5em]
        5 & $1900125/268435456$ & $\frac{25}{2}x^2 - \frac{275}{4}x + \frac{182463}{2252}$ \\[0.5em]
        6 & $-86477625/2147483648$ & $18x^2 - \frac{201}{2}x + \frac{460039}{3796}$
    \end{tabular}
    \caption{The first few $a_n$ and $q_n$. }
    \label{tab:an_and_qn}
\end{table}
Looking at $q_n(x)$, we easily find a pattern to the coefficients of $x$ and $x^2$, namely,
\begin{align}
    [x^2]q_n = \frac{n^2}{2}, \qquad  [x] q_n = -3n^2 + \frac{5}{4}n.
\end{align}
So we have
\begin{align}
    q_n (x) = \frac{n^2x^2}{2} + \qty(-3n^2+\frac{5}{4}n)x + q_n(0). 
\end{align}
On the other hand, the constants, 
\begin{align}
    q_n(0) = \qty{\frac{1}{4}, \frac{171}{20}, \frac{691}{28}, \frac{33809}{692}, \frac{182463}{2252}, \dots}
\end{align}
are the difficult part! They have no clear polynomial fits. However, we found they can once again be expressed in terms of Franel numbers,
\begin{align}
    q_n(0) = \frac{1}{4}-\frac{15n}{4}+\frac{49n^2}{12}-\frac{2n^2}{3}\frac{F_{n-1}}{F_n}, \label{eqn:qn0_ratio}
\end{align}
meaning the full expression is
\begin{subequations}
\begin{align}
    q_n(x) & = \frac{n^2x^2}{2} + \qty(-3n^2+\frac{5}{4}n)x + \frac{1}{4}-\frac{15n}{4}+\frac{49n^2}{12}-\frac{2n^2}{3}\frac{F_{n-1}}{F_n} \\
    & = \qty[\frac{(x-3)^2}{2}-\frac{5}{12}]n^2 + \frac{5}{4}(x-3)n + \frac{1}{4} - \frac{2n^2}{3}\frac{F_{n-1}}{F_n}.
\end{align}
\end{subequations}
Now recall 
\begin{align}
    R_0(w) = -1 + \sum_{n=1}^\infty a_nw^n,  \qquad a_n = 2(-1)^{n+1} \frac{(n!)^3F_n}{256^n}. 
\end{align}
Define 
\begin{align}
    D = w\frac{d}{dw}, \qquad \Rightarrow \qquad DR_0 = \sum_{n=1}^\infty na_nw^n, \qquad D^2R_0 = \sum_{n=1}^\infty n^2a_nw^n,
\end{align}
then from $R_1 = c + \sum_n q_n a_n  w^n$, we can write the polynomial part of $q_n$ in terms of Euler derivative,
\begin{align}
    R_1 = c + \qty[\frac{(x-3)^2}{2}-\frac{5}{12}]D^2R_0 + \frac{5}{4}(x-3)DR_0 + \frac{1}{4}(R_0+1) + T(w)
\end{align}
where we defined the remaining piece $T$ as
\begin{subequations}
\begin{align}
    T(w) & = -\frac{2}{3} \sum_{n=1}^\infty n^2 a_n \frac{F_{n-1}}{F_n}w^n \\
    & = -\frac{2}{3}\sum_{n=1}^\infty n^2 \cdot 2(-1)^{n+1}  \frac{(n!)^3F_n}{256^n} \frac{F_{n-1}}{F_n}w^n \\
    & = \frac{4}{3}\sum_{n=1}^\infty (-1)^n \frac{(n!)^3n^2}{256^n} F_{n-1}w^n,
\end{align}
\end{subequations}
and we'd like to express this in terms of $R_0$. First we'll try to write it as a single sum: Define
\begin{align}
    \mathcal R_0(w) \equiv \frac{1-R_0}{2} = \frac{1}{2}\qty(2-\sum_{n=1}^\infty a_nw^n) = 1-\sum_{n=1}^\infty (-1)^{n+1} \frac{(n!)^3F_n}{256^n}w^n \equiv \sum_{n=0}r_n w^n,
\end{align}
where we defined
\begin{align}
    r_n \equiv (-1)^n \frac{(n!)^3F_n}{256^n},
\end{align}
and note $r_0 = 1$. Now we see $T(w)$ can be written in terms of $r_{n-1}$:
\begin{subequations}
\begin{align}
    T(w) & = \frac{4}{3}\sum_{n=1}^\infty (-1)\frac{n^3\cdot n^2}{256} \times \underbracket{(-1)^{n-1} \frac{[(n-1)!]^3}{256^{n-1}}F_{n-1}}_{\mathclap{\text{$r_{n-1}$}}} w^n \\
    & = -\frac{4}{3} \sum_{n=1}^\infty \frac{n^5}{256} r_{n-1} w^n, 
\intertext{shift $n\rightarrow m=n-1$,}
    \Rightarrow T(w) & = -\frac{1}{192}\sum_{m=0}^\infty (m+1)^5 r_m w^{m+1} \\
    & = -\frac{w}{192} \sum_{m=0}^\infty (D+1)^5 r_m w^m \\
    & = -\frac{w}{192} (D+1)^5 \mathcal R_0(w) \\
    & = -\frac{w}{384} (D+1)^5 (1-R_0).
\end{align}
\end{subequations}
Combine everything, we have
\begin{align}
    R_1(w,x) = c + \qty[\frac{(x-3)^2}{2}-\frac{5}{12}]D^2R_0 + \frac{5}{4}(x-3)DR_0 + \frac{1}{4}(R_0+1)  -\frac{w}{384}(D+1)^5[1-R_0]. \label{eqn:R1_c}
\end{align}
It's tempting to set $c = -\frac{1}{4}$ to cancel the constant term by making some initial condition or asymptotic argument. Indeed $c = -\frac{1}{4}$ turns out to be correct, but let's consider carefully why it is the case. The constant is determined by matching the return profile with the depletion profile, order by order, we expect
\begin{align}
     C_0 \longrightarrow R_0, \qquad C_1 \longrightarrow R_1,
\end{align}
during the ``transition time''—which is $\log(4\alpha) < \chi < 3\log(4\alpha)$, after the first depletion but before the first return. But for all practical purposes, this is equivalent to saying 
\begin{align}
    z = e^{2(\chi-\log4\alpha)} \ \longrightarrow \ \infty, \qquad w = 256e^{2(\chi-3\log4\alpha)} \ \longrightarrow \ 0,
\end{align}
due to the exponential scaling. 

We can verify that the equality $C_0 = R_0$ is satisfied in this limit. Recall
\begin{align}
    C_0(z) = 2 \times \frac{e^{1/z}}{z}E_1(1/z) - 1,
\end{align}
where asymptotically \cite[Eq.~6.6.2]{NIST_DLMF}, we have
\begin{subequations}
\begin{gather}
    E_1(1/z) \longrightarrow \log z - \gamma \quad \text{ as } \quad z\rightarrow\infty, \\
    \Rightarrow \ C_0(z) \longrightarrow 2 \frac{e^{1/z}}{z}\log z - 2\frac{e^{1/z}}{z}\gamma - 1 \longrightarrow -1 \quad \text{ as } \quad z \rightarrow \infty,
\end{gather}
\end{subequations}
which just says the pump depletes to $-1$. Now check for $R_0$:
\begin{align}
    R_0(w) = -1 + \frac{w}{64} + \dots \ \longrightarrow \ -1 \quad \text{ as } \quad w\rightarrow0,
\end{align}
this is true by our definition of $R_0$. So the equality is satisfied at the 0th order. 

Now for the 1st order, we should have $C_1 = R_1$, and we shall use this to determine our constant $c$ in $R_1$. Recall from equation \eqref{eqn:C1_final} that $C_1$ is 
\begin{align}
    C_1 = \frac{1}{4} \qty[(2x^2-4x+3)D_z^2 + 5(x-1)D_z+1]C_0(z) \longrightarrow -\frac{1}{4}, \quad \text{ as } \quad z\rightarrow\infty,
\end{align}
meaning we should have
\begin{align}
    R_1 \longrightarrow -\frac{1}{4} \quad \text{ as } \quad w\rightarrow0,
\end{align}
looking back to equation \eqref{eqn:R1_c}, we see this is only satisfied if $c=-\frac{1}{4}$. And so we have fixed our constant. 

Then we have the final expression,
\begin{align}
    \Aboxed{R_1(w,x) & = \qty[\frac{(x-3)^2}{2}-\frac{5}{12}]D^2R_0 + \frac{5}{4}(x-3)DR_0 + \frac{1}{4}R_0  -\frac{w}{384}(D+1)^5[1-R_0]}. \label{eqn:R1_final}
\end{align}
The various derivatives are easily calculated from the Fourier transformed form of $R_0$, reproduced from equation \eqref{eqn:R0_FT}:
\begin{align}
    R_0(w) & = 1-2\int \frac{\abs*{\hat\phi(k)}^2}{1+wk^2/64} d^2k, \tag{\ref{eqn:R0_FT}}
\end{align}
let $\lambda = \frac{wk^2}{64}$, then
\begin{subequations}
\begin{align}
    DR_0 & = 2\int \frac{\abs*{\hat\phi(k)}^2\lambda}{(1+\lambda)^2}d^2k, \\
    D^2R_0 
    & = 2\int \abs*{\hat\phi(k)}^2 \frac{\lambda(1-\lambda)}{(1+\lambda)^3} d^2k, \\
    (D+1)^5 [1-R_0] & = 2\int d^2k \abs*{\hat\phi(k)}^2 \frac{\lambda^4-26\lambda^3+66\lambda^2-26\lambda+1}{(1+\lambda)^6}. 
\end{align}
\end{subequations}

\pagebreak

\section{2nd-order correlations}

We shall speed through the derivation of $\expval*{c^\dag c}$ and $\expval*{cc}$. While they can get complex in places, the derivations follow very closely the derivation for $\expval*{c}$, and no genuinely new ideas emerge. So instead of deriving every step, we will heavily cite figures directly from the Reference Sheets in the Supplementary Materials. 

Note these two correlators are enough to describe the system to 2nd order—$\expval*{c}, \expval*{c^\dag c}$ and $\expval*{cc}$ completely define all 1st- and 2nd-order moments and variances. Their precise relations can be found in Appendix \ref{app:moments_identities}.

\subsection{Pump photon number $\expval{c^\dag c}$}

\subsubsection{First depletion profile}

Call the expansion 
\begin{align}
    P(\alpha,\chi) \equiv \frac{\expval*{c^\dag c}}{\alpha^2} = 1 - \frac{\sinh^2\chi}{\alpha^2} + \frac{p_2(\chi)}{\alpha^4} + \frac{p_3(\chi)}{\alpha^6} + \dots
\end{align}
and each $p_m(\chi)$ is a finite exponential polynomial, 
\begin{align}
    p_m(\chi) = Z_m(\chi) + \sum_{j=1}^m \qty[e^{2j\chi}H_{m,j}^{(P)}(\chi) + e^{-2j\chi} H_{m,j}^{(P)}(-\chi)],
\end{align}
where $Z_m(\chi)$ is the $j=0$ term. The leading-order harmonics have the simple form
\begin{align}
    H_{j,j}^{(P)} = (-1)^j \frac{4j \cdot j!}{16^j}.
\end{align}
Compare this with the coefficients for $\expval*{c}$ as in equation \eqref{eqn:C0_coeff} (substitute $z = e^{2\chi}/16\alpha^2$), reproduced here:
\begin{gather}
    C_0(z) = 1+2\sum_{j=1}^\infty(-1)^j j!z^j, \label{eqn:C0_series} \qquad
    \Rightarrow \qquad H_{j,j}^{(C)} = (-1)^j \frac{2j!}{16^j}, 
\end{gather}
we have that 
\begin{align}
    H_{j,j}^{(P)} = 2j \ H_{j,j}^{(C)}.
\end{align}
The $j$ will become an Euler derivative. Explicitly, substituting $z = e^{2\chi}/16\alpha^2$ into $H_{j,j}^{(P)}$, we have the leading depletion profile is
\begin{align}
    P_\text{dep}^{(0)}  = 1-4z+16z^2-72z^3+384z^4-\dots = 1 + \sum_{j=1}^\infty 4j(-1)^jj!z^j, \label{eqn:P_dep_0}
\end{align}
whereas if we act the Euler derivative $D = z \frac{d}{dz}$ on $C_0(z)$, we have
\begin{align}
    DC_0(z) = 2\sum_{j=1}^\infty j(-1)^jj!z^j.
\end{align}
Compare the two, we have
\begin{align}
    \Aboxed{P_\text{dep}^{(0)} = 1+2DC_0(z)}.
\end{align}
\indent Now note the series for $P_\text{dep}^{(0)}$ is factorially divergent. Borel transform gives
\begin{align}
    \hat P_\text{dep}^{(0)}(zt) = 1+4\sum_{j=1}^\infty j(-zt)^j = 1-\frac{4zt}{(1+zt)^2} = \qty(\frac{1-zt}{1+zt})^2. 
\end{align}
So we have
\begin{align}
    P_\text{dep}^{(0)} = \int_0^\infty e^{-t} \qty(\frac{1-zt}{1+zt})^2 dt. 
\end{align}
We will make the same substitutions as for $C_0$, including
\begin{subequations}
\begin{gather}
    s_0 = \chi-\log(4\alpha), \qquad z=e^{2s_0}, \qquad t = e^{2u} \quad \Rightarrow \quad dt = 2e^{2u}du, \\
    \Rightarrow \ \frac{1-zt}{1+zt} = -\tanh(s_0+u), \\
    \Rightarrow \ \boxed{P_\text{dep}^{(0)}(s_0) = \int_{-\infty}^\infty 2e^{2u}e^{-e^{2u}} \tanh^2(s_0+u) \ du},
\end{gather}
\end{subequations}
with
\begin{align}
    \int 2e^{2u}e^{-e^{2u}} du = 1
\end{align}
as usual. 

\subsubsection{First and second depletion correction} \label{subsubsec:P_dep_1}
We shall simply claim the key answers. The relevant quadratic polynomial is
\begin{align}
    A_1(j,\chi) = 1+\frac{7}{4}(\chi-1)j+\frac{2\chi^2-4\chi+3}{4}j^2,
\end{align}
replacing $j\mapsto D$ gives
\begin{align}
    P_\text{dep}^{(1)} = \frac{1}{2} + A_1(D,\chi)(P_\text{dep}^{(0)}-1).
\end{align}
Or, more explicitly,
\begin{align}
    P_\text{dep}^{(1)} & = \frac{2\chi^2-4\chi+3}{4}D^2P_\text{dep}^{(0)} + \frac{7(\chi-1)}{4}DP_\text{dep}^{(0)}+P_\text{dep}^{(0)}-\frac{1}{2}. \label{eqn:P_dep_1}
\end{align}
Using $P_\text{dep}^{(0)} = 1+2DC_0$, this can also be written as
\begin{align}
    \Aboxed{P_\text{dep}^{(1)} = \frac{1}{2} + \frac{1}{2} \qty[(2\chi^2-4\chi+3)D^3+7(\chi-1)D^2+4D]C_0}. \label{eqn:P_dep_1_full}
\end{align}
\indent The second correction to depletion is
\begin{empheq}[box=\fbox]{align}
    P_\text{dep}^{(2)} & = \bigg[ \frac{(2\chi^2-4\chi+3)^2}{32}D^4 + \frac{76\chi^3-204\chi^2+213\chi-87}{96}D^3 + \\
    & \quad + \frac{78\chi^2-138\chi+107}{64}D^2 + \frac{9\chi-24}{64}D -\frac{1}{4} \bigg] P_\text{dep}^{(0)} + \frac{1}{8}-\frac{z}{32}-\frac{1}{64z}. \label{eqn:P_dep_2}
\end{empheq}

\subsubsection{First return profile}
The same exceptional scale appears, 
\begin{align}
    w = \epsilon^2z = \frac{e^{2\chi}}{16\alpha^6}.
\end{align}
The population exceptional terms emerge as (to read them from the Reference Table: Find the green cells in the ``4 Corrections'' tab, multiply them by the corresponding factors in the ``5 Indices'' tab)
\begin{align}
    P_\text{ret}^{(0)} \sim 1-\frac{w}{32} + \frac{5w^2}{512} - \frac{567w^3}{65536} + \frac{4671w^4}{262144} - \dots, \label{eqn:P_ret_0_series}
\end{align}
compare with the amplitude return profile,
\begin{align}
    R_0(w) \sim -1 + \frac{w}{64} - \frac{5w^2}{2048} + \frac{189w^3}{131072} - \frac{4671w^4}{2097152}+\dots , 
\end{align}
we see that (define $D = w\frac{d}{dw}$)
\begin{align}
    P_\text{ret}^{(0)} = 1-2DR_0. \label{eqn:P_ret_0_id}
\end{align}
In terms of Franel numbers,
\begin{align}
    P_\text{ret}^{(0)}(w) = 1+4\sum_{j=1}^\infty j(-1)^j (j!)^3 \frac{F_j}{256^j} w^j.
\end{align}
Or in terms of the resolvent operator, first rewrite $R_0$ as
\begin{align}
    R_0(w) = 1-2\expval{\qty(1+\frac{w}{4}Q^\dag Q)^{-1}} = \expval{\frac{\tilde\Delta-1}{\tilde\Delta+1}},
\end{align}
where we defined
\begin{align}
    \tilde\Delta \equiv \frac{w}{4}Q^\dag Q,
\end{align}
and we found the corresponding population return profile to be
\begin{align}
    P_\text{ret}^{(0)} = \expval{\qty(\frac{\tilde\Delta-1}{\tilde\Delta+1})^2}.
\end{align}
Now since the only $\chi$ dependence of $\tilde \Delta$ is $\tilde \Delta \propto e^{2\chi}$, taking the $\chi$ derivative gives
\begin{align}
    \frac{d}{d\chi}\qty(\frac{\tilde\Delta-1}{\tilde\Delta+1}) = 1-\qty(\frac{\tilde\Delta-1}{\tilde\Delta+1})^2.
\end{align}
Take the expectation value, we have
\begin{align}
    P_\text{ret}^{(0)} = 1-\frac{dR_0}{d\chi}. 
\end{align}
One can check that this is in fact the same statement as equation \eqref{eqn:P_ret_0_id}, i.e. 
\begin{align}
    \frac{dR_0}{d\chi} = 2DR_0.
\end{align}
Now recall $R_0 = \int \abs{h_1}^2 \tanh$, we simply have
\begin{align}
    \Aboxed{P_\text{ret}^{(0)}(s_1) = \int_{-\infty}^\infty \abs{h_1(u)}^2 \tanh^2(s_1+u) \ du}. 
\end{align}

\subsubsection{First return correction} \label{subsubsec:P_ret_1}
The first population return correction is
\begin{align}
    \Aboxed{P_\text{ret}^{(1)} = -\frac{dR_1}{d\chi}+\frac{3+P_\text{ret}^{(0)}}{8}},
\end{align}
where $R_1$ is calculated in Section \ref{subsec:R1}. But we can evaluate it more specifically. Now note the derivative here is the total derivative along $w=w(\chi)$, therefore,
\begin{align}
    \frac{dR_1}{d\chi} = \pdv{R_1}{\chi}\bigg\vert_w + \frac{dw}{d\chi}\pdv{R_1}{w} = \partial_\chi R_1 +2DR_1,
\end{align}
since $dw/d\chi = 2w$. Also substituting in 
\begin{align}
    P_\text{ret}^{(0)} = 1-2DR_0,
\end{align}
we have
\begin{align}
    P_\text{ret}^{(1)} = -\partial_\chi R_1 - 2DR_1 + \frac{1}{2} - \frac{1}{4}DR_0. 
\end{align}
From the expression for $R_1$ (equation \eqref{eqn:R1_final}), 
\begin{subequations}
\begin{gather}
    \partial_\chi\qty[\frac{(x-3)^2}{2}-\frac{5}{12}]=x-3, \qquad
    \partial_\chi\qty[\frac{5}{4}(x-3)] = \frac{5}{4},  \\
    \Rightarrow \ \partial_\chi R_1 = (x-3)D^2R_0 + \frac{5}{4}DR_0, \\
    \Rightarrow \ P_\text{ret}^{(1)} = \frac{1}{2} - \frac{3}{2}DR_0 - (x-3)D^2R_0 - 2DR_1. 
\end{gather}
\end{subequations}
We may also eliminate $R_1$. Note that
\begin{align}
    D\qty[w(D+1)^5(1-R_0)] = w(D+1)^6(1-R_0),
\end{align}
we have
\begin{align}
    \Aboxed{P_\text{ret}^{(1)} & = \frac{1}{2} - 2DR_0 - \frac{7}{2}(x-3)D^2R_0 - \qty[(x-3)^2-\frac{5}{6}]D^3R_0 + \frac{w}{192}(D+1)^6(1-R_0)}. \label{eqn:P_ret_1}
\end{align}

\subsubsection{The $m$-th passage and coherent stitching} \label{subsec:stitching_cDc}
There is no alternating sign for the population, we simply have 
\begin{align}
    \Aboxed{P_m^{(0)} (s_m) = \int_{-\infty}^\infty \abs{h_m(u)}^2 \tanh^2(s_m+u) \ du}. \label{eqn:P_m}
\end{align}
\indent Note this satisfies the relation
\begin{gather}
    P_m^{(0)} = 1-(-1)^{m+1}\frac{d\Phi_m^{(0)}}{ds_m},
\end{gather}
with $\Phi_m^{(0)}$ given in equation \eqref{eqn:Phi_m0}.

The additive stitching gives
\begin{align}
    P_\text{add} = 1+\sum_m\qty[P_m^{(0)}(s_m)-1] = 1-\sum_m \int \abs{h_m(u)}^2 \sech^2(s_m+u) \ du.
\end{align}
\indent In reality, we should use coherent stitching. Define the common coordinate,
\begin{align}
    y = s_m+u, \qquad \psi_m(y) = h_m(y-s_m), 
\end{align}
separate the even and odd passage sectors,
\begin{align}
    \Psi_e(y) = \sum_{m \text{ even}} \psi_m(y), \qquad \Psi_o(y) = \sum_{m \text{ odd}}\psi_m(y). 
\end{align}
With the amplitude, they appeared with opposite signs. But for the population, both sectors reduce the population during a passage, and they have the same sign
\begin{align}
    \Aboxed{P_\text{coh} = 1-\int(\abs{\Psi_\text{even}}^2 + \abs{\Psi_\text{odd}}^2)\sech^2y \ dy}. 
\end{align}
For example, expanding $\abs{\Psi_e}^2$ gives
\begin{align}
    \abs{\Psi_e}^2 = \sum_{m\text{ even}}\abs{\psi_m}^2 + \sum_{m<n}2\Re(\psi_m^*\psi_n) ,
\end{align}
and similarly for the odd sector. A pair contributes
\begin{align}
    2\Re(\psi_m^*\psi_n) = -2\Re\int h_m^*(y-s_m)h_n(y-s_n)\sech^2y \ dy, \qquad n-m>0\quad  \text{ even}. 
\end{align}

\subsubsection{Poisson averaging of clocks}\label{subsubsec:P_poisson}
The construction is nearly identical to the case for coherent amplitude, but this time, we define sector time to be
\begin{align}
    s_{m,N} (\chi) = \chi\frac{\sqrt{N+1}}{\alpha} - (2m+1) \log(4\sqrt{N+1}).
\end{align}
So the final result is
\begin{align}
    \Aboxed{P_\text{final} = 1- \mathbb E_N\qty[\frac{N}{\alpha^2}\int(\abs{\Psi_{\text{even},N}}^2 + \abs{\Psi_{\text{odd},N}}^2)\sech^2y \ dy]}. \label{eqn:P_poisson}
\end{align}

\subsection{2nd-order phase $\expval*{cc}$}
From the numerics, we could see immediately the shape of $\expval*{cc}$ closely resembles $\expval*{c^\dag c}$, indicating $\expval*{cc}$ may only differ from $\expval*{c^\dag c} = \int\abs{h}^2\tanh^2$ by a small perturbation. Indeed, we shall see the derivation for $\expval*{cc}$ is a lot more complex. We will heavily reference the Reference Table provided in the Supplementary Materials in this section.

\subsubsection{Formalism and first depletion profile}

Again we will use $x$ and $\chi$ interchangeably as the scaled time. In our convention of real pump and real generator, $\expval*{cc}$ is real. Set 
\begin{align}
    S \equiv \frac{\expval*{cc}}{\alpha^2},
\end{align}
we shall see its value is almost the same as population, 
\begin{align}
    P = \frac{\expval*{c^\dag c}}{\alpha^2}.
\end{align}
In their cosh and sinh forms,
\begin{subequations}
\begin{align}
    P & = 1 - \epsilon\sinh^2x+\sum_{N\geqslant1}\epsilon^{N+1}G_{k,P}(x), \\
    S & = 1-\epsilon\sinh^2x + \sum_{N\geqslant1}\epsilon^{N+1}G_{k,S}(x),
\end{align}
\end{subequations}
for example, 
\begin{subequations}
\begin{align}
    G_{2,S} & = -\frac{3}{8}-\frac{3}{8}x\sinh2x -\frac{1}{4}x^2\cosh2x+\frac{1}{8}\cosh4x+\frac{1}{4}\cosh2x-\frac{1}{2}x^2,\\
    G_{2,P} & = -\frac{1}{8}-\frac{3}{8}x\sinh2x-\frac{1}{4}x^2\cosh2x+\frac{1}{8}\cosh4x, \\
    G_{2,P} - G_{2,S} & = -\frac{\cosh2x-1-2x^2}{4} = - \frac{e^{2x}}{8} + \frac{1}{2}x^2 + \frac{1}{4} - \frac{e^{-2x}}{8}. \label{eqn:VY_demo}
\end{align}
\end{subequations}
We see that, because they have so many terms in common, it's natural to study their differences. To make this small deviation more visible, let's define the phase quadrature: 
\begin{subequations}
\begin{gather}
    Y = -i(c-c^\dag), \qquad \Rightarrow \qquad Y^2 = 2c^\dag c+1-c^2-c^{\dag2}, \\
    \Rightarrow \ V_Y \equiv \expval*{(\Delta Y)^2} = 1+2\alpha^2 (P-S), \qquad P = \frac{\expval*{c^\dag c}}{\alpha^2}, \\
    \text{or, } \quad S = P + \frac{1-V_Y}{2\alpha^2}. \label{eqn:S_P_VY}
\end{gather}
\end{subequations}
In fact, the supplied $\expval*{cc}$ not only shows $\expval*{cc}$, but also $V_Y$. For example, the terms in \eqref{eqn:VY_demo} are found in the ``cc 1 Harmonics'' tab (note the factor of 2 has already been included in the Reference Table) as:
\begin{table}[H]
\centering
\begin{tabular}{c|c|c|c|c}
\hline
\rowcolor[HTML]{EFEFEF} 
\cellcolor[HTML]{EFEFEF}                        & $\varepsilon^1$ & j = 1  & j = 0 & j = $-1$ \\ \hhline{>{\arrayrulecolor[HTML]{EFEFEF}}->{\arrayrulecolor{black}}|-|-|-|-}
\cellcolor[HTML]{EFEFEF}                        & \cellcolor[HTML]{EFEFEF}  $x^2$ & 0      & 1     & 0      \\ \hhline{>{\arrayrulecolor[HTML]{EFEFEF}}->{\arrayrulecolor{black}}|-|-|-|-}
\multirow{-3}{*}{\cellcolor[HTML]{EFEFEF}N = 2} & \cellcolor[HTML]{EFEFEF}  1        & $-1/4$ & 1/2   & $-1/4$ \\ \hline
\end{tabular}
\caption{Harmonic coefficients of the $\mathcal O(\epsilon)$ contribution to $V_Y-1$.}
\label{tab:cc_ref_harmonics}
\end{table}
And we notice only a lower harmonics and the zero-mode polynomial survive, whereas the $e^{\pm4x},x e^{\pm2x}$ and $x^2e^{\pm2x}$ terms are all cancelled. This reflects the smallness of the difference between $\expval*{c^\dag c}$ and $\expval*{cc}$.

Define
\begin{align}
    V_Y = 1+2\sum_{N\geqslant1}\epsilon^N (G_{N,P}-G_{N,S}) \equiv \sum_N \epsilon^N v_N.
\end{align}
This is the main ``innovation'' of $\expval*{cc}$ calculations—instead of analyzing the harmonics of $G_N$ directly, in this section we will focus on the harmonics of the difference $G_{N,P} - G_{N,S}$. However, the principle and key steps remain unchanged. 

$v_N$, in the harmonic form, can be written
\begin{align}
    v_N(x) = Z_N(x) + \sum_{j=1}^N \qty[e^{2jx} H_{N,j}(x) + e^{-2jx}H_{N,j}(-x)].
\end{align}
Again define the fixed diagonal index,
\begin{align}
    r = N-j, 
\end{align}
$r=0$ is the leading harmonics, responsible for the first depletion profile. $r=1$ is the sub-leading harmonics, responsible for the first correction, etc. The leading harmonics series is: 
\begin{align}
    -\frac{\epsilon}{4}e^{2x} + \frac{\epsilon^2}{16}e^{4x} - \frac{9\epsilon^3}{512}e^{6x} + \frac{3\epsilon^4}{512}e^{8x} - \dots,
\end{align}
this can be read from the ``0 Raw S harmonics'' tab in the Reference Table, $r=0$ column. Note they are equal to the leading harmonics of $\expval{c^\dag c}$,
\begin{align}
    H_{j,j} = (-1)^j\frac{4j \cdot j!}{16^j}, \quad \Rightarrow \quad V_0 = P_0
\end{align}
where $V_0$ is the leading-order profile of $V_Y$.

These positive harmonics can be written
\begin{align}
    \epsilon^{j+r} e^{2jx}H_{j+r,j} = \epsilon^r (16z)^jH_{j+r,j},
\end{align}
where $z = \epsilon e^{2x}/16$ as usual. The useful normalized coefficient is 
\begin{align}
    q_r(j,x) = (-1)^j\frac{16^j}{4j!}H_{j+r,j}.
\end{align}

\subsubsection{Formalizing the Euler derivatives}
In converting from normalized coefficients back to observables, we will often see terms of the form
\begin{align}
    4\sum_{j\geqslant1}(-1)^jj!j^kz^j,
\end{align}
so let's define the generating series $B$:
\begin{align}
    B(z) = 4\sum_{j\geqslant1}(-1)^jj!z^j, \label{eqn:B_generating_series}
\end{align}
in that case,
\begin{align}
    D^kB = 4\sum_{j\geqslant1}(-1)^jj!j^kz^j.
\end{align}
Therefore, if the normalized harmonic coefficient is the polynomial
\begin{align}
    q_r(j,x) = \sum_{k=0}^d a_{r,k}(x)j^k,
\end{align}
then the entire $r$-th harmonic series is
\begin{align}
    \sum_{j\geqslant1}4(-1)^jj!q_r(j,x)z^j = q_r(D,x)B(z),
\end{align}
where $D = z\frac{d}{dz}$ as usual. Here's an immediate example:

\subsubsection{First depletion correction} \label{subsubsec:VY_dep_1}
Since in the leading order, $V_0 = P_0$, we need to start with the sub-leading harmonics, $r=1$. The unnormalized and normalized coefficients are
\begin{table}[H]
    \centering
    \begin{tabular}{c|c|c}
        $j$ & $H_{j+1,j}$ & $q_1(j,x)$ \\\hline
        1 & $-\frac{5}{8}x^2 + \frac{7}{16}x$ & $\frac{5}{2}x^2-\frac{7}{4}x$ \\[0.5em]
        2 & $\frac{5}{16}x^2-\frac{3}{8}x+\frac{1}{8}$ & $10x^2-12x+4$ \\[0.5em]
        3 & $-\frac{153}{1024}x^2 + \frac{429}{2048}x - \frac{93}{1024}$ & $\frac{51}{2}x^2-\frac{143}{4}x + \frac{31}{2} $
    \end{tabular}
    \caption{The raw coefficients $H_{j+1,j}$ and the normalized $q_1(j,x)$. In the Reference Table, these are listed in the $r=1$ column in the ``cc 1 Harmonics'' tab and the ``cc 2 Normalized'' tab respectively.}
    \label{tab:tab:cc_ref_handq}
\end{table}
The good thing is all coefficients in this order are described exactly by some generic polynomials: 
\begin{align}
    q_1 (j,x) = (x-1) + p_1(x) j + p_2(x) j^2 + p_3(x) 
    j^3, 
\end{align}
where
\begin{subequations}
\begin{align}
    p_1(x) & = x^2-x+1, \\
    p_2(x) & = \frac{4x^2-3x-3}{4}, \\
    p_3(x) & = \frac{2x^2 - 4x+3}{4}, 
\end{align}
\end{subequations}
Replace $j$ with $D$:
\begin{align}
    q_1(D,x) B = (x-1)B + p_1DB + p_2D^2B + p_3D^3B. 
\end{align}
Note from equation \eqref{eqn:C0_series}, \eqref{eqn:P_dep_0} and \eqref{eqn:B_generating_series}, we have
\begin{align}
    DB = P_0-1, \qquad B = 2(C_0-1),
\end{align}
so we can write
\begin{align}
    q_1(D,x) B = 2(x-1)(C_0-1) + p_1(P_0-1) + p_2 DP_0 + p_3 D^2P_0.
\end{align}
The zero harmonic at this order is (the $N=2$ row in the Reference Table) 
\begin{align}
    Z_1(x) = x^2+\frac{1}{2}.
\end{align}
Combining, we have
\begin{align}
    \Aboxed{V_{Y,\text{ dep}}^{(1)} = \frac{2x^2-4x+3}{4}D^2P_0 + \frac{4x^2-3x-3}{4}DP_0+(x^2-x+1)P_0 +2(x-1)C_0 + \frac{3}{2}-x}, \label{eqn:VY_dep_1}
\end{align}
and therefore at $\mathcal O(\epsilon)$,
\begin{align}
    \Aboxed{V_Y = P_0 + \epsilon V_1 + \mathcal O(\epsilon^2)},
\end{align}
where $V_1$ is the first correction to $V_Y$, defined below.

What about the quantity we actually want to calculate, $S = \expval*{cc}/\alpha^2$? Starting from
\begin{align}
    S = P + \frac{\epsilon}{2}(1-V_Y),
\end{align}
expand out
\begin{subequations}
\begin{align}
    P & = P_0 + \epsilon P_1 + \epsilon^2 P_2 + \dots \\
    V_Y & = V_0 + \epsilon V_1 + \epsilon^2 V_2 + \dots \\
    S & = S_0 + \epsilon S_1 + \epsilon^2 S_2 + \dots,
\end{align} \label{eqn:PVS_expansion}
\end{subequations}
then we have the normalized $\expval*{cc}$ satisfying
\begin{align}
    S_0 = P_0, \qquad S_1 = P_1 + \frac{1-P_0}{2}.
\end{align}
All these relations are general. However, here, they're understood to apply in the first depletion region. Substituting in equation \eqref{eqn:P_dep_1} for $P_1$, we have
\begin{align}
    \Aboxed{S_\text{dep}^{(1)} = \frac{2x^2-4x+3}{4}D^2P_0 + \frac{7(x-1)}{4}DP_0 +\frac{1}{2}P_0}. \label{eqn:S_dep_1}
\end{align}

\subsubsection{First return profile and exceptional terms} \label{subsubsec:VY_dep_2}
For $r=2$, explicitly, the generic polynomial predicts (call these the ``regular'' terms)
\begin{align}
    q_2^\text{reg}(j,\chi) = \sum_{k=0}^5 a_k(x)j^k,
\end{align}
with
\begin{subequations}
\begin{align}
    a_5 & = \frac{(2x^2-4x+3)^2}{32}, \\
    a_4 & = \frac{x^4}{2}-\frac{35x^3}{24} + \frac{13x^2}{8} - \frac{21x}{32} - \frac{5}{32}, \\
    a_3 & = \frac{x^4}{2} - \frac{3x^3}{4} + \frac{19x^2}{32} - \frac{49x}{32} + \frac{107}{64}, \\
    a_2 & = \frac{5x^3}{4} - \frac{9x^2}{2} + \frac{253x}{64} - \frac{7}{8}, \\
    a_1 & = -\frac{5x^2}{4} + \frac{9x}{8} - \frac{3}{4}, \\
    a_0 & = \frac{3}{2} - \frac{13x}{8}.
\end{align}
\end{subequations}
These coefficients give the $r=2$ columns in the ``cc 3 Generic'' tab. The actual $j=1$ coefficient differs from the regular value by 
\begin{align}
    E_{2,1} = \frac{1}{128}. 
\end{align}
Its contribution is
\begin{align}
    4(-1)^11!E_{2,1}z = -\frac{z}{32}. 
\end{align}
The generic polynomial also does not fit the $j=0$ harmonics 
\begin{align}
    Z_2 = -\frac{3}{2}x^2 - \frac{1}{8},
\end{align}
as well as the $j=-1$ negative harmonics, 
\begin{align}
    \epsilon e^{-2x}\qty(-\frac{1}{4}) = -\epsilon^2 \frac{1}{64z}.
\end{align}
The complete second order correction is
\begin{align}
    \Aboxed{V_{Y,\text{dep}}^{(2)} = q_2^\text{reg}(D,x)B(z) -\frac{z}{32}-\frac{3x^2}{2}-\frac{1}{8}-\frac{1}{64z}}. \label{eqn:VY_dep_2}
\end{align}
The same principle can be used for every higher $r$. Refer to the ``4 Corrections'' tab, the first exceptional tower consists of the green terms.

The first return scale is, once again
\begin{align}
    w = \epsilon^2z = \frac{e^{2x}}{16\alpha^6} = 256e^{2s_1}, \qquad s_1 = \chi-3\log(4\alpha), 
\end{align}
for the first return, the relevant coefficients are (find the green cells in the ``4 Corrections'' tab, multiply them by the factors given in the green cells in the ``5 Indices'' tab)
\begin{align}
    V_{Y,\text{ret}}^{(0)} = 1 -\frac{w}{32} + \frac{5w^2}{512} - \frac{567w^3}{65536}+\frac{4671w^4}{262144} - \dots \label{eqn:V_ret_0}
\end{align}
we see they're in fact equal to the population-return coefficients, $P_\text{ret}^{(0)}$ as in equation \eqref{eqn:P_ret_0_series}. This gives us no new information, and we really need to calculate the first return correction.

For the first return correction, we will follow the general procedure laid out in Section \ref{subsec:R1}. Recall the general principle: Return terms were produced by $E_{r,j}$. The leading profile $R_0$ was produced by $E_{2n,n}$, whereas the first correction $R_1$ was produced by $E_{2n+1,n}$ We then took the ratio $q_n = \frac{E_{2n+1,n}}{E_{2n,n}}$, and we found that the ratios have simple polynomial patterns. This allowed us to find a general expression for $q_n \rightarrow E_{2n+1,n} \rightarrow R_1$. 

Here we will follow a similar procedure: We will similarly take the ratio, 
\begin{align}
    \tilde q_n = \frac{E_{2n+1,n}^{V_Y} - E_{2n+1,n}^P}{\beta_n}, \label{eqn:Qn_ratio}
\end{align}
where we defined $\beta\propto E_{2n,n}^{V_Y} = E_{2n,n}^P$ as the normalized coefficients of the first exceptional terms,
\begin{align}
    \beta_n \equiv \qty{-\frac{1}{32}, \frac{5}{512}, -\frac{567}{65536}, \frac{4671}{262144} \dots}, \label{eqn:betan_ratio}
\end{align}
exactly as they appear in the physical observables, in equation \eqref{eqn:V_ret_0}; and they can be read from the green terms in the Reference Table up to normalization given in the Index Guide. As for $E_{2n+1,n}^{V_Y} - E_{2n+1,n}^P$, these are the next-order corrections. In the Reference Tables, these are the difference between the corresponding blue terms. For example, 
\begin{subequations}
\begin{align}
    E_{3,1}^{V_Y} - E_{3,1}^P & = \qty[\qty(\frac{17}{128} - \frac{1}{128})x^2 + \qty(-\frac{167}{256} + \frac{5}{256})x + \qty(\frac{95}{128} + \frac{1}{128})]\times(-2) \\
    & = - \frac{x^2}{4} + \frac{81x}{64} -\frac{3}{2}, \\
    E_{5,2}^{V_Y} - E_{5,2}^P & = \qty[ \qty(\frac{109}{2048}-\frac{5}{1024})x^2 + \qty(-\frac{37}{128} + \frac{85}{4096})x + \qty(\frac{387}{1024} - \frac{63}{4096}) ] \times 4 \\
    & = \frac{99x^2}{512} - \frac{1099x}{1024} + \frac{1485}{1024}, \\
    E_{7,3}^{V_Y} - E_{7,3}^P & = \qty[\frac{16173-1701}{524288}x^2 + \frac{-181593+16443}{1048576}x + \frac{123687-7911}{524288} ]\times(-12) \\
    & = -\frac{5427x^2}{16384} + \frac{247725x}{131072} - \frac{5427}{2048}, \\
    & \dots
\end{align}
Now take the ratio to simplify the coefficients even further:
\begin{align}
    Q_1 &= \frac{E_{3,1}^{V_Y} - E_{3,1}^P}{\beta_1}  = \frac{-\frac{x^2}{4}+\frac{81x}{64}-\frac{3}{2}}{-\frac{1}{32}} = 8x^2 - \frac{81x}{2}+48, \\
    Q_2 &= \frac{E_{5,2}^{V_Y} - E_{5,2}^P}{\beta_2}  = \frac{\frac{99x^2}{512} - \frac{1099x}{1024}+\frac{1485}{1024}}{\frac{5}{512}} = \frac{99x^2}{5} - \frac{1099x}{10} + \frac{297}{2}, \\
    Q_3 &= \frac{E_{7,3}^{V_Y} - E_{7,3}^P}{\beta_3}  = \frac{-\frac{5427x^2}{16384} + \frac{247725x}{131072} - \frac{5427}{2048}}{-\frac{567}{65536}} = \frac{268x^2}{7} - \frac{9175x}{42} + \frac{2144}{7}, \\
    Q_4 & = \dots = \frac{10825x^2}{173} - \frac{250359x}{692} + \frac{357225}{692}, \\
    Q_5 & = \frac{52308x^2}{563} - \frac{3047939x}{5630} + \frac{2196936}{2815}, \\
    Q_6 & = \frac{122647x^2}{949} - \frac{4309727x}{5694} + \frac{2084999}{1898}.
\end{align}
\end{subequations}
and so forth. They have the form
\begin{align}
    \tilde q_n (x) = u_n x^2 + v_n x + c_n, 
\end{align}
A pattern emerges if we factor out $u_n$, and look at the quantities $(v_n+n/2)/u_n$ and $c_n/u_n$:
\begin{table}[H]
    \centering
    \begin{tabular}{c|c|c|c}
        $n$ & $u_n$ & $c_n/u_n$ & $(v_n+n/2)/u_n$  \\\hline
        1 & 8 & 6 & $-5$ \\
        2 & 99/5 & 15/2 & $-11/2$ \\
        3 & 268/7 & 8 & $-17/3$ \\
        4 & 10825/173 & 33/4 & $-23/4$ \\
        5 & 52308/563 & 42/5 & $-29/5$ \\
        6 & 122647/949 & 17/2 & $-35/6$
    \end{tabular}
    \caption{Coefficients and coefficient ratios in $\tilde q_n (x) = u_nx^2 + v_nx+c_n$.}
    \label{tab:un}
\end{table}
\noindent the last two columns are exactly
\begin{subequations}
\begin{align}
    \frac{v_n+n/2}{u_n} = -6+\frac{1}{n} \quad &\Rightarrow \quad v_n = u_n\qty(-6+\frac{1}{n}) - \frac{n}{2}, \\
    \frac{c_n}{u_n} = 9-\frac{3}{n}, \quad &\Rightarrow \quad c_n = u_n\qty(9-\frac{3}{n}),
\end{align}
\end{subequations}
Substituting this into $\tilde q_n(x) = u_nx^2 + v_nx + c_n$, we have
\begin{subequations}
\begin{align}
    \tilde q_n(x) & = u_nx^2 + \qty[u_n\qty(-6+\frac{1}{n}) - \frac{n}{2}] x + u_n\qty(9-\frac{3}{n}) \\
    & = u_n \qty[x^2 - 6x + \frac{x}{n}+9-\frac{3}{n}] -\frac{nx}{2} \\
    & = u_n \qty[(x-3)^2 + \frac{x-3}{n}] - \frac{nx}{2}.
\end{align}
\end{subequations}
So now the only thing we need to know is the sequence $u_n$. Here, we found that $u_n$ is closely related to the ratio $q_n(0)$ that we've seen when calculating the first return correction of $\expval*{c}$, in equation \eqref{eqn:qn0_ratio}:
\begin{align}
    q_n(0) = \frac{1}{4}-\frac{15n}{4}+\frac{49n^2}{12}-\frac{2n^2}{3}\frac{F_{n-1}}{F_n}, \tag{\ref{eqn:qn0_ratio}}
\end{align}
and it's related to $u_n$ by
\begin{align}
    u_n = 19n^2 -12n + 2 - 4q_n(0) = 1+3n+\frac{8n^2}{3}\qty(1+\frac{F_{n-1}}{F_n}).
\end{align}
\indent Recall once again our procedure is that we want to eventually calculate something like $\sum(E_{2n+1,n}^{V_Y} - E_{2n+1,n}^P)w^n$, and we found these coefficients are equal to $\tilde q_n \times \beta_n$ as defined in equation \eqref{eqn:Qn_ratio}, so this problem is reduced to calculating 
\begin{align}
    \tilde\Gamma_1 \equiv \sum_{n\geqslant1} \beta_n \tilde q_n(x) w^n. \label{eqn:Gamma_1_formal}
\end{align}
So we want to put $\tilde q_n$ in a series. Let's define 
\begin{align}
    \mathcal U(w) \equiv \sum_{n=1}^\infty a_n u_n w^n, \label{eqn:U_caligraphic_series}
\end{align}
substituting
\begin{subequations}
\begin{gather}
    u_n = 19n^2 -12n+2-4q_n(0): \\
    \Rightarrow \ \mathcal U = 19\sum_{n\geqslant1} n^2a_nw^n - 12\sum_{n\geqslant1} na_nw^n + 2\sum_{n\geqslant1}a_nw^n -4\sum_{n\geqslant1}a_nq_n(0)w^n,
\end{gather}
\end{subequations}
now recall our formal definitions for $R_0$ (equation \eqref{eqn:R0_formal_an}) and $R_1$ (equation \eqref{eqn:R1_formal}), namely,
\begin{align}
    R_0(w) = -1+\sum_{n\geqslant1} a_nw^n, \qquad R_1(w,x) = -\frac{1}{4} + \sum_{n\geqslant1}a_nq_n(x)w^n, 
\end{align}
we have
\begin{subequations}
\begin{align}
    \mathcal U(w) & = 19D^2R_0-12DR_0+2(R_0+1)-4(R_1(w,0)+\frac{1}{4})\\
    & = 19D^2R_0 - 12DR_0+2R_0-4R_1(w,0)+1.
\end{align}
\end{subequations}
Alternatively, we can express $R_1(w,0)$ in terms of $R_0$ only, starting from equation \eqref{eqn:R1_final}:
\begin{subequations}
\begin{align}
    R_1(w,0) & = \frac{49}{12}D^2R_0 - \frac{15}{4}DR_0 + \frac{1}{4}R_0 - \frac{w}{384}(D+1)^5(1-R_0), \\
    \Rightarrow \ \mathcal U(w) & =1+ R_0+ 3DR_0+ \frac{8}{3}D^2R_0 + \frac{w}{96}(D+1)^5(1-R_0). \label{eqn:U_caligraphic_final}
\end{align}
\end{subequations}
Now substitute into the formal series of the first return phase-population gap, $\tilde\Gamma_1$, defined in equation \eqref{eqn:Gamma_1_formal}:
\begin{subequations}
\begin{gather}
    \tilde q_n(x) = u_n\qty[(x-3)^2+\frac{x-3}{n}]-\frac{nx}{2}, \\
    \tilde\Gamma_1 = (x-3)^2\sum_{n\geqslant1}\beta_n u_n w^n + (x-3)\sum_{n\geqslant1} \frac{\beta_nu_n}{n}w^n - \frac{x}{2}\sum_{n\geqslant1}n\beta_nw^n.
\end{gather}
\end{subequations}
From Table \ref{tab:an_and_qn} for $a_n$ and equation \eqref{eqn:betan_ratio} for $\beta_n$, we see that $\beta_n = -2na_n$, therefore,
\begin{subequations}
\begin{align}
    \sum_{n\geqslant1}\beta_nu_nw^n & = -2\sum_{n\geqslant1}na_nu_nw^n = -2DU, \\
    \sum_{n\geqslant1}\frac{\beta_nu_n}{n}w^n & = -2\sum_{n\geqslant1}a_nu_nw^n= -2U, \\
    \sum_{n\geqslant1}n\beta_nw^n & = -2\sum_{n\geqslant1}n^2a_nw^n = -2D^2R_0,
\end{align}
\end{subequations}
substituting, we have
\begin{align}
    \tilde\Gamma_1(w,x) & =xD^2 R_0 -2\qty[(x-3)^2D + (x-3)]U(w).
\end{align}
$\tilde\Gamma_1$ is in fact not yet the full solution. Just like how the constant $-1/4$ in $R_1$ (section \ref{subsec:R1}) was not determined by the exceptional series, $\tilde V_{Y,\text{ ret}}^{(1)}$ also has a constant not yet determined, and must be found via asymptotic matching. 

From equation \eqref{eqn:P_dep_1_full} for $P_1$ and \eqref{eqn:VY_dep_1} for $V_1$, we have
\begin{align}
    V_{Y, \text{ dep}}^{(1)} - P_\text{dep}^{(1)} = \qty(x^2-\frac{5}{2}x+1)DP_0 + x(x-1)P_0 + 2(x-1)C_0 +2-x, \label{eqn:VY_P1_diff}
\end{align}
after the depletion but before the return, 
\begin{align}
    P_0 \longrightarrow1, \qquad C_0\longrightarrow-1, \qquad DP_0\longrightarrow0, \quad \text{ as } \quad z\rightarrow\infty,\ w\rightarrow0,
\end{align}
so 
\begin{align}
    V_{Y, \text{ dep}}^{(1)} - P_\text{dep}^{(1)} & \longrightarrow x(x-1)-2(x-1)+2-x = (x-2)^2, \quad \text{ as } \quad w\rightarrow0,
\end{align}
so the full expression is
\begin{gather}
    \Gamma_1 \longrightarrow(x-2)^2, \quad \tilde\Gamma_1 \longrightarrow0 \quad \text{ as } \quad w\rightarrow0, \label{eqn:Gamma1_asymptote} \\
    \Rightarrow \ \Gamma_1(w,x) = (x-2)^2+xD^2R_0-2\qty[(x-3)^2D+(x-3)]U(w), 
\end{gather}
or using the $R_0$-only expression for $U$, 
\begin{subequations}
\begin{align}
    \Gamma_1(w,x) & = (x-2)^2 + xD^2R_0 \\ 
    & \  - 2\qty[(x-3)^2D+(x-3)]\qty[1+R_0+3DR_0 + \frac{8}{3}D^2R_0 + \frac{w}{96}(D+1)^5(1-R_0)]. 
\end{align} \label{eqn:Gamma_1_final}
\end{subequations}
\indent Now for the actual observables: From $\Gamma_1 = V_{Y, \text{ ret}}^{(1)}-P_\text{ret}^{(1)}$, we have
\begin{align}
    \Aboxed{V_{Y,\text{ ret}}^{(1)} = P_\text{ret}^{(1)} + \Gamma_1}, \label{eqn:VY_ret_1}
\end{align}
where $\Gamma_1$ is given above in equation \eqref{eqn:Gamma_1_final}, and $P_\text{ret}^{(1)}$ is given in equation \eqref{eqn:P_ret_1}. 

Now let's convert this to $\expval*{cc}$. From the discussion around \eqref{eqn:PVS_expansion}, we have (use the identity $P_0 = 1-2DR_0$ in equation \eqref{eqn:P_ret_0_id})
\begin{align}
    S_\text{ret}^{(1)} = P_\text{ret}^{(1)} + \frac{1-P_\text{ret}^{(0)}}{2} = P_\text{ret}^{(1)} + DR_0,
\end{align}
and since we know $P_\text{ret}^{(1)}$ from equation \eqref{eqn:P_ret_1}, we have
\begin{align}
    \Aboxed{S_\text{ret}^{(1)} & = \frac{1}{2} - DR_0 - \frac{7}{2}(x-3)D^2R_0 - \qty[(x-3)^2-\frac{5}{6}]D^3R_0 + \frac{w}{192}(D+1)^6(1-R_0)}.
\end{align}
However, we see the phase-population gap $\Gamma_1$ we just calculated doesn't enter $S_\text{ret}^{(1)}$. In fact, it enters $S_\text{ret}^{(2)}$. At order $\epsilon^2$, we have
\begin{subequations}
\begin{align}
    S_2 = & P_2 - \frac{1}{2}V_1, \quad V_1 = P_1 +\Gamma_1, \\
    \Rightarrow \quad S_\text{ret}^{(2)} & = P_\text{ret}^{(2)}- \frac{1}{2}P_\text{ret}^{(1)} - \frac{1} {2}\Gamma_1 \\
    & = P_\text{ret}^{(2)} - \frac{1}{2}P_\text{ret}^{(1)} - \frac{1}{2}(x-2)^2 - \frac{x}{2}D^2R_0 + \qty[(x-3)^2D + (x-3)] U(w).
\end{align}
\end{subequations}
We will not calculate it explicitly here. If the goal, for example, is to observe $\tilde q_n$ and $\Gamma_1$, then $V_Y$ (and $P$) are the right target.

\subsubsection{The $m$-th passage and Poisson-clock} \label{subsubsec:VY_poisson}

We conjecture a general formula for the $m$-th passage here. We give a possible form here, but it is not yet proven for arbitrary $m$. In reality, the difference between $S$ and $P$ remains $\mathcal O(\epsilon)$ within our model. However, in the numerics, they deviate significantly for later time. It could be due to non-uniform sector-clock differences, or a missing exceptional tower, and this could be a topic of future research. 

We shall derive the possible form via the known relation between $P_1$ and $\Gamma_1$. Recall 
\begin{align}
    P_1(s_1) & = \int \rho_1(u) \tanh^2(s_1+u) \ du, 
\end{align}
where we have defined $\rho_1(u) = \abs*{h_1(u)}^2$, and (for example, from equation \eqref{eqn:Gamma_1_formal}) the coefficients of the phase-population gap are obtained by multiplying the population coefficients by a ratio $\tilde q_n$, that is,
\begin{subequations}
\begin{align}
    [e^{2ns_1}]\Gamma_1 & = \tilde q_n(x) [e^{2ns_1}]P_1, \\
    \text{or explicitly, } \quad \Gamma_1(s_1) & = g_0 + \sum_{n=1}^\infty \tilde q_n(x) \cdot \qty([e^{2ns_1}]P_1) \cdot e^{2ns_1}, \label{eqn:G1_Q_P1}\\
    \text{where } \quad \tilde q_n(x) & = u_n \qty[d_1^2 + \frac{d_1}{n}] - \frac{nx}{2},
\end{align}
\end{subequations}
where $d_1 = x-3$. Now, to find the coefficients of $P_1$, let $y=s_1+u$, and first try to expand $\tanh y$:
\begin{align}
    \tanh y = -1+2\sum_{n=1}^\infty (-1)^{n-1}e^{2ny}, 
\end{align}
and now expand $\tanh^2y$, we have (write $\xi = e^{2y}$ and $\tanh^2y = 1-\frac{4\xi}{(\xi+1)^2}$, then expand the bracket; alternatively, note $\tanh^\prime y = 1-\tanh^2y$)
\begin{align}
    \tanh^2y = 1+ 4\sum_{n=1}^\infty (-1)^n ne^{2ny},
\end{align}
then
\begin{subequations}
\begin{align}
    P_1(s_1) & = \int \rho_1 \tanh^2(s_1+u) \ du \\
    & = 1+4 \sum_{n\geqslant1} (-1)^n n \int du \  \rho_1 e^{2ny}.
\end{align}
\end{subequations}
Now let's write out equation \eqref{eqn:G1_Q_P1} explicitly:
\begin{subequations}
\begin{align}
    \Gamma_1 - g_0 & = \sum_{n=1}^\infty \qty(u_n\qty[d_1^2+\frac{d_1}{n}] - \frac{nx}{2}) \qty(4(-1)^n n \int du \ \rho_1 e^{2ny}) \\
    & = \int du \ \rho_1 \sum_{n=1}^\infty \qty(\underbracket{4d_1^2 (-1)^n n u_n e^{2ny}}_{\mathclap{\text{\ding{172}}}} + \underbracket{4d_1 (-1)^n u_n e^{2ny}}_{\mathclap{\text{\ding{173}}}} - \underbracket{2x(-1)^n n^2 e^{2ny}}_{\mathclap{\text{\ding{174}}}})
\end{align}
\end{subequations}
If we look closely at term \ding{172} and \ding{173}, we realize they ``almost'' contain $\tanh^2 y$ and $\tanh y$ respectively, we just need to handle the extra $u_n$ terms. We define a new type of weight, 
\begin{align}
    K_n = \int W_1(u) e^{2nu} \ du \equiv u_n \int \rho_1(u) e^{2nu}\ du, \quad n\geqslant1. \label{eqn:W1_intro}
\end{align}
One can check that 
\begin{align}
    K_0 = \int W_1(u) \ du = 1,
\end{align}
(explicitly, $W_1(u) = \abs*{h_1^\prime(u)}^2 - \frac{1}{4}\rho^{\prime\prime}_1(u)$.) Now terms \ding{172} and \ding{173} can be resummed:
\begin{subequations}
\begin{align}
    \text{\ding{173}} & = d_1 \int du \ W_1(u) 4\sum_{n=1}^\infty (-1)^n e^{2ny} \\
    & = -2d_1 \int du \ W_1(u) \qty[1+\tanh(s_1+u)], \\
    \text{\ding{172}} & = d_1^2 \int du \ W_1(u) 4\sum_{n=1}^\infty (-1)^n n e^{2ny} \\
    & = d_1^2 \int du \  W_1(u) \qty[\tanh^2(s_1+u)-1], \\
    \text{\ding{172}} + \text{\ding{173}} & = \int du \ W_1(u) \qty(-2d_1 - 2d_1\tanh y + d_1^2\tanh^2 y -d_1^2) \\
    & = \int du \ W_1(u) \ (1-d_1\tanh y)^2 - (1+d_1)^2.
\end{align}
\end{subequations}
Now for term \ding{174}, note that
\begin{subequations}
\begin{align}
    P_1(s_1) & = 1+4\sum_{n=1}^\infty (-1)^n n \int \rho_1  e^{2ny} \ du, \\
    \partial_{s_1}P_1(s_1) & = 8\sum_{n=1}^\infty (-1)^nn^2 \int \rho_1 e^{2ny} \ du, \\
    \Rightarrow \ \text{\ding{174}} & = -\frac{x}{4}\partial_{s_1}P_1(s_1). 
\end{align}
\end{subequations}
Putting everything together, we have
\begin{align}
    \Gamma_1 = g_0 - (1+d_1)^2 + \int W_1(u) \qty[1-d_1\tanh(s_1+u)]^2 \ du - \frac{x}{4}\partial_{s_1}P_1.
\end{align}
What about the constant $g_0$? We have to determine it via asymptote matching, namely,
\begin{align}
    \lim_{z\rightarrow\infty} (V_{Y,\text{dep}}^{(1)}-P_\text{dep}^{(1)}) = \lim_{w\rightarrow0}\Gamma_1(w,x) = g_0(x),
\end{align}
and we already calculated this quantity to be $(x-2)^2$ back in equation \eqref{eqn:Gamma1_asymptote}. Now recall $d_1 = x-3$, so $(1+d_1)^2 = (x-2)^2$ also, so $g_0 - (1+d_1)^2$ cancel. We are simply left with (re-instate $x=\chi$)
\begin{align}
    \Aboxed{\Gamma_1(\chi) = \int W_1(u) \qty[1-d_1\tanh(s_1+u)]^2 \ du - \frac{\chi}{4}\partial_{s_1}P_1}. \label{eqn:G1_integral}
\end{align}
\indent Now, since the population passage always appears to have the form
\begin{align}
    P_m(s_m) = \int \rho_m(u) \tanh^2(s_m+u) \ du, \quad s_m = \chi - (2m+1)\log4\alpha, \quad \rho_m(u) = \abs*{h_m(u)}^2,
\end{align}
it seems very probable that the form of $\Gamma_1(\chi)$ we just wrote down could be general. That is, for the $m$-th passage, (define $d_m =\chi-(2m+1)$) 
\begin{gather}
    \boxed{\Gamma_m(\chi) = \int W_m(u) \qty[1-d_m\tanh(s_m+u)]^2 \ du - \frac{\chi}{4}\partial_{s_m}P_m(s_m)}.
\end{gather}
This formula appears promising because $W_m(u)$ always has an explicit construction given by $W_m = \abs*{h_m^\prime}^2-\frac{1}{4}\rho_m^{\prime\prime}$, and the first six coefficients of the first return has been checked. However, it is not proven. For the $(1-d_m\tanh (s_m+u))^2$ term to be invariant, it would require that the coefficient ratio can always be written as
\begin{align}
    \tilde q_n^{(m)} = u_n^{(m)} \qty(d_m^2 + \frac{d_m}{n}) - \frac{n\chi}{2},
\end{align}
and it also assumes the constant matching condition $g_{m,0} = (1+d_m)^2$ always holds, as well as that the $\partial_{s_m}P_m$ term holds. 

If these conditions hold, together with the relation $S = P+\frac{1-V_Y}{2\alpha^2}$ as in equation \eqref{eqn:S_P_VY}, we have
\begin{align}
    \Aboxed{S = P+\frac{\epsilon}{2}(1-P)-\frac{\epsilon^2}{2}\Gamma_m + \mathcal O(\epsilon^3)},
\end{align}
or in terms of $\expval*{cc}$, 
\begin{align}
    \Aboxed{\expval*{cc} = \alpha^2P +\frac{1-P}{2}-\frac{\Gamma_m}{2\alpha^2} + \mathcal O(\alpha^{-4})}.
\end{align}
\indent Finally, the Poisson clock-averaging formula, written out explicitly, is (use $\alpha\rightarrow a_N =\sqrt{N+1}, \xi_N = \frac{x\sqrt{N+1}}{\alpha}$)
\begin{subequations}
\begin{empheq}[box=\fbox]{align}
    V_Y & = \mathbb E_N \bigg\{\frac{(N-\alpha^2)^2}{\alpha^2}\qty[1-\int (\abs*{\Psi_{\text{even},N}}^2 + \abs{\Psi_{\text{odd},N}}^2)\sech^2y \ dy ] \\
    & \quad + \frac{\alpha^2}{(N+1)^2} \bigg[\xi_N^2 -\sum_m (\xi_N-2m)^2 \\
    & \qquad + \sum_{m-n \text{ even}}\Re\int (1-d_m\tanh y)(1-d_n \tanh y)\psi_m^{\prime*}\psi_n^\prime \ dy\bigg] \\
    & \quad - \frac{2(N-\alpha^2)}{N+1} \sum_{m,n\text{ even}}\Re\int\tanh y (1-d_n\tanh y)\psi_m^*\psi_n^\prime \ dy \bigg\},
\end{empheq} \label{eqn:VY_poisson}
\end{subequations}
then $S$ can be obtained via $S = P+(1-V_Y)/2\alpha^2$.

\pagebreak
\section{3rd-order correlations}

Define
\begin{align}
    f_m(u) \equiv (-1)^{m+1}\tanh(s_m+u), \quad \rho_m(u) = \abs{h_m(u)}^2, \quad  s_m = \chi-(2m+1)\log4\alpha. 
\end{align}
So far, we found that the $m$-th passage of $\expval*{c}$ has the leading profile given by:
\begin{align}
    \Phi_m^{(0)} = \int \rho_m f_m \ du, 
\end{align}
and the leading profile for $\expval*{c^\dag c}$ and $\expval*{cc}$ is
\begin{align}
    P_m^{(0)} = \int \rho_m f_m^2 \ du. 
\end{align}
It's natural to ask, what about the quantity
\begin{align}
    \int \rho_m f_m^3 \ du = (-1)^{m+1} \int \rho_m \tanh^3(s_m+u) \ du,
\end{align}
does it represent any correlation functions? It turns out this is unsurprisingly the leading profile of $\expval*{c^\dag cc}$ and $\expval*{ccc}$. But let us try to derive it nonetheless.

\subsection{3rd-order mixed correlation $\expval*{c^\dag cc}$} 
\subsubsection{Formalism and a key identity} \label{subsec:cDcc_id}
Let our quantity of interest be 
\begin{align}
    U \equiv \frac{\expval*{c^\dag cc}}{\alpha^3}.
\end{align}
Also let
\begin{align}
    F(\alpha,\tau) = \expval*{c}, \qquad \tau = gt,
\end{align}
and define the conserved number,
\begin{align}
    M = n_a + n_c.
\end{align}
The Heisenberg equations of motion are
\begin{align}
    \frac{da}{d\tau} \equiv \dot a = b^\dag c, \quad \dot b^\dag = ac^\dag, \quad \dot c = -ab,
\end{align}
and differentiate the last equation once more gives
\begin{subequations}
\begin{align}
    \ddot c & = -(2n_a+1)c = -(2M+1)c + 2c^\dag cc, \\
    \Rightarrow \ \ddot F & = -2\expval*{Mc} - F + 2\expval*{c^\dag cc}, \\
    \Rightarrow \ \expval*{c^\dag cc} & = \expval*{Mc} + \frac{1}{2}F + \frac{1}{2}\ddot F.
\end{align}
\end{subequations}
$\expval*{Mc}$ can be converted to an $\alpha$-derivative. For real $\alpha$:
\begin{align}
    \alpha\partial_\alpha\ket\alpha = (n_c-\alpha^2)\ket\alpha, \qquad \alpha\partial_\alpha\bra\alpha = \bra\alpha(n_c-\alpha^2), 
\end{align}
differentiating both the bra and ket at fixed time $\tau$—note this means evaluating the expectation value in the Heisenberg picture, and applying the above differentiation identity in the Schr\"odinger picture. This means the $n_c$ in the identity is that of the initial state at $\tau=0$, $n_c(0)$; and note $n_a(0) = 0$:
\begin{subequations}
\begin{align}
    \alpha\partial_\alpha F\big\vert_\tau & = (\alpha\partial_\alpha\bra\alpha) c(\tau)\ket\alpha + \bra\alpha c(\tau)(\alpha\partial_\alpha\ket\alpha) \\
    & = \bra\alpha(n_c(0)-\alpha^2) c(\tau)\ket\alpha + \bra\alpha c(\tau)(n_c(0)-\alpha^2)\ket\alpha \\
    & = \expval*{(n_c(0) + n_a(0))c(\tau)} + \expval*{c(\tau)(n_c(0)+n_a(0))} - 2\alpha^2F \\
    & = \expval*{Mc+cM}-2\alpha^2F \\
    \Rightarrow \ \alpha\partial_\alpha F\big\vert_\tau & = 2\expval*{Mc} + (1-2\alpha^2)F,
\end{align}
\end{subequations}
where we used $[M,c] = -c$ in the last step. Substitute this back to the equation for $\expval*{c^\dag cc}$, we have
\begin{align}
    \expval*{c^\dag cc} & = \alpha^2 F + \frac{\alpha}{2}F_\alpha\big\vert_\tau + \frac{1}{2}\ddot F.
\end{align}
Recall that 
\begin{align}
    F = \alpha C(\epsilon,\chi), \qquad \epsilon=\alpha^{-2}, \qquad \chi=\alpha\tau,
\end{align}
we have 
\begin{align}
    \alpha\partial_\alpha\big\vert_\tau = \chi\partial_\chi-2\epsilon\partial_\epsilon, \qquad \partial_\tau\big\vert_\alpha = \alpha\partial_\chi, \qquad \partial^2_\tau\big\vert_\alpha = \alpha^2\partial_\chi^2,
\end{align}
and at fixed $\tau$, 
\begin{align}
    \alpha\partial_\alpha C\big\vert_\tau = \chi C_\chi - 2\epsilon C_\epsilon, \qquad \ddot F = \alpha^3C_{\chi\chi},
\end{align}
dividing both sides of the $\expval*{c^\dag cc}$ equation by $\alpha^3$, we have
\begin{align}
    \Aboxed{U = C + \frac{1}{2}C_{\chi\chi} + \frac{\epsilon}{2}\qty(C+\chi C_\chi -2\epsilon C_\epsilon)}. \label{eqn:U_id}
\end{align}
No approximations have yet been made, this is an exact identity that shows $\expval*{c^\dag cc}$ is entirely determined by $C$. But note this does not mean the numerical value of $U$ is determined by the numerical value of $C$. This identity requires us to know the precise dependence of $C$ on $\chi$ and $\epsilon$. 

Write out their formal series,
\begin{align}
    C(\epsilon,\chi) & = \sum_{n\geqslant0}\epsilon^nc_n(\chi), \qquad c_0 = 1, \\
    U(\epsilon,\chi) & = \sum_{n\geqslant0}\epsilon^n u_n(\chi), 
\end{align}
note that
\begin{align}
    C_\chi = \sum_{n\geqslant0}\epsilon^n c_n^\prime, \quad C_{\chi\chi} = \sum_{n\geqslant0} \epsilon^n c_n^{\prime\prime}, \quad C_\epsilon = \sum_{n\geqslant1} n\epsilon^{n-1}c_n = \sum_{n\geqslant0} n\epsilon^{n-1}c_n,
\end{align}
substituting into the identity, we have
\begin{align}
    \sum_{n} \epsilon^n u_n  & = \sum_n \qty(\epsilon^n c_n + \frac{1}{2}\epsilon^nc_n^{\prime\prime}) + \frac{\epsilon}{2}\sum_n\qty(\epsilon^n c_n + \chi\epsilon^n c_n^\prime - 2\epsilon n\epsilon^{n-1}c_n),
\end{align}
look at the last bracketed terms, it is
\begin{align}
    \Rightarrow \ \frac{1}{2}\sum_n \epsilon^{n+1}\qty((1-2n)c_n + \chi c_n^\prime) = \frac{1}{2}\sum_n \epsilon^n\qty((3-2n)c_{n-1}+\chi c_{n-1}^\prime),
\end{align}
so we can equate the terms in the identity,
\begin{align}
    u_n = c_n + \frac{1}{2}c_n^{\prime\prime} + \frac{1}{2}\qty[(3-2n)c_{n-1}+\chi c_{n-1}^\prime], \quad n\geqslant1.
\end{align}
\indent Now recall in going into harmonic analysis, $C$ becomes a function of $z$ and $\chi$,
\begin{align}
    C = C_0(z) + \epsilon C_1(z,\chi) + \epsilon^2 C_2(z,\chi) + \dots, \qquad z = \frac{\epsilon e^{2\chi}}{16}.
\end{align}
Define the total derivative in $\chi$, holding $\epsilon$ fixed, as
\begin{align}
    \mathcal L \equiv \frac{d}{d\chi} = \pdv{\chi} + \pdv{z}{\chi}\pdv{z} = \pdv{\chi} + 2z\pdv{z} = \partial_\chi+ 2D,
\end{align}
so we can take $\chi$ derivatives while holding $z$ constant. Also note $\epsilon\partial_\epsilon z=z$, so we have
\begin{align}
    \epsilon\partial_\epsilon[\epsilon^rC_r(z,\chi)] = \epsilon^r(r+D)C_r,
\end{align}
substitute back into identity \eqref{eqn:U_id}, we have
\begin{subequations}
\begin{align}
    \sum_r \epsilon^r U_r & = \sum_r \epsilon^r \qty(1 + \frac{1}{2}\mathcal L^2) C_r   + \sum_r \frac{\epsilon^{r+1}}{2} \qty( 1+ \chi \mathcal L - 2(r+D) ) C_r \\
    \Rightarrow \ \Aboxed{U_r & = \qty(1+\frac{1}{2}\mathcal L^2)C_r + \frac{1}{2}\qty[(3-2r) + \chi\mathcal L - 2D]C_{r-1}}, \label{eqn:Ur_id}
\end{align}
\end{subequations}
for example, for $r=0$, 
\begin{align}
    \Aboxed{U_0 = \qty(1+\frac{1}{2}\mathcal L^2)C_0}.
\end{align}

\subsubsection{First depletion profile}
Let's analyze the leading harmonics. Once again we use the notation that
\begin{align}
    U_n(x) = Z_n(x) + \sum_{j=1}^n \qty[e^{2jx} \tilde H_{n,j}(x) + e^{-2jx}\tilde H_{n,j}(-x)],
\end{align}
where harmonic order $r$ is defined as $r = n-j$. Both $\expval*{c^\dag cc}$ and $\expval*{ccc}$ have the same leading coefficients, 
\begin{align}
    \tilde H_{j,j} = (-1)^j \frac{2(2j^2+1)j!}{16^j}, 
\end{align}
the first few values are
\begin{align}
    \tilde H_{j,j} = \qty{-\frac{3}{8}, \frac{9}{64}, -\frac{57}{1024}, \frac{99}{4096}, \dots}
\end{align}
as can be easily read from the Reference Tables. Note that this means
\begin{align}
    U_0 = 1+2\sum_{j\geqslant1} (-1)^j j! (1+2j^2) z^j = (1+2D^2)C_0,
\end{align}
where as we recall,
\begin{align}
    C_0(z) = 1+2\sum_{j=1}^\infty(-1)^j j! z^j. \tag{\ref{eqn:C0_coeff}}
\end{align}
Also recall
\begin{subequations}
\begin{gather}
    C_0(z) = \int_0^\infty e^{-t} \frac{1-zt}{1+zt}dt, \tag{\ref{eqn:C0_fraction}}\\
    \text{note } \ (1+2D^2)\qty(\frac{1-zt}{1+zt}) = \qty(\frac{1-zt}{1+zt})^3, \\
    \text{so } \ \boxed{U_0 = \int_0^\infty e^{-t} \qty(\frac{1-zt}{1+zt})^3 dt}. 
\end{gather}
\end{subequations}

\subsubsection{First depletion correction} \label{subsubsec:U_dep_1}
We know that
\begin{align}
    C_1 = \frac{1}{4} \qty[(2x^2-4x+3)D^2 + 5(x-1)D + 1]C_0, \tag{\ref{eqn:C1_final}}
\end{align}
so 
\begin{align}
    U_\text{dep}^{(1)} = \qty(1+\frac{1}{2}\mathcal L^2) C_1 + \frac{1}{2}\qty[1+x\mathcal L -2D]C_0,
\end{align}
the necessary ingredients we need to evaluate this equation are:
\begin{subequations}
\begin{gather}
    \text{let } \quad \hat P \equiv (2x^2-4x+3)D^2 + 5(x-1)D + 1, \\
    \partial_x\hat P = 4(x-1)D^2 + 5D, \quad \partial_x^2 \hat P = 4D^2, \\
    1+\frac{1}{2}\mathcal L^2 = 1+2D^2+2D\partial_x + \frac{1}{2}\partial_x^2, \\
    (1+\frac{1}{2}\mathcal L^2)C_1 = \qty[\frac{1+2D^2}{4}\hat P + 2(x-1)D^3 + 3D^2]C_0,  \\
    \frac{1}{2}(1+x\mathcal L - 2D) C_0 = \qty((x-1)D + \frac{1}{2})C_0, \\
    \Rightarrow \ U_\text{dep}^{(1)} = \frac{1+2D^2}{4}\hat P C_0 + \qty[(1+2D^2)\qty((x-1)D+\frac{1}{2})+2D^2]C_0 \\
    \Rightarrow \boxed{U_\text{dep}^{(1)} = \frac{1+2D^2}{4} \qty[(2x^2-4x+3)D^2 + 9(x-1)D+3]C_0 + 2D^2C_0}. \label{eqn:U_dep_1}
\end{gather}
\end{subequations}
Or, grouping terms by powers of $x$, we have
\begin{subequations}
\begin{align}
    U_\text{dep}^{(1)} & = P_{U_1}(D,x)C_0, \\
    P_{U_1}(D,x) & = \qty(\frac{3}{4} - \frac{9}{4}D+\frac{17}{4}D^2-\frac{9}{2}D^3 + \frac{3}{2}D^4) \\
    & \quad + x\qty(\frac{9}{4}D-D^2+\frac{9}{2}D^3-2D^4) + x^2\qty(\frac{1}{2}D^2+D^4).
\end{align}
\end{subequations}
We obtain the normalized harmonic polynomial simply by replacing $D$ with $j$. For example, at $j=1$, we have the normalized polynomial 
\begin{align}
    B_{r=1,j=1}(1,x) = -\frac{1}{4}+\frac{15}{4}x+\frac{3}{2}x^2,
\end{align}
the physical $e^{2x}$ coefficient is therefore
\begin{align}
    \tilde H_{2,1} = \frac{1}{32}-\frac{15}{32}x - \frac{3}{16}x^2,
\end{align}
as can be easily read from the Reference Table.

\subsubsection{Second correction and the first return profile} \label{subsubsec:U_dep_2} 
Starting from $r=2$, the generic polynomial no longer describes every normalized coefficient. The first exceptional term is $-\frac{3}{128z}$, and the second correction to the first depletion is
\begin{align}
    \Aboxed{U_\text{dep}^{(2)} = P_{U_2}(D,x)C_0 + \frac{1-x}{4} + \frac{3z}{64} - \frac{3}{128z}}, \label{eqn:U_dep_2}
\end{align}
where
\begin{subequations}
\begin{align}
    P_{U_2}(D,x) & = p_0(D) + xp_1(D) + x^2p_2(D) + x^3p_3(D) + x^4p_4(D), \\
    p_0 & = -\frac{19}{32} + \frac{51}{64}D + \frac{39}{32}D^2 - \frac{31}{8}D^3 + \frac{181}{32}D^4 - \frac{41}{16}D^5 + \frac{9}{16}D^6,\\
    p_1 & = \frac{1}{4} - \frac{23}{64}D - \frac{101}{32}D^2 + \frac{47}{8}D^3 - \frac{145}{16}D^4 + \frac{99}{16}D^5 - \frac{3}{2}D^6, \\
    p_2 & = -\frac{1}{8}-\frac{1}{8}D + \frac{51}{32}D^2 - \frac{25}{8}D^3 + \frac{89}{16}D^4 - \frac{23}{4}D^5 + \frac{7}{4}D^6, \\
    p_3 &= \frac{25}{24}D^3 - \frac{1}{2}D^4 + \frac{25}{12}D^5 - D^6, \\
    p_4 &=\frac{1}{8}D^4 + \frac{1}{4}D^6.
\end{align}
\end{subequations}
\indent Now focus on the first tower of exceptional terms (shaded green in the Reference Table):
\begin{align}
    E_{r=2n,j=n} = \qty{-\frac{3}{128}, -\frac{45}{8192}, -\frac{1197}{524288}, \dots}.
\end{align}
The physical contribution is
\begin{align}
    \epsilon^r2(-1)^j j! z^j E_{r,j} = \epsilon^{r-2j}w^j \qty[2(-1)^jj!E_{r,j}],
\end{align}
so for example, the contribution of $E_{2,1}$ is 
\begin{align}
    2(-1)\qty(-\frac{3}{128})w = \frac{3w}{64}. 
\end{align}
\indent Now recall
\begin{align}
    R_0(w) = -1 + \frac{w}{64} - \frac{5w^2}{2048} + \frac{189w^3}{131072} - \frac{4671w^4}{2097152} + \dots
\end{align}
is the leading amplitude return profile. We found that
\begin{align}
    U_\text{ret}^{(0)} = -1+\frac{3w}{64} - \frac{45w^2}{2048} + \frac{3591w^3}{131072} - \frac{154143w^4}{2097152} + \dots,
\end{align}
and the coefficients satisfy
\begin{subequations}
\begin{align}
    [w^j]U_\text{ret}^{(0)} & = (1+2j^2) [w^j] R_0, \\
    \Rightarrow \Aboxed{U_\text{ret}^{(0)} & = (1+2D_w^2) R_0}. 
\end{align}
\end{subequations}

\subsubsection{First return correction} \label{subsubsec:U_ret_1}
If we define
\begin{align}
    R_0 = -1+\sum_{j\geqslant1}a_jw^j, \qquad R_1 = \sum_{j\geqslant1}r_{1,j}w^j, \qquad U_\text{ret}^{(1)} = \sum_{j\geqslant1}\mu_j w^j,
\end{align}
then from the Reference Tables, it is easily extracted that
\begin{align}
    (1+2j^2) \qty[r_{1,j}(x) + j(x-3)a_j + \frac{1}{2}a_j] + 2j^2a_j,
\end{align}
in other words,
\begin{align}
    \Aboxed{U_\text{ret}^{(1)} & = (1+2D_w^2) \qty[R_1 + (x-3)D_wR_0 + \frac{1}{2}R_0] + 2D_w^2R_0}.\label{eqn:U_ret_1}
\end{align}

\subsubsection{The $m$-th passage}
Define 
\begin{align}
    f_m(u) = (-1)^{m+1}\tanh(s_m+u), \quad s_m = x - (2m+1)\log4\alpha, \quad \rho_m(u) = \abs{h_m(u)}^2,
\end{align}
then we know the leading amplitude profile is
\begin{align}
    C_m(s_m) = \int \rho_m(u) f_m(u) \ du,
\end{align}
while the leading 3rd-order mixed profile is
\begin{align}
    B_m(s_m) = \int\rho_m(u)f_m(u)^3 \ du.
\end{align}
And note that
\begin{align}
    f_m^3 = f_m + \frac{1}{2}\partial_{s_m}^2 f_m,
\end{align}
we have
\begin{align}
    \Aboxed{B_m = C_m + \frac{1}{2}\partial_{s_m}^2 C_m}.
\end{align}

\subsubsection{Coherent stitching and Poisson clock averaging} \label{subsubsec:U_poisson}
We do not need to write out explicitly the coherent sum or the $N$-sector decomposition. Since $U$ is determined by $C$ via equation \eqref{eqn:U_id}, we can directly apply our existing results for $C$. 

Recall the Hilbert space decomposes into different $N$ sectors, which really need to be treated differently. With the amplitude $C$, this was done by the substitution $\alpha\rightarrow a_N, x \rightarrow\xi_N$, where 
\begin{align}
    C_\text{total} = \mathbb E_N F(a_N, \xi_N), \qquad a_N = \sqrt{N+\frac{3}{2}}, \qquad \xi_N = \frac{x a_N}{\alpha} = xa_N\sqrt\epsilon,
\end{align}
the $\mathbb E_N$ here represents a Poisson-weighted sum,
\begin{align}
    \mathbb E_N[X_N] = \sum_{N=0}^\infty p_N(\epsilon) X_N, \quad \text{ where } \quad p_N(\epsilon) = e^{-1/\epsilon} \frac{\epsilon^{-N}}{N!},
\end{align}
Now we want to calculate $U$ from our identity between $U$ and $C$, reproduced here:
\begin{align}
    U = C + \frac{1}{2}C_{xx} + \frac{\epsilon}{2}\qty(C+x C_x -2\epsilon C_\epsilon), \tag{\ref{eqn:U_id}}
\end{align}
and we see we need to know $xC_x, C_{xx},$ and $-2\epsilon C_\epsilon$: 
\begin{subequations}
\begin{align}
    C_x & = \mathbb E_N \qty[a_N \sqrt\epsilon F_\xi(a_N,\xi_N)], \\
    xC_x & = \mathbb E_N \qty[\xi_N F_\xi(a_N,\xi_N)], \\
    \frac{1}{2}C_{xx} & = \mathbb E_N \qty[\frac{1}{2}\epsilon a_N^2F_{\xi\xi}(a_N,\xi_N)], 
\end{align}
\end{subequations}
and noting that both $p_N$ and $F$ depend on $\epsilon$:
\begin{subequations}
\begin{gather}
    -2\epsilon\partial_\epsilon p_N(\epsilon) = 2\qty(N-\frac{1}{\epsilon})p_N(\epsilon), \\
    -2\epsilon\partial_\epsilon F(a_N, x a_N\sqrt\epsilon) = -2\epsilon\qty(\frac{1}{2}xa_N\epsilon^{-1/2})F_\xi(a_N,\xi_N) = -\xi_NF_\xi(a_N,\xi_N), \\
    \Rightarrow \ -2\epsilon C_\epsilon = \mathbb E_N \qty[2\qty(N-\frac{1}{\epsilon})F(a_N,\xi_N) - \xi_NF_\xi(a_N,\xi_N)].
\end{gather}
\end{subequations}
Combine all terms, we have
\begin{subequations}
\begin{align}
    U & = \mathbb E_N\qty[F + \frac{1}{2}\epsilon a_N^2 F_{\xi\xi} + \qty(\epsilon N + \frac{\epsilon}{2}-1)F] \\
    & = \epsilon \mathbb E_N \qty[a_N^2 \qty(F+\frac{1}{2}F_{\xi\xi})-F] \\
    \Rightarrow U & = \mathbb E_N \qty[\frac{(N+3/2)(F+\frac{1}{2}F_{\xi\xi})-F}{\alpha^2}].
\end{align}
\end{subequations}
Or, written in the more explicit form,
\begin{align}
    \Aboxed{U =  \frac{1}{\alpha^2} \mathbb E_N \qty[(N+\frac{3}{2})\qty{1-\int (\abs{\Psi_{\text{even,}N}}^2 - \abs{\Psi_{\text{odd,}N}}^2 )(1+\tanh^3y) dy } - C_N(\chi)]}. \label{eqn:U_poisson}
\end{align}

\subsection{3rd-order phase $\expval*{ccc}$}
Recall our definitions for $C$ and $U$, and now we introduce $T$:
\begin{align}
    C = \frac{\expval*{c}}{\alpha}, \qquad U = \frac{\expval*{c^\dag cc}}{\alpha^3}, \qquad T = \frac{\expval*{ccc}}{\alpha^3}.
\end{align}
We shall see that just like with $\expval*{c^\dag c}$ and $\expval*{cc}$, the difference between $U$ and $T$ is also of $\mathcal O(\epsilon^2)$. For example, based on our BCH expansions,
\begin{subequations}
\begin{gather}
    C = 1-\frac{\epsilon}{2}\sinh^2x + \dots, \\
    U = 1-\frac{3\epsilon}{2}\sinh^2x + \dots ,\qquad T = 1-\frac{3\epsilon}{2}\sinh^2x +\dots \\
    U - T = \epsilon^2(x^2-\sinh^2x) + \dots, 
\end{gather}
\end{subequations}
this motivates us to study the difference, $U-T$. First define the quadratures and the symmetric product,
\begin{subequations}
\begin{gather}
    X = c+c^\dag , \qquad Y = -i(c-c^\dag), \\
    YXY  = c^\dag cc + c^{\dag2}c -c^3 -c^{\dag3} +c+c^\dag, \\
    \Rightarrow \ \expval*{YXY} = 2 \qty(\expval*{c^\dag cc}+\expval*{c}-\expval*{ccc}),
\end{gather}
\end{subequations}
therefore just like how we defined $V_Y = 1+2\alpha^2(P-S)$ from the operator relation $Y^2 = 1+ 2c^\dag c - c^2 - c^{\dag2}$, we shall define
\begin{align}
    \Aboxed{Q = 
    \frac{\expval*{YXY}}{2\alpha}= C+ \alpha^2(U-T)}. 
\end{align}
Or, 
\begin{align}
    \Aboxed{T = U + \epsilon(C-Q)}, \label{eqn:T_def}
\end{align}
Note this suggests $Q$ is a highly useful quantity to calculate—knowing $C$ and $Q$ to the first correction allows us to determine the difference $T-U$ to its second order. As such, $Q$ is the quantity we will supply in our Reference Sheets. To first order, 
\begin{align}
    \Rightarrow \ Q & = 1+\epsilon\qty(x^2-\frac{3}{2}\sinh^2x) + \dots.
\end{align}
Or more generally, write out the perturbative expansions of $C,U,T$,
\begin{subequations}
\begin{align}
    C & = C_0 + \epsilon C_1 + \dots , \\
    U & = U_0 + \epsilon U_1 + \epsilon^2 U_2 + \dots, \\
    T & = T_0 + \epsilon T_1 + \epsilon^2 T_2 + \dots,
\end{align}
\end{subequations}
and the definition of $Q$ gives
\begin{subequations}
\begin{align}
    Q & = C_0 + \epsilon_1 C_1 + \frac{1}{\epsilon}\cancel{(U_0 - T_0)} + U_1 - T_1 + \dots, \\
    \Rightarrow \ Q_0 & = C_0 + U_1 - T_1, \\
    Q_1 & = C_1 + U_2 - T_2. \label{eqn:Q1}
\end{align}
\end{subequations}

\subsubsection{First depletion and return}

From the Reference Sheets, we see $\expval*{ccc}$ and $\expval*{c^\dag cc}$ have the same leading order coefficients for both depletion and return. This means
\begin{gather}
    T_0 = (1+2D_z^2)C_0 = U_0, \\
    \quad \boxed{T_0(z) = U_0(z) = \int_0^\infty e^{-t} \qty(\frac{1-zt}{1+zt})^3 dt}, \\
    \text{or } \ \boxed{T_0(s_0) = -\int_{-\infty}^\infty \rho_0(u) \tanh^3(s_0+u) \ du}.
\end{gather}
Same for the return profile:
\begin{align}
    T_\text{ret}^{(0)} &= U_\text{ret}^{(0)} = (1+2D_w^2)R_0 = (1+\partial_{s_1}^2) R_0, \\
    \Aboxed{T_\text{ret}^{(0)} &= \int_{-\infty}^\infty \rho_1(u) \tanh^3(s_1+u) \ du},
\end{align}

\subsubsection{First depletion correction} \label{subsubsec:T_dep_1}
The first correction, $T_\text{dep,1}$ is completely defined by the polynomials, 
\begin{align}
    \Aboxed{T_\text{dep}^{(1)}(z,x) = \frac{1+2D_z^2}{4} \qty[(2x^2-4x+3)D_z^2 + 9(x-1)D_z+3] C_0(z)}, \label{eqn:T_dep_1}
\end{align}
and one can check that by replacing $D$ with $j$ and substituting $j$, the polynomial expression returns the normalized coefficients. Also, if we compare with $U_1$ in equation \eqref{eqn:U_dep_1}, we find
\begin{align}
    T_1 = U_1 - 2D_z^2C_0. 
\end{align}
Note from $U_0=(1+2D_z^2)C_0$, this is equal to 
\begin{align}
    T_1 = U_1 + C_0 - U_0, \label{eqn:T1_id}
\end{align}
and also note that from $Q_0 = C_0 + U_1 - T_1$, this means
\begin{align}
    Q_0 = U_0 = T_0.
\end{align}
This also suggests we can study a different quantity,
\begin{subequations}
\begin{align}
    U-T & = \epsilon(Q-C) \\
    & = \epsilon(Q_0 -C_0) + \epsilon^2 (Q_1 - C_1) + \dots \\
    & = \epsilon(U_0-C_0) + \epsilon^2 (U_1 - C_1 +G) + \dots,
\end{align}
\end{subequations}
where we defined
\begin{align}
    G \equiv Q_1 - U_1. \label{eqn:Gap_def}
\end{align}
The point is that both $(U_0 - C_0)$ is known, and $(U_1 - C_1)$ is also known. Only the difference $G = Q_1 - U_1$ is unknown. This suggests we can study the difference $U-T$ purely by studying the smaller difference $G = Q_1 - U_1$. This also mean we may write equation \eqref{eqn:Q1} as
\begin{subequations}
\begin{align}
    T_2 &= U_2 +C_1 - U_1 - G, \label{eqn:T2_G}\\
    \Rightarrow \ T & = U_0 + \epsilon(U_1+C_0-U_0) +\epsilon^2(U_2+C_1-U_1-G).
\end{align}
\end{subequations}
In fact, we can do better—we can omit $U_2$ altogether, and simply calculate
\begin{align}
    T & \simeq U_0 + \epsilon(U_1+C_0-U_0) +\epsilon^2(C_1-U_1-G).
\end{align}
This is because $U$ also has a $\epsilon^2 U_2$ term, which cancels with the $\epsilon^2 U_2$ in $T$ when we take their difference.

\subsubsection{Second correction to the first depletion} \label{subsubsec:T_dep_2}
We can read off the spreadsheet both the generic polynomials and the exceptional terms. There are three corrections: 
\begin{subequations}
\begin{align}
    \text{constant: } &\quad \frac{9}{4}(1-x), \\
    \text{positive harmonic: } &\quad E_{2,1} = -\frac{3}{128}, \quad \xlongrightarrow{\text{contributing}} \quad \frac{3z}{64}, \\
    \text{negative harmonic: } &\quad -\frac{3}{8}\epsilon e^{-2x}, \quad \xlongrightarrow{\text{contributing}} \quad -\epsilon^2 \frac{3}{128z}, 
\end{align}
\end{subequations}
summing them up,
\begin{align}
    \Aboxed{T_\text{dep}^{(2)} = P_{T_2}(D,x)C_0 + \frac{9}{4}(1-x)+\frac{3z}{64}-\frac{3}{128z}}. \label{eqn:T_dep_2}
\end{align}
and $P_{T_2}$ is almost identical to $P_{U_2}$:
\begin{subequations}
\begin{align}
    P_{T_2}(D,x) & = p_0(D) + xp_1(D) + x^2p_2(D) + x^3p_3(D) + x^4p_4(D), \\
    p_0 & = -\frac{99}{32} + \frac{243}{64}D + \frac{23}{32}D^2 - \frac{11}{8}D^3 + \frac{133}{32}D^4 - \frac{41}{16}D^5 + \frac{9}{16}D^6,\\
    p_1 & = \frac{9}{4} - \frac{183}{64}D - \frac{101}{32}D^2 + \frac{51}{8}D^3 - \frac{113}{16}D^4 + \frac{99}{16}D^5 - \frac{3}{2}D^6, \\
    p_2 & = -\frac{9}{8}-\frac{9}{8}D - \frac{13}{32}D^2 - \frac{41}{8}D^3 + \frac{73}{16}D^4 - \frac{23}{4}D^5 + \frac{7}{4}D^6, \\
    p_3 &= \frac{25}{24}D^3 - \frac{1}{2}D^4 + \frac{25}{12}D^5 - D^6, \\
    p_4 &=\frac{1}{8}D^4 + \frac{1}{4}D^6.
\end{align}
\end{subequations}
\indent Now we have all the ingredients to calculate $G_\text{dep} = U_2 + C_1 - U_1 - T_2$ as in equation \eqref{eqn:T2_G}, putting it all together (equation \eqref{eqn:C1_final} for $C_1$, \eqref{eqn:U_dep_1} for $U_1$,and \eqref{eqn:U_dep_2} for $U_2$), we have
\begin{subequations}
\begin{align}
    G_\text{dep}(z,x) & = (x^2-2x+2)C_0 + \qty(x^2+\frac{3}{2}x-2)D_zC_0 \\
    & \quad + (2x^2-3)D_z^2C_0 + (2x^2-5x+2)D_z^3C_0 + 2(x-1). 
\end{align}
\end{subequations}
Two limits are useful: Before depletion,
\begin{subequations}
\begin{gather}
    C_0 \longrightarrow1,\qquad  D_z^k C_0 \longrightarrow0, \\
    \Rightarrow \ G_\text{dep} \longrightarrow x^2 - 2x+2+2(x-1) = x^2.
\end{gather}
\end{subequations}
After depletion, 
\begin{align}
    C_0 \longrightarrow-1, \qquad D^k_z C_0 \longrightarrow0,
\end{align}
giving
\begin{align}
    G_\text{dep} \longrightarrow -(x^2-2x+2)+2(x-1) = -(x-2)^2. \label{eqn:G_dep_lim_after} 
\end{align}

\subsubsection{First return correction} \label{subsubsec:T_ret_1}
We will follow closely the procedure leading to equation \eqref{eqn:Ur_id}. Note the total derivative is
\begin{align}
    \mathcal L \equiv \frac{d}{dx} = \partial_x + 2D_w \equiv \partial_x + 2D,
\end{align}
now let the local return expansion be
\begin{align}
    R(\epsilon,x)=R_0(w) + \epsilon R_1(w,x) + \dots ,
\end{align}
and since $\epsilon\partial_\epsilon w = 3w$, 
\begin{align}
    \epsilon\partial_\epsilon [\epsilon^r R_r(w,x)] = \epsilon^r (r+3D)R_r.
\end{align}
completely analogous to equation \eqref{eqn:Ur_id}; except now we have
\begin{align}
    U_\text{ret}^{(1)} & = \qty(1+\frac{1}{2}\mathcal L^2)R_1 + \frac{1}{2}(1+x\mathcal L-6D) R_0.
\end{align}
Note 
\begin{align}
    \mathcal L R_0 = 2DR_0,
\end{align}
since $R_0$ has no explicit $x$-dependence. And also (now analogous to the calculations leading up to \eqref{eqn:U_dep_1}),
\begin{subequations}
\begin{gather}
    \mathcal L^2 R_1 = \partial_x^2 R_1 + 4D\partial_x R_1 + 4D^2R_1, \\
    \Rightarrow \ U_\text{ret}^{(1)} = (1+2D^2)R_1 + 2D\partial_x R_1 + \frac{1}{2}\partial_x^2 R_1 + \frac{1}{2}R_0 + (x-3)DR_0.
\end{gather}
\end{subequations}
Now use our expression \eqref{eqn:R1_final} for $R_1$:
\begin{subequations}
\begin{gather}
    R_1(w,x) = \qty[\frac{(x-3)^2}{2}-\frac{5}{12}]D^2R_0 + \frac{5}{4}(x-3)DR_0 + \frac{1}{4}R_0  - \frac{w}{384}(D+1)^5[1-R_0], \tag{\ref{eqn:R1_final}}\\
    \Rightarrow \ \partial_x R_1 = (x-3)D^2R_0 + \frac{5}{4}DR_0, \\
    \partial_x^2 R_1 = D^2 R_0, \\
    \Rightarrow \ 2D\partial_xR_1 + \frac{1}{2}\partial_x^2 R_1 = 2(x-3)D^3R_0 + 3D^2R_0.
\end{gather}
\end{subequations}
Substituting back into the expression for $U_\text{ret}^{(1)}$, 
\begin{subequations}
\begin{align}
    \Rightarrow \ U_\text{ret}^{(1)} & = (1+2D^2) R_1 + (x-3) (D+2D^3) R_0 + 3D^2 R_0 + \frac{1}{2}R_0 \\
    \Rightarrow \ U_\text{ret}^{(1)} & = (1+2D^2) \qty[R_1 + (x-3)DR_0 + \frac{1}{2}R_0] + 2D^2 R_0.
\end{align}
\end{subequations}
Now applying the identity \eqref{eqn:T1_id} but for the return profile (replace $C_0 \rightarrow R_0$):
\begin{subequations}
\begin{align}
    T_\text{ret}^{(1)} & = U_\text{ret}^{(1)} + R_0 - U_\text{ret}^{(0)} = U_\text{ret}^{(1)} - 2D^2R_0, \\
    \Rightarrow \ \Aboxed{T_\text{ret}^{(1)} & = (1+2D^2) \qty[R_1 + (x-3)DR_0 + \frac{1}{2}R_0]}. \label{eqn:T1_ret_penultimate}
\end{align}
\end{subequations}
Or substitute equation \eqref{eqn:R1_final} to eliminate $R_1$, we have
\begin{align}
    \Aboxed{T_\text{ret}^{(1)} & = (1+2D^2) \qty{\qty[\frac{(x-3)^2}{2}-\frac{5}{12}]D^2R_0 + \frac{9}{4}(x-3)DR_0 + \frac{3}{4}R_0 - \frac{w}{384}(D+1)^5(1-R_0)}}. \label{eqn:T1_ret_final}
\end{align}
We can check equation \eqref{eqn:T1_ret_penultimate} immediately. Write
\begin{align}
    R_1 = -\frac{1}{4} + \sum_{j\geqslant1} a_j q_j(x) w^j,
\end{align}
then equation \eqref{eqn:T1_ret_penultimate} becomes
\begin{align}
    [w^j] T_\text{ret}^{(1)} & = (1+2j^2)a_j \qty[q_j(x) + j(x-3) + \frac{1}{2}].
\end{align}
Recall, for example, from Table \ref{tab:an_and_qn}, 
\begin{align}
    q_1(x) = \frac{x^2}{2}-\frac{7x}{4}+ \frac{1}{4},
\end{align}
so for $j=1$,
\begin{align}
    [w]T_\text{ret}^{(1)} & = \frac{3}{64} \qty[\frac{x^2}{2} - \frac{7x}{4} + \frac{1}{4}+x-3+\frac{1}{2}] = \frac{3x^2}{128}-\frac{9x}{256}-\frac{27}{256},
\end{align}
which agrees with the first blue term in the $\expval*{ccc}$ Reference Sheet (under the ``4 Corrections'' tab), up to normalization (given in the ``5 Indices'' tab).

The asymptotes also match. As $w\rightarrow0$, 
\begin{align}
    R_0 \longrightarrow-1, \quad R_1 \longrightarrow-\frac{1}{4}, \quad D_wR_0 \longrightarrow 0, 
\end{align}
and therefore from equation \eqref{eqn:T1_ret_penultimate},
\begin{align}
    T_\text{ret}^{(1)} \longrightarrow -\frac{1}{4}-\frac{1}{2}=-\frac{3}{4},
\end{align}
which agrees with the $z\rightarrow\infty$ limit of $T_\text{dep}^{(1)}$ given in equation \eqref{eqn:T_dep_1}. 

Now let's calculate the phase gap at return. Define
\begin{align}
    G_\text{ret} & = Q_\text{ret}^{(1)} - U_\text{ret}^{(1)} = R_1 + U_\text{ret}^{(2)} - T_\text{ret}^{(2)} - U_\text{ret}^{(1)},
\end{align}
where we used equation \eqref{eqn:Q1}, now adopted for the return case. 

Let's calculate the first term of $G_\text{ret}$ explicitly. The coefficients of $w$ are (see Tab 4 of the Amplitude table for $R_1$, the $\expval*{c^\dag cc}$ table for $U_\text{ret}^{(1)}$, and the $\expval*{ccc}$ table for $T_\text{ret}^{(2)}$)
\begin{subequations}
\begin{align}
    [w] R_1 & = \frac{2x^2-7x+1}{256}, \\
    [w]U_\text{ret}^{(1)} & = \frac{6x^2-9x-19}{256}, \\
    [w]U_\text{ret}^{(2)} & = \frac{24x^4-88x^3-206x^2 + 699x + 341}{4096}, \\
    [w]T_\text{ret}^{(2)} & = \frac{24x^4-88x^3 -1806x^2+7995x-7467}{4096},  \\
    \Rightarrow \ [w]G_\text{ret} & = \frac{1536x^2-7264x+8128}{4096} = \frac{3}{8}x^2 - \frac{227}{128}x + \frac{127}{64}.
\end{align}
\end{subequations}
And we know $a_1 = \frac{1}{64}$. Define the ratio $g_j(x)$:
\begin{align}
    g_j(x)\equiv \frac{[w]G_\text{ret}}{a_j}, \qquad g_1(x) = 24x^2 - \frac{227}{2}x + 127.
\end{align}
The first few $g_j(x)$ are, for example,
\begin{table}[H]
    \centering
    \begin{tabular}{c|c}
        $j$ & $g_j(x)$ \\\hline
        1 & $24x^2-\frac{227}{2}x+127$ \\[0.5em]
        2 & $\frac{891}{5}x^2 - \frac{4599}{5}x + \frac{5722}{5}$ \\[0.5em]
        3 & $\frac{5092}{7}x^2 - \frac{55071}{14}x + \frac{36385}{7}$
    \end{tabular}
    \caption{The first three coefficient ratios $g_j$ for the third-order return gap.}
    \label{tab:g_j}
\end{table}
We might find the coefficients for $x^2$ strongly resemble $u_j$ in Table \ref{tab:un} that we've seen in the $\expval*{cc}$ calculations,
\begin{gather}
    u_j = 19j^2 -12j+2-4q_j(0) = \qty{8, \frac{99}{5}, \frac{268}{7}, \dots}.
\end{gather}
Indeed, we can relate $g_j(x)$ to $u_j$ by (define $d= x-3$)
\begin{align}
    g_j(x) & = u_j \qty[1+4jd+(1+2j^2)d^2] - \frac{jx}{2}(1+2j^2) - j^2.  \label{eqn:aj_uj_series}
\end{align}
We want to put this into a series—but note that we already defined such a series at least for the $u_j$ terms in equation \eqref{eqn:U_caligraphic_series} and \eqref{eqn:U_caligraphic_final}:
\begin{subequations}
\begin{align}
    \mathcal U(w) & = \sum_{j\geqslant1} a_j u_j w^j \label{eqn:U_series} \\
    \Rightarrow \mathcal U(w) & = 1+ R_0+ 3D_wR_0+ \frac{8}{3}D_w^2R_0 + \frac{w}{96}(D_w+1)^5(1-R_0) \label{eqn:U_caligraphic_final_repeat}
\end{align}
\end{subequations}
where $a_j$ are the coefficients of $R_0$. Summing $g_j(x)$, the term containing $u_j$ gives
\begin{align}
    \sum_{j\geqslant1} a_j u_j \qty[1+4jd+(1+2j^2)d^2]w^j = (1+d^2)\mathcal U + 4dD_w\mathcal U + 2d^2D_w^2\mathcal U. 
\end{align}
The remaining terms give
\begin{align}
    -\frac{x}{2}\sum_{j\geqslant1} j(1+2j^2)a_jw^j - \sum_{j\geqslant1}j^2a_jw^j = -\frac{x}{2}(D_w+2D_w^3)R_0 - D_w^2 R_0.
\end{align}
So $G_\text{ret}$, up to a constant independent of $w$, is
\begin{align}
    \tilde G_\text{ret} = \qty[(1+d^2)+4dD_w+2d^2D_w^2]\mathcal U - \frac{x}{2}(D_w+2D_w^3)R_0 - D_w^2 R_0.
\end{align}
Note every term here vanishes as $w\rightarrow0$, the asymptotic behavior of $G_\text{ret}$ is entirely determined by the constant, which determined back in equation \eqref{eqn:G_dep_lim_after} as $-(x-2)^2$. Therefore,
\begin{align}
    \Aboxed{G_\text{ret} & = -(x-2)^2 + \qty[1+(x-3)^2+4(x-3)D_w + 2(x-3)^2D_w^2] \mathcal U(w) - \frac{x}{2}(D_w+2D_w^3)R_0 - D_w^2R_0},
\end{align}
with $\mathcal U(w)$ given above in equation \eqref{eqn:U_caligraphic_final_repeat}. 

We can also derive its integral form using the weight $W_1$ already introduced for $\expval*{cc}$ in equation \eqref{eqn:W1_intro}. We shall follow closely the derivation leading up to equation \eqref{eqn:G1_integral}. Let $y=s_1+u$, then define the expansions for tanh's. 
\begin{subequations}
\begin{align}
    \tanh y & = -1 + 2\sum_{j\geqslant1} (-1)^{j-1}e^{2jy}, \\
    \tanh^2 y & = 1+4\sum_{j\geqslant1} (-1)^j je^{2jy}, \label{eqn:tanh2_series} \\
    \tanh^3y & = -1 + 2\sum_{j\geqslant1}(-1)^{j-1}(1+2j^2)e^{2jy},
\end{align}
\end{subequations}
with the last equation coming from $\tanh^3y = \tanh y + \frac{1}{2}\frac{d^2}{dy^2}\tanh y$. Already we notice these 3 series should be able to fit into the $u_j$ terms in equation \eqref{eqn:aj_uj_series}, but let us show it explicitly. 

Write $d = x-3$, then $G_\text{ret}$ written as a series is
\begin{subequations}
\begin{align}
    G_\text{ret} & = -(1+d)^2 \\
    & \quad + \underbracket{\sum_{j\geqslant1} a_j u_j \qty[1+4jd+(1+2j^2)d^2]w^j}_{\mathclap{\text{\ding{172}}}} \\
    & \quad - \underbracket{\frac{x}{2}(D_w+2D_w^3)R_0 - D_w^2R_0}_{\mathclap{\text{\ding{173}}}}, 
\end{align}
\end{subequations}
write out the expansion for $R_0$:
\begin{subequations}
\begin{align}
    R_0(w) & = -1 + \sum_{j\geqslant1} a_j w^j = \int \rho_1(u) \tanh(s_1+u) \ du \\
    & = -1 + 2\sum_{j\geqslant1} (-1)^{j-1} e^{2js_1} \int \rho_1(u) e^{2ju} \ du \\
    & = -1 + \sum_{j\geqslant1} \qty[\frac{2(-1)^{j-1}}{256^j} \int \rho_1(u) e^{2ju} \ du]\  w^j,
\end{align}
\end{subequations}
matching coefficients, we see
\begin{align}
    a_j = \frac{2(-1)^{j-1}}{256^j} \int \rho_1(u) e^{2ju} \ du. 
\end{align}
Now let's go back to term \ding{172} in $G_\text{ret}$. First note that the $u_j$ is the same $u_j$ used in defining the weight function $W_1(u)$ back in the $\expval*{cc}$ calculation, in equation \eqref{eqn:W1_intro}:
\begin{align}
    K_j = \int W_1(u) e^{2ju} \ du \equiv u_j \int \rho_1(u) e^{2ju}\ du, \quad j\geqslant1, \tag{\ref{eqn:W1_intro}}
\end{align}
therefore
\begin{subequations}
\begin{align}
    a_j u_j w^j & = 2(-1)^{j-1} e^{2js_1} u_j \int \rho_1(u) e^{2ju} \ du, \\
    \Rightarrow \ a_j u_j w^j & = 2(-1)^{j-1} \int W_1(u) e^{2jy} \ du. 
\end{align}
\end{subequations}
Now we can evaluate \ding{172} in full,
\begin{subequations}
\begin{align}
    \text{\ding{172}} & = \sum_{j\geqslant1} a_j u_j \qty[1+4jd+(1+2j^2)d^2]w^j \\
    & = \sum_{j\geqslant1} a_j u_j w^j + 4d\sum_{j\geqslant1} ja_j u_jw^j + d^2\sum_{j\geqslant1} (1+2j^2)a_ju_jw^j . 
\end{align}
\end{subequations}
For the first sum, use
\begin{subequations}
\begin{gather}
    2\sum_{j\geqslant1} (-1)^{j-1}e^{2jy} = 1+\tanh y, \\
    \Rightarrow \ \sum_{j\geqslant1} a_j u_j w^j = \int W_1(u) [1+\tanh y]\ du.
\end{gather}
\end{subequations}
In particular, this derives the integral form of $\mathcal U(w)$ as in equation \eqref{eqn:U_series}:
\begin{align}
    \mathcal U (w) = 1+\int W_1(u) \tanh y \ du. 
\end{align}
For the second sum, use the expansion for $\tanh^2y$, we have
\begin{subequations}
\begin{gather}
    2\sum_{j\geqslant1} (-1)^{j-1}je^{2jy} = \frac{1}{2}(1-\tanh^2y),  \\
    \Rightarrow \ 4d\sum_{j\geqslant1} ja_j u_jw^j = 2d \int W_1(u) [1-\tanh^2y] \ du.
\end{gather}
\end{subequations}
For the third sum, use the expansion for $\tanh^3y$, we have
\begin{align}
    d^2\sum_{j\geqslant1} (1+2j^2)a_ju_jw^j = d^2 \int W_1(u) [1+\tanh^3y]\ du.
\end{align}
Combine all three terms, we have
\begin{subequations}
\begin{align}
    \text{\ding{172}} & = \int W_1(u) \qty[1+\tanh y + 2d(1-\tanh^2y) + d^2(1+\tanh^3y)] \ du \\
    & = \int W_1(u) \qty((1+d)^2 + \tanh y[1-d\tanh y]^2) \ du \\
    & = (1+d)^2 + \int W_1(u) \tanh y [1-d\tanh y]^2 \ du.
\end{align}
\end{subequations}
Note that $(1+d)^2$ term cancels with 
the $-(1+d)^2$ in $G_\text{ret}$. 

Now let's turn our attention to term \ding{173}. First, note that since 
\begin{align}
    w = 256e^{2s_1} \ \Rightarrow \ D_w = \frac{1}{2}\partial_{s_1}.
\end{align}
Denote the leading 3rd-order return profile by 
\begin{align}
    B_1(s_1) \equiv U_\text{ret}^{(0)} = T_\text{ret}^{(0)} = (1+2D_w^2)R_0 = \int \rho_1(u) \tanh^3(s_1+u)\ du,
\end{align}
then the first term in \ding{173} is
\begin{subequations}
\begin{gather}
    (D_w+2D_w^3)R_0 = D_w(1+2D_w^2)R_0 = D_wB_1 = \frac{1}{2}\partial_{s_1}B_1, \\
    \Rightarrow \ -\frac{x}{2}(D_w+2D_w^3)R_0 = -\frac{x}{4}\partial_{s_1}B_1.
\end{gather}
\end{subequations}
As for the final term in \ding{173}, differentiate:
\begin{subequations}
\begin{align}
    \partial_{s_1}R_0 &= \int \rho_1(u) \qty[1-\tanh^2(s_1+u)] \ du = 1-P_1(s_1). \\
    \partial_{s_1}^2R_0 &= -\partial_{s_1}P_1, \\
    \Rightarrow \ -D_w^2 R_0 & = -\frac{1}{4}\partial_{s_1}^2R_0 = \frac{1}{4}\partial_{s_1}P_1. 
\end{align}
\end{subequations}
Putting everything together, we have
\begin{subequations}
\begin{empheq}[box=\fbox]{align}
    G_\text{ret}(s_1,x) & = \int W_1(u) \tanh(s_1+u) \qty[1-(x-3)\tanh(s_1+u)]^2 \ du \\
    & \quad - \frac{x}{4}\partial_{s_1}B_1(s_1) + \frac{1}{4}\partial_{s_1}P_1(s_1). 
\end{empheq}
\end{subequations}
One can check its asymptotic limits—as the derivatives vanish either before or after the return passage, the integral gives
\begin{subequations}
\begin{align}
    G_\text{ret} &\longrightarrow -(1+d)^2 = -(x-2)^2  &\text{ as } \quad s_1 \rightarrow -\infty, \\
    G_\text{ret} &\longrightarrow + (1-d)^2 = (x-4)^2  &\text{ as } \quad s_1 \rightarrow+\infty. 
\end{align}\label{eqn:gap_final}
\end{subequations}
\indent Finally, connect back to the observables, our definition in equation \eqref{eqn:T_def} and \eqref{eqn:Gap_def} give
\begin{subequations}
\begin{gather}
    Q = U+\epsilon G_\text{ret} + \mathcal O(\epsilon^2), \qquad T = U + \epsilon(C-Q), \\
    \Rightarrow \ \boxed{T = U+\epsilon(C-U)-\epsilon^2 G_\text{ret}+\mathcal O(\epsilon^3),}
\end{gather}
\end{subequations}
or in the unnormalized form,
\begin{align}
    \expval*{ccc} & = \qty(1-\frac{1}{\alpha^2})\expval*{c^\dag cc} + \expval*{c} - \frac{G}{\alpha} + \dots,
\end{align}
which makes the sizes of the different contributions clear.

\subsubsection{The $m$-th passage and Poisson clock averaging} \label{subsubsec:T_poisson}
Use the established definitions,
\begin{align}
    s_m = x-(2m+1)\log4\alpha, \qquad \rho_m = \abs{h_m}^2, \qquad f_m(u) = (-1)^{m+1} \tanh(s_m+u), 
\end{align}
the leading amplitude and cubic profiles are
\begin{align}
    C_m^{(0)} = \int \rho_m f_m \ du, \qquad B_m^{(0)} = \int \rho_m f_m^3 \ du. 
\end{align}
Once again, we have
\begin{align}
    f_m + \frac{1}{2}\partial_{s_m}^2 f_m = f_m^3, 
\end{align}
integration then gives
\begin{align}
    T_m^{(0)} = U_m^{(0)} = B_m^{(0)} = C_m^{(0)} + \frac{1}{2}\partial_{s_m}^2C_m^{(0)}.
\end{align}
The extension of equation \eqref{eqn:gap_final} suggests
\begin{subequations}
\begin{empheq}[box=\fbox]{align}
    G_m & = \int W_m(u) f_m(u) \qty[1-(x-(2m+1))\tanh(s_m+u)]^2 \ du \\
    & \quad - \frac{x}{4}\partial_{s_m} B_m^{(0)} - \frac{1}{4}\partial_{s_m}^2 C_m^{(0)}.
\end{empheq}
\end{subequations}
\indent The simplest way to stitch the passages together is adding them directly,
\begin{align}
    T_\text{add}^{(0)} & = 1 + \sum_{m=0}^{M-1} (-1)^{m+1} \int \rho_m(u) \qty[1+\tanh^3(s_m+u)] \ du.
\end{align}
\indent As for coherent stitching, set $\psi_m(y) = h_m(y-s_m)$, and let $\Psi_\text{even}(y) = \sum_{m \text{ even}} \psi_m(y)$, and similarly for $\Psi_\text{odd}$, we have
\begin{align}
    T_\text{coh}^{(0)} = 1- \int \qty(\abs{\Psi_\text{even}}^2 - \abs{\Psi_\text{odd}}^2) (1+\tanh^3y) \ dy. 
\end{align}
\indent For Poisson averaging, substitute in $a_N = \sqrt{N+1}$, we have
\begin{subequations}
\begin{empheq}[box=\fbox]{align}
    Q & = \mathbb E_N \bigg\{ \frac{(N-\alpha^2)^2}{\alpha^2} \qty[1-\int (\abs*{\Psi_{\text{even},N}}^2 + \abs{\Psi_{\text{odd},N}}^2)\sech^2y \ dy ] \\
    & \quad + \frac{\alpha^2}{(N+1)^2} \bigg[\xi_N^2+\sum_m (-1)^{m+1}(\xi_N-2m)^2 \\
    & \qquad + \sum_{m-n\text{ even}}(-1)^{m+1} \Re \int\tanh y(1-d_m\tanh y)(1-d_n\tanh y) \psi_m^{\prime*}\psi_n^\prime \ dy \bigg] \\
    & \quad - \frac{2(N-\alpha^2)}{N+1}\sum_{m-n\text{ even}}(-1)^{m+1} \Re\int\tanh^2y(1-d_n\tanh y)\psi_m^*\psi_n^\prime \ dy\bigg\},
\end{empheq} \label{eqn:Q_poisson}
\end{subequations}
then we may use $T = U+(C-Q)/\alpha^2$ to calculate $T$.

\pagebreak
\section{4th-order correlations through the first depletion} \label{sec:4th}

\subsection{Identities}
There are a number of observables in an experimental setting, such as photon number statistics, that require 4th-order correlations. In this section we list the key 4th-order results, but only in the experimentally relevant first-depletion window. Define the normalized correlation functions as
\begin{align}
    q_{22} = \frac{\expval*{c^{\dag2}c^2}}{\alpha^4}, \quad q_{13} = \frac{\expval*{c^\dag c^3}}{\alpha^4}, \quad q_{04} = \frac{\expval*{c^4}}{\alpha^4}, \quad R = \frac{\expval*{ab\,n_c}}{\alpha^4}.
\end{align}
We shall see the first 3 share the same leading profiles. $R$ is the difference of two known leading profiles. Finally, $q_{22}$ is entirely determined by $P = \expval*{c^\dag c}/\alpha^2$:
\begin{align}
    \Aboxed{q_{22} = \frac{P_{\chi\chi}+8P-2}{6} + \frac{\epsilon}{3}(2P+2\chi P_\chi - 4\epsilon P_\epsilon -1)}. 
\end{align}
\indent All other 4th-order correlations can be calculated from these existing 4 using the relations
\begin{table}[H]
    \centering
    \begin{tabular}{c|c}
        Correlators & Normalized identities \\\hline
        $\expval*{n_ac^2}$ & $S-q_{13}+\frac{\epsilon\chi}{2}S_\chi-\epsilon^2S_\epsilon$ \\[0.5em]
        $\expval*{n_an_c}$ & $P-q_{22}+\frac{\epsilon\chi}{2}P_\chi-\epsilon^2P_\epsilon$ \\[0.5em]
        $\expval*{n_an_b}$ & $1-2P+q_{22}+\epsilon(1-P-\chi P_\chi)+2\epsilon^2P_\epsilon$ \\[0.5em]
        $\expval*{a^{\dag2}a^2}$ & $1-2P+q_{22}-\epsilon\chi P_\chi+2\epsilon^2 P_\epsilon$ \\[0.5em] 
        $\expval*{abc^2}$ & $-T_\chi/3$ \\[0.5em]
        $\expval*{a^\dag b^\dag c^2}$ & $-U_\chi-2R$ \\[0.5em]
        $\expval*{n_a ab}$ & $-C_\chi-\frac{\epsilon}{2}(C_\chi + \chi C_{\chi\chi} - 2\epsilon C_{\chi\epsilon} - R$ \\[0.5em]
        $\expval*{a^2b^2}$ & $\frac{1}{2}S_{\chi\chi} - 2q_{13} + 2S + \epsilon(S+\chi S_\chi-2\epsilon S_\epsilon) $
    \end{tabular}
    \caption{4-th order correlation function identities. Note every quantity has been divided by $\alpha^4$ in the second column. Other quantities can be obtained by conjugating and by swapping $a\leftrightarrow b$.}
    \label{tab:4th_order}
\end{table}

We will study quantities such as intensity correlations and covariances:
\begin{align}
    g_c^{(2)} = \frac{g_{22}}{P}, \quad g_a^{(2)} = \frac{d_a}{(1-P)^2}, \quad g_{ab}^{(2)} = \frac{d_{ab}}{(1-P)^2}, \quad \frac{\text{Cov}(n_a,n_c)}{\alpha^4} = d_{ac} - P(1-P), 
\end{align}
where
\begin{align}
    d_a & = 1-2P + q_{22} - \epsilon\chi P_\chi + 2\epsilon^2 P_\epsilon , \\
    d_{ab} & = 1-2P + q_{22} + \epsilon(1-P-\chi P_\chi) + 2\epsilon^2P_\epsilon, \\
    d_{ac} & = P-q_{22} + \frac{\epsilon\chi}{2}P_\chi - \epsilon^2 P_\epsilon. 
\end{align}
Also the pump photon number variance is
\begin{align}
    \frac{\text{Var}(n_c)}{\alpha^4} = q_{22} - P^2 + \epsilon P.
\end{align}
For pump quadratures $X = c+c^\dag, Y = -i(c-c^\dag)$,
\begin{align}
    \frac{\expval*{X^4}}{\alpha^4} & = 2q_{04} + 8q_{13} + 6q_{22} + 12\epsilon (P+S) + 3\epsilon^2, \\
    \frac{\expval*{Y^4}}{\alpha^4} & = 2q_{04} - 8q_{13} + 6q_{22} + 12\epsilon (P-S) + 3\epsilon^2. \label{eqn:4_quadrature_Y}
\end{align}

\subsection{Leading profiles to the first depletion}
The leading profile has the same tanh profile, raised to the 4th power,
\begin{align}
    Q_m^{(0)} (s_m) = \int \abs*{h_m(u)}^2 \tanh^4(s_m+u) \ du,
\end{align}
with $q_{22,m}^{(0)} = q_{13,m}^{(0)} = q_{04,m}^{(0)} = Q_m^{(0)}$. Since we are only interested in the first depletion, let's define
\begin{align}
    f_z(u) = \frac{1-zu}{1+zu}, \qquad \mathbb E_z[F] = \int_0^\infty e^{-u} F(u) \ du, 
\end{align}
then the previous $m=0$ profiles are simply
\begin{align}
    C_0 = \mathbb E_z[f_z], \qquad P_0 = S_0 = \mathbb E_z[f_z^2], \qquad U_0 = T_0 = \mathbb E_z[f_z^3],
\end{align}
and the 4th order depletion profile has the same pattern, 
\begin{align}
    \Aboxed{Q_0^{(0)} = \mathbb E_z[f_z^4] = 1 + \qty(\frac{8}{3}D+\frac{4}{3}D^3)C_0 = \frac{2D^2P_0 + 4P_0 -1}{3}}.
\end{align}
However, the depletion profile of $R$ has the form
\begin{align}
    \Aboxed{R_0 = P_0 - Q_0 = \mathbb E_z[f_z^2(1-f_z^2)] = -\qty(\frac{2}{3}D+\frac{4}{3}D^3)C_0}. 
\end{align}

\subsection{First correction to the first depletion} \label{subsec:4th_corrections}
Write 
\begin{align}
    q_{ij} = q_{ij,0}+\epsilon Q_{ij,1} + \mathcal O(\epsilon^2), \qquad R = R_0 + \epsilon R_1+\mathcal O(\epsilon^2), 
\end{align}
the correction to $q_{22}$ is
\begin{align}
    Q_{22,1} = 2\qty[2+\frac{11(\chi-1)}{3}D + \frac{2\chi^2-4\chi+9}{3}D^2+\frac{11(\chi-1)}{6}D^3 + \frac{2\chi^2-4\chi+3}{6}D^4]P_0-1,
\end{align}
and
\begin{align}
    Q_{13,1} & = Q_{22,1} + \frac{1}{2}(P_0-Q_0), \\
    Q_{04,1} & = Q_{22,1} + 2(P_0-Q_0) \\
    & = \qty[\frac{4}{3}+\frac{11(\chi-1)}{3}D + \frac{2\chi^2-4\chi+5}{3}D^2+\frac{11(\chi-1)}{6}D^3 + \frac{2\chi^2-4\chi+3}{6}D^4]P_0-\frac{1}{3},
\end{align}
and finally,
\begin{subequations}
\begin{align}
    R_1 & = \bigg[-\frac{11}{12}D + \frac{11}{6}D^2 - \frac{13}{3}D^3+\frac{11}{3}D^4-D^5 \\
    & \qquad + \chi\qty(-\frac{11}{6}D^2+\frac{2}{3}D^3-\frac{11}{3}D^4+\frac{4}{3}D^5)-\chi^2\qty(\frac{1}{3}D^3+\frac{2}{3}D^5)\bigg]C_0.
\end{align}
\end{subequations}
We see the relations
\begin{gather}
    q_{13}-q_{22} = \frac{\epsilon}{2}R_0 + \mathcal O(\epsilon^2), \qquad q_{04} - q_{22} = 2\epsilon R_0 + \mathcal O(\epsilon^2), \\
    \Rightarrow \ q_{04} - 4q_{13} + 3q_{22} = \mathcal O(\epsilon^2),
\end{gather}
note the last relation appears in equation \eqref{eqn:4_quadrature_Y}—these equally large quantities nearly cancel, and therefore the expression should be calculated analytically before floating-point evaluation.

\pagebreak
\section{Squeezed pump analysis through the first depletion} \label{sec:squeezed}

What if the pump starts with a displaced squeezed state? We will study this system using the standard single-mode Gaussian-state formalism \cite{OLIVARES_2012,MANDEL_1995,BARNETT_2002}.
The BCH coefficients derived above remain unchanged,  only the expectation values of the starting states must be replaced.


\subsection{Squeezing theory}
Define the squeezing operator, use the convention
\begin{align}
    S(\zeta) = \exp[\frac{1}{2}(\zeta^*c^2-\zeta c^{\dag2})], \qquad \zeta = re^{i\theta}.
\end{align}
Define
\begin{align}
    \mu = \cosh r, \qquad \kappa = e^{i\theta}\sinh r,
\end{align}
BCH expansion shows the Bogoliubov transformations,
\begin{align}
    S^\dag c S = \mu c - \kappa c^\dag, \qquad ScS^\dag =\mu c + \kappa c^\dag,
\end{align}
which means the squeezed vacuum $\ket{0_\zeta} = S(\zeta)\ket0$ is annihilated by
\begin{align}
    (\mu c + \kappa c^\dag)\ket{0_\zeta} = 0, 
\end{align}
while its number-state expansion is
\begin{align}
    \ket{0_\zeta} = \frac{1}{\sqrt{\cosh r}} \sum_{k=0}^\infty (-e^{i\theta} \tanh r )^k \frac{\sqrt{(2k)!}}{2^kk!}\ket{2k},
\end{align}
\indent The displacement operator is
\begin{subequations}
\begin{gather}
    D(\alpha) = \exp(\alpha c^\dag - \alpha^*c), \qquad \ket\alpha = D(\alpha)\ket0, \\
    D^\dag(\alpha)cD(\alpha) = c+\alpha,\qquad D(\alpha) c D^\dag(\alpha) = c - \alpha.
\end{gather}
\end{subequations}
\indent The displaced squeezed state is
\begin{align}
    \ket{\alpha,\zeta}  = D(\alpha)S(\zeta)\ket0, 
\end{align}
satisfying
\begin{gather}
    \expval*{c}{\alpha,\zeta} = \alpha, \qquad \expval*{c^\dag c} = \abs{\alpha}^2 + \sinh^2r, \qquad \expval*{cc} = \alpha^2 -e^{i\theta} \sinh r \cosh r .
\end{gather}
\indent Now we want to derive action of $c$ and $c^\dag$. Define the squeezed-mode annihilator,
\begin{align}
    b = D(\alpha) S(\zeta) c S^\dag(\zeta) D^\dag(\alpha) = \mu(c-\alpha) + \kappa(c^\dag-\alpha^*), 
\end{align}
then
\begin{align}
    b\ket{\alpha,\zeta} = 0.
\end{align}
Let 
\begin{align}
    \ket{1_b} = b^\dag\ket{\alpha,\zeta} = D(\alpha)S(\zeta) \ket1,
\end{align}
direct substitutions show
\begin{align}
    c-\alpha = \mu b - \kappa b^\dag, \qquad c^\dag -\alpha^* = -\kappa^* b + \mu b^\dag, 
\end{align}
therefore
\begin{align}
    c\ket{\alpha,\zeta} = \alpha \ket{\alpha,\zeta} -\kappa\ket{1_b}, \quad c^\dag\ket{\alpha,\zeta} = \alpha^*\ket{\alpha,\zeta} +\mu\ket{1_b}, 
\end{align}
so a squeezed state is the eigenstate of the combination
\begin{align}
    (\mu c+\kappa c^\dag) \ket{\alpha,\zeta} = (\mu\alpha+\kappa\alpha^*)\ket{\alpha,\zeta}.
\end{align}
\indent Now let's study the fluctuations. Let
\begin{align}
    \delta c = c-\alpha,
\end{align}
then for a squeezed state,
\begin{gather}
    \expval*{\delta c} = 0, \qquad \nu \equiv \expval*{\delta c^\dag \delta c} = \sinh^2r, \\
    \eta \equiv \expval*{\delta c \delta c} = -e^{i\theta} \sinh r \cosh r, \qquad \expval*{\delta c^{\dag2}} = \eta^*, \qquad \abs{\eta}^2 = \nu(\nu+1).
\end{gather}
\indent For the quadratures
\begin{align}
    X = c+c^\dag, \qquad Y = -i(c-c^\dag), 
\end{align}
the variances $V_A = \expval*{(A-\expval*{A})^2}$ are
\begin{align}
    V_X = 1+2\nu +2\Re\eta, \qquad V_Y = 1+2\nu -2\Re\eta.
\end{align}
So for example, 
\begin{subequations}
\begin{align}
    \theta = 0: \ V_X = e^{-2r}, \ &\Rightarrow \ X \text{ is squeezed}, \\
    \theta = \pi: \ V_Y = e^{-2r}, \ &\Rightarrow \ Y \text{ is squeezed}. 
\end{align}
\end{subequations}
\indent Some numerical sanity checks: we should recover the coherent results at $r=0$, regardless of what $\theta$ is. Also, at the starting time, we should have
\begin{subequations}
\begin{align}
    V_X(0) & = \cosh 2r - \cos\theta \sinh 2r, \\
    V_Y(0) & = \cosh 2r + \cos\theta \sinh 2r, \\
    \expval*{n_c}(0) & = \alpha^2+\nu, \\
    \text{Var}(n_c)(0) & = \alpha^2V_X(0) +2\nu(\nu+1).
\end{align}
\end{subequations}
\indent Define the normal generating function which allows for calculations of expectation values of normally-ordered operator products \cite[Sec.~11.9.1]{MANDEL_1995}, 
\begin{align}
    F(u,v) = \expval*{e^{uc^\dag}e^{vc}}, \qquad \Rightarrow \qquad M_{pq} \equiv \expval*{(c^\dagger)^p c^q} = \pdv[p]{u}\pdv[q]{v} F(u,v) \bigg\vert_{u=v=0},
\end{align}
and note that 
\begin{subequations}
\begin{align}
    F(u,v) & = \expval*{e^{uc^\dag}e^{vc}} = e^{u\alpha^*+v\alpha} \expval*{e^{u\delta c^\dag} e^{v\delta c}} \\
    & = e^{u\alpha^*+v\alpha - \frac{1}{2}uv} \expval*{e^{u\delta c^\dag+v\delta c}} \\
    & = e^{u\alpha^*+v\alpha - \frac{1}{2}uv} \exp[\frac{1}{2}\qty(u^2\delta c^{\dag2} + v^2\delta c^2 + 2uv\delta c^\dag \delta c + uv)] \\
    & = \exp[u\alpha^* + v\alpha + \frac{\eta^* u^2 + \eta v^2}{2}+\nu uv], 
\end{align}
\end{subequations}
where we used $\expval*{e^X} = \exp(\frac{1}{2}\expval*{X^2})$ when taking expectation value over zero-mean Gaussian state in going into the 3rd line. Now define the exponent as
\begin{align}
    f(u,v) = u\alpha^* + v\alpha + \frac{\eta^* u^2 + \eta v^2}{2}+\nu uv.
\end{align}
And we may calculate the expectation values as
\begin{subequations}
\begin{align}
    F(0,0) & = 1, \\
    M_{01} & = \pdv{F}{v} = F \pdv{f}{v} = F(\alpha + \eta v + \nu u) \xlongrightarrow{u=v=0} \alpha, \\
    M_{11} & = \pdv{u} \qty[F(\alpha+\eta v + \nu u)] = \pdv{F}{u}(\alpha+\eta v + \nu u) + F\nu \xlongrightarrow{u=v=0} \abs{\alpha}^2 + \nu, \\
    M_{02} & = \pdv{v}\qty[F(\alpha+\eta v + \nu u)] \xlongrightarrow{u=v=0} \alpha^2+\eta, \\
    M_{12} & = \pdv{u} \qty[\pdv{F}{v}\, (\alpha+\eta v + \nu u)+F\eta] = M_{11}\alpha + M_{01}\nu + \alpha^*\eta = \alpha^*\alpha^2 + \alpha^*\eta + 2\alpha \nu, \\
    M_{03} & = \alpha^3 + 3\alpha\eta, \\
    M_{23} & = \alpha^5 + \alpha^3(\eta^*+3\eta+6\nu) + \alpha(3\abs*{\eta}^2+6\nu\eta+6\nu^2), \\
    M_{22}&= \alpha^4 + \alpha^2(4\nu+\eta+\eta^*) + 2\nu^2+|\eta|^2, \\
    M_{13} &= \alpha^4+3\alpha^2(\nu+\eta)+3\nu\eta, \\
    M_{04} &= \alpha^4+6\alpha^2\eta+3\eta^2.
\end{align}
\end{subequations}
The general formula for $M_{pq}$ is
\begin{subequations}
\begin{align}
    M_{pq} & = \expval*{(\alpha^*+\delta c^\dag)^p(\alpha+\delta c)^q} \\
    & = \sum_{i=0}^p \sum_{j=0}^q \mqty(p \\ i) \mqty(q \\ j) (\alpha^*)^{p-i} \alpha^{q-j} \expval*{(\delta c^\dag)^i (\delta c)^j},
\end{align}
\end{subequations}
but it is easier to use the recursive relations,
\begin{subequations}
\begin{align}
    M_{p+1,q} & = \alpha^* M_{pq} + p\eta^* M_{p-1,q} + q\nu M_{p,q-1}, \\
    M_{p,q+1} & = \alpha M_{pq} + q\eta M_{p,q-1} + p\nu M_{p-1,q}. 
\end{align}
\end{subequations}

\subsection{BCH coefficient replacement}
Crucially, the BCH coefficients are unchanged:
\begin{subequations}
\begin{align}
    \expval*{c(\tau)} & = \expval*{c} - \frac{\tau^2}{2!} \expval*{c}+ \frac{\tau^4}{4!}\qty(\expval*{c}-4\expval*{c^\dag c^2}) + \frac{\tau^6}{6!}\qty(-\expval*{c}+60\expval*{c^\dag c^2} - 16 \expval*{(c^\dag)^2c^3}) - \dots \\
    & = M_{01} - \frac{\tau^2}{2!} M_{01} + \frac{\tau^4}{4!} \qty(M_{01}-4M_{12}) + \frac{\tau^6}{6!} \qty(-M_{01}+60M_{12}-16M_{23}) - \dots.
\end{align}
\end{subequations}
For a coherent pump, the substitution is easy, with $M_{n,n+1} = \alpha^{2n+1}$, and this retrieves our previous results. For a squeezed pump, 
\begin{align}
    \expval*{c(\tau)}_\zeta = \alpha - \frac{\alpha\tau^2}{2} + \frac{\tau^4}{4!} \qty[\alpha-4\alpha^3-4\alpha(2\nu+\eta)] + \frac{\tau^6}{6!} \qty[-\alpha+60(\alpha^3+\alpha(2\nu+\eta))-16M_{23}] + \dots
\end{align}
Now first define $C_\zeta = \expval*{c}_\zeta/\alpha$. Then substitute 
\begin{align}
    x=\alpha\tau, \qquad \epsilon=\frac{1}{\alpha^2}, 
\end{align}
and we have the above terms become, for example,
\begin{subequations}
\begin{align}
    -\frac{\tau^2}{2} & = -\epsilon \frac{x^2}{2}, \\ 
    -\frac{4\alpha^2\tau^4}{4!} = -\epsilon \frac{x^4}{6}, \qquad   -&\frac{4(\eta+2\nu)\tau^4}{4!} = -\epsilon^2 \frac{\eta+2\nu}{6}x^4, \\
    -\frac{16\alpha^4\tau^6}{6!} & = -\epsilon \frac{x^6}{45}, \\
    -\frac{16\alpha^2(\eta^*+3\eta+6\nu)\tau^6}{6!} & = -\epsilon^2 \frac{\eta^*+3\eta +6\nu}{45}x^6, \\
    -\frac{16(3\abs*{\eta}^2+6\eta\nu+6\nu^2)\tau^6}{6!} & = -\epsilon^3 \frac{3\abs*{\eta}^2+6\eta\nu+6\nu^2}{45}x^6, 
\end{align}
\end{subequations}
where the last three equations are in the $-16M_{23}$ term. So our result becomes
\begin{subequations}
\begin{align}
    C_\zeta & = 1 - \epsilon \qty[\frac{x^2}{2}+\frac{x^4}{6}+\frac{x^6}{45}+\dots] \\
    & \qquad + \epsilon^2 \qty[\frac{x^4}{24} + \frac{x^6}{12} + \dots - \frac{\eta+2\nu}{6}x^4 - \frac{\eta^*+3\eta+6\nu}{45}x^6 - \dots ] \\
    & \qquad - \epsilon^3 \qty[\frac{x^6}{720} - \frac{\eta+2\nu}{12}x^6 + \frac{3\abs*{\eta}^2+6\eta\nu + 6\nu^2}{45}x^6 + \dots ] + \dots  \\
    & \equiv 1-\epsilon G_1^\zeta + \epsilon^2 G_2^\zeta - \epsilon^3 G_3^\zeta + \dots ,
\end{align} \label{eqn:C_zeta}
\end{subequations}
where we've seen that
\begin{align}
    G_1^\zeta = G_1
\end{align}
is unchanged, and
\begin{align}
    G_2^\zeta = G_2 - \frac{\eta+2\nu}{6}x^4 - \frac{\eta^* +3\eta +6\nu}{45}x^6-\dots .
\end{align}
\indent Now, how to do this systematically? How to identify the entire sequence and find all the coefficients, and not just the first few terms? Could we find a pattern to the correction terms involving $\eta$ and $\nu$, and resum them into some closed form expressions the same way we've done with all the other corrections so far? The answer is we can! First, let's study the replacement rules in more detail. 

In our main analysis, we assumed $\alpha$ to be real and left everything proportional to $\alpha^n$. How do we work backwards to $M_{ij}$? For example, if we see $\alpha^5$, how do we know if it came from $\alpha^*\alpha^4$ (i.e. replace it as $M_{14}$), or $\alpha^{*2}\alpha^3$ (replace it as $M_{23}$), or others? 

Thankfully, it turns out we do not need to go back and re-evaluate the BCH commutation relations. The replacement rules are completely predictable and depend on which correlation function we're dealing with. For example, with $\expval*{c}$, we notice that the only operator combinations that contribute to the expectation values are of the form $(c^\dag)^n c^{n+1}$. In more detail, the rules are as follows:
\begin{subequations}
\begin{align}
    \expval*{c}: & & \alpha^{2j+1} & \longrightarrow M_{j,j+1}, \\
    \expval*{c^\dag c}: & & \alpha^{2j} & \longrightarrow M_{j,j}, \\
    \expval*{c^2}: & & \alpha^{2j+2} & \longrightarrow \ M_{j,j+2}, \\
    \expval*{c^\dag c^2}: & & \alpha^{2j+1} & \longrightarrow M_{j,j+1}, && j\geqslant1, \\
    \expval*{c^3}: & & \alpha^{2j+3} & \longrightarrow M_{j,j+3},  \\
    \expval*{c^{\dagger 2}c^2}:  && \alpha^{2j} &\longrightarrow M_{j,j}, && j\geqslant2, \\
    \expval*{c^\dagger c^3}: && \alpha^{2j+2} &\longrightarrow M_{j,j+2}, && j\geqslant1,    \\
    \expval*{c^4}: && \alpha^{2j+4} &\longrightarrow M_{j,j+4}, 
    \\
    \expval*{ab\,n_c}: && \alpha^{2j+1} &\longrightarrow M_{j,j+1}, && j\geqslant1.
\end{align}
\end{subequations}

\subsection{Resummation into cosh and sinh}

Now we are in a position to find $G_2^\zeta$. First, expand the coherent result $G_1^\zeta$:
\begin{align}
    G_1(x) = \frac{1}{2}\sinh^2x = \frac{\cosh2x-1}{4} = \sum_{m=0}^\infty 4^m\frac{x^{2m+2}}{(2m+2)!}, 
\end{align}
so the $m$-th term in $-\epsilon G_1(x)$ is
\begin{align}
    -\epsilon 4^m  \frac{x^{2m+2}}{(2m+2)!} = -\frac{4^m\tau^{2m+2}}{(2m+2)!}\alpha^{2m},
\end{align}
since $C_\zeta = \expval*{c}/\alpha$, the $\alpha^{2m}$ term in fact comes from 
\begin{align}
    \frac{\alpha^{2m+1}}{\alpha} \longrightarrow \frac{1}{\alpha}M_{m,m+1}. 
\end{align}
From the explicit formula for $M_{m,m+1}$ or just by inferring the pattern, we have that to first order/to first factor of $\eta$ and $\nu$:
\begin{subequations}
\begin{align}
    \frac{M_{m,m+1}}{\alpha} & = \alpha^{2m}  + \alpha^{2m-2} \qty[\frac{\eta^*}{2}m(m-1) + \frac{\eta}{2}m(m+1) + \nu m(m+1)] + \mathcal O(\alpha^{2m-4}) \\
    & = \alpha^{2m} + \alpha^{2m-2} \mathcal S(m) + \mathcal O(\alpha^{2m-4}), 
\end{align}\label{eqn:M_m_m+1}
\end{subequations}
where we defined
\begin{align}
    \mathcal S(m) & \equiv \frac{\eta^*}{2}m(m-1)+\frac{\eta}{2}m(m+1)+\nu m(m+1)
\end{align}
as the squeezing corrections. For example, the first few terms are
\begin{table}[H]
    \centering
    \begin{tabular}{c|c}
         $m$ & Squeezed correction \\\hline
         0 & 0 \\ 
         1 & $\eta+2\nu$ \\
         2 & $\eta^*+3\eta+6\nu$ \\
         3 & $3\eta^*+6\eta+12\nu$ \\
         4 & $6\eta^*+10\eta+20\nu$ 
    \end{tabular}
    \caption{The first few squeezed correction terms $\mathcal S(m)$ multiplied to the coherent results.}
    \label{tab:squeezed_correction}
\end{table}
One can check that the first two terms appear in the explicit equation for $C_\zeta$, in equation \eqref{eqn:C_zeta}.

The contribution of the first-order corrections to $C_\zeta$ is
\begin{subequations}
\begin{align}
    -\frac{4^m\tau^{2m+2}}{(2m+2)!}\frac{M_{m,m+1}}{\alpha} & = -\epsilon 4^m \frac{x^{2m+2}}{(2m+2)!} \\
    & \quad - \epsilon^2 4^m \frac{x^{2m+2}}{(2m+2)!} \qty[\frac{\eta^*}{2}m(m-1) + \frac{\eta}{2}m(m+1) + \nu m(m+1)] + \dots
\end{align}
\end{subequations}
and we can define
\begin{subequations}
\begin{gather}
    G_2^\zeta = G_2 - \Delta G_2, \quad  \text{where } \\
   \Delta G_2(x) = \sum_{m=0}^\infty 4^m \frac{x^{2m+2}}{(2m+2)!} \qty[\frac{\eta^*}{2}m(m-1) + \frac{\eta}{2}m(m+1) + \nu m(m+1)].
\end{gather}
\end{subequations}
Once again, use the Euler derivative trick to replace the $m$'s:
\begin{subequations}
\begin{gather}
    \qty(\frac{x}{2}\frac{d}{dx}-1)x^{2m+2} = mx^{2m+2},  \\
    \Rightarrow \ \Delta G_2 = \qty[\frac{\eta^*}{2} \qty(\frac{x}{2}\frac{d}{dx}-1) \qty(\frac{x}{2}\frac{d}{dx}-2) + \qty(\frac{\eta}{2}+\nu)\qty(\frac{x}{2}\frac{d}{dx}-1)\qty(\frac{x}{2}\frac{d}{dx})]G_1,
\end{gather}
\end{subequations}
and substitute $G_1 = \frac{1}{4}(\cosh2x-1)$, we have
\begin{subequations}
\begin{gather}
    \frac{x}{2}\frac{dG_1}{dx} = \frac{x}{4}\sinh2x, \\
    \qty(\frac{x}{2}\frac{d}{dx})^2G_1 = \frac{x^2}{4}\cosh 2x + \frac{x}{8}\sinh 2x, \\
    \Rightarrow \ \qty(\frac{x}{2}\frac{d}{dx}-1)\qty(\frac{x}{2}\frac{d}{dx}-2)G_1 = \qty[\qty(\frac{x}{2}\frac{d}{dx})^2 - \frac{3x}{2}\frac{d}{dx}+2]G_1 = \frac{x^2+2}{4}\cosh 2x -\frac{5x}{8}\sinh 2x - \frac{1}{2}, \\
    \qty(\frac{x}{2}\frac{d}{dx}-1)\qty(\frac{x}{2}\frac{d}{dx})G_1 = \frac{x^2}{4}\cosh2x - \frac{x}{8}\sinh2x.
\end{gather}
\end{subequations}
So the fully resummed result is
\begin{subequations}
\begin{empheq}[box=\fbox]{align}
    \Delta G_2(x) & = \eta^* \qty[\frac{x^2+2}{8}\cosh2x - \frac{5x}{16}\sinh2x - \frac{1}{4}] \\
    & \quad + \eta\qty[\frac{x^2}{8}\cosh2x - \frac{x}{16}\sinh2x] \\
    & \quad + \nu \qty[\frac{x^2}{4}\cosh2x - \frac{x}{8}\sinh2x].
\end{empheq} \label{eqn:Delta_G2}
\end{subequations}
Expanding the expression in $x$ gives
\begin{align}
    \Delta G_2 = \frac{\eta+2\nu}{6}x^4 + \frac{\eta^*+3\eta+6\nu}{45}x^6 + \frac{3\eta^*+6\eta+12\nu}{630}x^8 + \dots
\end{align}
and we can retrieve the original sequence. And so we have
\begin{align}
    \Aboxed{C_\zeta = 1-\epsilon G_1 + \epsilon^2(G_2-\Delta G_2) + \mathcal O(\epsilon^3)}.
\end{align}
Now, we will also define,
\begin{subequations}
\begin{align}
    C_\text{coh} & = C_0 + \epsilon C_1 + \mathcal O(\epsilon^2), \\
    C_\zeta & = C_0 + \epsilon C_{1,\zeta} + \mathcal O(\epsilon^2), \\
    \Rightarrow \ \Delta C_1 & \equiv C_{1,\zeta} - C_1, \\
    \Rightarrow \ C_\zeta & = C_0 + \epsilon (C_1 + \Delta C_1) + \mathcal O(\epsilon^2).
\end{align}
\end{subequations}
So $\epsilon \Delta C_1$ is the change in the first depletion profile due to squeezing. What part of $\Delta C_1$ is caused by $\Delta G_2$? This can be answered by expanding the cosh and sinh into their harmonics. However, even if we had done that, with only $G_2$ calculated, we would've only gotten the first term in the infinite series. We can only obtain the later terms by expanding $G_3, G_4,\dots$. So we shall delay the harmonic expansion for now, and first study the general form for $G_N^\zeta$.

\subsection{General form for $G_{N}^\zeta$}

Recall from equation \eqref{eqn:M_m_m+1} that we had a simple replacement rule to find squeezing to first order, 
\begin{subequations}
\begin{align}
    \alpha^{2m} \longrightarrow\frac{\alpha^{2m+1}}{\alpha} & \longrightarrow \frac{M_{m,m+1}}{\alpha} = \alpha^{2m} + \epsilon\alpha^{2m}\mathcal S(m) + \dots, \\
    \text{where } \ \mathcal S(m) & = \frac{\eta^*}{2}m(m-1)+\frac{\eta}{2}m(m+1)+\nu m(m+1) \\
    & = \qty[\nu+\frac{\eta+\eta^*}{2}]m^2 + \qty[\nu+\frac{\eta-\eta^*}{2}]m.
\end{align}
\end{subequations}
With the replacement rule in mind, we can find the general form for $G_{N+1}^\zeta$. Recall the power series expansion of $G_2$ begins as
\begin{subequations}
\begin{align}
    G_2(x) & = \frac{x^2}{24} + \frac{x^6}{12} + \frac{97x^8}{2520} + \dots, \\
    \Rightarrow \ \epsilon^2 G_2 & = \epsilon^2 \frac{x^2}{24} + \epsilon^2 \frac{x^6}{12} + \epsilon^2\frac{97x^8}{2520} + \dots.
\end{align}
\end{subequations}
The $x^4$ term is
\begin{align}
    \epsilon^2x^4 = \frac{1}{\alpha^4}(\alpha\tau)^4 = \tau^4,
\end{align}
this contains $\alpha^0$, and so it corresponds to $m=0$ term—it has no squeezed correction at this order. The next term:
\begin{align}
    \epsilon^2 x^6 = \frac{1}{\alpha^4}(\alpha\tau)^6 = \alpha^2\tau^6,
\end{align}
is replaced by
\begin{subequations}
\begin{gather}
    \alpha^2\longrightarrow \frac{M_{12}}{\alpha} = \alpha^2 + \eta + 2\nu, \\
    \Rightarrow \ \epsilon^2 \frac{x^6}{12} = \frac{\alpha^2 \tau^6}{12}\longrightarrow \frac{\alpha^2\tau^6}{12} + \frac{(\eta+2\nu)\tau^6}{12} = \epsilon^2 \frac{x^6}{12} + \epsilon^3 \frac{\eta+2\nu}{12}x^6,
\end{gather}
\end{subequations}
where we used $\tau^6 = \epsilon^3 x^6$. Similarly, for the $x^8$ term, we have
\begin{subequations}
\begin{gather}
\epsilon^2x^8 = \alpha^4\tau^8, \\
    \alpha^4 \longrightarrow \frac{M_{23}}{\alpha} = \alpha^4 + \alpha^2(\eta^*+3\eta+6\nu) + \mathcal O((\eta,\eta^*,\nu)^2), \\
    \Rightarrow \ \epsilon^2 \frac{97x^8}{2520} \longrightarrow \epsilon^2 \frac{97x^8}{2520} + \epsilon^3 \frac{97}{2520}(\eta^*+3\eta+6\nu) x^8 + \dots
\end{gather}
\end{subequations}
So, term by term, we have
\begin{align}
    \epsilon^2 G_2 \longrightarrow \epsilon^2G_2 + \epsilon^3 \qty[\frac{\eta+2\nu}{12}x^6+\frac{97}{2520}(\eta^*+3\eta+6\nu)x^8 + \dots].
\end{align}
Looking at the terms inside the square bracket, they are clearly related to $\mathcal S(m)$, and it seems like we are calculating the squeezing corrections again. But note that it is multiplied by $\epsilon^3$. We are in fact, calculating the squeezing correction to $G_3$, not $G_2$. The correction $\mathcal S(m)$ is general—it is linear in squeezing order, but applies to all $\epsilon$ order. Now let's find the precise relations between the terms in the square bracket and $\mathcal S(m)$. Consider the differential operator
\begin{align}
    \frac{x}{2}\frac{d}{dx}-2,
\end{align}
acting on the first three powers, we have
\begin{subequations}
\begin{align}
    \qty(\frac{x}{2}\frac{d}{dx}-2)x^4 & = (2-2)x^4 = 0x^4, \\
    \qty(\frac{x}{2}\frac{d}{dx}-2)x^6 & = (3-2)x^6 = 1x^6, \\
    \qty(\frac{x}{2}\frac{d}{dx}-2)x^8 & = (4-2)x^8 = 2x^8,
\end{align}
\end{subequations}
so it reproduces the sequence $m=0,1,2\dots$, and
\begin{subequations}
\begin{align}
    \mathcal S\qty(\frac{x}{2}\frac{d}{dx}-2)G_2 & = \frac{\mathcal S(0)}{24}x^4 + \frac{\mathcal S(1)}{12}x^6 + \frac{97\mathcal S(2)}{2520} x^8 + \dots \\
    & = \frac{\eta+2\nu}{12}x^6 + \frac{97}{2520}(\eta^*+3\eta+6\nu)x^8 + \dots,
\end{align}
\end{subequations}
so we have
\begin{align}
    \epsilon^2 G_2 \longrightarrow \epsilon^2 G_2 + \epsilon^3 \mathcal S \qty(\frac{x}{2}\frac{d}{dx}-2)G_2 + \dots.
\end{align}
The $\epsilon^3$ term in fact contributes to $-\epsilon^3 G_3$, so the $\epsilon^3$ term is
\begin{subequations}
\begin{gather}
    -\epsilon^3 G_3 + \epsilon^3 \mathcal S \qty(\frac{x}{2}\frac{d}{dx}-2)G_2 = -\epsilon^3 \qty[G_3 - \mathcal S \qty(\frac{x}{2}\frac{d}{dx}-2)G_2], \\
    \Rightarrow \ G_3^\zeta = G_3 - \mathcal S \qty(\frac{x}{2}\frac{d}{dx}-2)G_2 + \mathcal O((\eta,\eta^*,\nu)^2).
\end{gather}
\end{subequations}
This pattern is in fact general to all $G_N$, by replacing $2$ with $N$. Since every term of $G_N$ has the form
\begin{align}
    g_{N,m} x^{2m+2N}, 
\end{align}
we have
\begin{align}
    \qty(\frac{x}{2}\frac{d}{dx}-N)x^{2m+2N} = mx^{2m+2N},
\end{align}
so 
\begin{align}
    \mathcal S \qty(\frac{x}{2}\frac{d}{dx}-N)G_N = \sum_{m=0}^\infty g_{N,m} \mathcal S(m) x^{2m+2N}.
\end{align}
And we've seen that the squeezed correction carries an additional $\epsilon$, so when acting on $G_N$, it in fact contributes to $G_{N+1}^\zeta$. And taking into account the minus sign, we have
\begin{align}
    \Aboxed{G_{N+1}^\zeta = G_{N+1} - \mathcal S \qty(\frac{x}{2}\frac{d}{dx}-N)G_N + \mathcal O((\eta,\eta^*,\nu)^2)}.
\end{align}

\subsection{Squeezed depletion profile}

Recall back in equation \eqref{eqn:Delta_G2}, we expressed $\Delta G_2(x)$ as an expansion in $\epsilon$ and fixed $x$. With the eventual goal of obtaining the depletion profile, we do as before and expand the sinh and cosh in terms of their harmonics,
\begin{subequations}
\begin{align}
    \Delta G_2(x) & = e^{2x} \qty[\eta^*\frac{2x^2-5x+4}{32} + \eta\frac{2x^2-x}{32} + \nu\frac{2x^2-x}{16}] \\
    & \quad - \frac{\eta^*}{4} \\
    & \quad + e^{-2x} \qty[\eta^*\frac{2x^2+5x+4}{32} + \eta\frac{2x^2+x}{32} + \nu\frac{2x^2+x}{16}] \\
    & \equiv e^{2x} A_+(x) - \frac{\eta^*}{4} + e^{-2x} A_-(x),
\end{align}
\end{subequations}
where we defined
\begin{align}
    A_\pm(x) = \eta^*\frac{2x^2\mp5x+4}{32} + \eta\frac{2x^2\mp x}{32} + \nu\frac{2x^2\mp x}{16},
\end{align}
and substituting 
\begin{align}
    z = \frac{\epsilon e^{2x}}{16} \quad \Rightarrow \quad e^{2x} = \frac{16z}{\epsilon}, \qquad e^{-2x} = \frac{\epsilon}{16z},
\end{align}
we have
\begin{align}
    -\epsilon^2\Delta G_2 & = -16\epsilon z A_+(x) + \epsilon^2 \frac{\eta^*}{4} - \frac{\epsilon^3}{16z} A_-(x).
\end{align}
The first term, proportional to $\epsilon$, is the first depletion correction. Recall the leading terms in $C_0$ are
\begin{align}
    C_0(z) = 1-2z+4z^2-12z^3 + 48z^4 - \dots = 1+2\sum_{j=1}^\infty (-1)^j j!z^j,
\end{align}
matching the $j=1$ coefficient which is $-2z$, we write:
\begin{align}
    -16\epsilon z A_+(x) = \epsilon(-2z) [8A_+(x)] \equiv \epsilon(-2z)\Delta B_1(1,x),
\end{align}
where we defined the normalized coefficient
\begin{subequations}
\begin{align}
    \Delta B_1(1,x) &= 8A_+(x) = \eta^*\qty(\frac{x^2}{2}-\frac{5x}{4}+1) + \eta\qty(\frac{x^2}{2}-\frac{x}{4})+\nu\qty(x^2-\frac{x}{2}).
\end{align}
\end{subequations}
So $\Delta G_2$ gives the squeezed correction multiplying the $-2z$ term of $C_0$. 

More generally, define 
\begin{align}
    \Delta B_1(j,x) = B_{1,\zeta}(j,x) - B_1(j,x),
\end{align}
and the normalization constant for general $j$ in $\Delta B_1(j,x)$ are exactly as before, just like
\begin{align}
    B_1(j,x) = -\frac{16^j}{2j!} [e^{2jx}]G_{j+1}(x)
\end{align}
in the coherent analysis, we now have
\begin{align}
    \Delta B_1(j,x) = \frac{16^j}{2j!}[e^{2jx}] \mathcal S\qty(\frac{x}{2}\frac{d}{dx}-j)G_j.
\end{align}
\indent Now let's compute the $j=2$ term, which is a correction to $G_3^\zeta$:
\begin{align}
    G_3^\zeta = G_3 - \mathcal S\qty(\frac{x}{2}\frac{d}{dx}-2)G_2 + \dots.
\end{align}
(And recall the first few $G_N$ were listed all the way back in equation \eqref{eqn:harmonics_full}.) The leading harmonic in $G_2$ is
\begin{align}
    H_2 (x) \equiv \frac{e^{4x}}{64},
\end{align}
introduce the shorthand
\begin{align}
    Q_2 = \frac{x}{2}\frac{d}{dx}-2, 
\end{align}
then
\begin{subequations}
\begin{align}
    Q_2H_2 & = 2(x-1)H_2, \\
    Q_2^2H_2 & = (4x^2-7x+4)H_2, \\
    \Rightarrow \ Q_2(Q_2-1) H_2 & = (4x^2-9x+6)H_2, \\
    Q_2(Q_2+1)H_2 & = (4x^2-5x+2)H_2, \\
    \Rightarrow \ \mathcal S(Q_2)H_2 = \bigg[\frac{\eta^*}{2} (4x^2-9x + &6) + \qty(\frac{\eta}{2}+\nu)(4x^2-5x+2)\bigg]H_2,
\end{align}
\end{subequations}
the normalized coefficient is (note $16^2/2(2!)$ exactly cancels with the 1/64 by design)
\begin{subequations}
\begin{align}
    \Delta B_1(2,x) & = \frac{\eta^*}{2}(4x^2-9x+6) + \qty(\frac{\eta}{2}+\nu)(4x^2-5x+2) \\
    & = \eta^*\qty(2x^2-\frac{9}{2}x+3) + (\eta+2\nu)\qty(2x^2-\frac{5}{2}x+1), 
\end{align}
\end{subequations}
and its contribution to $C_\zeta$ is
\begin{align}
    -\epsilon^3 \qty[-H_2\Delta B_1(2,x)] = \epsilon^3 \frac{e^{4x}}{64}\Delta B_1(2,x) = \epsilon(4z^2)\Delta B_1(2,x) \xlongrightarrow{\text{contributing to}} \epsilon \Delta C_1.
\end{align}
\indent Now let's calculate $\Delta B_1(3,x)$, we have for $j=3$,
\begin{align}
    G_4^\zeta = G_4 - \mathcal S\qty(\frac{x}{2}\frac{d}{dx}-3)G_3 + \dots,
\end{align}
and the leading harmonics is 
\begin{align}
    H_3(x) = \frac{3e^{6x}}{1024},
\end{align}
define
\begin{align}
    Q_3 = \frac{x}{2}\frac{d}{dx}-3,
\end{align}
we have
\begin{subequations}
\begin{align}
    Q_3H_3 & = 3(x-1)H_3, \\
    Q_3^2H_3 & = \qty(\frac{3x}{2}+9(x-1)^2)H_3 = \qty(9x^2 - \frac{33x}{2}+9)H_3, \\
    \Rightarrow \ Q_3(Q_3-1) H_3 & = \qty(9x^2 - \frac{39x}{2}+12)H_3, \\
    Q_3(Q_3+1) H_3 & = \qty(9x^2 - \frac{27x}{2}+6)H_3, \\
    \Rightarrow \ \mathcal S(Q_3) H_3  = \bigg[\frac{\eta^*}{2}\bigg(& 9x^2- \frac{39x}{2}+12\bigg)+ \qty(\frac{\eta}{2}+\nu)\qty(9x^2 - \frac{27x}{2}+6)\bigg]H_3 ,
\end{align}
\end{subequations}
the normalized coefficient is
\begin{subequations}
\begin{align}
    \Delta B_1(3,x) & = \frac{\eta^*}{2}\qty(9x^2-\frac{39x}{2}+12) + \qty(\frac{\eta}{2}+\nu)\qty(9x^2-\frac{27x}{2}+6) \\
    & = \eta^*\qty(\frac{9}{2}x^2 - \frac{39}{4}x+6) + (\eta+2\nu)\qty(\frac{9}{2}x^2 - \frac{27}{4}x+3),
\end{align}
\end{subequations}
and its contribution to $C_\zeta$ is
\begin{align}
    \epsilon \Delta C_1 \sim \epsilon(-12z^3)\Delta B_1(3,x).
\end{align}
\indent We can keep going, but with $\Delta B_1(j,x)$ for $j=1,2,3$, a general pattern can already be inferred. We have that
\begin{subequations}
\begin{align}
    \Delta B_1(j,x) & = \eta^* \qty[\frac{j^2}{2}x^2 - \qty(j^2+\frac{j}{4})x + \frac{j(j+1)}{2}] + (\eta+2\nu) \qty[\frac{j^2x^2}{2} - \qty(j^2-\frac{3j}{4})x + \frac{j(j-1)}{2}] \\
    & = \eta^* \qty[\frac{j^2}{2}(x-1)^2 + \frac{j}{4}(2-x)] + (\eta+2\nu) \qty[\frac{j^2}{2}(x-1)^2 + \frac{j}{4}(3x-2)].
\end{align}
\end{subequations}
\indent A resummation is now possible. In summary, we have found
\begin{align}
    \Delta C_1(z,x) = (-2z)\Delta B_1(1,x) + (4z^2) \Delta B_1(2,x) + (-12z^3)\Delta B_z(3,x) + \dots,
\end{align}
where the terms in the brackets are the terms in 
\begin{align}
    C_0(z)-1 = -2z + 4z^2 - 12z^3 + \dots = 2\sum_{j=1}^\infty (-1)^j j! z^j, 
\end{align}
and we can use the $j\rightarrow D$ trick, to write
\begin{subequations}
\begin{empheq}[box=\fbox]{align}
    \Delta C_1(z,x) & = \qty{\eta^* \qty[\frac{(x-1)^2}{2}D^2 + \frac{2-x}{4}D] + (\eta+2\nu) \qty[\frac{(x-1)^2}{2}D^2 + \frac{3x-2}{4}D]}C_0(z) \\
    & = \frac{\eta^* + \eta + 2\nu}{2}(x-1)^2 D^2 C_0 + \frac{\eta^*(2-x)+(\eta+2\nu)(3x-2)}{4} DC_0,
\end{empheq}
\end{subequations}
and once again we can use the Borel summed $C_0$,
\begin{subequations}
\begin{align}
    C_0(z) & = \int_0^\infty e^{-t} \frac{1-zt}{1+zt}dt, \\
    DC_0(z) & = -2\int e^{-t} \frac{zt}{(1+zt)^2} dt, \\
    D^2C_0(z) & = -2\int e^{-t} \frac{zt(1-zt)}{(1+zt)^2} dt,
\end{align}
\end{subequations}
and the squeezed result through the first depletion correction is
\begin{align}
    \Aboxed{C_\zeta = C_0 + \epsilon (C_1 + \Delta C_1) + \mathcal O(\epsilon^2)}.
\end{align}
\indent Higher-order squeezing corrections, as well as corrections to other correlation functions can be similarly evaluated—by replacing $\alpha^n \rightarrow M_{ij}$, which produces multiple infinite series involving $\eta, \eta^*, \nu$, then finding the pattern for each series. In the supplied numerical package, squeezing correction can be calculated to up to $\mathcal O(\epsilon^3)$.

\pagebreak
\section{Reference data and book-keeping} \label{sec:book-keeping}

\subsection{1st-3rd moments definitions and identities} \label{app:moments_identities}

Note $\expval*{c}$, $\expval*{c^\dag c}$, $\expval*{cc}$ and their derivatives completely define all 1st- and 2nd-order moments. Together with $\expval*{c^\dag cc}$ and $\expval*{ccc}$, these 5 correlators and their derivatives completely define all 2nd- and 3rd-order moments. 

Define the unnormalized pump moments:
\begin{subequations}
\begin{gather}
    f \equiv \expval*{c}, \qquad n \equiv \expval*{a^\dag a} = \expval*{b^\dag b}\\
    p \equiv \expval*{c^\dag c}, \qquad s \equiv \expval*{cc},  \\
    u \equiv \expval*{c^\dag cc}, \qquad t \equiv \expval*{ccc} ,
\end{gather}
\end{subequations}
then the other correlators can be constructed as
\begin{table}[H]
    \centering
    \begin{tabular}{c|c}
        Correlator & Relation \\\hline
        $n$ & $\alpha^2-p$ \\[0.5em]
       $\expval*{ab}$  & $i\alpha f^\prime$ \\[0.5em]
       $\expval*{abc}$ & $i\alpha s^\prime/2$ \\[0.5em]
       $\expval*{abc^\dag}$ & $i\alpha p^\prime/2$ \\[0.5em]
       $\expval*{a^\dag a c} = \expval*{b^\dag bc}$ & $-\frac{1}{2}(\alpha^2f^{\prime\prime}+f)$
    \end{tabular}
    \caption{Moment identities for the real-generator convention.}
    \label{tab:moments_id}
\end{table}
Now define the quadratures as 
\begin{gather}
    X_j = j + j^\dag, \qquad Y_j = -i(j-j^\dag), \qquad \delta A = A-\expval*{A}, \\
    \Rightarrow \ [X_j,Y_j] = 2i, \qquad \text{Var}(A) = \expval*{(\delta A)^2}, 
\end{gather}
where $j = a, b, c$, $A, B = X, Y$. Then
\begin{table}[H]
    \centering
    \begin{tabular}{c|c}
        Definition & Relations \\\hline
        $V_{xc}=\expval*{(\delta X_c)^2}$ & $1+2p+2s-4f^2$ \\[0.5em]
        $V_{pc}=\expval*{(\delta Y_c)^2}$ & $1+2p-2s$ \\[0.5em]
        $\begin{cases}
            V_{xj} = \text{Var }X_j \\
            V_{pj} = \text{Var }Y_j
        \end{cases}$ & $1+2n$ \\[1em]
        $V_{xp}^\pm = \text{Var}[(X_a\pm Y_b)/\sqrt2]$ & $1\pm2n\pm2\Im \expval*{ab}$ \\[0.5em]
        $V_{xp}^+V_{xp}^-$ & $(1+2n)^2-4(\Im \expval*{ab})^2$ \\[0.5em]
        $V_{abc} = \expval*{X_aY_b\delta X_c}$ & $\alpha[p^\prime+s^\prime-4ff^\prime]$ \\
    \end{tabular}
    \caption{Quadrature variances and a third-order cross moment.}
    \label{tab:quadratures}
\end{table}
In our provided codes, all possible 2nd- and 3rd-order moments are automatically calculated.

\pagebreak

\subsection{Programming packages guide}

The leading-order results are extremely simple and can be readily implemented. They provide $\sim$ 99\% accuracy ($10^{-2}$ absolute error) at $\alpha=25$, and the error scales as $\alpha^{-2}$ as discussed in Section \ref{subsec:error_scaling}. For those who aim for absolute errors of $10^{-12}$ or smaller during the first depletion, they should use the supplied MATLAB package for implementing our analytic models. 

The packages come with all the possible corrections already implemented—namely, all available higher-order corrections to the first depletion, coherent interference between passages, and Poisson clocks of different sectors. Furthermore, they automatically calculate all correlations routinely used in an experiment setting, such as all 2nd- and 3rd-order variances and moments, all of which can be obtained via the 5 correlators we already studied. 

Evaluating the first depletion profile already suffices for most research purposes. We recommend the default setting:
\begin{itemize}
    \item Passage-interference turned OFF. It provides only marginal accuracy improvement, at the cost of longer evaluation time.
    \item Everything else is turned ON by default, including all perturbative corrections and Poisson clocks. It appears adding more correction orders does improve accuracy, but this is not proven. The user should experiment with different settings to find the optimal one.
    \item The reader is also free to reduce end time $\chi_\text{end}$ and the number of passages $M$ to some appropriate range, for example, set $\chi_\text{end} = 2\log4\alpha$ and $M = 1$ to capture the first depletion only. However, the computation speedup is negligible.
\end{itemize}
\indent If there are any situations where an even higher accuracy than $10^{-12}$ is required, there is a way to obtaining arbitrarily high precision—but restricted to the first depletion only—by using Pad\'e approximants (see Section \ref{subsec:pade}):
\begin{enumerate}
    \item Use the supplied Python code to generate BCH coefficients of an observable to some high $1/\alpha^2$ order. 
    \item Find the closed-form expressions $G_n(\chi)$, again using the Python code supplied. 
    \item Evaluate $G_n$ numerically for some range of $\chi$ covering the first depletion. At each $\chi$ value, apply Pad\'e approximants.
\end{enumerate}

\pagebreak
\subsection{Coherent amplitude raw data}

\footnotesize
As an explicit example, we will show here the complete raw data for the first few perturbative orders in $\expval*{c}$. For the other correlations and for a more complete raw data, we refer the readers to the Supplementary Materials. Let us first give a review of the formalism. Recall that ultimately, what we want to calculate is
\begin{align}
    C = \frac{\expval*{c}}{\alpha} = 1 - \epsilon G_1 + \epsilon^2 G_2 - \epsilon^3 G_3 + \epsilon^4 G_4 - \dots...
\end{align}
where $G_N(\chi)$ are functions of scaled time, $\chi = \alpha\tau$. Each $G_N$ is an infinite sum that, when summed over, we get,
\begin{subequations}
\begin{align}
    G_1(\chi) & = \frac{\sinh^2\chi}{2} = \frac{1}{4}(\cosh2\chi-1), \\
    G_2(\chi) & = \frac{-3-4\chi^2+2\cosh(2\chi)-4\chi^2\cosh(2\chi)+\cosh(4\chi)-2\chi\sinh(2\chi)}{32}, \\
    G_3(\chi) &= \frac{1}{1536} \Big(-108-240 \chi^2+3 (45-52 \chi^2+16 \chi^4) \cosh(2 \chi) -12 (3+8 \chi^2) \cosh(4 \chi) + \\
    & \quad \dots  + 9 \cosh(6 \chi)+30 \chi \sinh(2 \chi)+16 \chi^3 \sinh(2 \chi)+72 \chi \sinh(4 \chi)\Big), \\
    G_4(\chi) & = \frac{1}{6144} \bigg( -633-1080 \chi^2+80 \chi^4+(45-696 \chi^2+232 \chi^4-32 \chi^6) \cosh(2 \chi) \\
    & \dots +24 (29+32 \chi^2+16 \chi^4)+ \cosh(4 \chi)-117 \cosh(6 \chi)-162 \chi^2 \cosh(6 \chi)  \\
    & \dots +9 \cosh(8 \chi) -426 \chi \sinh(2 \chi)-376 \chi^3 \sinh(2 \chi) +16 \chi^5 \sinh(2 \chi) \\
    & \dots -738 \chi \sinh(4 \chi)-704 \chi^3 \sinh(4 \chi)+189 \chi \sinh(6 \chi)\bigg), \\
    G_5(\chi) & = \frac{1}{2949120} \bigg(10 (111510-142605 \chi^2+3588 \chi^4-1952 \chi^6+192 \chi^8) \cosh(2 \chi) \\
    & \dots +3 (-127320-492480 \chi^2+32640 \chi^4-160 (2055+4047 \chi^2+1568 \chi^4+256 \chi^6) \cosh(4 \chi) \\
    & \dots +90 (1049+2326 \chi^2+648 \chi^4) \cosh(6 \chi) -10440 \cosh(8 \chi)-11520 \chi^2 \cosh(8 \chi) \\
    & \dots +450 \cosh(10 \chi)+212835 \chi \sinh(2 \chi)+84360 \chi^3 \sinh(2 \chi) +29088 \chi^5 \sinh(2 \chi) \\
    & \dots -1280 \chi^7 \sinh(2 \chi)+604200 \chi \sinh(4 \chi)+398720 \chi^3 \sinh(4 \chi)+133120 \chi^5 \sinh(4 \chi) \\
    & \dots -175410 \chi \sinh(6 \chi)-149040 \chi^3 \sinh(6 \chi)+15840 \chi \sinh(8 \chi))\bigg). 
\end{align}
\end{subequations}
Expand in terms of harmonics (we'll use $x$ instead of $\chi$ from now on):
\begin{subequations}
\begin{align}
    G_1(x) & = \frac{1}{8}e^{2x} - \frac{1}{4} + \frac{1}{8}e^{-2x}, \\
    G_2(x) & = \frac{1}{64}e^{4x} + \qty(-\frac{x^2}{16}- \frac{x}{32}+\frac{1}{32})e^{2x} - \frac{x^2}{8}-\frac{3}{32} + \text{(negative harmonics}, x \rightarrow-x), \\
    G_3(x) & = \frac{3}{1024}e^{6x} + \qty(-\frac{x^2}{32} + \frac{3x}{128} - \frac{3}{256}) e^{4x}  + \qty(\frac{x^4}{64} + \frac{x^3}{192} - \frac{13x^2}{256} + \frac{5x}{512} + \frac{45}{1024})e^{2x}   \\
    & \quad - \qty(\frac{5x^2}{32} + \frac{9}{128}) + \text{(negative harmonics}, x \rightarrow-x), \\
    G_4(x) & = \frac{3}{4096}e^{8x} + \qty(-\frac{27x^2}{2048}+\frac{63x}{4096}-\frac{39}{4096})e^{6x} + \qty(\frac{x^4}{32} - \frac{11x^3}{192} + \frac{x^2}{16} - \frac{123x}{2048} + \frac{29}{512})e^{4x} \\
    & \quad + \qty(-\frac{x^6}{384}+\frac{x^5}{768}+\frac{29x^4}{1536}-\frac{47x^3}{1536}-\frac{29x^2}{512}-\frac{71x}{2048}+\frac{15}{4096})e^{2x} \\
    & \quad + \qty(\frac{5x^4}{384}-\frac{45x^2}{256}-\frac{211}{2048}) + \text{(negative harmonics}, x \rightarrow-x), \\
    G_5(x) & = \frac{15}{65536}e^{10x} + \qty(-\frac{3x^2}{512}+\frac{33x}{4096}-\frac{87}{16384})e^{8x} + \\
    & \quad + \qty(\frac{243x^4}{8192} -\frac{621x^3}{8192}+\frac{3489x^2}{32768} - \frac{5847x}{65536}+\frac{3147}{65536})e^{6x} \\
    & \quad + \qty(-\frac{x^6}{48} + \frac{13x^5}{192} - \frac{49x^4}{384} + \frac{623x^3}{3072}-\frac{1349x^2}{4096} + \frac{5035x}{16384} - \frac{685}{4096})e^{4x} \\
    & \quad + \qty(\frac{x^8}{3072} - \frac{x^7}{1536} - \frac{61x^6}{18432} + \frac{303x^5}{20480} + \frac{299x^4}{49152} + \frac{703x^3}{16384}-\frac{15845x^2}{65536} + \frac{14189x}{131072} + \frac{6195}{32768})e^{2x} \\
    & \quad + \qty(\frac{17x^4}{512} - \frac{513x^2}{1024}-\frac{1061}{8192}) + (\text{negative harmonics}, x \rightarrow-x).
\end{align}
\end{subequations}
Now let's put it into table form: 

\subsubsection{The raw harmonics / $[e^{2jx}]G_{N}$ factors}

\begin{table}[H]
\centering
\footnotesize
\renewcommand{\arraystretch}{1.5}
\begin{tabular}{|c|c|c|c|c|}
\hline
$N$ & $N-j = 0$ & $N-j = 1$ & $N-j = 2$ & $N-j = 3$ \\
\hline
$1$ & $\frac{1}{8}$ & $-\frac{1}{4}$ & (negative harmonics) & $0$ \\\hline
$2$ & $\frac{1}{64}$ & $\begin{array}{c} -\frac{x^2}{16} - \frac{x}{32} \\ + \frac{1}{32} \end{array}$ & $\begin{array}{c} -\frac{x^2}{8} \\ - \frac{3}{32} \end{array}$ & (negative harmonics) \\\hline
$3$ & $\frac{3}{1024}$ & $\begin{array}{c} -\frac{x^2}{32} + \frac{3x}{128} \\ - \frac{3}{256} \end{array}$ & $\begin{array}{c} \frac{x^4}{64} + \frac{x^3}{192} \\ - \frac{13x^2}{256} + \frac{5x}{512} \\ + \frac{45}{1024} \end{array}$ & $\begin{array}{c} -\frac{5x^2}{32} \\ - \frac{9}{128} \end{array}$ \\\hline
$4$ & $\frac{3}{4096}$ & $\begin{array}{c} -\frac{27x^2}{2048} + \frac{63x}{4096} \\ - \frac{39}{4096} \end{array}$ & $\begin{array}{c} \frac{x^4}{32} - \frac{11x^3}{192} \\ + \frac{x^2}{16} - \frac{123x}{2048} \\ + \frac{29}{512} \end{array}$ & $\begin{array}{c} -\frac{x^6}{384} + \frac{x^5}{768} \\ + \frac{29x^4}{1536} - \frac{47x^3}{1536} \\ - \frac{29x^2}{512} - \frac{71x}{2048} \\ + \frac{15}{4096} \end{array}$ \\\hline
$5$ & $\frac{15}{65536}$ & $\begin{array}{c} -\frac{3x^2}{512} + \frac{33x}{4096} \\ - \frac{87}{16384} \end{array}$ & $\begin{array}{c} \frac{243x^4}{8192} - \frac{621x^3}{8192} \\ + \frac{3489x^2}{32768} - \frac{5847x}{65536} \\ + \frac{3147}{65536} \end{array}$ & $\begin{array}{c} -\frac{x^6}{48} + \frac{13x^5}{192} \\ - \frac{49x^4}{384} + \frac{623x^3}{3072} \\ - \frac{1349x^2}{4096} + \frac{5035x}{16384} \\ - \frac{685}{4096} \end{array}$ \\\hline
$6$ & $\frac{45}{524288}$ & $\begin{array}{c} -\frac{375x^2}{131072} + \frac{1125x}{262144} \\ - \frac{765}{262144} \end{array}$ & $\begin{array}{c} \frac{3x^4}{128} - \frac{35x^3}{512} \\ + \frac{3381x^2}{32768} - \frac{5601x}{65536} \\ + \frac{5025}{131072} \end{array}$ & $\begin{array}{c} -\frac{729x^6}{16384} + \frac{6075x^5}{32768} \\ - \frac{27999x^4}{65536} + \frac{43893x^3}{65536} \\ - \frac{205131x^2}{262144} + \frac{312525x}{524288} \\ - \frac{64251}{262144} \end{array}$ \\\hline
$7$ & $\frac{315}{8388608}$ & $\begin{array}{c} -\frac{405x^2}{262144} + \frac{2565x}{1048576} \\ - \frac{3555}{2097152} \end{array}$ & $\begin{array}{c} \frac{9375x^4}{524288} - \frac{29375x^3}{524288} \\ + \frac{183765x^2}{2097152} - \frac{304785x}{4194304} \\ + \frac{126645}{4194304} \end{array}$ & $\begin{array}{c} -\frac{x^6}{16} + \frac{37x^5}{128} \\ - \frac{2893x^4}{4096} + \frac{36215x^3}{32768} \\ - \frac{154881x^2}{131072} + \frac{418833x}{524288} \\ - \frac{145965}{524288} \end{array}$ \\\hline
$8$ & $\frac{315}{16777216}$ & $\begin{array}{c} -\frac{15435x^2}{16777216} + \frac{50715x}{33554432} \\ - \frac{35595}{33554432} \end{array}$ & $\begin{array}{c} \frac{3645x^4}{262144} - \frac{23895x^3}{524288} \\ + \frac{76365x^2}{1048576} - \frac{1016325x}{16777216} \\ + \frac{203265}{8388608} \end{array}$ & $\begin{array}{c} -\frac{78125x^6}{1048576} + \frac{765625x^5}{2097152} \\ - \frac{3850375x^4}{4194304} + \frac{6025525x^3}{4194304} \\ - \frac{24804075x^2}{16777216} + \frac{31437435x}{33554432} \\ - \frac{4962615}{16777216} \end{array}$ \\\hline
$9$ & $\frac{2835}{268435456}$ & $\begin{array}{c} -\frac{315x^2}{524288} + \frac{8505x}{8388608} \\ - \frac{48195}{67108864} \end{array}$ & $\begin{array}{c} \frac{756315x^4}{67108864} - \frac{2557065x^3}{67108864} \\ + \frac{16582545x^2}{268435456} - \frac{27669915x}{536870912} \\ + \frac{10835055}{536870912} \end{array}$ & $\begin{array}{c} -\frac{10935x^6}{131072} + \frac{222345x^5}{524288} \\ - \frac{1137645x^4}{1048576} + \frac{14249655x^3}{8388608} \\ - \frac{57472785x^2}{33554432} + \frac{140253525x}{134217728} \\ - \frac{2602935}{8388608} \end{array}$ \\\hline
$10$ & $\frac{14175}{2147483648}$ & $\begin{array}{c} -\frac{229635x^2}{536870912} + \frac{790965x}{1073741824} \\ - \frac{564165}{1073741824} \end{array}$ & $\begin{array}{c} \frac{315x^4}{32768} - \frac{8715x^3}{262144} \\ + \frac{7310205x^2}{134217728} - \frac{6115095x}{134217728} \\ + \frac{9455355}{536870912} \end{array}$ & $\begin{array}{c} -\frac{12353145x^6}{134217728} + \frac{128825655x^5}{268435456} \\ - \frac{666828015x^4}{536870912} + \frac{1044619485x^3}{536870912} \\ - \frac{4161162915x^2}{2147483648} + \frac{4956208425x}{4294967296} \\ - \frac{707069475}{2147483648} \end{array}$ \\
\hline
\end{tabular}%
\caption{Polynomial coefficients for harmonics $e^{2jx}$, columns $N-j \in \{0, 1, 2, 3\}$}
\end{table}

\begin{table}[H]
\centering
\footnotesize
\renewcommand{\arraystretch}{1.5}
\resizebox{\textwidth}{!}{%
\begin{tabular}{|c|c|c|c|}
\hline
$N$ & $N-j = 4$ & $N-j = 5$ & $N-j = 6$ \\
\hline
$4$ & $\begin{array}{c} \frac{5x^4}{384} - \frac{45x^2}{256} \\ - \frac{211}{2048} \end{array}$ & (negative harmonics) & (negative harmonics) \\\hline
$5$ & $\begin{array}{c} \frac{x^8}{3072} - \frac{x^7}{1536} \\ - \frac{61x^6}{18432} + \frac{303x^5}{20480} \\ + \frac{299x^4}{49152} + \frac{703x^3}{16384} \\ - \frac{15845x^2}{65536} + \frac{14189x}{131072} \\ + \frac{6195}{32768} \end{array}$ & $\begin{array}{c} \frac{17x^4}{512} - \frac{513x^2}{1024} \\ - \frac{1061}{8192} \end{array}$ & (negative harmonics) \\\hline
$6$ & $\begin{array}{c} \frac{x^8}{96} - \frac{5x^7}{96} \\ + \frac{43x^6}{288} - \frac{1747x^5}{5120} \\ + \frac{769x^4}{1024} - \frac{10523x^3}{8192} \\ + \frac{215373x^2}{131072} - \frac{846925x}{524288} \\ + \frac{504371}{524288} \end{array}$ & $\begin{array}{c} -\frac{x^{10}}{30720} + \frac{5x^9}{36864} \\ + \frac{5x^8}{18432} - \frac{53x^7}{15360} \\ + \frac{1393x^6}{294912} - \frac{8397x^5}{327680} \\ + \frac{100927x^4}{393216} - \frac{175633x^3}{393216} \\ + \frac{147787x^2}{524288} - \frac{554827x}{1048576} \\ - \frac{12825}{32768} \end{array}$ & $\begin{array}{c} -\frac{61x^6}{46080} + \frac{1513x^4}{12288} \\ - \frac{8109x^2}{16384} - \frac{47321}{65536} \end{array}$ \\\hline
$7$ & $\begin{array}{c} \frac{6561x^8}{131072} - \frac{19683x^7}{65536} \\ + \frac{261225x^6}{262144} - \frac{6096573x^5}{2621440} \\ + \frac{8822633x^4}{2097152} - \frac{12223991x^3}{2097152} \\ + \frac{50112699x^2}{8388608} - \frac{70239369x}{16777216} \\ + \frac{3323181}{2097152} \end{array}$ & $\begin{array}{c} -\frac{x^{10}}{240} + \frac{17x^9}{576} \\ - \frac{137x^8}{1152} + \frac{8503x^7}{23040} \\ - \frac{31969x^6}{30720} + \frac{102469x^5}{40960} \\ - \frac{320713x^4}{65536} + \frac{137177x^3}{16384} \\ - \frac{12122407x^2}{1048576} + \frac{45702751x}{4194304} \\ - \frac{11538621}{2097152} \end{array}$ & $\begin{array}{c} \frac{x^{12}}{368640} - \frac{7x^{11}}{368640} \\ + \frac{13x^{10}}{1474560} + \frac{12631x^9}{26542080} \\ - \frac{5561x^8}{2949120} + \frac{33571x^7}{3440640} \\ - \frac{1272421x^6}{11796480} + \frac{727439x^5}{1572864} \\ - \frac{64675x^4}{65536} + \frac{4484749x^3}{1572864} \\ - \frac{6227059x^2}{1048576} + \frac{7763047x}{2097152} \\ + \frac{29280423}{8388608} \end{array}$ \\\hline
$8$ & $\begin{array}{c} \frac{x^8}{8} - \frac{13x^7}{16} \\ + \frac{8639x^6}{3072} - \frac{133887x^5}{20480} \\ + \frac{5818877x^4}{524288} - \frac{7283743x^3}{524288} \\ + \frac{26157981x^2}{2097152} - \frac{123594255x}{16777216} \\ + \frac{9356493}{4194304} \end{array}$ & $\begin{array}{c} -\frac{59049x^{10}}{1310720} + \frac{190269x^9}{524288} \\ - \frac{105705x^8}{65536} + \frac{6685173x^7}{1310720} \\ - \frac{266013459x^6}{20971520} + \frac{1064913489x^5}{41943040} \\ - \frac{684702915x^4}{16777216} + \frac{875982595x^3}{16777216} \\ - \frac{3406623381x^2}{67108864} + \frac{4510542375x}{134217728} \\ - \frac{48163857}{4194304} \end{array}$ & $\begin{array}{c} \frac{x^{12}}{720} - \frac{19x^{11}}{1440} \\ + \frac{17x^{10}}{240} - \frac{118949x^9}{414720} \\ + \frac{5833x^8}{5760} - \frac{806369x^7}{258048} \\ + \frac{8053163x^6}{983040} - \frac{25023647x^5}{1310720} \\ + \frac{62599451x^4}{1572864} - \frac{854183849x^3}{12582912} \\ + \frac{1499019311x^2}{16777216} - \frac{5625587651x}{67108864} \\ + \frac{352891071}{8388608} \end{array}$ \\\hline
$9$ & $\begin{array}{c} \frac{1953125x^8}{8388608} - \frac{6640625x^7}{4194304} \\ + \frac{281434375x^6}{50331648} - \frac{434364375x^5}{33554432} \\ + \frac{2856700725x^4}{134217728} - \frac{3382584825x^3}{134217728} \\ + \frac{11239867575x^2}{536870912} - \frac{11998589115x}{1073741824} \\ + \frac{1595098275}{536870912} \end{array}$ & $\begin{array}{c} -\frac{x^{10}}{5} + \frac{41x^9}{24} \\ - \frac{2993x^8}{384} + \frac{124839x^7}{5120} \\ - \frac{56393347x^6}{983040} + \frac{274922731x^5}{2621440} \\ - \frac{314049373x^4}{2097152} + \frac{690712903x^3}{4194304} \\ - \frac{2238533553x^2}{16777216} + \frac{9646374333x}{134217728} \\ - \frac{327794091}{16777216} \end{array}$ & $\begin{array}{c} \frac{177147x^{12}}{5242880} - \frac{1830519x^{11}}{5242880} \\ + \frac{41616423x^{10}}{20971520} - \frac{337956381x^9}{41943040} \\ + \frac{2171675043x^8}{83886080} - \frac{20065001949x^7}{293601280} \\ + \frac{37745494337x^6}{251658240} - \frac{46189818641x^5}{167772160} \\ + \frac{112607733761x^4}{268435456} - \frac{137847481073x^3}{268435456} \\ + \frac{511768962711x^2}{1073741824} - \frac{646946609373x}{2147483648} \\ + \frac{52432596969}{536870912} \end{array}$ \\\hline
$10$ & $\begin{array}{c} \frac{98415x^8}{262144} - \frac{688905x^7}{262144} \\ + \frac{4924395x^6}{524288} - \frac{181826289x^5}{8388608} \\ + \frac{73308075x^4}{2097152} - \frac{2684500455x^3}{67108864} \\ + \frac{33979275585x^2}{1073741824} - \frac{68021399115x}{4294967296} \\ + \frac{2081749365}{536870912} \end{array}$ & $\begin{array}{c} -\frac{9765625x^{10}}{16777216} + \frac{517578125x^9}{100663296} \\ - \frac{1194453125x^8}{50331648} + \frac{925203125x^7}{12582912} \\ - \frac{44965636875x^6}{268435456} + \frac{155991093375x^5}{536870912} \\ - \frac{414880481715x^4}{1073741824} + \frac{416695044885x^3}{1073741824} \\ - \frac{1209469079025x^2}{4294967296} + \frac{1145520309285x}{8589934592} \\ - \frac{67264380315}{2147483648} \end{array}$ & $\begin{array}{c} \frac{4x^{12}}{15} - \frac{43x^{11}}{15} \\ + \frac{15817x^{10}}{960} - \frac{566827x^9}{8640} \\ + \frac{49039187x^8}{245760} - \frac{221858647x^7}{458752} \\ + \frac{3985990317x^6}{4194304} - \frac{63765125579x^5}{41943040} \\ + \frac{4114896321x^4}{2097152} - \frac{266867603725x^3}{134217728} \\ + \frac{402178397001x^2}{268435456} - \frac{1616300403105x}{2147483648} \\ + \frac{25466548047}{134217728} \end{array}$ \\
\hline
\end{tabular}%
}
\caption{Polynomial coefficients for harmonics $e^{2jx}$, columns $N-j \in \{4, 5, 6\}$}
\end{table}

\begin{table}[H]
\centering
\footnotesize
\renewcommand{\arraystretch}{1.5}
\resizebox{\textwidth}{!}{%
\begin{tabular}{|c|c|c|}
\hline
$N$ & $N-j = 7$ & $N-j = 8$ \\
\hline
$7$ & $\begin{array}{c} -\frac{299x^6}{61440} + \frac{15539x^4}{49152} - \frac{381681x^2}{65536} \\ + \frac{354837}{262144} \end{array}$ & (negative harmonics) \\
\hline
$8$ & $\begin{array}{c} -\frac{x^{14}}{5160960} + \frac{x^{13}}{491520} - \frac{53x^{12}}{8847360} \\ - \frac{1001x^{11}}{26542080} + \frac{36803x^{10}}{106168320} - \frac{3577309x^9}{1486356480} \\ + \frac{1547923x^8}{55050240} - \frac{16243319x^7}{82575360} + \frac{38194009x^6}{47185920} \\ - \frac{104127527x^5}{31457280} + \frac{311294257x^4}{25165824} - \frac{670327015x^3}{25165824} \\ + \frac{1263330463x^2}{33554432} - \frac{1974221611x}{67108864} - \frac{649774335}{33554432} \end{array}$ & $\begin{array}{c} \frac{277x^8}{2064384} - \frac{55681x^6}{1474560} + \frac{1405441x^4}{786432} \\ + \frac{6422049x^2}{524288} - \frac{221091951}{8388608} \end{array}$ \\
\hline
$9$ & $\begin{array}{c} -\frac{x^{14}}{2520} + \frac{7x^{13}}{1440} - \frac{289x^{12}}{8640} \\ + \frac{35519x^{11}}{207360} - \frac{68599x^{10}}{92160} + \frac{7308647x^9}{2580480} \\ - \frac{96552761x^8}{10321920} + \frac{189255277x^7}{6881280} - \frac{194050803x^6}{2621440} \\ + \frac{185536867x^5}{1048576} - \frac{5923289x^4}{16384} + \frac{20725951389x^3}{33554432} \\ - \frac{110402369991x^2}{134217728} + \frac{405065115663x}{536870912} - \frac{2999507337}{8388608} \end{array}$ & $\begin{array}{c} \frac{x^{16}}{82575360} - \frac{11x^{15}}{61931520} + \frac{247x^{14}}{247726080} \\ + \frac{103x^{13}}{212336640} - \frac{96359x^{12}}{2548039680} + \frac{2387393x^{11}}{5945425920} \\ - \frac{626845039x^{10}}{118908518400} + \frac{2411500279x^9}{47563407360} - \frac{6774565079x^8}{21139292160} \\ + \frac{18875337631x^7}{10569646080} - \frac{19228435621x^6}{2013265920} + \frac{51106951869x^5}{1342177280} \\ - \frac{355292896151x^4}{3221225472} + \frac{279672349221x^3}{1073741824} - \frac{1818275581639x^2}{4294967296} \\ + \frac{2462914829575x}{8589934592} + \frac{91494840543}{536870912} \end{array}$ \\
\hline
$10$ & $\begin{array}{c} -\frac{1594323x^{14}}{73400320} + \frac{5845851x^{13}}{20971520} - \frac{82255257x^{12}}{41943040} \\ + \frac{413848197x^{11}}{41943040} - \frac{330367977x^{10}}{8388608} + \frac{153178900983x^9}{1174405120} \\ - \frac{858401890419x^8}{2348810240} + \frac{294064841621x^7}{335544320} - \frac{4850372287851x^6}{2684354560} \\ + \frac{17054886545397x^5}{5368709120} - \frac{10015662179485x^4}{2147483648} + \frac{11870323674579x^3}{2147483648} \\ - \frac{42775864367247x^2}{8589934592} + \frac{52176349548117x}{17179869184} - \frac{2011212684801}{2147483648} \end{array}$ & $\begin{array}{c} \frac{x^{16}}{10080} - \frac{23x^{15}}{15120} + \frac{197x^{14}}{15120} \\ - \frac{17063x^{13}}{207360} + \frac{269459x^{12}}{622080} - \frac{23071163x^{11}}{11612160} \\ + \frac{824367769x^{10}}{103219200} - \frac{3035676559x^9}{106168320} + \frac{1932830807x^8}{20643840} \\ - \frac{185349532849x^7}{660602880} + \frac{285326738711x^6}{377487360} - \frac{304066266437x^5}{167772160} \\ + \frac{6070155426625x^4}{1610612736} - \frac{20742698763919x^3}{3221225472} + \frac{72824309976181x^2}{8589934592} \\ - \frac{264363437953141x}{34359738368} + \frac{7690881172173}{2147483648} \end{array}$ \\
\hline
\end{tabular}%
}
\caption{Polynomial coefficients for harmonics $e^{2jx}$, columns $N-j \in \{7, 8\}$}
\end{table}

\begin{table}[H]
\centering
\footnotesize
\renewcommand{\arraystretch}{1.5}
\resizebox{\textwidth}{!}{%
\begin{tabular}{|c|c|c|}
\hline
$N$ & $N-j = 9$ & $N-j = 10$ \\
\hline
$9$ & $\begin{array}{c} \frac{25591x^8}{41287680} - \frac{192311x^6}{1966080} + \frac{3359605x^4}{1048576} - \frac{489074787x^2}{2097152} \\ + \frac{7136434671}{33554432} \end{array}$ & (negative harmonics) \\
\hline
$10$ & $\begin{array}{c} -\frac{x^{18}}{1486356480} + \frac{13x^{17}}{990904320} - \frac{37x^{16}}{330301440} + \frac{3781x^{15}}{11147673600} \\ + \frac{76271x^{14}}{35672555520} - \frac{3314243x^{13}}{71345111040} + \frac{536522597x^{12}}{713451110400} - \frac{2201286947x^{11}}{237817036800} \\ + \frac{152754259757x^{10}}{1902536294400} - \frac{445405367143x^9}{761014517760} + \frac{10643297017x^8}{2642411520} - \frac{247578589271x^7}{10569646080} \\ + \frac{3499336106141x^6}{32212254720} - \frac{9087105412447x^5}{21474836480} + \frac{11315596105639x^4}{8589934592} - \frac{25213714845973x^3}{8589934592} \\ + \frac{151903466310057x^2}{34359738368} - \frac{213206071586793x}{68719476736} - \frac{3209841913527}{2147483648} \end{array}$ & $\begin{array}{c} -\frac{50521x^{10}}{3715891200} + \frac{559345x^8}{66060288} \\ - \frac{41685619x^6}{31457280} + \frac{438258185x^4}{8388608} \\ + \frac{120422753631x^2}{67108864} - \frac{704420899383}{268435456} \end{array}$ \\
\hline
\end{tabular}%
}
\caption{Polynomial coefficients for harmonics $e^{2jx}$, columns $N-j \in \{9, 10\}$}
\end{table}

\pagebreak
\subsubsection{The normalized coefficients $B_r(j,s)$}

The normalization is $B_r(j,x) = (-1)^r\frac{16^j}{2j!}[e^{2jx}]G_{j+1}(x)$, where $r = N-j$. With this normalization many patterns become obvious. 

\begin{table}[H]
\centering
\footnotesize
\renewcommand{\arraystretch}{1.5}
\begin{tabular}{|c|c|c|c|c|}
\hline
$N$ & $N-j=0$ & $N-j=1$ & $N-j=2$ & $N-j=   3$ \\
\hline
$1$ & $1$ & $\frac{1}{4}$ & (negative harmonics) & 0 \\\hline
$2$ & $1$ & $\begin{array}{c} \frac{x^{2}}{2} + \frac{x}{4} \\ - \frac{1}{4} \end{array}$ & $-\frac{x^{2}}{8} - \frac{3}{32}$ & (negative harmonics) \\\hline
$3$ & $1$ & $\begin{array}{c} 2x^{2} - \frac{3x}{2} \\ + \frac{3}{4} \end{array}$ & $\begin{array}{c} \frac{x^{4}}{8} + \frac{x^{3}}{24} \\ - \frac{13x^{2}}{32} + \frac{5x}{64} \\ + \frac{45}{128} \end{array}$ & $\frac{5x^{2}}{32} + \frac{9}{128}$ \\\hline
$4$ & $1$ & $\begin{array}{c} \frac{9x^{2}}{2} - \frac{21x}{4} \\ + \frac{13}{4} \end{array}$ & $\begin{array}{c} 2x^{4} - \frac{11x^{3}}{3} \\ + 4x^{2} - \frac{123x}{32} \\ + \frac{29}{8} \end{array}$ & $\begin{array}{c} \frac{x^{6}}{48} - \frac{x^{5}}{96} \\ - \frac{29x^{4}}{192} + \frac{47x^{3}}{192} \\ + \frac{29x^{2}}{64} + \frac{71x}{256} \\ - \frac{15}{512} \end{array}$ \\\hline
$5$ & $1$ & $\begin{array}{c} 8x^{2} - 11x \\ + \frac{29}{4} \end{array}$ & $\begin{array}{c} \frac{81x^{4}}{8} - \frac{207x^{3}}{8} \\ + \frac{1163x^{2}}{32} - \frac{1949x}{64} \\ + \frac{1049}{64} \end{array}$ & $\begin{array}{c} \frac{4x^{6}}{3} - \frac{13x^{5}}{3} \\ + \frac{49x^{4}}{6} - \frac{623x^{3}}{48} \\ + \frac{1349x^{2}}{64} - \frac{5035x}{256} \\ + \frac{685}{64} \end{array}$ \\\hline
$6$ & $1$ & $\begin{array}{c} \frac{25x^{2}}{2} - \frac{75x}{4} \\ + \frac{51}{4} \end{array}$ & $\begin{array}{c} 32x^{4} - \frac{280x^{3}}{3} \\ + \frac{1127x^{2}}{8} - \frac{1867x}{16} \\ + \frac{1675}{32} \end{array}$ & $\begin{array}{c} \frac{243x^{6}}{16} - \frac{2025x^{5}}{32} \\ + \frac{9333x^{4}}{64} - \frac{14631x^{3}}{64} \\ + \frac{68377x^{2}}{256} - \frac{104175x}{512} \\ + \frac{21417}{256} \end{array}$ \\\hline
$7$ & $1$ & $\begin{array}{c} 18x^{2} - \frac{57x}{2} \\ + \frac{79}{4} \end{array}$ & $\begin{array}{c} \frac{625x^{4}}{8} - \frac{5875x^{3}}{24} \\ + \frac{12251x^{2}}{32} - \frac{20319x}{64} \\ + \frac{8443}{64} \end{array}$ & $\begin{array}{c} \frac{256x^{6}}{3} - \frac{1184x^{5}}{3} \\ + \frac{2893x^{4}}{3} - \frac{36215x^{3}}{24} \\ + \frac{51627x^{2}}{32} - \frac{139611x}{128} \\ + \frac{48655}{128} \end{array}$ \\\hline
$8$ & $1$ & $\begin{array}{c} \frac{49x^{2}}{2} - \frac{161x}{4} \\ + \frac{113}{4} \end{array}$ & $\begin{array}{c} 162x^{4} - 531x^{3} \\ + \frac{1697x^{2}}{2} - \frac{22585x}{32} \\ + \frac{4517}{16} \end{array}$ & $\begin{array}{c} \frac{15625x^{6}}{48} - \frac{153125x^{5}}{96} \\ + \frac{770075x^{4}}{192} - \frac{1205105x^{3}}{192} \\ + \frac{1653605x^{2}}{256} - \frac{2095829x}{512} \\ + \frac{330841}{256} \end{array}$ \\\hline
$9$ & $1$ & $\begin{array}{c} 32x^{2} - 54x \\ + \frac{153}{4} \end{array}$ & $\begin{array}{c} \frac{2401x^{4}}{8} - \frac{24353x^{3}}{24} \\ + \frac{52643x^{2}}{32} - \frac{87841x}{64} \\ + \frac{34397}{64} \end{array}$ & $\begin{array}{c} 972x^{6} - 4941x^{5} \\ + \frac{25281x^{4}}{2} - \frac{316659x^{3}}{16} \\ + \frac{1277173x^{2}}{64} - \frac{3116745x}{256} \\ + \frac{57843}{16} \end{array}$ \\\hline
$10$ & $1$ & $\begin{array}{c} \frac{81x^{2}}{2} - \frac{279x}{4} \\ + \frac{199}{4} \end{array}$ & $\begin{array}{c} 512x^{4} - \frac{5312x^{3}}{3} \\ + \frac{23207x^{2}}{8} - \frac{19413x}{8} \\ + \frac{30017}{32} \end{array}$ & $\begin{array}{c} \frac{117649x^{6}}{48} - \frac{1226911x^{5}}{96} \\ + \frac{6350743x^{4}}{192} - \frac{9948757x^{3}}{192} \\ + \frac{13210041x^{2}}{256} - \frac{15733995x}{512} \\ + \frac{2244665}{256} \end{array}$ \\\hline
\end{tabular}%
\caption{Polynomial coefficients for scaled harmonics $B_1(j,x)$, columns $N-j \in \{0, 1, 2, 3\}$}
\end{table}

\begin{table}[H]
\centering
\footnotesize
\renewcommand{\arraystretch}{1.5}
\begin{tabular}{|c|c|c|c|}
\hline
$N$ & $N-j = 4$ & $N-j = 5$ & $N-j = 6$ \\
\hline
$4$ & $\begin{array}{c} \frac{5x^{4}}{384} - \frac{45x^{2}}{256} \\ - \frac{211}{2048} \end{array}$ & (negative harmonics) & (negative harmonics) \\\hline
$5$ & $\begin{array}{c} \frac{x^{8}}{384} - \frac{x^{7}}{192} \\ - \frac{61x^{6}}{2304} + \frac{303x^{5}}{2560} \\ + \frac{299x^{4}}{6144} + \frac{703x^{3}}{2048} \\ - \frac{15845x^{2}}{8192} + \frac{14189x}{16384} \\ + \frac{6195}{4096} \end{array}$ & $\begin{array}{c} -\frac{17x^{4}}{512} + \frac{513x^{2}}{1024} \\ + \frac{1061}{8192} \end{array}$ & (negative harmonics) \\\hline
$6$ & $\begin{array}{c} \frac{2x^{8}}{3} - \frac{10x^{7}}{3} \\ + \frac{86x^{6}}{9} - \frac{1747x^{5}}{80} \\ + \frac{769x^{4}}{16} - \frac{10523x^{3}}{128} \\ + \frac{215373x^{2}}{2048} - \frac{846925x}{8192} \\ + \frac{504371}{8192} \end{array}$ & $\begin{array}{c} \frac{x^{10}}{3840} - \frac{5x^{9}}{4608} \\ - \frac{5x^{8}}{2304} + \frac{53x^{7}}{1920} \\ - \frac{1393x^{6}}{36864} + \frac{8397x^{5}}{40960} \\ - \frac{100927x^{4}}{49152} + \frac{175633x^{3}}{49152} \\ - \frac{147787x^{2}}{65536} + \frac{554827x}{131072} \\ + \frac{12825}{4096} \end{array}$ & $\begin{array}{c} -\frac{61x^{6}}{46080} + \frac{1513x^{4}}{12288} \\ - \frac{8109x^{2}}{16384} - \frac{47321}{65536} \end{array}$ \\\hline
$7$ & $\begin{array}{c} \frac{2187x^{8}}{128} - \frac{6561x^{7}}{64} \\ + \frac{87075x^{6}}{256} - \frac{2032191x^{5}}{2560} \\ + \frac{8822633x^{4}}{6144} - \frac{12223991x^{3}}{6144} \\ + \frac{16704233x^{2}}{8192} - \frac{23413123x}{16384} \\ + \frac{1107727}{2048} \end{array}$ & $\begin{array}{c} \frac{4x^{10}}{15} - \frac{17x^{9}}{9} \\ + \frac{137x^{8}}{18} - \frac{8503x^{7}}{360} \\ + \frac{31969x^{6}}{480} - \frac{102469x^{5}}{640} \\ + \frac{320713x^{4}}{1024} - \frac{137177x^{3}}{256} \\ + \frac{12122407x^{2}}{16384} - \frac{45702751x}{65536} \\ + \frac{11538621}{32768} \end{array}$ & $\begin{array}{c} \frac{x^{12}}{46080} - \frac{7x^{11}}{46080} \\ + \frac{13x^{10}}{184320} + \frac{12631x^{9}}{3317760} \\ - \frac{5561x^{8}}{368640} + \frac{33571x^{7}}{430080} \\ - \frac{1272421x^{6}}{1474560} + \frac{727439x^{5}}{196608} \\ - \frac{64675x^{4}}{8192} + \frac{4484749x^{3}}{196608} \\ - \frac{6227059x^{2}}{131072} + \frac{7763047x}{262144} \\ + \frac{29280423}{1048576} \end{array}$ \\\hline
$8$ & $\begin{array}{c} \frac{512x^{8}}{3} - \frac{3328x^{7}}{3} \\ + \frac{34556x^{6}}{9} - \frac{44629x^{5}}{5} \\ + \frac{5818877x^{4}}{384} - \frac{7283743x^{3}}{384} \\ + \frac{8719327x^{2}}{512} - \frac{41198085x}{4096} \\ + \frac{3118831}{1024} \end{array}$ & $\begin{array}{c} \frac{19683x^{10}}{1280} - \frac{63423x^{9}}{512} \\ + \frac{35235x^{8}}{64} - \frac{2228391x^{7}}{1280} \\ + \frac{88671153x^{6}}{20480} - \frac{354971163x^{5}}{40960} \\ + \frac{228234305x^{4}}{16384} - \frac{875982595x^{3}}{49152} \\ + \frac{1135541127x^{2}}{65536} - \frac{1503514125x}{131072} \\ + \frac{16054619}{4096} \end{array}$ & $\begin{array}{c} \frac{4x^{12}}{45} - \frac{38x^{11}}{45} \\ + \frac{68x^{10}}{15} - \frac{118949x^{9}}{6480} \\ + \frac{5833x^{8}}{90} - \frac{806369x^{7}}{4032} \\ + \frac{8053163x^{6}}{15360} - \frac{25023647x^{5}}{20480} \\ + \frac{62599451x^{4}}{24576} - \frac{854183849x^{3}}{196608} \\ + \frac{1499019311x^{2}}{262144} - \frac{5625587651x}{1048576} \\ + \frac{352891071}{131072} \end{array}$ \\\hline
$9$ & $\begin{array}{c} \frac{390625x^{8}}{384} - \frac{1328125x^{7}}{192} \\ + \frac{56286875x^{6}}{2304} - \frac{28957625x^{5}}{512} \\ + \frac{190446715x^{4}}{2048} - \frac{225505655x^{3}}{2048} \\ + \frac{749324505x^{2}}{8192} - \frac{799905941x}{16384} \\ + \frac{106339885}{8192} \end{array}$ & $\begin{array}{c} \frac{4096x^{10}}{15} - \frac{20992x^{9}}{9} \\ + \frac{95776x^{8}}{9} - \frac{166452x^{7}}{5} \\ + \frac{56393347x^{6}}{720} - \frac{274922731x^{5}}{1920} \\ + \frac{314049373x^{4}}{1536} - \frac{690712903x^{3}}{3072} \\ + \frac{746177851x^{2}}{4096} - \frac{3215458111x}{32768} \\ + \frac{109264697}{4096} \end{array}$ & $\begin{array}{c} \frac{59049x^{12}}{5120} - \frac{610173x^{11}}{5120} \\ + \frac{13872141x^{10}}{20480} - \frac{112652127x^{9}}{40960} \\ + \frac{723891681x^{8}}{81920} - \frac{6688333983x^{7}}{286720} \\ + \frac{37745494337x^{6}}{737280} - \frac{46189818641x^{5}}{491520} \\ + \frac{112607733761x^{4}}{786432} - \frac{137847481073x^{3}}{786432} \\ + \frac{170589654237x^{2}}{1048576} - \frac{215648869791x}{2097152} \\ + \frac{17477532323}{524288} \end{array}$ \\\hline
$10$ & $\begin{array}{c} 4374x^{8} - 30618x^{7} \\ + 109431x^{6} - \frac{20202921x^{5}}{80} \\ + \frac{4887205x^{4}}{12} - \frac{178966697x^{3}}{384} \\ + \frac{755095013x^{2}}{2048} - \frac{1511586647x}{8192} \\ + \frac{46261097}{1024} \end{array}$ & $\begin{array}{c} \frac{1953125x^{10}}{768} - \frac{103515625x^{9}}{4608} \\ + \frac{238890625x^{8}}{2304} - \frac{185040625x^{7}}{576} \\ + \frac{2997709125x^{6}}{4096} - \frac{10399406225x^{5}}{8192} \\ + \frac{27658698781x^{4}}{16384} - \frac{27779669659x^{3}}{16384} \\ + \frac{80631271935x^{2}}{65536} - \frac{76368020619x}{131072} \\ + \frac{4484292021}{32768} \end{array}$ & $\begin{array}{c} \frac{16384x^{12}}{45} - \frac{176128x^{11}}{45} \\ + \frac{1012288x^{10}}{45} - \frac{36276928x^{9}}{405} \\ + \frac{49039187x^{8}}{180} - \frac{221858647x^{7}}{336} \\ + \frac{1328663439x^{6}}{1024} - \frac{63765125579x^{5}}{30720} \\ + \frac{1371632107x^{4}}{512} - \frac{266867603725x^{3}}{98304} \\ + \frac{134059465667x^{2}}{65536} - \frac{538766801035x}{524288} \\ + \frac{8488849349}{32768} \end{array}$ \\\hline
\end{tabular}%
\caption{Polynomial coefficients for scaled harmonics $B_1(j,x)$, columns $N-j \in \{4, 5, 6\}$}
\end{table}

\begin{table}[H]
\centering
\footnotesize
\renewcommand{\arraystretch}{1.5}
\resizebox{\textwidth}{!}{%
\begin{tabular}{|c|c|c|}
\hline
$N$ & $N-j = 7$ & $N-j = 8$ \\
\hline
$7$ & $\begin{array}{c} \frac{299x^{6}}{61440} - \frac{15539x^{4}}{49152} + \frac{381681x^{2}}{65536} \\ - \frac{354837}{262144} \end{array}$ & (negative harmonics) \\\hline
$8$ & $\begin{array}{c} \frac{x^{14}}{645120} - \frac{x^{13}}{61440} + \frac{53x^{12}}{1105920} \\ + \frac{1001x^{11}}{3317760} - \frac{36803x^{10}}{13271040} + \frac{3577309x^{9}}{185794560} \\ - \frac{1547923x^{8}}{6881280} + \frac{16243319x^{7}}{10321920} - \frac{38194009x^{6}}{5898240} \\ + \frac{104127527x^{5}}{3932160} - \frac{311294257x^{4}}{3145728} + \frac{670327015x^{3}}{3145728} \\ - \frac{1263330463x^{2}}{4194304} + \frac{1974221611x}{8388608} + \frac{649774335}{4194304} \end{array}$ & $\begin{array}{c} \frac{277x^{8}}{2064384} - \frac{55681x^{6}}{1474560} + \frac{1405441x^{4}}{786432} \\ + \frac{6422049x^{2}}{524288} - \frac{221091951}{8388608} \end{array}$ \\\hline
$9$ & $\begin{array}{c} \frac{8x^{14}}{315} - \frac{14x^{13}}{45} + \frac{289x^{12}}{135} \\ - \frac{35519x^{11}}{3240} + \frac{68599x^{10}}{1440} - \frac{7308647x^{9}}{40320} \\ + \frac{96552761x^{8}}{161280} - \frac{189255277x^{7}}{107520} + \frac{194050803x^{6}}{40960} \\ - \frac{185536867x^{5}}{16384} + \frac{5923289x^{4}}{256} - \frac{20725951389x^{3}}{524288} \\ + \frac{110402369991x^{2}}{2097152} - \frac{405065115663x}{8388608} + \frac{2999507337}{131072} \end{array}$ & $\begin{array}{c} \frac{x^{16}}{10321920} - \frac{11x^{15}}{7741440} + \frac{247x^{14}}{30965760} \\ + \frac{103x^{13}}{26542080} - \frac{96359x^{12}}{318504960} + \frac{2387393x^{11}}{743178240} \\ - \frac{626845039x^{10}}{14863564800} + \frac{2411500279x^{9}}{5945425920} - \frac{6774565079x^{8}}{2642411520} \\ + \frac{18875337631x^{7}}{1321205760} - \frac{19228435621x^{6}}{251658240} + \frac{51106951869x^{5}}{167772160} \\ - \frac{355292896151x^{4}}{402653184} + \frac{279672349221x^{3}}{134217728} - \frac{1818275581639x^{2}}{536870912} \\ + \frac{2462914829575x}{1073741824} + \frac{91494840543}{67108864} \end{array}$ \\\hline
$10$ & $\begin{array}{c} \frac{531441x^{14}}{71680} - \frac{1948617x^{13}}{20480} + \frac{27418419x^{12}}{40960} \\ - \frac{137949399x^{11}}{40960} + \frac{110122659x^{10}}{8192} - \frac{51059633661x^{9}}{1146880} \\ + \frac{286133963473x^{8}}{2293760} - \frac{294064841621x^{7}}{983040} + \frac{1616790762617x^{6}}{2621440} \\ - \frac{5684962181799x^{5}}{5242880} + \frac{10015662179485x^{4}}{6291456} - \frac{3956774558193x^{3}}{2097152} \\ + \frac{14258621455749x^{2}}{8388608} - \frac{17392116516039x}{16777216} + \frac{670404228267}{2097152} \end{array}$ & $\begin{array}{c} \frac{2x^{16}}{315} - \frac{92x^{15}}{945} + \frac{788x^{14}}{945} \\ - \frac{17063x^{13}}{3240} + \frac{269459x^{12}}{9720} - \frac{23071163x^{11}}{181440} \\ + \frac{824367769x^{10}}{1612800} - \frac{3035676559x^{9}}{1658880} + \frac{1932830807x^{8}}{322560} \\ - \frac{185349532849x^{7}}{10321920} + \frac{285326738711x^{6}}{5898240} - \frac{304066266437x^{5}}{2621440} \\ + \frac{6070155426625x^{4}}{25165824} - \frac{20742698763919x^{3}}{50331648} + \frac{72824309976181x^{2}}{134217728} \\ - \frac{264363437953141x}{536870912} + \frac{7690881172173}{33554432} \end{array}$ \\\hline
\end{tabular}%
}
\caption{Polynomial coefficients for scaled harmonics $B_1(j,x)$, columns $N-j \in \{7, 8\}$}
\end{table}

\begin{table}[H]
\centering
\footnotesize
\renewcommand{\arraystretch}{1.5}
\resizebox{\textwidth}{!}{%
\begin{tabular}{|c|c|c|}
\hline
$N$ & $N-j = 9$ & $N-j = 10$ \\
\hline
$9$ & $\begin{array}{c} -\frac{25591x^{8}}{41287680} + \frac{192311x^{6}}{1966080} - \frac{3359605x^{4}}{1048576} + \frac{489074787x^{2}}{2097152} \\ - \frac{7136434671}{33554432} \end{array}$ & (negative harmonics) \\\hline
$10$ & $\begin{array}{c} \frac{x^{18}}{185794560} - \frac{13x^{17}}{123863040} + \frac{37x^{16}}{41287680} - \frac{3781x^{15}}{1393459200} \\ - \frac{76271x^{14}}{4459069440} + \frac{3314243x^{13}}{8918138880} - \frac{536522597x^{12}}{89181388800} + \frac{2201286947x^{11}}{29727129600} \\ - \frac{152754259757x^{10}}{237817036800} + \frac{445405367143x^{9}}{95126814720} - \frac{10643297017x^{8}}{330301440} + \frac{247578589271x^{7}}{1321205760} \\ - \frac{3499336106141x^{6}}{4026531840} + \frac{9087105412447x^{5}}{2684354560} - \frac{11315596105639x^{4}}{1073741824} + \frac{25213714845973x^{3}}{1073741824} \\ - \frac{151903466310057x^{2}}{4294967296} + \frac{213206071586793x}{8589934592} + \frac{3209841913527}{268435456} \end{array}$ & $\begin{array}{c} -\frac{50521x^{10}}{3715891200} + \frac{559345x^{8}}{66060288} \\ - \frac{41685619x^{6}}{31457280} + \frac{438258185x^{4}}{8388608} \\ + \frac{120422753631x^{2}}{67108864} - \frac{704420899383}{268435456} \end{array}$ \\\hline
\end{tabular}%
}
\caption{Polynomial coefficients for scaled harmonics $B_1(j,x)$, columns $N-j \in \{9, 10\}$}
\end{table}

\pagebreak
\subsubsection{The generic polynomial $P$'s}
\sethlcolor{red}

Recall we found the $r=1$ sub-leading harmonics can be fit exactly to a polynomial expression. However, starting from the $r=2$ harmonics, the first few terms cannot fit into polynomials. The polynomial fit only works for large $j$'s. 

So we will use the large $j$ data to find our polynomial expressions, and record the predictions in the following tables. 

The \hl{terms highlighted red} are ones where the predicted values are different to the actual coefficients, $B$. The differences give rise to the ``boundary'' terms (which give some constant corrections that are of no interests to us) and the ``exceptional'' terms (which are of interest and used to calculate later-time profiles).

\begin{table}[H]
\centering
\footnotesize
\renewcommand{\arraystretch}{1.5}
\begin{tabular}{|c|c|c|c|c|}
\hline
$N$ & $N-j = 0$ & $N-j = 1$ & $N-j = 2$ & $N-j = 3$ \\
\hline
$1$ & $1$ & $\frac{1}{4}$ & (negative harmonics) & $0$ \\\hline
$2$ & $1$ & $\frac{x^{2}}{2} + \frac{x}{4} - \frac{1}{4}$ & $\text{$-\frac{x^{2}}{8}$} \text{\hl{$+ \frac{x}{4}$}} \text{\hl{$- \frac{11}{32}$}}$ & (negative harmonics) \\\hline
$3$ & $1$ & $2x^{2} - \frac{3x}{2} + \frac{3}{4}$ & $\begin{array}{c} \frac{x^{4}}{8} + \frac{x^{3}}{24} - \frac{13x^{2}}{32} \\ + \frac{5x}{64} \text{\hl{$+ \frac{46}{128}$}} \end{array}$ & $\frac{5x^{2}}{32}$ $\text{\hl{$- \frac{9x}{32}$}} \text{\hl{$+ \frac{33}{128}$}}$ \\\hline
$4$ & $1$ & $\frac{9x^{2}}{2} - \frac{21x}{4} + \frac{13}{4}$ & $\begin{array}{c} 2x^{4} - \frac{11x^{3}}{3} + 4x^{2} \\ - \frac{123x}{32} + \frac{29}{8} \end{array}$ & $\begin{array}{c} \frac{x^{6}}{48} - \frac{x^{5}}{96} - \frac{29x^{4}}{192} \\ + \frac{47x^{3}}{192} \text{\hl{$+ \frac{117x^{2}}{256}$}} \text{\hl{$+ \frac{135x}{512}$}} \\ \text{\hl{$- \frac{7}{256}$}} \end{array}$ \\\hline
$5$ & $1$ & $8x^{2} - 11x + \frac{29}{4}$ & $\begin{array}{c} \frac{81x^{4}}{8} - \frac{207x^{3}}{8} + \frac{1163x^{2}}{32} \\ - \frac{1949x}{64} + \frac{1049}{64} \end{array}$ & $\begin{array}{c} \frac{4x^{6}}{3} - \frac{13x^{5}}{3} + \frac{49x^{4}}{6} \\ - \frac{623x^{3}}{48} + \frac{1349x^{2}}{64} - \frac{5035x}{256} \\ + \frac{685}{64} \end{array}$ \\\hline
$6$ & $1$ & $\frac{25x^{2}}{2} - \frac{75x}{4} + \frac{51}{4}$ & $\begin{array}{c} 32x^{4} - \frac{280x^{3}}{3} + \frac{1127x^{2}}{8} \\ - \frac{1867x}{16} + \frac{1675}{32} \end{array}$ & $\begin{array}{c} \frac{243x^{6}}{16} - \frac{2025x^{5}}{32} + \frac{9333x^{4}}{64} \\ - \frac{14631x^{3}}{64} + \frac{68377x^{2}}{256} - \frac{104175x}{512} \\ + \frac{21417}{256} \end{array}$ \\\hline
$7$ & $1$ & $18x^{2} - \frac{57x}{2} + \frac{79}{4}$ & $\begin{array}{c} \frac{625x^{4}}{8} - \frac{5875x^{3}}{24} + \frac{12251x^{2}}{32} \\ - \frac{20319x}{64} + \frac{8443}{64} \end{array}$ & $\begin{array}{c} \frac{256x^{6}}{3} - \frac{1184x^{5}}{3} + \frac{2893x^{4}}{3} \\ - \frac{36215x^{3}}{24} + \frac{51627x^{2}}{32} - \frac{139611x}{128} \\ + \frac{48655}{128} \end{array}$ \\\hline
$8$ & $1$ & $\frac{49x^{2}}{2} - \frac{161x}{4} + \frac{113}{4}$ & $\begin{array}{c} 162x^{4} - 531x^{3} + \frac{1697x^{2}}{2} \\ - \frac{22585x}{32} + \frac{4517}{16} \end{array}$ & $\begin{array}{c} \frac{15625x^{6}}{48} - \frac{153125x^{5}}{96} + \frac{770075x^{4}}{192} \\ - \frac{1205105x^{3}}{192} + \frac{1653605x^{2}}{256} - \frac{2095829x}{512} \\ + \frac{330841}{256} \end{array}$ \\\hline
$9$ & $1$ & $32x^{2} - 54x + \frac{153}{4}$ & $\begin{array}{c} \frac{2401x^{4}}{8} - \frac{24353x^{3}}{24} + \frac{52643x^{2}}{32} \\ - \frac{87841x}{64} + \frac{34397}{64} \end{array}$ & $\begin{array}{c} 972x^{6} - 4941x^{5} + \frac{25281x^{4}}{2} \\ - \frac{316659x^{3}}{16} + \frac{1277173x^{2}}{64} - \frac{3116745x}{256} \\ + \frac{57843}{16} \end{array}$ \\\hline
$10$ & $1$ & $\frac{81x^{2}}{2} - \frac{279x}{4} + \frac{199}{4}$ & $\begin{array}{c} 512x^{4} - \frac{5312x^{3}}{3} + \frac{23207x^{2}}{8} \\ - \frac{19413x}{8} + \frac{30017}{32} \end{array}$ & $\begin{array}{c} \frac{117649x^{6}}{48} - \frac{1226911x^{5}}{96} + \frac{6350743x^{4}}{192} \\ - \frac{9948757x^{3}}{192} + \frac{13210041x^{2}}{256} - \frac{15733995x}{512} \\ + \frac{2244665}{256} \end{array}$ \\
\hline
\end{tabular}%
\caption{Polynomial fits for scaled harmonics $B_1(j,x)$, columns $N-j \in \{0, 1, 2, 3\}$}
\end{table}

\begin{table}[H]
\centering
\footnotesize
\renewcommand{\arraystretch}{1.5}
\begin{tabular}{|c|c|c|c|}
\hline
$N$ & $N-j = 4$ & $N-j = 5$ & $N-j = 6$ \\
\hline
$4$ & $\frac{5x^{4}}{384} \text{\hl{$- \frac{5x^{3}}{96}$}} \text{\hl{$- \frac{5x^{2}}{256}$}} \text{\hl{$+ \frac{37x}{512}$}} \text{\hl{$- \frac{515}{2048}$}}$ & (negative harmonics) & (negative harmonics) \\\hline
$5$ & $\begin{array}{c} \frac{x^{8}}{384} - \frac{x^{7}}{192} \\ - \frac{61x^{6}}{2304} + \frac{303x^{5}}{2560} \\ \text{\hl{$+ \frac{305x^{4}}{6144}$}} \text{\hl{$+ \frac{2063x^{3}}{6144}$}} \\ \text{\hl{$- \frac{15743x^{2}}{8192}$}} \text{\hl{$+ \frac{14271x}{16384}$}} \\ \text{\hl{$+ \frac{12461}{8192}$}} \end{array}$ & $\text{$-\frac{17x^{4}}{512}$} \text{\hl{$+ \frac{77x^{3}}{768}$}} \text{\hl{$+ \frac{345x^{2}}{1024}$}} \text{\hl{$- \frac{4047x}{4096}$}} \text{\hl{$+ \frac{8629}{8192}$}}$ & (negative harmonics) \\\hline
$6$ & $\begin{array}{c} \frac{2x^{8}}{3} - \frac{10x^{7}}{3} \\ + \frac{86x^{6}}{9} - \frac{1747x^{5}}{80} \\ + \frac{769x^{4}}{16} - \frac{10523x^{3}}{128} \\ + \frac{215373x^{2}}{2048} - \frac{846925x}{8192} \\ \text{\hl{$+ \frac{63047}{1024}$}} \end{array}$ & $\begin{array}{c} \frac{x^{10}}{3840} - \frac{5x^{9}}{4608} \\ - \frac{5x^{8}}{2304} + \frac{53x^{7}}{1920} \\ \text{\hl{$- \frac{1387x^{6}}{36864}$}} \text{\hl{$+ \frac{24941x^{5}}{122880}$}} \\ \text{\hl{$- \frac{100561x^{4}}{49152}$}} \text{\hl{$+ \frac{175547x^{3}}{49152}$}} \\ \text{\hl{$- \frac{148913x^{2}}{65536}$}} \text{\hl{$+ \frac{554777x}{131072}$}} \\ \text{\hl{$+ \frac{101485}{32768}$}} \end{array}$ & $\begin{array}{c} -\frac{61x^{6}}{46080} \text{\hl{$+ \frac{61x^{5}}{7680}$}} \\ \text{\hl{$+ \frac{1025x^{4}}{12288}$}} \text{\hl{$- \frac{4385x^{3}}{12288}$}} \\ \text{\hl{$+ \frac{13059x^{2}}{16384}$}} \text{\hl{$- \frac{280413x}{131072}$}} \\ \text{\hl{$+ \frac{87055}{65536}$}} \end{array}$ \\\hline
$7$ & $\begin{array}{c} \frac{2187x^{8}}{128} - \frac{6561x^{7}}{64} \\ + \frac{87075x^{6}}{256} - \frac{2032191x^{5}}{2560} \\ + \frac{8822633x^{4}}{6144} - \frac{12223991x^{3}}{6144} \\ + \frac{16704233x^{2}}{8192} - \frac{23413123x}{16384} \\ + \frac{1107727}{2048} \end{array}$ & $\begin{array}{c} \frac{4x^{10}}{15} - \frac{17x^{9}}{9} \\ + \frac{137x^{8}}{18} - \frac{8503x^{7}}{360} \\ + \frac{31969x^{6}}{480} - \frac{102469x^{5}}{640} \\ + \frac{320713x^{4}}{1024} - \frac{137177x^{3}}{256} \\ \text{\hl{$+ \frac{12122427x^{2}}{16384}$}} \text{\hl{$- \frac{45703131x}{65536}$}} \\ \text{\hl{$+ \frac{1442349}{4096}$}} \end{array}$ & $\begin{array}{c} \frac{x^{12}}{46080} - \frac{7x^{11}}{46080} \\ + \frac{13x^{10}}{184320} + \frac{12631x^{9}}{3317760} \\ \text{\hl{$- \frac{11107x^{8}}{737280}$}} \text{\hl{$+ \frac{66827x^{7}}{860160}$}} \\ \text{\hl{$- \frac{70507x^{6}}{81920}$}} \text{\hl{$+ \frac{908501x^{5}}{245760}$}} \\ \text{\hl{$- \frac{2073493x^{4}}{262144}$}} \text{\hl{$+ \frac{18010255x^{3}}{786432}$}} \\ \text{\hl{$- \frac{50269279x^{2}}{1048576}$}} \text{\hl{$+ \frac{64421519x}{2097152}$}} \\ \text{\hl{$+ \frac{14219249}{524288}$}} \end{array}$ \\\hline
$8$ & $\begin{array}{c} \frac{512x^{8}}{3} - \frac{3328x^{7}}{3} \\ + \frac{34556x^{6}}{9} - \frac{44629x^{5}}{5} \\ + \frac{5818877x^{4}}{384} - \frac{7283743x^{3}}{384} \\ + \frac{8719327x^{2}}{512} - \frac{41198085x}{4096} \\ + \frac{3118831}{1024} \end{array}$ & $\begin{array}{c} \frac{19683x^{10}}{1280} - \frac{63423x^{9}}{512} \\ + \frac{35235x^{8}}{64} - \frac{2228391x^{7}}{1280} \\ + \frac{88671153x^{6}}{20480} - \frac{354971163x^{5}}{40960} \\ + \frac{228234305x^{4}}{16384} - \frac{875982595x^{3}}{49152} \\ + \frac{1135541127x^{2}}{65536} - \frac{1503514125x}{131072} \\ + \frac{16054619}{4096} \end{array}$ & $\begin{array}{c} \frac{4x^{12}}{45} - \frac{38x^{11}}{45} \\ + \frac{68x^{10}}{15} - \frac{118949x^{9}}{6480} \\ + \frac{5833x^{8}}{90} - \frac{806369x^{7}}{4032} \\ + \frac{8053163x^{6}}{15360} - \frac{25023647x^{5}}{20480} \\ \text{\hl{$+ \frac{62599481x^{4}}{24576}$}} \text{\hl{$- \frac{854186209x^{3}}{196608}$}} \\ \text{\hl{$+ \frac{1499030431x^{2}}{262144}$}} \text{\hl{$- \frac{5625660735x}{1048576}$}} \\ \text{\hl{$+ \frac{44112387}{16384}$}} \end{array}$ \\\hline
$9$ & $\begin{array}{c} \frac{390625x^{8}}{384} - \frac{1328125x^{7}}{192} \\ + \frac{56286875x^{6}}{2304} - \frac{28957625x^{5}}{512} \\ + \frac{190446715x^{4}}{2048} - \frac{225505655x^{3}}{2048} \\ + \frac{749324505x^{2}}{8192} - \frac{799905941x}{16384} \\ + \frac{106339885}{8192} \end{array}$ & $\begin{array}{c} \frac{4096x^{10}}{15} - \frac{20992x^{9}}{9} \\ + \frac{95776x^{8}}{9} - \frac{166452x^{7}}{5} \\ + \frac{56393347x^{6}}{720} - \frac{274922731x^{5}}{1920} \\ + \frac{314049373x^{4}}{1536} - \frac{690712903x^{3}}{3072} \\ + \frac{746177851x^{2}}{4096} - \frac{3215458111x}{32768} \\ + \frac{109264697}{4096} \end{array}$ & $\begin{array}{c} \frac{59049x^{12}}{5120} - \frac{610173x^{11}}{5120} \\ + \frac{13872141x^{10}}{20480} - \frac{112652127x^{9}}{40960} \\ + \frac{723891681x^{8}}{81920} - \frac{6688333983x^{7}}{286720} \\ + \frac{37745494337x^{6}}{737280} - \frac{46189818641x^{5}}{491520} \\ + \frac{112607733761x^{4}}{786432} - \frac{137847481073x^{3}}{786432} \\ + \frac{170589654237x^{2}}{1048576} - \frac{215648869791x}{2097152} \\ \text{\hl{$+ \frac{8738766193}{262144}$}} \end{array}$ \\\hline
$10$ & $\begin{array}{c} 4374x^{8} - 30618x^{7} \\ + 109431x^{6} - \frac{20202921x^{5}}{80} \\ + \frac{4887205x^{4}}{12} - \frac{178966697x^{3}}{384} \\ + \frac{755095013x^{2}}{2048} - \frac{1511586647x}{8192} \\ + \frac{46261097}{1024} \end{array}$ & $\begin{array}{c} \frac{1953125x^{10}}{768} - \frac{103515625x^{9}}{4608} \\ + \frac{238890625x^{8}}{2304} - \frac{185040625x^{7}}{576} \\ + \frac{2997709125x^{6}}{4096} - \frac{10399406225x^{5}}{8192} \\ + \frac{27658698781x^{4}}{16384} - \frac{27779669659x^{3}}{16384} \\ + \frac{80631271935x^{2}}{65536} - \frac{76368020619x}{131072} \\ + \frac{4484292021}{32768} \end{array}$ & $\begin{array}{c} \frac{16384x^{12}}{45} - \frac{176128x^{11}}{45} \\ + \frac{1012288x^{10}}{45} - \frac{36276928x^{9}}{405} \\ + \frac{49039187x^{8}}{180} - \frac{221858647x^{7}}{336} \\ + \frac{1328663439x^{6}}{1024} - \frac{63765125579x^{5}}{30720} \\ + \frac{1371632107x^{4}}{512} - \frac{266867603725x^{3}}{98304} \\ + \frac{134059465667x^{2}}{65536} - \frac{538766801035x}{524288} \\ + \frac{8488849349}{32768} \end{array}$ \\\hline
\end{tabular}%
\caption{Polynomial fits for scaled harmonics $B_1(j,x)$, columns $N-j \in \{4, 5, 6\}$}
\end{table}

\begin{table}[H]
\centering
\footnotesize
\renewcommand{\arraystretch}{1.5}
\resizebox{\textwidth}{!}{%
\begin{tabular}{|c|c|c|}
\hline
$N$ & $N-j = 7$ & $N-j = 8$ \\
\hline
7 & $\begin{array}{c} \frac{299x^{6}}{61440} \text{\hl{$- \frac{83x^{5}}{4096}$}} \\ \text{\hl{$- \frac{13243x^{4}}{49152}$}} \text{\hl{$+ \frac{23857x^{3}}{32768}$}} \text{\hl{$+ \frac{399489x^{2}}{65536}$}} \\ \text{\hl{$- \frac{22106655x}{1048576}$}} \text{\hl{$+ \frac{6275955}{262144}$}} \end{array}$ & (negative harmonics) \\\hline
8 & $\begin{array}{c} \frac{x^{14}}{645120} - \frac{x^{13}}{61440} \\ + \frac{53x^{12}}{1105920} + \frac{1001x^{11}}{3317760} \text{\hl{$- \frac{4597x^{10}}{1658880}$}} \\ \text{\hl{$+ \frac{1784087x^{9}}{92897280}$}} \text{\hl{$- \frac{4634669x^{8}}{20643840}$}} \text{\hl{$+ \frac{540971x^{7}}{344064}$}} \\ \text{\hl{$- \frac{152863643x^{6}}{23592960}$}} \text{\hl{$+ \frac{417468553x^{5}}{15728640}$}} \text{\hl{$- \frac{625667279x^{4}}{6291456}$}} \\ \text{\hl{$+ \frac{452319895x^{3}}{2097152}$}} \text{\hl{$- \frac{2590642675x^{2}}{8388608}$}} \text{\hl{$+ \frac{4152280099x}{16777216}$}} \\ \text{\hl{$+ \frac{305815017}{2097152}$}} \end{array}$ & $\begin{array}{c} \frac{277x^{8}}{2064384} \text{\hl{$- \frac{277x^{7}}{258048}$}} \\ \text{\hl{$- \frac{14867x^{6}}{491520}$}} \text{\hl{$+ \frac{62491x^{5}}{327680}$}} \text{\hl{$+ \frac{546961x^{4}}{786432}$}} \\ \text{\hl{$- \frac{4669845x^{3}}{1048576}$}} \text{\hl{$+ \frac{20807193x^{2}}{524288}$}} \text{\hl{$- \frac{2280994719x}{16777216}$}} \\ \text{\hl{$+ \frac{1282341297}{8388608}$}} \end{array}$ \\\hline
9 & $\begin{array}{c} \frac{8x^{14}}{315} - \frac{14x^{13}}{45} \\ + \frac{289x^{12}}{135} - \frac{35519x^{11}}{3240} + \frac{68599x^{10}}{1440} \\ - \frac{7308647x^{9}}{40320} + \frac{96552761x^{8}}{161280} - \frac{189255277x^{7}}{107520} \\ \text{\hl{$+ \frac{582152509x^{6}}{122880}$}} \text{\hl{$- \frac{556611211x^{5}}{49152}$}} \text{\hl{$+ \frac{379091785x^{4}}{16384}$}} \\ \text{\hl{$- \frac{20726097329x^{3}}{524288}$}} \text{\hl{$+ \frac{110403719435x^{2}}{2097152}$}} \text{\hl{$- \frac{405073233483x}{8388608}$}} \\ \text{\hl{$+ \frac{749898345}{32768}$}} \end{array}$ & $\begin{array}{c} \frac{x^{16}}{10321920} - \frac{11x^{15}}{7741440} \\ + \frac{247x^{14}}{30965760} + \frac{103x^{13}}{26542080} \text{\hl{$- \frac{19261x^{12}}{63700992}$}} \\ \text{\hl{$+ \frac{2383487x^{11}}{743178240}$}} \text{\hl{$- \frac{625889329x^{10}}{14863564800}$}} \text{\hl{$+ \frac{803175851x^{9}}{1981808640}$}} \\ \text{\hl{$- \frac{6775274347x^{8}}{2642411520}$}} \text{\hl{$+ \frac{2699987357x^{7}}{188743680}$}} \text{\hl{$- \frac{19284345853x^{6}}{251658240}$}} \\ \text{\hl{$+ \frac{154237225567x^{5}}{503316480}$}} \text{\hl{$- \frac{359333379475x^{4}}{402653184}$}} \text{\hl{$+ \frac{284782012145x^{3}}{134217728}$}} \\ \text{\hl{$- \frac{1872440011667x^{2}}{536870912}$}} \text{\hl{$+ \frac{2639988194963x}{1073741824}$}} \text{\hl{$+ \frac{83598027681}{67108864}$}} \end{array}$ \\\hline
10 & $\begin{array}{c} \frac{531441x^{14}}{71680} - \frac{1948617x^{13}}{20480} \\ + \frac{27418419x^{12}}{40960} - \frac{137949399x^{11}}{40960} + \frac{110122659x^{10}}{8192} \\ - \frac{51059633661x^{9}}{1146880} + \frac{286133963473x^{8}}{2293760} - \frac{294064841621x^{7}}{983040} \\ + \frac{1616790762617x^{6}}{2621440} - \frac{5684962181799x^{5}}{5242880} + \frac{10015662179485x^{4}}{6291456} \\ - \frac{3956774558193x^{3}}{2097152} \text{\hl{$+ \frac{14258621460285x^{2}}{8388608}$}} \text{\hl{$- \frac{17392116562911x}{16777216}$}} \\ \text{\hl{$+ \frac{335202117243}{1048576}$}} \end{array}$ & $\begin{array}{c} \frac{2x^{16}}{315} - \frac{92x^{15}}{945} \\ + \frac{788x^{14}}{945} - \frac{17063x^{13}}{3240} + \frac{269459x^{12}}{9720} \\ - \frac{23071163x^{11}}{181440} + \frac{824367769x^{10}}{1612800} - \frac{3035676559x^{9}}{1658880} \\ \text{\hl{$+ \frac{7731323753x^{8}}{1290240}$}} \text{\hl{$- \frac{185349621049x^{7}}{10321920}$}} \text{\hl{$+ \frac{285327208231x^{6}}{5898240}$}} \\ \text{\hl{$- \frac{912202283651x^{5}}{7864320}$}} \text{\hl{$+ \frac{6070198347073x^{4}}{25165824}$}} \text{\hl{$- \frac{20742942259007x^{3}}{50331648}$}} \\ \text{\hl{$+ \frac{72825611955197x^{2}}{134217728}$}} \text{\hl{$- \frac{264370022342909x}{536870912}$}} \text{\hl{$+ \frac{3845571588623}{16777216}$}} \end{array}$ \\\hline
\end{tabular}%
}
\caption{Polynomial fits for scaled harmonics $B_1(j,x)$, columns $N-j \in \{7, 8\}$}
\end{table}

\begin{table}[H]
\centering
\footnotesize
\renewcommand{\arraystretch}{1.5}
\resizebox{\textwidth}{!}{%
\begin{tabular}{|c|c|c|}
\hline
$N$ & $N-j = 9$ & $N-j = 10$ \\
\hline
9 & $\begin{array}{c} -\frac{25591x^{8}}{41287680} \text{\hl{$+ \frac{11059x^{7}}{3440640}$}} \text{\hl{$+ \frac{35155x^{6}}{393216}$}} \\ \text{\hl{$- \frac{513193x^{5}}{2621440}$}} \text{\hl{$- \frac{5222789x^{4}}{1048576}$}} \text{\hl{$+ \frac{56952761x^{3}}{8388608}$}} \text{\hl{$+ \frac{717717099x^{2}}{2097152}$}} \\ \text{\hl{$- \frac{174235602447x}{134217728}$}} \text{\hl{$+ \frac{55140168369}{33554432}$}} \end{array}$ & (negative harmonics) \\\hline
10 & $\begin{array}{c} \frac{x^{18}}{185794560} - \frac{13x^{17}}{123863040} + \frac{37x^{16}}{41287680} \\ - \frac{3781x^{15}}{1393459200} \text{\hl{$- \frac{76217x^{14}}{4459069440}$}} \text{\hl{$+ \frac{662017x^{13}}{1783627776}$}} \text{\hl{$- \frac{535861727x^{12}}{89181388800}$}} \\ \text{\hl{$+ \frac{2199627037x^{11}}{29727129600}$}} \text{\hl{$- \frac{152737609837x^{10}}{237817036800}$}} \text{\hl{$+ \frac{148577842037x^{9}}{31708938240}$}} \text{\hl{$- \frac{28434506509x^{8}}{880803840}$}} \\ \text{\hl{$+ \frac{165640800503x^{7}}{880803840}$}} \text{\hl{$- \frac{3521237174257x^{6}}{4026531840}$}} \text{\hl{$+ \frac{9176530286827x^{5}}{2684354560}$}} \text{\hl{$- \frac{11486224623307x^{4}}{1073741824}$}} \\ \text{\hl{$+ \frac{25826881598841x^{3}}{1073741824}$}} \text{\hl{$- \frac{157983616535637x^{2}}{4294967296}$}} \text{\hl{$+ \frac{231623458705557x}{8589934592}$}} \text{\hl{$+ \frac{2807658855477}{268435456}$}} \end{array}$ & $\begin{array}{c} -\frac{50521x^{10}}{3715891200} \\ \text{\hl{$+ \frac{50521x^{9}}{371589120}$}} \text{\hl{$+ \frac{797519x^{8}}{110100480}$}} \\ \text{\hl{$- \frac{10023061x^{7}}{165150720}$}} \text{\hl{$- \frac{80171753x^{6}}{94371840}$}} \\ \text{\hl{$+ \frac{1773318167x^{5}}{251658240}$}} \text{\hl{$- \frac{99954205x^{4}}{25165824}$}} \\ \text{\hl{$- \frac{39069039005x^{3}}{402653184}$}} \text{\hl{$+ \frac{220708486207x^{2}}{67108864}$}} \\ \text{\hl{$- \frac{110121627474205x}{8589934592}$}} \text{\hl{$+ \frac{4540578668177}{268435456}$}} \end{array}$ \\\hline
\end{tabular}%
}
\caption{Polynomial fits for scaled harmonics $B_1(j,x)$, columns $N-j \in \{9, 10\}$}
\end{table}

\pagebreak
\sethlcolor{customGreen}

\subsubsection{The exceptional terms $E$}
The exceptional terms are defined as the differences between the generic polynomials and the actual values. 
\begin{align}
E_r(j,x):=B_r(j,x)-P_r(j,x),\qquad j=N-r\ge 0.
\end{align}
At the first-return scale,
\begin{align}
  w=\epsilon^2z,
\end{align}
so an order-$r$ correction in $z$ becomes an order-$(3r-2N)$ correction in $w$:
\begin{align}
  \epsilon^rz^j=\epsilon^q w^j,
  \qquad q=r-2j=3r-2N.
\end{align}
This becomes most obvious when viewed on a table. There are many towers of exceptional terms, but here we only highlight two of them with colours:
\begin{itemize}
    \item The \hl{terms coloured green} contribute to the first return profile, $R_0$ (the $q=0$ case).
    \sethlcolor{customBlue}
    \item The \hl{terms coloured blue} contribute to the first return profile correction, $R_1$ (the $q=1$ case). 
\end{itemize}
For the complete table, see Supplementary Materials. 

\begin{table}[H]
\centering
\footnotesize
\renewcommand{\arraystretch}{1.5}
\begin{tabular}{|c|c|c|c|c|}
\hline
$N$ & $N-j=0$ & $N-j=1$ & $N-j=2$ & $N-j=3$ \\
\hline
$2$ & $0$ & $0$ & $-\frac{1}{4}x + \frac{1}{4}$ & --- \\\hline
$3$ & $0$ & $0$ & \cellcolor[HTML]{ddeee7} $-\frac{1}{128}$ & $\frac{9}{32}x - \frac{3}{16}$ \\\hline
$4$ & $0$ & $0$ & $0$ & \cellcolor[HTML]{e0eaf7} $-\frac{1}{256}x^{2} + \frac{7}{512}x - \frac{1}{512}$ \\\hline
\end{tabular}%
\caption{Exceptional differences $E_r(j,x)=B_r(j,x)-P_r(j,x)$ for $N-j\in\{0, 1, 2, 3\}$. Rows containing only zeros have been omitted.}
\end{table}

\begin{table}[H]
\centering
\footnotesize
\renewcommand{\arraystretch}{1.5}
\begin{tabular}{|c|c|c|c|}
\hline
$N$ & $N-j=4$ & $N-j=5$ & $N-j=6$ \\
\hline
$4$ & $\begin{array}{c} \frac{5}{96}x^{3} - \frac{5}{32}x^{2} - \frac{37}{512}x \\ + \frac{19}{128} \end{array}$ & --- & --- \\\hline
$5$ & $\begin{array}{c} -\frac{1}{1024}x^{4} + \frac{23}{3072}x^{3} - \frac{51}{4096}x^{2} \\ - \frac{41}{8192}x - \frac{71}{8192} \end{array}$ & $\begin{array}{c} -\frac{77}{768}x^{3} + \frac{21}{128}x^{2} + \frac{4047}{4096}x \\ - \frac{473}{512} \end{array}$ & --- \\\hline
$6$ &\cellcolor[HTML]{ddeee7} $-\frac{5}{8192}$ & $\begin{array}{c} -\frac{1}{6144}x^{6} + \frac{25}{12288}x^{5} - \frac{61}{8192}x^{4} \\ + \frac{43}{24576}x^{3} + \frac{563}{32768}x^{2} + \frac{25}{65536}x \\ + \frac{1115}{32768} \end{array}$ & $\begin{array}{c} -\frac{61}{7680}x^{5} + \frac{61}{1536}x^{4} + \frac{4385}{12288}x^{3} \\ - \frac{1323}{1024}x^{2} + \frac{280413}{131072}x - \frac{16797}{8192} \end{array}$ \\\hline
$7$ & $0$ & \cellcolor[HTML]{e0eaf7}$-\frac{5}{4096}x^{2} + \frac{95}{16384}x - \frac{171}{32768}$ & $\begin{array}{c} -\frac{1}{49152}x^{8} + \frac{3}{8192}x^{7} - \frac{659}{294912}x^{6} \\ + \frac{3191}{983040}x^{5} + \frac{3893}{262144}x^{4} - \frac{23753}{262144}x^{3} \\ + \frac{452807}{1048576}x^{2} - \frac{2317143}{2097152}x + \frac{841925}{1048576} \end{array}$ \\\hline
$8$ & $0$ & $0$ & $\begin{array}{c} -\frac{5}{4096}x^{4} + \frac{295}{24576}x^{3} - \frac{695}{16384}x^{2} \\ + \frac{18271}{262144}x - \frac{8025}{131072} \end{array}$ \\\hline
$9$ & $0$ & $0$ & \cellcolor[HTML]{ddeee7}$-\frac{63}{524288}$ \\\hline
\end{tabular}%
\caption{Exceptional differences $E_r(j,x)=B_r(j,x)-P_r(j,x)$ for $N-j\in\{4, 5, 6\}$. Rows containing only zeros have been omitted.}
\end{table}

\begin{table}[H]
\centering
\footnotesize
\renewcommand{\arraystretch}{1.5}
\begin{tabular}{|c|c|c|}
\hline
$N$ & $N-j=7$ & $N-j=8$ \\
\hline
$7$ & $\begin{array}{c} \frac{83}{4096}x^{5} - \frac{287}{6144}x^{4} - \frac{23857}{32768}x^{3} \\ - \frac{1113}{4096}x^{2} + \frac{22106655}{1048576}x - \frac{828849}{32768} \end{array}$ & --- \\\hline
$8$ & $\begin{array}{c} -\frac{1}{491520}x^{10} + \frac{29}{589824}x^{9} - \frac{65}{147456}x^{8} \\ + \frac{2027}{1474560}x^{7} + \frac{87607}{23592960}x^{6} - \frac{191689}{3145728}x^{5} \\ + \frac{1026255}{2097152}x^{4} - \frac{16305655}{6291456}x^{3} + \frac{63981749}{8388608}x^{2} \\ - \frac{203836877}{16777216}x + \frac{38144301}{4194304} \end{array}$ & $\begin{array}{c} \frac{277}{258048}x^{7} - \frac{277}{36864}x^{6} - \frac{62491}{327680}x^{5} \\ + \frac{17885}{16384}x^{4} + \frac{4669845}{1048576}x^{3} - \frac{1798143}{65536}x^{2} \\ + \frac{2280994719}{16777216}x - \frac{46982289}{262144} \end{array}$ \\\hline
$9$ & $\begin{array}{c} -\frac{5}{6144}x^{6} + \frac{305}{24576}x^{5} - \frac{1289}{16384}x^{4} \\ + \frac{36485}{131072}x^{3} - \frac{337361}{524288}x^{2} + \frac{2029455}{2097152}x \\ - \frac{86043}{131072} \end{array}$ & $\begin{array}{c} -\frac{1}{5898240}x^{12} + \frac{31}{5898240}x^{11} - \frac{1517}{23592960}x^{10} \\ + \frac{140909}{424673280}x^{9} + \frac{25331}{94371840}x^{8} - \frac{6143467}{330301440}x^{7} \\ + \frac{2329593}{10485760}x^{6} - \frac{22909249}{12582912}x^{5} + \frac{1010120831}{100663296}x^{4} \\ - \frac{1277415731}{33554432}x^{3} + \frac{13541107507}{134217728}x^{2} - \frac{44268341347}{268435456}x \\ + \frac{3948406431}{33554432} \end{array}$ \\\hline
$10$ &\cellcolor[HTML]{e0eaf7} $-\frac{567}{1048576}x^{2} + \frac{5859}{2097152}x - \frac{6219}{2097152}$ & $\begin{array}{c} -\frac{5}{12288}x^{8} + \frac{35}{4096}x^{7} - \frac{5869}{73728}x^{6} \\ + \frac{174217}{393216}x^{5} - \frac{27943}{16384}x^{4} + \frac{15218443}{3145728}x^{3} \\ - \frac{162747377}{16777216}x^{2} + \frac{823048721}{67108864}x - \frac{262005073}{33554432} \end{array}$ \\\hline
\end{tabular}%
\caption{Exceptional differences $E_r(j,x)=B_r(j,x)-P_r(j,x)$ for $N-j\in\{7, 8\}$. Rows containing only zeros have been omitted.}
\end{table}

\begin{table}[H]
\centering
\footnotesize
\renewcommand{\arraystretch}{1.5}
\begin{tabular}{|c|c|c|}
\hline
$N$ & $N-j=9$ & $N-j=10$ \\
\hline
$9$ & $\begin{array}{c} -\frac{11059}{3440640}x^{7} + \frac{689}{81920}x^{6} + \frac{513193}{2621440}x^{5} \\ + \frac{116449}{65536}x^{4} - \frac{56952761}{8388608}x^{3} - \frac{28580289}{262144}x^{2} \\ + \frac{174235602447}{134217728}x - \frac{1946143845}{1048576} \end{array}$ & --- \\\hline
$10$ & $\begin{array}{c} -\frac{1}{82575360}x^{14} + \frac{11}{23592960}x^{13} - \frac{1049}{141557760}x^{12} \\ + \frac{23713}{424673280}x^{11} - \frac{7433}{106168320}x^{10} - \frac{41019871}{11890851840}x^{9} \\ + \frac{157143391}{2642411520}x^{8} - \frac{1765222967}{2642411520}x^{7} + \frac{5475267029}{1006632960}x^{6} \\ - \frac{4471243719}{134217728}x^{5} + \frac{42657129417}{268435456}x^{4} - \frac{153291688217}{268435456}x^{3} \\ + \frac{1520037556395}{1073741824}x^{2} - \frac{4604346779691}{2147483648}x + \frac{201091529025}{134217728} \end{array}$ & $\begin{array}{c} -\frac{50521}{371589120}x^{9} + \frac{50521}{41287680}x^{8} + \frac{10023061}{165150720}x^{7} \\ - \frac{2805319}{5898240}x^{6} - \frac{1773318167}{251658240}x^{5} + \frac{176841095}{3145728}x^{4} \\ + \frac{39069039005}{402653184}x^{3} - \frac{3133929143}{2097152}x^{2} + \frac{110121627474205}{8589934592}x \\ - \frac{655624945945}{33554432} \end{array}$ \\\hline
\end{tabular}%
\caption{Exceptional differences $E_r(j,x)=B_r(j,x)-P_r(j,x)$ for $N-j\in\{9, 10\}$. Rows containing only zeros have been omitted.}
\end{table}

\subsubsection{Index reference table: $(N,r,j)$ and return order}

Recall we mostly worked with normalized coefficients like $B$ where underlying patterns become clear. But these coefficients can obscure the physics because they do not directly contribute to the observable $C = \expval*{c}/\alpha$. The following table shows the overall factor to multiply in order to obtain the physical amplitude. 

Find the corresponding coefficients in the table for $B$ or for $E$, multiply them by the corresponding factors here to get the actual contribution to $C$. 

For example, the cells marked ``$E\neq0$'' are those that contribute to $R_n$—the $n$-th order correction to the return profile. The green terms are those contributing to the first return profile $R_0$. From the previous section, these are:
\begin{align}
    N = 3, \ r = 2: && & -\frac{1}{128}, \\
    N = 6, \ r = 4: && & -\frac{5}{8192}, \\
    N = 9, \ r = 6: && & -\frac{63}{524288}, \dots
\end{align}
and in the Index table below, we can read off the corresponding factors
\begin{align}
    N = 3, \ r = 2: && & -2z, \\
    N = 6, \ r = 4: && & +2\cdot 2!z^2, \\
    N = 9, \ r = 6: && & -2\cdot 3!z^3, \dots
\end{align}
also recall how $N$ was defined, 
\begin{align}
    C = \sum_N \epsilon^N G_N,
\end{align}
meaning for each $N$, we need a factor of $\epsilon^N$. Multiply everything together, we have
\begin{align}
    N = 3, \ r = 2: && & +\frac{1}{64}\epsilon^3z, \\
    N = 6, \ r = 4: && & -\frac{5}{2048}\epsilon^6z^2, \\
    N = 9, \ r = 6: && & +\frac{189}{131072}\epsilon^9z^3, \dots
\end{align}
and recall $w = \epsilon^2z$, this gives us the precise series for $R_0$: 
\begin{align}
    R_0 = \frac{w}{64} - \frac{5w^2}{2048} + \frac{189w^3}{131072} - \dots
\end{align}
\indent Note that the ``boundary'' terms and the ``$Z-P_n$ terms are finite corrections to the generic polynomials, they do not contribute to return terms and are of no interests to us.

\newcommand{\IndexPos}[1]{%
  $\begin{array}{c}
    j=#1\\
    2\cdot #1!\cdot z^{#1}
  \end{array}$%
}

\newcommand{\IndexNeg}[1]{%
  $\begin{array}{c}
    j=#1\\
    -2\cdot #1!\cdot z^{#1}
  \end{array}$%
}

\newcommand{\IndexPosReturn}[2]{%
  $\begin{array}{c}
    j=#1\\
    2\cdot #1!\cdot z^{#1}\\
    R_{#2}\quad E\ne 0
  \end{array}$%
}

\newcommand{\IndexNegReturn}[2]{%
  $\begin{array}{c}
    j=#1\\
    -2\cdot #1!\cdot z^{#1}\\
    R_{#2}\quad E\ne 0
  \end{array}$%
}

\newcommand{\IndexZero}[1]{%
  $\begin{array}{c}
    j=0\\
    Z-P_{#1}(0,\chi)\\
    \Delta_0\ne 0
  \end{array}$%
}

\newcommand{\IndexZeroFirst}{%
  $\begin{array}{c}
    j=0\\
    Z-P_1(0,\chi)
  \end{array}$%
}

\newcommand{\IndexBoundary}[2]{%
  $\begin{array}{c}
    j=-#1\\
    L/(16z)^{#1}\\
    R_{#2}\ \mathrm{boundary}
  \end{array}$%
}


\begin{table}[H]
\centering
\scriptsize
\renewcommand{\arraystretch}{1.45}
\setlength{\tabcolsep}{4pt}
\begin{tabular}{|c|c|c|c|c|}
\hline
$N$ & $r=0$ & $r=1$ & $r=2$ & $r=3$ \\
\hline
$1$
& \IndexNeg{1}
& \IndexZeroFirst
& \IndexBoundary{1}{4}
& \textemdash
\\ \hline

$2$
& \IndexPos{2}
& \IndexNeg{1}
& \IndexZero{2}
& \IndexBoundary{1}{5}
\\ \hline

$3$
& \IndexNeg{3}
& \IndexPos{2}
& \cellcolor[HTML]{ddeee7}\IndexNegReturn{1}{0}
& \IndexZero{3}
\\ \hline

$4$
& \IndexPos{4}
& \IndexNeg{3}
& \IndexPos{2}
& \cellcolor[HTML]{e0eaf7}\IndexNegReturn{1}{1}
\\ \hline

$5$
& \IndexNeg{5}
& \IndexPos{4}
& \IndexNeg{3}
& \IndexPos{2}
\\ \hline

$6$
& \IndexPos{6}
& \IndexNeg{5}
& \IndexPos{4}
& \IndexNeg{3}
\\ \hline

$7$
& \IndexNeg{7}
& \IndexPos{6}
& \IndexNeg{5}
& \IndexPos{4}
\\ \hline

$8$
& \IndexPos{8}
& \IndexNeg{7}
& \IndexPos{6}
& \IndexNeg{5}
\\ \hline

$9$
& \IndexNeg{9}
& \IndexPos{8}
& \IndexNeg{7}
& \IndexPos{6}
\\ \hline

$10$
& \IndexPos{10}
& \IndexNeg{9}
& \IndexPos{8}
& \IndexNeg{7}
\\ \hline
\end{tabular}
\caption{Index map for $r=0,\ldots,3$.}
\end{table}


\begin{table}[H]
\centering
\scriptsize
\renewcommand{\arraystretch}{1.45}
\setlength{\tabcolsep}{4pt}
\begin{tabular}{|c|c|c|c|}
\hline
$N$ & $r=4$ & $r=5$ & $r=6$ \\
\hline
$2$
& \IndexBoundary{2}{8}
& \textemdash
& \textemdash
\\ \hline

$3$
& \IndexBoundary{1}{6}
& \IndexBoundary{2}{9}
& \IndexBoundary{3}{12}
\\ \hline

$4$
& \IndexZero{4}
& \IndexBoundary{1}{7}
& \IndexBoundary{2}{10}
\\ \hline

$5$
& \IndexNegReturn{1}{2}
& \IndexZero{5}
& \IndexBoundary{1}{8}
\\ \hline

$6$
& \cellcolor[HTML]{ddeee7}\IndexPosReturn{2}{0}
& \IndexNegReturn{1}{3}
& \IndexZero{6}
\\ \hline

$7$
& \IndexNeg{3}
& \cellcolor[HTML]{e0eaf7}\IndexPosReturn{2}{1}
& \IndexNegReturn{1}{4}
\\ \hline

$8$
& \IndexPos{4}
& \IndexNeg{3}
& \IndexPosReturn{2}{2}
\\ \hline

$9$
& \IndexNeg{5}
& \IndexPos{4}
& \cellcolor[HTML]{ddeee7}\IndexNegReturn{3}{0}
\\ \hline

$10$
& \IndexPos{6}
& \IndexNeg{5}
& \IndexPos{4}
\\ \hline
\end{tabular}
\caption{Index map for $r=4,\ldots,6$.}
\end{table}

\begin{table}[H]
\centering
\scriptsize
\renewcommand{\arraystretch}{1.45}
\setlength{\tabcolsep}{4pt}

\begin{minipage}[t]{0.48\textwidth}
\centering
\begin{tabular}{|c|c|c|}
\hline
$N$ & $r=7$ & $r=8$ \\
\hline
$4$
& \IndexBoundary{3}{13}
& \IndexBoundary{4}{16}
\\ \hline

$5$
& \IndexBoundary{2}{11}
& \IndexBoundary{3}{14}
\\ \hline

$6$
& \IndexBoundary{1}{9}
& \IndexBoundary{2}{12}
\\ \hline

$7$
& \IndexZero{7}
& \IndexBoundary{1}{10}
\\ \hline

$8$
& \IndexNegReturn{1}{5}
& \IndexZero{8}
\\ \hline

$9$
& \IndexPosReturn{2}{3}
& \IndexNegReturn{1}{6}
\\ \hline

$10$
& \cellcolor[HTML]{e0eaf7}\IndexNegReturn{3}{1}
& \IndexPosReturn{2}{4}
\\ \hline
\end{tabular}
\caption{Index map for $r=7,8$.}
\end{minipage}%
\hfill
\begin{minipage}[t]{0.48\textwidth}
\centering
\begin{tabular}{|c|c|c|}
\hline
$N$ & $r=9$ & $r=10$ \\
\hline
$5$
& \IndexBoundary{4}{17}
& \IndexBoundary{5}{20}
\\ \hline

$6$
& \IndexBoundary{3}{15}
& \IndexBoundary{4}{18}
\\ \hline

$7$
& \IndexBoundary{2}{13}
& \IndexBoundary{3}{16}
\\ \hline

$8$
& \IndexBoundary{1}{11}
& \IndexBoundary{2}{14}
\\ \hline

$9$
& \IndexZero{9}
& \IndexBoundary{1}{12}
\\ \hline

$10$
& \IndexNegReturn{1}{7}
& \IndexZero{10}
\\ \hline
\end{tabular}
\caption{Index map for $r=9,10$.}
\end{minipage}

\end{table}

\normalsize
\printbibliography[title={References}]